UiO : University of Oslo

Lars Musland

# Theory and calculations of thermoelectric transport in heterostructures

**Thesis submitted for the degree of Philosophiae Doctor**

Department of Physics
Faculty of Mathematics and Natural Sciences

**2019**

# Preface

The work behind this thesis was done as a part of the THELMA project, which was a five year long Norwegian effort to research the thermoelectric effect, funded by the Norwegian research council (project number 228854). The thermoelectric effect is a generic term for effects that couple heat and charge transport in materials, and which if sufficiently large, can be utilized to create cooling elements, heat pumps, and even electrical generators without moving parts. My own section of the project was concerned with the thermoelectric properties of heterostructure materials, which have been predicted, and to some extent demonstrated to be beneficial for thermoelectric applications.

The overarching goal of my own project in particular, was the implementation of a general purpose numerical solver capable of simulating heat- and charge transport in heterostructure materials, so that the thermoelectric properties of such materials could be calculated from structural information. While this task is in itself to involved to finish in a single PhD-project, it has served as the overlying motivation for all of my work, which has mostly been focused on laying the ground work for the implementation by studying various formalisms and methods, in order to find the ones most appropriate for the particular problem.

Before starting to implement a solver, one must determine the theoretical formalism in which to formulate the problem, and the method by which to solve the involved equations. The appropriate choices are determined by the accuracy of the results, and by the computational requirements of the calculations. Thus, when selecting a method, one must consider to which extent it results in an accurate solution, how much computation time is required for the calculation, and also whether an unreasonable amount of memory is required. Further, when selecting the theoretical formalism itself, one must consider whether it is theoretically justified for the particular problem, whether it might still give reasonable answers even if this is not the case, and finally, whether there are actually any methods capable of solving the relevant equations within a reasonable amount of time.

My investigation into these questions has taken two separate forms. Partially, it has consisted of a purely theoretical study, where I have studied the literature of the various formalisms, and put some effort into trying to understand the regimes in which they are each justified. Partially, my investigation has also consisted of the implementation of a series of small scale numerical tests, intended to study the computational efficiency of the different methods, and to some extent also to study whether there are notable differences between results of the different formalisms. The latter point concerns in particular differences between the semiclassical Boltzmann equation and approaches based on quantum transport.

Finally, having determined that the NEGF formalism is as an accurate starting point for these calculations, and that most methods for solving the equations involved are to computationally demanding, I have spent some time trying to develop a method to speed up the solution of these equations by utilizing Monte Carlo calculations

## Acknowledgments

First of all, I would like to thank Espen-Flage Larsen, Joakim Bergli and Øystein Prytz for accepting the task of being my supervisors during my work as a PhD student, and for always providing as much support and help as they have been able to. Particularly I would like to thank Espen, with whom I have had the largest cooperation during this work, and who always remains positive, even when things are not working as well as planned.

I would also like to thank the remainder of my colleagues and fellow students at the structure physics group at the University of Oslo, for providing a highly enjoyable environment in which to write a PhD thesis, and for many interesting discussions and entertaining Friday evenings.

Finally, I would like to thank my parents, my grand parents, my sister, the remainder of my family, and my friends for always being supportive.

**Lars Musland**
Oslo, May 2019

# Contents

# Chapter 1

# Introduction

## 1.1 Thermoelectric effects, and thermoelectric transport coefficients

The thermoelectric effect, is a generic term for effects that couple the two major transport modes in materials: transport of heat, and transport of electrical charge. To understand the meaning of this, imagine some device, or material slab connected to two contacts, between which there is a voltage difference $V$, and a temperature difference $\Delta T$. In the commonly presented picture, the flow of heat and charge are independent, meaning that the electrical current $I$ is found from Ohm's law

$$V = RI, \tag{1.1}$$

$R$ being the resistance of the device, while the heat current $q$ is found from the law of heat conduction as

$$q = k\Delta T, \tag{1.2}$$

$k$ being the heat conductance, or thermal conductance. However, in almost all materials there is a small amount of coupling between these transport modes, meaning that voltages can induce heat currents, while temperature gradients can induce electrical currents. Thus, more generally, one must make a modification to Ohm's law as

$$V = RI - \alpha\Delta T, \tag{1.3}$$

where $\alpha$ is known as the Seebeck coefficient. Similarly, the law of heat conduction must also be generalized, and is typically written as

$$q = k\Delta T + \Pi I, \tag{1.4}$$

where $\Pi$ is referred to as the Peltier coefficient.

With these generalizations, we can identify two important effects. First, a temperature difference may induce currents or voltages. This is referred to as the Seebeck effect. In open circuit conditions, where no current is allowed to flow, the Seebeck effect will induce a voltage $V = -\alpha\Delta T$. Secondly, even in the absence of a temperature difference, there may still be a heat flow $q = \Pi I$. If two devices with different Peltier coefficients are connected, the heat flow $q$ will not be conserved, and accordingly this effect can be used to create heat sources and heat sinks. This is known as the Peltier effect.

The coefficients $R$, $k$, $\alpha$ and $\Pi$ are referred to as transport coefficients. In this thesis I will in particular refer to $k$, $\alpha$ and $\Pi$ as thermoelectric transport coefficients, while the purely electrical case will usually be considered separately. One can also define transport coefficients as material properties. At the macroscopic scale, one may define coefficients $\overleftrightarrow{\sigma}$, $\overleftrightarrow{\alpha}$, $\overleftrightarrow{\kappa}$ and $\overleftrightarrow{\pi}$, such that the relations

$$\boldsymbol{j} = -\overleftrightarrow{\sigma}\,(\boldsymbol{\nabla} V + \overleftrightarrow{\alpha}\boldsymbol{\nabla} T) \quad \text{and,} \tag{1.5}$$

$$\boldsymbol{\phi}_Q = -\overleftrightarrow{\kappa}\boldsymbol{\nabla} T + \overleftrightarrow{\pi}\boldsymbol{j}, \tag{1.6}$$

apply at each point of the device. Here $\boldsymbol{j}$ and $\boldsymbol{\phi}_Q$ are respectively the electrical current density and the heat flux density, while $\boldsymbol{\nabla}$ denotes the gradient operation. $T$ and $V$ denotes temperature and voltage as before. The coefficients $\overleftrightarrow{\sigma}$, $\overleftrightarrow{\alpha}$, $\overleftrightarrow{\kappa}$ and $\overleftrightarrow{\pi}$ are material properties, and are respectively referred to as the electrical conductivity, Seebeck coefficient, thermal conductivity and Peltier coefficient of the material. In general these quantities are not scalar, but tensors. This means they correspond to general linear transformations, which can be represented as matrices.

Another thermoelectric effect which is sometimes mentioned, is the Thomson effect, which can be stated as

$$\dot{q} = \boldsymbol{j} \cdot \overleftrightarrow{\mathcal{K}}\boldsymbol{\nabla} T, \tag{1.7}$$

where $\dot{q}$ is the local rate of heat generation, and $\overleftrightarrow{\mathcal{K}}$ is the Thomson coefficient. However, the Thomson effect is a second order effect, and can be derived from the thermoelectric relations presented above. In fact, it can be regarded as a continuum limit of the Peltier effect, and one can show that

$$\overleftrightarrow{\mathcal{K}} = T\frac{\partial \overleftrightarrow{\alpha}}{\partial T}, \tag{1.8}$$

although we will not do that here.

The thermal conductivity and conductance are typically decomposed in two contributions as $\overleftrightarrow{\kappa} = \overleftrightarrow{\kappa}_e + \overleftrightarrow{\kappa}_l$ and $k = k_e + k_l$, where subscripts $e$ and $l$ respectively denotes contributions to the heat flux from electrons and from lattice vibrations. The electron contribution $\kappa_e$ is intimately related to the electronic conductivity $\sigma$, and one defines a coefficient of proportionality $L = \kappa_e/\sigma T$, which is known as the Lorenz coefficient. One can also define a device Lorenz coefficient as $L = Rk_e/T$. In metals the Lorenz coefficient usually has a value approximately equal to $2.44 \cdot 10^{-8}\,\mathrm{V}^2/\mathrm{K}^2$ [9].

## 1.2 Applications of the thermoelectric effect

One of the most common applications of the thermoelectric effect today, is the use of Peltier elements, which are electrical devices that utilize the Peltier effect to transport heat from one side of the device to the other. Peltier elements are usually made use of for cooling purposes, but in principle they can also be used as heat pumps. However, as of yet their low efficiency is limiting the latter application.

While Peltier elements can not compete with compressor based technologies in terms of efficiency, they have several other advantages. First of all, they have no moving parts, which means they are noiseless, and have an extremely long durability, requiring as good as no maintenance. Secondly, they have an extremely short response time, meaning that the cooling rate can be adjusted almost instantaneously. The latter fact is highly advantageous in those cooling applications where the temperature must be kept close to some specific value. This is the main application of Peltier elements today.

The Seebeck effect is also made use of in a few technological applications. First of all, many heat sensors and temperature measurement devices make use of the Seebeck effect, since temperature differences can be read off as voltages. This is probably the application of the Seebeck effect that has the biggest consequence in our daily lives. However, an application which is perhaps more interesting, is that the Seebeck effect can be used to construct electrical generators without moving parts. However, just like thermoelectric heat pumps, such thermoelectric generators are limited by their low efficiency. Because of this, their application is restricted to niche cases. An example of such niche cases would be space probes, where the absence of maintenance requirements trumps any efficiency concern.

The efficiency of thermoelectric heat pumps and generators are both determined by a single device parameter

$$ZT = \frac{\alpha\Pi}{Rk} = \frac{\alpha^2}{Rk}T, \tag{1.9}$$

which is known as the figure of merit. The equivalence of the two expressions will be shown later. The efficiency of a thermoelectric generator in particular, is given by the formula

$$\eta = \frac{\chi - 1}{\chi + T_1/T_2}\eta_C, \tag{1.10}$$

where $\chi = \sqrt{ZT + 1}$, $T_1$ and $T_2$ are respectively the temperatures on the cold and hot sides of the device, and $\eta_C$ is the Carnot efficiency $\eta_C = 1 - T_1/T_2$, which is the maximal possible efficiency of a heat engine. In the limit $ZT \to \infty$, we can approximate the efficiency formula as $\eta = \left(1 - 1/\sqrt{ZT}\right)\eta_C$. Thus, the efficiency approaches $\eta_C$ as $ZT$ is increased, but very slowly.

In addition to the device figure of merit $ZT$, one defines a similar material figure of merit

$$zT = \frac{\sigma\alpha^2}{\kappa}T, \tag{1.11}$$

where the transport coefficients and material parameters are assumed to either be scalar, or referring to a single well defined direction in the material. The device figure of merit $ZT$ will often lie close to the smallest value of $zT$ in the materials from which the device is constructed, so accordingly the material figure of merit $zT$ is a crucial parameter when choosing materials for thermoelectric applications.

## 1.3 Heterostructure materials, and related nomenclature

As can to some extent be discerned from the name, a heterostructure material is defined as a material where different sections of the material have different compositions or structure, and where this variation occurs at a mesoscopic size scale (nm-$\mu$m range). Thus, the composition or structure of the material varies as a function of position. An important special case is that of quasi-one-dimensional heterostructures, which are particularly important since in an attempt to limit the scope of the project, all of my work has been limited to such structures. The term quasi-one-dimensional simply means that all of the mesoscale structural variation occurs along a single dimension, while the material is homogeneous along the two remaining dimensions. Thus, one can introduce a coordinate system $xyz$ such that the composition and structure of the material depends only on the $z$-coordinate, and is independent of $x$ and $y$.

A particularly important subclass of quasi-one-dimensional heterostructures is made up by the superlattices. A superlattice is a quasi-one-dimensional heterostructure which is periodic also in the direction of mesoscale variation. Thus, in the coordinate system introduced above, there is some distance $d$ such that the material composition at coordinate $(x, y, z)$ is always the same as at coordinate $(x, y, z + d)$. In my work, the smallest possible value of $d$ will be referred to as the superlattice period. In order for the material to classify as an actual heterostructure, and not as some bulk material, the superlattice period should be at least a few nanometers. In addition, the literature often reserves the term superlattice to structures where the period is quite small, typically no longer than a few tens of nm. However, as there is not really any good term for quasi-one-dimensional heterostructures in the more general case, I will in fact use the term superlattice also when the period is arbitrarily large.

Since a superlattice is periodic along all three dimensions, it can be constructed as a repeating pattern of a single small block of material. This block will be referred to as the supercell, and sometimes as the superlattice supercell. The term unit cell on the other hand, will in this work be reserved for the unit cells of the underlying materials from which the heterostructure is composed. For instance, in a superlattice consisting of repeating layers of HgTe and CdTe, the term unit cell will always refer to the unit cell of HgTe or CdTe, and never to the supercell of the superlattice.

In the simplest configuration, a superlattice consists simply of a repeating pattern of two different layer types, with differing composition or structure. Often the material in the two layers will also have differing electron band edges. If electron transport occurs in the conduction band, the layers with higher conduction band minima is typically referred to as barriers, while the layers with lower band minima are referred to as wells. The difference between the two conduction band minima is called the barrier height. If electron transport occurs in the valence band, one will often adopt the opposite convention, where the layers with lower valence band maxima are referred to as barriers.

A second important subclass of quasi-one-dimensional heterostructures that should be introduced, is the thin film. A thin film consists of a mesoscopically dimensioned layer suspended between two macroscopic bulk regions. The thin layer must have a composition or structure different from both bulk regions, and may or may not also contain internal variations. In addition, one or both bulk regions may consist of vacuum.

## 1.4 Advantages of heterostructuring for thermoelectric applications

There are multiple reasons why heterostructures are interesting for thermoelectric applications. One important effect, which has already been utilized with considerable success, is the fact that including structures at a mesoscopic size scale will typically reduce the mean free path of phonons in the material. Since phonons carry heat but not charge, they will in most cases contribute only to $\kappa_l$ and $k_l$, and not to the other transport coefficients introduced above. An exception to this occurs in the presence of phonon drag effects, which are normally ignored in standard materials. In the absence of such effects, it is then easy to see that the material figure of merit

$$zT = \frac{\sigma\alpha^2}{\kappa_e + \kappa_l} T \tag{1.12}$$

can only be positively impacted by the reduction of phonon transport.

Another important consequence of heterostructuring a material, is an effect known as energy filtering, where electrons are freely moving only when their energy lies in some limited range, and where the motion of electrons outside of this range is inhibited. For instance, by including small barriers consisting of materials with higher conduction band minima, the motion of conduction band electrons would be inhibited in the range below the barrier height. Earlier work has predicted energy filtering to have a positive impact on a central trade-off in thermoelectric materials: As a function of doping concentration, the conductivity of the material tends to increase, much as would be expected, but the Seebeck coefficient of the material tends to drop. Thus, since the figure of merit is proportional both to $\sigma$ and $\alpha^2$, there is a trade-off between these coefficients when selecting the optimal doping concentration. Energy filtering has been predicted to reduce this trade-off, which means the product $\sigma\alpha^2$ is increased at the optimum.

In addition, some work has indicated that energy filtering may also reduce the Lorenz coefficient. Writing the figure of merit as

$$zT = \frac{\alpha^2}{L}\left(1 + \frac{\kappa_l}{\kappa_e}\right)^{-1} \tag{1.13}$$

shows that this would also be highly beneficial.

These were the effects that motivated our research project to spend some resources on studyin heterostructure materials, and in particular to begin preparatory work to implement a transport solver capable of calculating the thermo-electric transport coeeficients of such materials. The latter task has served as motivation for all work presented in this thesis.

## 1.5 Outline of the thesis

This thesis is divided into two major parts. Part I provides an overview of some of the most important formalisms of transport theory, while Part II provides a presentation and discussion of my own work. The order in which these parts are read is not important, as long as the reader is prepared to accept that part II references a few topics in part I. Published works, and work submitted for publication are included at the end of part II.

Chapter 2, the first chapter of part I, is an introduction to the Landauer-Büttiker theory of transport. I open with this subject, since it is in a sense the simplest transport formalism. Also, it provides a conceptually simple and intuitive way of thinking about transport in general. However, while always useful as a conceptual tool, the Landauer-Büttiker theory is not very useful for making quantitative predictions, except when the transport problem can be regarded as ballistic, i.e. free from scattering. Various scattering models can be included in the theory, but these will always either be phenomenological or taken from another formalism.

In order to rigorously obtain predictions where scattering is included, it is better to make use of formalisms derived from first principles. Such formalisms are the subject of the remaining four chapters of part I. Chapter 3 deals with the path integral, or field integral, which is the most general of the formalisms, since it is in fact equivalent to the many particle Schrödinger equation. While rarely directly applicable in practice, the field integral forms an important theoretical starting point, from which most of the remaining theory is derived.

Chapter 4 deals with transport in the linear regime, i.e. the regime where perturbations are small enough for all responses to be simple linear functions of the stimuli. This is an important topic in thermoelectric theory, since at least those thermoelectric applications that are thermally driven will almost always be in the linear regime. Some of the expressions in Chapter 4 are derived from the field integral, but the most important ones, known as the Kubo relations, are derived using the standard Hilbert space operator formalism. Some discussion is also provided for how to convert these expressions into field integrals.

Chapter 5 deals with the perturbative expansion of the field integral. The discussion is limited to the fermionic sector, and to the expansion of two-point correlation functions. We accordingly end up with the Non-equilibrium Green's function, or NEGF formalism, which is the first of the rigorous formalisms that is practically applicable to any major extent. However, application requires additional approximations, which may or may not be severe, depending on the particular problem. Conveniently, transport expressions in the NEGF formalism can be formulated in a language similar to that of the Landauer-Büttiker theory.

The final chapter of part I, Chapter 6, discusses two remaining, and very important transport formalisms. First, by introducing a Markov approximation to the NEGF formalism, we end up with the master equation approach. Secondly, by introducing an additional assumption of slow spatial variations, we obtain the semiclassical Boltzmann equation, which is probably the most commonly applied of all transport formalisms. I also discuss the multi band version of the Boltzmann equation, which must be justified in a somewhat different manner than the single band version. While not particularly important in bulk systems, the last point may be of major importance when modeling superlattices.

Part I serves three different purposes in the thesis. Firstly, it serves as a reference for many of the theories and equations I have already applied in my work. Thus, part II will commonly make reference to equations and discussions in part I, but mostly to Chapters 2, 5 and 6. A second purpose served by part I, is as a useful overview and work of reference for future work. This applies particularly to Chapters 4 and 6, the subjects of which I have in retrospect realized would have been a better starting point for my own work than the NEGF formalism. This will be more thoroughly discussed later. In any case, these subjects should be useful for future improvements of my implementations, and Chapters 4 and 6 serve as useful references.

The third and final purpose served by part I, is as a documentation of the more theoretical and self serving work I have done during my PhD. In the process of determining the formalism most appropriate for my implementations, I have spent a considerable time reviewing various literature, and on trying to understand the various formalisms, their ranges of applicability, and how they all relate to each other. Part I documents this work, and presents an overview of what I have learned and understood from it.

The more practical aspects of my work are presented in Part II, which is divided into two chapters. Chapter 7 serves as an introduction to this part of my work. There, I discuss the motivation behind my work, our path towards determining the formalism and method of our implementation, our overall progress towards this implementation, and finally how the papers included at the end fit into this picture.

Chapter 8 contains a discussion of various sub-projects that did not result in the submission of a manuscript, as well as some additional results related to the topic of our third paper, Paper III. Section 8.1 describes some experiments with the Boltzmann Monte Carlo method, which was the first method of transport calculations I pursued. Monte Carlo methods are methods where the desired calculation is performed by averaging a set of random results. In particular, the Boltzmann Monte Carlo method obtains a solution to the Boltzmann equation by explicitly simulating the movement of particles, according to physical forces acting upon them, and a sequence of random scattering events. Per today, this is the most efficient method to solve the Boltzmann equation under general conditions.

Sections 8.2 and 8.3 are both concerned with methods for reducing the computational burden associated with the solution of the NEGF equations. Section 8.2 in particular, is concerned with the calculation of the retarded Green's function, a demanding task which must be performed multiple times during a transport calculation. While we eventually landed on the RGF method[6] as the most efficient method for doing these calculations, Section 8.2 investigates an alternative based on perturbation theory.

Section 8.3 is concerned with a NEGF Monte Carlo method. Motivated by the success of the Boltzmann Monte Carlo method, we have attempted to develop a Monte Carlo method which solves the NEGF equations. This method is the subject of our third paper, Paper III, and Section 8.2 merely includes some additional results, which were omitted from the paper.

Finally, Chapter 9 contains an overview of what I consider to be the most important conclusions I have arrived at during my work.

Part I

# An overview of selected topics from transport theory, relevant to the calculation of thermoelectric coefficients

# Chapter 2

# The Landauer-Büttiker theory of transport

In this chapter, I give an introduction to transport theory, in the form of Landauer-Büttiker theory, which is a very intuitive conceptualization of transport. It is just as useful for conceptualizing semiclassical transport as quantum transport[7, 22, 17]. However, the theory is purely phenomenological, and it is also most useful when scattering is assumed elastic. Thus, for predictive calculations, more rigorous formalisms are required.

## 2.1 Landauer-Büttiker expressions

### 2.1.1 Derivation

Consider some electronic device or material sample $S$, and a set $\{p\}$ of leads connected to $S$. For each lead $p$ we define the direction pointing away from $S$ as the positive $z$ axis. We assume that $p$ is translationally invariant along the $z$ axis with some period $a$, and we ignore many particle interactions in the leads. Electron motion in the leads is then determined by the Schrödinger equation

$$\hat{H}\psi(\boldsymbol{x}) = E\psi(\boldsymbol{x}), \tag{2.1}$$

here in its time independent form. Here $\hat{H}$ is the Hamiltonian operator, $\psi(\boldsymbol{x})$ is the wave function, and $E$ is the energy. From Bloch's theorem[11] it follows that the solutions can be written as

$$\psi(\boldsymbol{x}) = \psi_m(k, \boldsymbol{x}) = \phi_m(k, \boldsymbol{x})e^{ikz}, \text{ with} \tag{2.2}$$

$$E = E_m(k). \tag{2.3}$$

Such Bloch states are discussed more thoroughly in Section 6.3.1. The index $m$ refers to the transversal mode of the lead, and the functions $\phi_m(k, \boldsymbol{x})$ are periodic in $z$.

In the Landauer-Büttiker formalism, the system $S$ is dealt with entirely in terms of a transmission function $\mathcal{T}$, which describes probabilities of transfer between different leads and energies. As we shall see, all the physics relevant to electronic transport calculations is captured by the transmission function. The formalism is not concerned with exactly how $\mathcal{T}$ is to be calculated, so this must be extracted from one of the formalisms to be described below. The following derivation follows Datta[7].

In the most general version of the formalism, the transmission function takes the form $\mathcal{T}_{qp}^{nm}(E, E')$, which is defined as the probability that an electron entering $S$ from lead $p$, with energy $E$ and in transversal mode $n$, exits in lead $q$, with energy $E'$ and in transversal mode $m$. However, in reality the Landauer-Büttiker formalism is rarely useful unless the scattering mechanism in $S$ is elastic, so that the exiting energy $E'$ must equal the incoming energy $E$. In that case the transmission function simplifies to $\mathcal{T}_{qp}^{nm}(E)$, which is defined as before, except that the electron now also exits with energy $E$.

Consider the total electron flux $\Phi_q$ in lead $q$. This can be separated in two contribution classes: the incoming contributions from lead $q$ itself, and the outgoing contributions having been transmitted through $S$ from all leads $p$. Thus, the total electron flux in $q$ can be written

$$\Phi_q = \sum_{nk} \Phi_q^n(k) - \sum_{pmnk} \mathcal{T}_{qp}^{nm}(E_m(k))\Phi_p^n(k), \tag{2.4}$$

where the sum is only over values of $k$ with negative flux (moving towards $S$).

We assume each lead to be internally close to equilibrium, and well described by a chemical potential $\mu_p$ and a temperature $T_p$. Since electrons are fermions their occupation is described by the Fermi function[7, 11]

$$f_p(E) = \frac{1}{e^{\beta_p(E-\mu_p)} + 1}. \tag{2.5}$$

If the leads are assumed to have length $L$, the density of electrons in some state $\psi_m(k, \boldsymbol{x})$ will be $f(E_m(k))/L$, and if the group velocity

$$v_m(k) = \frac{1}{\hbar}\frac{\mathrm{d}E}{\mathrm{d}k}, \tag{2.6}$$

of the state is negative, then the flux hitting $S$ from that state will be $|\Phi_p^m(k)| = f(E_m(k))|v_m(k)|/L$.

For a continuing flux to make sense, we must take the limit $L \to \infty$. (2.4) then becomes

$$\begin{aligned}
\Phi_q &= \sum_n \int \frac{\mathrm{d}k}{2\pi} f_q(E_n(k))v_n(k) - \sum_{pmn} \int \frac{\mathrm{d}k}{2\pi} \mathcal{T}_{qp}^{nm}(E_m(k)) f_p(E_m(k)) v_m(k) \\
&= \sum_n \int \frac{\mathrm{d}k}{2\pi}\frac{1}{\hbar}\frac{\mathrm{d}E}{\mathrm{d}k} f_q(E_n(k)) - \sum_{pmn} \int \frac{\mathrm{d}k}{2\pi}\frac{1}{\hbar}\frac{\mathrm{d}E}{\mathrm{d}k} \mathcal{T}_{qp}^{nm}(E_m(k)) f_p(E_m(k)) \\
&= -\frac{1}{h}\sum_n \int \mathrm{d}E\, f_q(E) + \frac{1}{h}\sum_{pmn} \int \mathrm{d}E\, \mathcal{T}_{qp}^{nm}(E) f_p(E).
\end{aligned} \tag{2.7}$$

In the last expression, the indexes $m$ and $n$ no longer represents only the transversal modes, but possibly also a discrete set of values $k$ where the bands intersect the energy $E$, and where $v(k)$ is positive.

The total probability of moving from a mode $n$ to any new mode in any lead must be one, so

$$\sum_{pm} \mathcal{T}_{pq}^{mn}(E) = 1. \tag{2.8}$$

Inserting this in (2.7), we get

$$\Phi_q = \frac{1}{h} \sum_{pmn} \int \mathrm{d}E \left( \mathcal{T}_{qp}^{nm}(E) f_p(E) - \mathcal{T}_{pq}^{mn}(E) f_q(E) \right). \tag{2.9}$$

At this point we rewrite the expressions in terms of $i(E)$, the flux contribution from energy $E$, meaning that

$$\Phi_q = \int \mathrm{d}E \, i_q(E). \tag{2.10}$$

The current in lead $q$ is then

$$I_q = -e\Phi_q = -e \int \mathrm{d}E \, i_q(E), \tag{2.11}$$

where $e$ is the elementary charge. Since the energy flux contribution from energy $E$ must be $Ei(E)$, we can also calculate the total heat flux in lead $q$ as

$$q_q = \int \mathrm{d}E \, (E - \mu_q) i_q(E). \tag{2.12}$$

A more detailed argument for why this is the case is given in Section 3.4.3.

From (2.9) we can express $i_q(E)$ as

$$i_q(E) = \frac{1}{h} \sum_{p} \left( \bar{\mathcal{T}}_{qp}(E) f_p(E) - \bar{\mathcal{T}}_{pq}(E) f_q(E) \right) \text{ where,} \tag{2.13}$$

$$\bar{\mathcal{T}}_{qp}(E) = \sum_{mn} \mathcal{T}_{qp}^{nm}(E). \tag{2.14}$$

Using more sophisticated formalisms, one can show that under certain conditions

$$\sum_{p} \bar{\mathcal{T}}_{pq}(E) = \sum_{p} \bar{\mathcal{T}}_{qp}(E), \tag{2.15}$$

as is done for instance in the appendix of our third paper, Paper III. Inserting this in (2.13), we finally find

$$i_q(E) = \frac{1}{h} \sum_{p} \bar{\mathcal{T}}_{qp}(E) \left( f_p(E) - f_q(E) \right). \tag{2.16}$$

### 2.1.2 Linear limit and Landauer formula

In the linear limit, the currents $I_q$ are linear in the lead voltages $V_{pq} = V_p - V_q$. We define the conductances $G_{qp}$ through the relation $I_q = \sum_p G_{qp} V_{pq}$. Let us expand (2.16) to the first order in the potential differences $\Delta\mu_p = \mu_p - \mu$, where $\mu$ is the equilibrium potential. We get

$$i_q(E) = \frac{1}{h} \sum_{p} \bar{\mathcal{T}}_{qp}(E) \frac{\partial f}{\partial \mu}(E) (\Delta\mu_p - \Delta\mu_q) = \frac{1}{h} \sum_{p} \bar{\mathcal{T}}_{qp}(E) \mathrm{Th}(E) \Delta\mu_{pq}, \tag{2.17}$$

where $\Delta\mu_{pq} = \mu_p - \mu_q = \Delta\mu_p - \Delta\mu_q$, and

$$\mathrm{Th}(E) = -\frac{\partial f}{\partial E}(E) = \frac{\beta}{4\cosh^2 \beta(E-\mu)/2}. \tag{2.18}$$

Inserting (2.17) in (2.11), and utilizing the fact that $\Delta\mu_{qp} = -eV_{qp}$, we get

$$I_q = \frac{e^2}{h}\sum_p \int \mathrm{d}E\, \bar{\mathcal{T}}_{qp}(E)\mathrm{Th}(E)V_{pq}. \tag{2.19}$$

Thus, from the definition of $G_{qp}$,

$$G_{qp} = \frac{e^2}{h}\int \mathrm{d}E\, \bar{\mathcal{T}}_{qp}(E)\mathrm{Th}(E). \tag{2.20}$$

A particularly important special case is that of a device $S$ with only two leads. In that case there is only one conductance $G = G_{12} = G_{21}$, and one transmission function $\bar{\mathcal{T}}(E) = \bar{\mathcal{T}}_{12}(E) = \bar{\mathcal{T}}_{21}(E)$. The relation between $G$ and $\bar{\mathcal{T}}$ is found by removing the indices $p$ and $q$ from (2.20). Clearly, by (1.1) $R = 1/G$. An interesting limit is obtained for $T \to 0$. The function $\mathrm{Th}(E)$ then becomes a delta function, and we get

$$G = \frac{e^2}{h}\bar{\mathcal{T}}(\mu). \tag{2.21}$$

If scattering in $S$ can be ignored, then $\mathcal{T}_{12}^{mn} \sim \delta_{mn}$, and (2.21) becomes

$$G = \frac{2e^2}{h}M, \tag{2.22}$$

where the factor of 2 comes from spin degeneracy, and $M$ is the number of transversal modes per lead and spin. (2.22) is the well known Landauer formula for the quantization of conductance[7, 15].

### 2.1.3 Thermoelectric coefficients

Following for instance Ref. [17], we can also find linear expressions for currents due to temperature differences $\Delta T_{pq} = T_p - T_q$. Expanding (2.16) to the first order in $\Delta T_p = T_p - T$, we get

$$i_q(E) = -\frac{1}{h}\sum_p \bar{\mathcal{T}}_{qp}(E)\frac{\partial f}{\partial \beta}(E)\frac{\Delta T_p - \Delta T_q}{T^2} = \frac{1}{hT}\sum_p \bar{\mathcal{T}}_{qp}(E)\mathrm{Th}(E)(E-\mu)\Delta T_{pq}. \tag{2.23}$$

Thus, we find from (2.11) that we can write $I_q = \sum_p A_{qp}\Delta T_{pq}$, where

$$A_{qp} = -\frac{e}{hT}\int \mathrm{d}E\, \bar{\mathcal{T}}_{qp}(E)\mathrm{Th}(E)(E-\mu). \tag{2.24}$$

In a situation where there are both voltage and temperature differences between the leads, the lead currents will be $I_q = \sum_p (G_{qp}V_{pq} + A_{qp}\Delta T_{pq})$. Similarly, we can write the lead heat currents as $q_q = \sum_p (B_{qp}V_{pq} + C_{qp}\Delta T_{pq})$. Inserting (2.17) and (2.23) in (2.12), it is easy to see that

$$B_{qp} = -\frac{e}{h}\int \mathrm{d}E\,\bar{\mathcal{T}}_{qp}(E)\mathrm{Th}(E)(E-\mu), \text{ and} \tag{2.25}$$

$$C_{qp} = \frac{1}{hT}\int \mathrm{d}E\,\bar{\mathcal{T}}_{qp}(E)\mathrm{Th}(E)(E-\mu)^2. \tag{2.26}$$

If we limit the situation to two leads, the relations above simplify to $I = GV + A\Delta T$ and $q = BV + C\Delta T$, which are merely reexpressions of the relations (1.3) and (1.4). $A$, $B$ and $C$ are found by removing the indices $p$ and $q$ from (2.24) to (2.26). Expressions for the thermoelectric transport coefficients $\alpha$ and $k_e$ are most easily obtained by considering the special case where the current $I = 0$. We then have $GV + A\Delta T = 0$, which can be solved to find the voltage $V = -\alpha\Delta T$, with $\alpha = A/G$. Further, the heat current becomes $q = -B\alpha\Delta T + C\Delta T = k_e\Delta T$, with $k_e = C - B\alpha$. Finally considering the heat flux in the general case, we see that it can be written $k_e\Delta T + \Pi I$, with $\Pi = B/G$. Comparing (2.24) to (2.25), we also find $\Pi = \alpha T$.

Summarizing, we have obtained the expressions

$$\alpha = A/G, \tag{2.27}$$

$$k_e = C - AB/G, \quad \text{and} \tag{2.28}$$

$$\Pi = B/G = \alpha T, \tag{2.29}$$

allowing us to rewrite the expressions for $I$ and $q$ as

$$I = G(V + \alpha\Delta T), \text{ and} \tag{2.30}$$

$$q = k\Delta T + \Pi I, \tag{2.31}$$

in agreement with (1.3) and (1.4).

The Seebeck coefficient $\alpha$ can be rewritten in an interesting way by introducing the probability distribution

$$P(E) = \frac{e^2}{hG}\bar{\mathcal{T}}(E)\mathrm{Th}(E). \tag{2.32}$$

This function is positive, and integrates to 1 by (2.20). From (2.24), we get

$$\alpha = \frac{A}{G} = -\frac{e}{hGT}\int \mathrm{d}E\,\bar{\mathcal{T}}(E)\mathrm{Th}(E)(E-\mu) = -\frac{1}{eT}\langle E-\mu\rangle = -\frac{k_B}{e}\langle\chi\rangle, \tag{2.33}$$

where the brackets denote the expectation value with respect to the probability distribution $P$, $k_B$ is the Boltzmann constant, and the dimensionless quantity $\chi = \beta(E-\mu)$.

As mentioned, the electronic heat conductance $k_e$ is often expressed as $k_e = GLT$, where $L$ is the Lorenz number or Lorenz coefficient. $L$ can also be expressed in terms of $P$ and $\chi$, as

$$\begin{aligned} L = \frac{C - B\alpha}{GT} &= \frac{1}{hGT^2} \int \mathrm{d}E \, \bar{\mathcal{T}}(E) \mathrm{Th}(E)(E-\mu)^2 - \frac{\Pi\alpha}{T} \\ &= \frac{\left\langle (E-\mu)^2 \right\rangle}{e^2 T^2} - \alpha^2 = \frac{k_B^2}{e^2} \left\langle \chi^2 \right\rangle - \frac{k_B^2}{e^2} \langle \chi \rangle^2 = \frac{k_B^2}{e^2} \left\langle (\chi - \langle \chi \rangle)^2 \right\rangle = \frac{k_B^2}{e^2} \mathrm{Var}\, \chi, \end{aligned} \tag{2.34}$$

where Var denotes variance taken with respect to $P$.

In a metallic system we can approximate $\bar{\mathcal{T}}(E)$ as constant over an energy range of a few $k_B T$. Thus, $P(E) \approx \mathrm{Th}(E)$, and we can calculate $L$ explicitly. since $\mathrm{Th}(E)$ is a symmetric function, $\langle \chi \rangle = 0$. Accordingly

$$\mathrm{Var}\, \chi = \int \mathrm{d}E \, \mathrm{Th}(E) \chi^2 = \int \mathrm{d}\chi \, \frac{\chi^2}{4 \cosh^2 \chi/2} = \frac{\pi^2}{3}, \tag{2.35}$$

which is found by contour integrating the function $z^3/4/\cosh^2 z/2$ around the rectangle $[-\infty, \infty] \times [0, 2\pi i]$. By (2.34) we find

$$L = \frac{\pi^2 k_B^2}{3e^2} = 2.44 \cdot 10^{-8} \, \mathrm{V}^2/\mathrm{K}^2, \tag{2.36}$$

the standard value of the Lorenz number[9].

## 2.2 Discussion of the leads

In the previous section, drastic approximations were introduced in the physics of the leads. They were assumed to be infinitely long and perfectly periodic, and they were assumed to be free of scattering, so that the electrons are described by a single particle Hamiltonian.

Traditionally, these approximations are justified by the assumption that the leads are highly conductive compared to the device $S$ [7]. If this is the case, then most of the changes in $\mu$ and $T$ will happen inside $S$, so that the leads are close to equilibrium. In addition, the leads will then only give minor contributions to the resistance, and to other measurable properties of the combined system. This means that the physics of the leads will not significantly affect measurable quantities, so that whether or not the leads are described realistically becomes irrelevant.

If the assumption that most of the restiance resides in $S$ does not hold, then the leads of the previous section can be considered a purely theoretical tool. That is, instead of taking the leads to be the real physical leads, we instead include these physical leads in the description of $S$ itself. The leads of the previous section are then merely theoretical abstractions that we use to create non-equilibrium conditions in the theoretical description of $S$, and to pick up the resulting currents. In the same way as an experimenter will choose the physical leads in the most convenient way in order to do his measurements, theorists may choose these theoretical leads in as convenient a manner as possible. This is done by keeping the lead physics maximally simple.

## 2.3 Coherent and incoherent regimes

In this section we briefly discuss the behavior of the transmission functions $\mathcal{T}(E)$. To do this in the general case requires a more sophisticated formalism, but we can produce some meaningful results in two simplified regimes. These are the coherent and incoherent regimes, which are determined by how the transport length scale compares to the coherence or phase relaxation length $L_\phi$ [7]. Coherent transport occurs over length scales much shorter than $L_\phi$, and is characterized by carriers taking on the character of classical waves. Incoherent transport occurs over length scales much larger than $L_\phi$, and is characterized by carriers taking on a character more like classical particles.

### 2.3.1 Incoherent regime

In the incoherent regime, interference effects due to the wavelike nature of particles can be ignored. We can then think of the particles as classical, and describe their motion using simple probabilities. Consider a device $S$ with two leads, and bisect it in two parts $A$ and $B$, $A$ being closer to lead 1 and $B$ closer to lead 2. Assume we know the transmission coefficients of each part to be $\mathcal{T}_A$ and $\mathcal{T}_B$ respectively, and that transport occurs in a single mode in each part.

Then, again following Datta[7], we can find the transmission function of $S$ by the following considerations: Assume lead 1 contains a flux $\Phi^+$ of particles moving towards $A$, and a flux $\Phi^-$ moving away from $A$. Assume further that at the boundary between parts $A$ and $B$, there is a flux $\Phi_A^+$ entering $B$ from $A$, and a flux $\Phi_A^-$ entering $A$ from $B$. Clearly these fluxes are related by the transmission function of $A$, so that

$$\Phi_A^+ = \mathcal{T}_A\Phi^+ + \mathcal{R}_A\Phi_A^-, \text{ and} \tag{2.37}$$

$$\Phi^- = \mathcal{T}_A\Phi_A^- + \mathcal{R}_A\Phi^+. \tag{2.38}$$

Here the reflection coefficient $\mathcal{R}$ is given by $\mathcal{R} = 1 - \mathcal{T}$. Solving (2.38) for $\Phi_A^-$ we get

$$\Phi_A^- = \frac{1}{\mathcal{T}_A}\Phi^- - \frac{\mathcal{R}_A}{\mathcal{T}_A}\Phi^+, \tag{2.39}$$

and inserting that in (2.37) we get

$$\Phi_A^+ = \mathcal{T}_A\Phi^+ + \frac{\mathcal{R}_A}{\mathcal{T}_A}\Phi^- - \frac{\mathcal{R}_A^2}{\mathcal{T}_A}\Phi^+. \tag{2.40}$$

This can be conveniently written in matrix form as

$$\begin{bmatrix}\Phi_A^+\\ \Phi_A^-\end{bmatrix} = \begin{bmatrix}\mathcal{T}_A - \mathcal{R}_A^2/\mathcal{T}_A & \mathcal{R}_A/\mathcal{T}_A\\ -\mathcal{R}_A/\mathcal{T}_A & 1/\mathcal{T}_A\end{bmatrix}\begin{bmatrix}\Phi^+\\ \Phi^-\end{bmatrix}. \tag{2.41}$$

Consider now the fluxes in lead 2. We refer to the flux exiting $B$ as $\Phi_B^+$, and the flux moving toward $B$ as $\Phi_B^-$. Clearly these fluxes are related to the fluxes

$\Phi_A^+$ and $\Phi_A^-$ in a manor completely analogous to the above relations, and we thus get

$$\begin{bmatrix}\Phi_B^+\\ \Phi_B^-\end{bmatrix} = \begin{bmatrix}\mathcal{T}_B - \mathcal{R}_B^2/\mathcal{T}_B & \mathcal{R}_B/\mathcal{T}_B\\ -\mathcal{R}_B/\mathcal{T}_B & 1/\mathcal{T}_B\end{bmatrix}\begin{bmatrix}\Phi_A^+\\ \Phi_A^-\end{bmatrix}. \tag{2.42}$$

Combining (2.41) and (2.42) we get

$$\begin{bmatrix}\Phi_B^+\\ \Phi_B^-\end{bmatrix} = \begin{bmatrix}\mathcal{T}_A - \mathcal{R}_A^2/\mathcal{T}_A & \mathcal{R}_A/\mathcal{T}_A\\ -\mathcal{R}_A/\mathcal{T}_A & 1/\mathcal{T}_A\end{bmatrix}\begin{bmatrix}\mathcal{T}_B - \mathcal{R}_B^2/\mathcal{T}_B & \mathcal{R}_B/\mathcal{T}_B\\ -\mathcal{R}_B/\mathcal{T}_B & 1/\mathcal{T}_B\end{bmatrix}\begin{bmatrix}\Phi^+\\ \Phi^-\end{bmatrix}, \tag{2.43}$$

and since considerations similar to those above could also have been made for the entire device $S$ with transmission $\mathcal{T}$, we must in fact have

$$\begin{bmatrix}\mathcal{T} - \mathcal{R}^2/\mathcal{T} & \mathcal{R}/\mathcal{T}\\ -\mathcal{R}/\mathcal{T} & 1/\mathcal{T}\end{bmatrix} = \begin{bmatrix}\mathcal{T}_A - \mathcal{R}_A^2/\mathcal{T}_A & \mathcal{R}_A/\mathcal{T}_A\\ -\mathcal{R}_A/\mathcal{T}_A & 1/\mathcal{T}_A\end{bmatrix}\begin{bmatrix}\mathcal{T}_B - \mathcal{R}_B^2/\mathcal{T}_B & \mathcal{R}_B/\mathcal{T}_B\\ -\mathcal{R}_B/\mathcal{T}_B & 1/\mathcal{T}_B\end{bmatrix}. \tag{2.44}$$

The relations above are easily generalized to a case where we divide $S$ into a sequence of parts $S_1 \cdots S_N$. Since the fluxes on the boundaries are in each case given by expression similar to (2.41), (2.44) generalizes to

$$\begin{bmatrix}\mathcal{T} - \mathcal{R}^2/\mathcal{T} & \mathcal{R}/\mathcal{T}\\ -\mathcal{R}/\mathcal{T} & 1/\mathcal{T}\end{bmatrix} = \prod_{n=1}^{N}\begin{bmatrix}\mathcal{T}_n - \mathcal{R}_n^2/\mathcal{T}_n & \mathcal{R}_n/\mathcal{T}_n\\ -\mathcal{R}_n/\mathcal{T}_n & 1/\mathcal{T}_n\end{bmatrix}. \tag{2.45}$$

In fact, evaluating the lower right corner of the matrix product in (2.44), we get

$$\frac{1}{\mathcal{T}} = -\frac{\mathcal{R}_A\mathcal{R}_B}{\mathcal{T}_A\mathcal{T}_B} + \frac{1}{\mathcal{T}_A\mathcal{T}_B} = \frac{1-(1-\mathcal{T}_A)(1-\mathcal{T}_B)}{\mathcal{T}_A\mathcal{T}_B} = \frac{1}{\mathcal{T}_A} + \frac{1}{\mathcal{T}_B} - 1, \tag{2.46}$$

and by a simple argument of induction it is easy to see that this generalizes to the case of (2.45) as

$$\frac{1}{\mathcal{T}} - 1 = \sum_{n=1}^{N}\left(\frac{1}{\mathcal{T}_n} - 1\right). \tag{2.47}$$

In particular, if $S$ is divided into $N$ identical pieces, all having transmission $\mathcal{T}_1$ and length $L_1$, then

$$\frac{1}{\mathcal{T}} - 1 = N\left(\frac{1}{\mathcal{T}_1} - 1\right). \tag{2.48}$$

We define the back scattering mean free path $\lambda$ as

$$\lambda = \frac{L_1}{1/\mathcal{T}_1 - 1} = \frac{\mathcal{T}_1 L_1}{1-\mathcal{T}_1}. \tag{2.49}$$

Note that by (2.48) this is the ratio of two quantities both proportional to $N$. Thus, if we rather than $L_1$ and $\mathcal{T}_1$, insert the transmission and length of some larger set of $N$ parts, (2.49) gives the same value of $\lambda$. Accordingly, $\lambda$ is to some extent independent of the subdivision of $S$. However, this is only true as long as the lengths $L_1 \gg L_\phi$. Using (2.49), we can write (2.48) as

$$\mathcal{T} = \frac{\lambda}{\lambda + L}, \tag{2.50}$$

in agreement with the literature[7, 22, 17].

## 2.3.2 Coherent regime

In the coherent regime, we can not think of electrons as classical particles, but we can think of them as classical waves. Consider the same situation as above, where $S$ is subdivided in two parts $A$ and $B$. The fluxes $\Phi^+$, $\Phi^-$, $\Phi_A^+$, $\Phi_A^-$, $\Phi_B^+$ and $\Phi_B^-$ are then associated with amplitudes $\phi^+$, $\phi^-$, $\phi_A^+$, $\phi_A^-$, $\phi_B^+$ and $\phi_B^-$, where in each case $\Phi = |\phi|^2$. The flux amplitudes entering and exiting a subsystem are linearly related, and we can thus write their relationship in a manor similar to (2.37) and (2.38), as[7]

$$\phi_A^+ = t_A^+\phi^+ + r_A^+\phi_A^-, \text{ and} \tag{2.51}$$

$$\phi^- = t_A^-\phi_A^- + r_A^-\phi^+, \tag{2.52}$$

where $|t_A^+|^2 = |t_A^-|^2 = \mathcal{T}_A$ and $|r_A^+|^2 = |r_A^-|^2 = \mathcal{R}_A$. Completely analogously to the incoherent case, these equations can also be written in matrix form as

$$\begin{bmatrix} \phi_A^+ \\ \phi_A^- \end{bmatrix} = \begin{bmatrix} t_A^+ - r_A^+ r_A^-/t_A^- & r_A^+/t_A^- \\ -r_A^-/t_A^- & 1/t_A^- \end{bmatrix} \begin{bmatrix} \phi^+ \\ \phi^- \end{bmatrix}, \tag{2.53}$$

and similarly for $\phi_B^+$ and $\phi_B^-$. Thus, (2.43) also has an analogous version in the coherent case, as

$$\begin{bmatrix} \phi_B^+ \\ \phi_B^- \end{bmatrix} = \begin{bmatrix} t_A^+ - r_A^+ r_A^-/t_A^- & r_A^+/t_A^- \\ -r_A^-/t_A^- & 1/t_A^- \end{bmatrix} \begin{bmatrix} t_B^+ - r_B^+ r_B^-/t_B^- & r_B^+/t_B^- \\ -r_B^-/t_B^- & 1/t_B^- \end{bmatrix} \begin{bmatrix} \phi^+ \\ \phi^- \end{bmatrix}, \tag{2.54}$$

and in fact, in the general case we get an equation analogous to (2.45) as

$$\begin{bmatrix} t^+ - r^+ r^-/t^- & r^+/t^- \\ -r^-/t^- & 1/t^- \end{bmatrix} = \prod_{n=1}^{N} \begin{bmatrix} t_n^+ - r_n^+ r_n^-/t_n^- & r_n^+/t_n^- \\ -r_n^-/t_n^- & 1/t_n^- \end{bmatrix}. \tag{2.55}$$

In the special case where all the parts $S_1 \cdots S_N$ are identical, this simplifies to

$$\begin{bmatrix} t^+ - r^+ r^-/t^- & r^+/t^- \\ -r^-/t^- & 1/t^- \end{bmatrix} = \begin{bmatrix} t_1^+ - r_1^+ r_1^-/t_1^- & r_1^+/t_1^- \\ -r_1^-/t_1^- & 1/t_1^- \end{bmatrix}^N. \tag{2.56}$$

The matrices in this expression are sometimes referred to as transfer matrices.

## 2.4 Macroscopic devices

Landauer formalism is usually applied to the study of transport in the microscopic regime. However, following Ref. [17], we can also use it to form a reasonable picture of transport in macroscopic devices. Consider a macroscopic device $S$ with two leads. Assume both $S$ and the leads to be composed of a crystalline material, thus being repetitions of some fundamental cell. In a macroscopic system, boundary conditions have relatively small effects, and we may thus assume periodic boundary conditions in the directions orthogonal to the leads. Then the transversal modes $m$ may be indexed by the transversal component of the Bloch vector, $\boldsymbol{k}_\perp$. Thus, we may write (2.14) as

$$\bar{\mathcal{T}}(E) = 2 \sum_{\boldsymbol{k}_\perp \boldsymbol{k}'_\perp} \mathcal{T}^{\boldsymbol{k}_\perp \boldsymbol{k}'_\perp}(E), \tag{2.57}$$

where the additional factor of 2 comes from spin degeneracy. Let us simplify the expression by making the unjustified assumption that $\mathcal{T}^{\boldsymbol{k}_\perp \boldsymbol{k}'_\perp}(E) \sim \delta_{\boldsymbol{k}_\perp \boldsymbol{k}'_\perp}$. Then

$$\bar{\mathcal{T}}(E) = 2 \sum_{\boldsymbol{k}_\perp} \mathcal{T}^{\boldsymbol{k}_\perp}(E) \approx \frac{2A}{(2\pi)^2} \int \mathrm{d}\boldsymbol{k}_\perp \, \mathcal{T}(\boldsymbol{k}_\perp, E), \tag{2.58}$$

where $A$ is the cross sectional area of $S$ and the leads, and where we have used the macroscopic size to justify switching to an integral.

Since $S$ is macroscopic, we are definitely in the incoherent regime, and since $S$ is also periodic we may thus substitute $\mathcal{T}$ from (2.50). However, in the macroscopic regime we also have $L \gg \lambda$, so that in fact we may take $\mathcal{T} \approx \lambda/L$. Inserting this in (2.58) we get

$$\bar{\mathcal{T}}(E) = \frac{2A}{L} \int \frac{\mathrm{d}\boldsymbol{k}_\perp}{(2\pi)^2} \lambda(\boldsymbol{k}_\perp, E). \tag{2.59}$$

Inserting (2.59) in (2.20), we get

$$G = \frac{A}{L} \frac{2e^2}{h} \int \frac{\mathrm{d}\boldsymbol{k}_\perp}{(2\pi)^2} \int \mathrm{d}E \, \lambda(\boldsymbol{k}_\perp, E) \mathrm{Th}(E). \tag{2.60}$$

Similar expressions can be obtained also for the transport coefficients $A$ $B$ and $C$.

We also know that for macroscopic devices, the conductance should be $G = A\sigma/L$, where $\sigma$ is the conductivity. Comparing this with (2.60), we get

$$\sigma = \frac{2e^2}{h} \int \frac{\mathrm{d}\boldsymbol{k}_\perp}{(2\pi)^2} \int \mathrm{d}E \, \lambda(\boldsymbol{k}_\perp, E) \mathrm{Th}(E), \tag{2.61}$$

which is an expression that is independent of the dimensions of the device. Intuitively, this relation should hold at least approximately also beyond the approximation $\mathcal{T}^{\boldsymbol{k}_\perp \boldsymbol{k}'_\perp}(E) \sim \delta_{\boldsymbol{k}_\perp \boldsymbol{k}'_\perp}$, but then with more complex expressions for the back scattering mean free path $\lambda$. In fact, it is not hard to see that (2.61) is equivalent to the relaxation time approximation of the Boltzmann equation[17].

The Seebeck and Lorenz coefficients can still be found from (2.33) and (2.34), but with the probability $P$ replaced with

$$P(E) = \frac{2e^2}{\sigma h} \int \frac{\mathrm{d}\boldsymbol{k}_\perp}{(2\pi)^2} \lambda(\boldsymbol{k}_\perp, E) \mathrm{Th}(E). \tag{2.62}$$

These expressions are then also independent of device dimensions, and should be regarded as approximate estimates of the material coefficients.

## Chapter 3

# The Non-equilibrium field integral

This chapter contains a derivation of transport expressions within the formalism of field-integrals or path-integrals. The theory is mostly taken from Ref. [2]. In Section 3.1 we introduce and derive the general field integral, while in Section 3.2 we specialize to transport calculations in a device of the type described in the previous chapter. In Section 3.3 we integrate over the lead degrees of freedom to obtain a description in terms of the system $S$ alone. Finally, in Section 3.4 we derive expressions for the expectation values of electrical current and the heat current.

## 3.1 General field integral

Consider any quantum system $S$, and any observable $A$. In general, the state of $S$ is described by a density operator $\rho$, and the expectation value of $A$ is given by the expression[21, 18] $\langle A\rangle = \mathrm{Tr}\,\hat{A}\rho$. In particular, the expectation value at time $t$ is $\langle A(t)\rangle = \mathrm{Tr}\,\hat{A}\rho(t)$, where

$$\rho(t) = U(t, t_0)\rho_0 U^\dagger(t, t_0). \tag{3.1}$$

Here $\rho_0 = \rho(t_0)$ represents the state at time $t_0$ and $U(t, t_0)$ is the propagator, or evolution operator[21], evolving the state from $t_0$ to $t$. The propagator satisfies the Schrödinger equation

$$\hbar i \frac{\mathrm{d}}{\mathrm{d}t} U(t, t_0) = \hat{H}(t) U(t, t_0), \tag{3.2}$$

where $\hat{H}$ is the Hamiltonian. If $\hat{H}$ is not time dependent, then $U$ depends only on the difference between the time arguments, and we can write $U(t, t_0) = U(t - t_0)$. The propagator $U(t)$ satisfies[21]

$$U(-t) = U^\dagger(t), \qquad \text{and} \tag{3.3}$$

$$U(t + t') = U(t)U(t'). \tag{3.4}$$

Still assuming a time independent Hamiltonian, the general expression for $\langle A\rangle$ becomes

$$\begin{aligned} \langle A(t)\rangle &= \mathrm{Tr}\,\hat{A}\,U(t - t_0)\rho_0 U^\dagger(t - t_0) \\ &= \mathrm{Tr}\,U^\dagger(t - t_0)\hat{A}U(t - t_0)\rho_0 \\ &= \mathrm{Tr}\,\hat{A}(t)\rho_0, \end{aligned} \tag{3.5}$$

where we have defined $\hat{A}(t) = U^\dagger(t - t_0)\hat{A}U(t - t_0)$. Generalizing this expression, we define

$$\langle \hat{A}_1(t_1)\hat{A}_2(t_2)\cdots\hat{A}_n(t_n)\rangle = \mathrm{Tr}\,\hat{A}_1(t_1)\hat{A}_2(t_2)\cdots\hat{A}_n(t_t)\rho_0, \tag{3.6}$$

where $\hat{A}_k(t) = U^\dagger(t-t_0)\hat{A}_k U(t-t_0)$. In particular, we define the two time correlation function $\langle \hat{A}(t)\hat{B}(t')\rangle = \operatorname{Tr}\hat{A}(t)\hat{B}(t')\rho_0$. We adopt the convention of writing these expectation values with hats over the observables, in order to distinguish them from another type of expectation value to be introduced below.

### 3.1.1 Field integral expression for the propagator

A detailed derivation of the field integral is given in Ref. [2], where Chapters 3 and 4 deal with the field integral in imaginary time, and a generalization to real time is briefly covered in chapter 11. Here we will only cover the gist of the argument.

Making use of (3.4) and defining $\Delta t = (t_f - t_0)/N$, we can write $U(t_f - t_0)$ as

$$U(t_f - t_0) = U(\Delta t)^N. \tag{3.7}$$

For $N \to \infty$, $\Delta t$ becomes arbitrarily small, and accordingly we can obtain an approximate solution of (3.2) as

$$U(\Delta t) \approx I - \frac{i}{\hbar}\hat{H}\Delta t. \tag{3.8}$$

Inserting this in (3.7) we get

$$U(t_f - t_0) \approx \left(I - \frac{i}{\hbar}\hat{H}\Delta t\right)^N. \tag{3.9}$$

We assume that the system $S$ consists of a set of bosonic fields that we denote $\hat{\phi}$, and a set of fermionic fields that we denote $\hat{\psi}$. We introduce single particle bases $\{i\}$ and $\{j\}$ for the fields $\hat{\phi}$, and $\hat{\psi}$ respectively, and write the corresponding field operators as $\hat{\phi}_i$ and $\hat{\psi}_j$. Finally, we organize these operators in vectors $\hat{\boldsymbol{\phi}} = [\hat{\phi}_i]$ and $\hat{\boldsymbol{\psi}} = [\hat{\psi}_j]$.

The identity operator of the system can then be written as[2]

$$I = \int \mathrm{d}(\boldsymbol{\phi},\boldsymbol{\psi}) e^{-|\boldsymbol{\phi}|^2 - \bar{\boldsymbol{\psi}}\boldsymbol{\psi}} |\boldsymbol{\phi},\boldsymbol{\psi}\rangle\langle\boldsymbol{\phi},\bar{\boldsymbol{\psi}}|, \tag{3.10}$$

where we have introduced coherent states

$$|\boldsymbol{\phi},\boldsymbol{\psi}\rangle = e^{\hat{\boldsymbol{\phi}}^\dagger\boldsymbol{\phi} + \hat{\boldsymbol{\psi}}^\dagger\boldsymbol{\psi}}|0\rangle, \tag{3.11}$$

and defined the integration measure $\mathrm{d}(\boldsymbol{\phi},\boldsymbol{\psi})$ as

$$\int \mathrm{d}(\boldsymbol{\phi},\boldsymbol{\psi}) = \iint \frac{\mathbf{d}\boldsymbol{\phi}^\dagger \mathbf{d}\boldsymbol{\phi}}{\pi^d} \iint \mathbf{d}\bar{\boldsymbol{\psi}}\mathbf{d}\boldsymbol{\psi} = \iint \prod_i \frac{\mathrm{d}\phi_i^\star \mathrm{d}\phi_i}{\pi} \iint \prod_j \mathrm{d}\bar{\psi}_j \mathrm{d}\psi_j \tag{3.12}$$

In these expressions $\psi_j$ and $\bar{\psi}_j$ are Grassmann numbers[2].

We insert $N+1$ identity operators on the form of (3.10) in (3.9), such that there is one identity operator on each side of each factor $(I - i\hat{H}\Delta t/\hbar)$. Further, we assume that the Hamiltonian of the system is expressed on normal ordered form as

$$\hat{H} = H(\hat{\boldsymbol{\phi}}^\dagger, \hat{\boldsymbol{\phi}}, \hat{\boldsymbol{\psi}}^\dagger, \hat{\boldsymbol{\psi}}), \tag{3.13}$$

where $H$ is an analytic function, and the operators are ordered such that $\hat{\phi}_i$ and $\hat{\psi}_j$ are always to the right of the adjoint fields. The benefit of the coherent state representation is that $\hat{\phi}_i|\boldsymbol{\phi}, \boldsymbol{\psi}\rangle = \phi_i|\boldsymbol{\phi}, \boldsymbol{\psi}\rangle$ and $\hat{\psi}_j|\boldsymbol{\phi}, \boldsymbol{\psi}\rangle = \psi_j|\boldsymbol{\phi}, \boldsymbol{\psi}\rangle$. Thus, since the Hamiltonian is normal ordered, we have

$$\begin{aligned}\langle \boldsymbol{\phi}, \bar{\boldsymbol{\psi}}| \left(I - \frac{i}{\hbar}\hat{H}\Delta t\right)|\boldsymbol{\phi}', \boldsymbol{\psi}\rangle &= \left(1 - \frac{i}{\hbar}H(\boldsymbol{\phi}^\dagger, \boldsymbol{\phi}', \bar{\boldsymbol{\psi}}, \boldsymbol{\psi})\Delta t\right)\langle \boldsymbol{\phi}, \boldsymbol{\psi}|\boldsymbol{\phi}, \boldsymbol{\psi}\rangle \\ &\approx e^{-\frac{i}{\hbar}H(\boldsymbol{\phi}^\dagger, \boldsymbol{\phi}', \bar{\boldsymbol{\psi}}, \boldsymbol{\psi})\Delta t + \boldsymbol{\phi}^\dagger\boldsymbol{\phi}' + \bar{\boldsymbol{\psi}}\boldsymbol{\psi}},\end{aligned} \tag{3.14}$$

where we have also used the formula $\langle \boldsymbol{\phi}, \bar{\boldsymbol{\psi}}|\boldsymbol{\phi}', \boldsymbol{\psi}\rangle = e^{\boldsymbol{\phi}^\dagger\boldsymbol{\phi}' + \bar{\boldsymbol{\psi}}\boldsymbol{\psi}}$ for the inner product of coherent states[2].

Making use of these steps, we can after some rearrangement write (3.9) as

$$U(t_f - t_0) = \int D(\phi, \psi) e^{iS^+[\phi,\psi]/\hbar} |\boldsymbol{\phi}_N, \boldsymbol{\psi}_N\rangle\langle \boldsymbol{\phi}_0, \bar{\boldsymbol{\psi}}_0|, \tag{3.15}$$

where we have defined the field integral

$$\int D(\phi, \psi) = \iint \prod_{n=0}^{N} \mathrm{d}(\boldsymbol{\phi}_n, \boldsymbol{\psi}_n). \tag{3.16}$$

and the action $S^+$ as

$$\begin{aligned}&\frac{i}{\hbar}S^+[\phi, \psi] = -|\boldsymbol{\phi}_0|^2 - \bar{\boldsymbol{\psi}}_0\boldsymbol{\psi}_0 + \\ &\frac{i}{\hbar}\sum_{n=0}^{N-1}\left\{\hbar i\boldsymbol{\phi}^\dagger_{n+1}(\boldsymbol{\phi}_{n+1} - \boldsymbol{\phi}_n) + \hbar i\bar{\boldsymbol{\psi}}_{n+1}(\boldsymbol{\psi}_{n+1} - \boldsymbol{\psi}_n) - H(\boldsymbol{\phi}^\dagger_{n+1}, \boldsymbol{\phi}_n, \bar{\boldsymbol{\psi}}_{n+1}, \boldsymbol{\psi}_n)\Delta t\right\}.\end{aligned} \tag{3.17}$$

Since we are interested in the limit $\Delta t \to 0$, we can formally approximate the sum as

$$\begin{aligned}&\sum_{n=0}^{N-1}\Delta t\left\{\hbar i\boldsymbol{\phi}^\dagger_{n+1}\tfrac{\boldsymbol{\phi}_{n+1}-\boldsymbol{\phi}_n}{\Delta t} + \hbar i\bar{\boldsymbol{\psi}}_{n+1}\tfrac{\boldsymbol{\psi}_{n+1}-\boldsymbol{\psi}_n}{\Delta t} - H(\boldsymbol{\phi}^\dagger_{n+1}, \boldsymbol{\phi}_n, \bar{\boldsymbol{\psi}}_{n+1}, \boldsymbol{\psi}_n)\right\} \\ &\approx \int_{t_0}^{t_f}\mathrm{d}t\,\left(\hbar i\boldsymbol{\phi}^\dagger\dot{\boldsymbol{\phi}} + \hbar i\bar{\boldsymbol{\psi}}\dot{\boldsymbol{\psi}} - H(\boldsymbol{\phi}^\dagger, \boldsymbol{\phi}, \bar{\boldsymbol{\psi}}, \boldsymbol{\psi})\right),\end{aligned} \tag{3.18}$$

where $\boldsymbol{\phi}_n = \boldsymbol{\phi}(n\Delta t)$ and $\boldsymbol{\psi}_n = \boldsymbol{\psi}(n\Delta t)$. This integral must however be thought of as a purely symbolic notation, since in reality the limit $N \to \infty$ is not well defined for the field integral expressions above, and since the derivative of Grassmann numbers is in any case not meaningful.

We will also require a field integral expression for $U^\dagger$. Taking the adjoint of (3.9), we get

$$U^\dagger(t_f - t_0) = \left(I + \frac{i}{\hbar}\hat{H}\Delta t\right)^N . \tag{3.19}$$

then simply repeating the argument above, we eventually end up with the expression

$$U^\dagger(t_f - t_0) = \int D(\phi,\psi) e^{iS^-[\phi,\psi]/\hbar} |\boldsymbol{\phi}_0, \boldsymbol{\psi}_0\rangle\langle \boldsymbol{\phi}_N, \bar{\boldsymbol{\psi}}_N|, \tag{3.20}$$

where the action $S^-$ is given by

$$\begin{aligned} \frac{i}{\hbar}S^-[\phi,\psi] &= -|\boldsymbol{\phi}_N|^2 - \bar{\boldsymbol{\psi}}_N\boldsymbol{\psi}_N \qquad (3.21)\\ -\frac{i}{\hbar}\sum_{n=0}^{N-1} &\Big\{\hbar i \boldsymbol{\phi}_n^\dagger(\boldsymbol{\phi}_{n+1} - \boldsymbol{\phi}_n) + \hbar i \bar{\boldsymbol{\psi}}_n(\boldsymbol{\psi}_{n+1} - \boldsymbol{\psi}_n) - H(\boldsymbol{\phi}_n^\dagger, \boldsymbol{\phi}_{n+1}, \bar{\boldsymbol{\psi}}_n, \boldsymbol{\psi}_{n+1})\Delta t\Big\}. \end{aligned}$$

Again, the sum can formally be approximated as

$$\begin{aligned} &\sum_{n=0}^{N-1} \Delta t \left\{\hbar i \boldsymbol{\phi}_n^\dagger \frac{\boldsymbol{\phi}_{n+1} - \boldsymbol{\phi}_n}{\Delta t} + \hbar i \bar{\boldsymbol{\psi}}_n \frac{\boldsymbol{\psi}_{n+1} - \boldsymbol{\psi}_n}{\Delta t} - H(\boldsymbol{\phi}_n^\dagger, \boldsymbol{\phi}_{n+1}, \bar{\boldsymbol{\psi}}_n, \boldsymbol{\psi}_{n+1})\right\} \\ &\approx \int_{t_0}^{t_f} \mathrm{d}t \, \Big(\hbar i \boldsymbol{\phi}^\dagger \dot{\boldsymbol{\phi}} + \hbar i \bar{\boldsymbol{\psi}} \dot{\boldsymbol{\psi}} - H(\boldsymbol{\phi}^\dagger, \boldsymbol{\phi}, \bar{\boldsymbol{\psi}}, \boldsymbol{\psi})\Big) . \qquad (3.22) \end{aligned}$$

### 3.1.2 Field integral expressions for expectation values

Single particle fermionic and bosonic observables respectively take the form

$$\hat{A} = \sum_{ij} A_{ij} \hat{\psi}_i^\dagger \hat{\psi}_j, \text{ and} \tag{3.23}$$

$$\hat{B} = \sum_{ij} B_{ij} \hat{\phi}_i^\dagger \hat{\phi}_j, \tag{3.24}$$

where $A_{ij} = \langle i|\hat{A}|j\rangle$ and $B_{ij} = \langle i|\hat{B}|j\rangle$ for single particle states $|i\rangle$ and $|j\rangle$. Thus, we have

$$\langle A\rangle = \sum_{ij} A_{ij} \langle \hat{\psi}_i^\dagger \hat{\psi}_j\rangle, \tag{3.25}$$

and similarly for the bosonic case. Accordingly, all single particle expectation values can be found by calculating expectation values on the form $\langle \hat{\psi}_i^\dagger \hat{\psi}_j\rangle$ and $\langle \hat{\phi}_i^\dagger \hat{\phi}_j\rangle$.

By (3.6) we have

$$\begin{aligned}\langle\hat{\psi}_i^\dagger(t)\hat{\psi}_j(t')\rangle &= \operatorname{Tr} U^\dagger(t-t_0)\hat{\psi}_i^\dagger U(t-t_0)U^\dagger(t'-t_0)\hat{\psi}_j U(t'-t_0)\rho_0 \\ &= \operatorname{Tr} U^\dagger(t-t_0)\hat{\psi}_i^\dagger U^\dagger(t_f-t)U(t_f-t')\hat{\psi}_j U(t'-t_0)\rho_0 \\ &= \operatorname{Tr} U^\dagger(t_f-t_0)U(t_f-t)\hat{\psi}_i^\dagger U(t-t')\hat{\psi}_j U(t'-t_0)\rho_0 \\ &= \operatorname{Tr} U^\dagger(t-t_0)\hat{\psi}_i^\dagger U^\dagger(t'-t)\hat{\psi}_j U^\dagger(t_f-t')U(t_f-t_0)\rho_0,\end{aligned} \tag{3.26}$$

where we have introduced an arbitrary final time $t_f > t, t'$, and made use of (3.3) and (3.4). The product $U(t_f - t')\hat{\psi}_j U(t' - t_0)$ in the second line can be written as a field integral by the same procedure as in the previous section, where we approximate the propagators as products of linear factors, and then insert $N + 1$ identity operators given by (3.10). However, we must take care to place the appropriate identity operator to the right of $\hat{\psi}_j$, so that we can make use of the relation $\hat{\psi}_j|\boldsymbol{\phi}, \boldsymbol{\psi}\rangle = \psi_j|\boldsymbol{\phi}, \boldsymbol{\psi}\rangle$ to get rid of this operator.

Clearly, we end up with an expression identical to (3.15), except for an additional factor of $\psi_{mj} = \psi_j(t')$. As long as we assume the Hamiltonian to be even in the fermionic fields, this factor can be commuted to the left in the expression, so that we end up with

$$U(t_f - t')\hat{\psi}_j U(t' - t_0) = \int D(\phi, \psi)\psi_j(t')e^{iS^+[\phi,\psi]/\hbar}|\boldsymbol{\phi}_N, \boldsymbol{\psi}_N\rangle\langle\boldsymbol{\phi}_0, \bar{\boldsymbol{\psi}}_0|. \tag{3.27}$$

Proceeding in a precisely analogous manor, we also find the relation

$$U^\dagger(t - t_0)\hat{\psi}_i^\dagger U^\dagger(t_f - t) = \int D(\phi, \psi)\bar{\psi}_i(t)e^{iS^-[\phi,\psi]/\hbar}|\boldsymbol{\phi}_0, \boldsymbol{\psi}_0\rangle\langle\boldsymbol{\phi}_N, \bar{\boldsymbol{\psi}}_N|. \tag{3.28}$$

Inserting (3.27) and (3.28) in the second line of (3.26), we obtain after some rearrangement

$$\langle\hat{\psi}_i^\dagger(t)\hat{\psi}_j(t')\rangle = \int D(+,-)\bar{\psi}_i^-(t)\psi_j^+(t')e^{iS[\phi^+,\psi^+,\phi^-,\psi^-]/\hbar}, \tag{3.29}$$

where we have introduced the short notation

$$\int D(+,-) = \int D(\phi^+, \psi^+, \phi^-, \psi^-) = \int D(\phi^+, \psi^+) \int D(\phi^-, \psi^-), \tag{3.30}$$

as well as the total action $S$, defined by the expression

$$\begin{aligned}\frac{i}{\hbar}S[\phi^+, \psi^+, \phi^-, \psi^-] = \frac{i}{\hbar}\left(S^+[\phi^+, \psi^+] + S^-[\phi^-, \psi^-]\right) + \boldsymbol{\phi}_N^{-\dagger}\boldsymbol{\phi}_N^+ + \bar{\boldsymbol{\psi}}_N^-\boldsymbol{\psi}_N^+ \\ + \ln\langle\boldsymbol{\phi}_0^+, \bar{\boldsymbol{\psi}}_0^+|\rho_0|\boldsymbol{\phi}_0^-, -\boldsymbol{\psi}_0^-\rangle,\end{aligned} \tag{3.31}$$

where $S^+$ and $S^-$ are given respectively by (3.17) and (3.21), and where the negative sign in front of $\boldsymbol{\psi}_0^-$ comes from the Grassmann anti-commutation relations.

Motivated by this result, we make the general definition

$$\langle A_1(t_1)\cdots A_n(t_n)\rangle = \int D(+,-)A_1(t_1)\cdots A_n(t_n)e^{iS[\phi^+,\psi^+,\phi^-,\psi^-]/\hbar}, \quad (3.32)$$

which allows us to write (3.29) simply as $\langle\hat{\psi}_i^\dagger(t)\hat{\psi}_j(t')\rangle = \langle\bar{\psi}_i^-(t)\psi_j^+(t')\rangle$. (3.32) defines the general non-equilibrium field integral, also known as the Keldysh field integral.[2]

As long as $t > t'$, we can make use of the same procedure as above to write also the product $U(t_f - t)\hat{\psi}_i^\dagger U(t-t')\hat{\psi}_j U(t'-t_0)$ from the third line of (3.26) as a field integral. Combining the resulting expression with (3.20), we obtain $\langle\hat{\psi}_i^\dagger(t)\hat{\psi}_j(t')\rangle = \langle\bar{\psi}_i^+(t)\psi_j^+(t')\rangle$. Finally, $U^\dagger(t-t_0)\hat{\psi}_i^\dagger U^\dagger(t'-t)\hat{\psi}_j U^\dagger(t_f-t')$ from the fourth line of (3.26) can be written as a field integral in the same manner, as long as we require $t < t'$. Combining the resulting expression with (3.15), we obtain $\langle\hat{\psi}_i^\dagger(t)\hat{\psi}_j(t')\rangle = \langle\bar{\psi}_i^-(t)\psi_j^-(t')\rangle$. Summarizing, we have found the following relations between operator and field integral expectation values:

$$\begin{aligned}\langle\hat{\psi}_i^\dagger(t)\hat{\psi}_j(t')\rangle &= \langle\bar{\psi}_i^-(t)\psi_j^+(t')\rangle && (3.33)\\ &= \langle\bar{\psi}_i^+(t)\psi_j^+(t')\rangle \qquad t > t' \\ &= \langle\bar{\psi}_i^-(t)\psi_j^-(t')\rangle \qquad t < t'.\end{aligned}$$

The opposite product $\langle\hat{\psi}_i(t)\hat{\psi}_j^\dagger(t')\rangle$ can be found in a similar fashion, by simply exchanging the operators $\hat{\psi}_j$ and $\hat{\psi}_i^\dagger$ in (3.26), renaming the indices, and then repeating the calculations above. In the end, the results are

$$\begin{aligned}\langle\hat{\psi}_i(t)\hat{\psi}_j^\dagger(t')\rangle &= \langle\psi_i^-(t)\bar{\psi}_j^+(t')\rangle && (3.34)\\ &= \langle\psi_i^+(t)\bar{\psi}_j^+(t')\rangle \qquad t \geq t' \\ &= \langle\psi_i^-(t)\bar{\psi}_j^-(t')\rangle \qquad t \leq t'.\end{aligned}$$

Clearly, the same procedure can also be applied to the bosonic expressions. This yields the near identical expressions

$$\begin{aligned}\langle\phi_i^\dagger(t)\phi_j(t')\rangle &= \langle\phi_i^{-\star}(t)\phi_j^+(t')\rangle && (3.35)\\ &= \langle\phi_i^{+\star}(t)\phi_j^+(t')\rangle \qquad t > t' \\ &= \langle\phi_i^{-\star}(t)\phi_j^-(t')\rangle \qquad t < t' \\ \langle\phi_i(t)\phi_j^\dagger(t')\rangle &= \langle\phi_i^-(t)\phi_j^{+\star}(t')\rangle && (3.36)\\ &= \langle\phi_i^+(t)\phi_j^{+\star}(t')\rangle \qquad t \geq t' \\ &= \langle\phi_i^-(t)\phi_j^{-\star}(t')\rangle \qquad t \leq t'.\end{aligned}$$

## 3.2 Transport field integral

### 3.2.1 Definition of the model

Let us now return to the system described in Section 2.1, where a system $S$ is connected to a set of leads $p$. We will assume that the degrees of freedom in $S$ are a bosonic field $\boldsymbol{\phi}$, and fermionic field $\boldsymbol{\psi}$, and that these are controlled by a Hamiltonian on the form of (3.13). We also assume that lead $p$ contains a fermionic field $\boldsymbol{\psi}_p$, and a bosonic field $\boldsymbol{\phi}_p$. Following the discussion of Section 2.2, we take the only purpose of the leads to be to drive $S$ out of equilibrium, and accordingly choose the lead physics from considerations of simplicity rather than realism. Thus, we assume the lead Hamiltonians to be quadratic in the fields, and write the Hamiltonian of lead $p$ as

$$\hat{H}_p = \hat{\boldsymbol{\psi}}_p^\dagger H_p^F \hat{\boldsymbol{\psi}}_p + \hat{\boldsymbol{\phi}}_p^\dagger H_p^B \hat{\boldsymbol{\phi}}_p, \tag{3.37}$$

where $H_p^F$ and $H_p^B$ are matrix representations of the fermionic and bosonic single particle Hamiltonians respectively.

An other simplification we will make, is to assume the leads to be internally in equilibrium at $t = t_0$. This can be justified from an assumption that the leads are not in contact with $S$ before this time. Accordingly, this assumption is more plausible than the one made in Chapter 2, where we assumed the leads to be in equilibrium at all times. The density operator $\rho_p(t_0)$ can then be expressed as[18]

$$\rho_p(t_0) = \frac{1}{Z_p} e^{-\beta_p(\hat{H}_p - \mu_p \hat{N}_p^F)}, \tag{3.38}$$

where $T_p = 1/k_B\beta_p$ is the temperature, $\mu_p$ the chemical potential of the fermion field, and $\hat{N}_p^F = \hat{\boldsymbol{\psi}}_p^\dagger \hat{\boldsymbol{\psi}}_p$ is the fermion number operator.

We assume that the interactions between the leads and the system $S$ consists of simple single particle hopping, so that the interaction terms are also quadratic in the fields. Thus, we can write the total Hamiltonian as

$$\hat{H} = \hat{H}_S + \sum_p \left( \hat{H}_p + \hat{\boldsymbol{\phi}}_p^\dagger b_p \hat{\boldsymbol{\phi}} + \hat{\boldsymbol{\phi}}^\dagger b_p^\dagger \hat{\boldsymbol{\phi}}_p + \hat{\boldsymbol{\psi}}_p^\dagger t_p \hat{\boldsymbol{\psi}} + \hat{\boldsymbol{\psi}}^\dagger t_p^\dagger \hat{\boldsymbol{\psi}}_p \right), \tag{3.39}$$

where $\hat{H}_S$ is given by (3.13), $\hat{H}_p$ by (3.37), and $b_p$ and $t_p$ are matrices describing single particle hopping between the leads and $S$, for bosons and fermions respectively.

On the other hand, the total density operator $\rho_0 = \rho(t_0)$ will be assumed to be on the product form

$$\rho_0 = \rho_S(t_0) \prod_p \rho_p(t_0), \tag{3.40}$$

where all the factors are assumed to commute. This form can again be justified from the assumption that the leads are first brought into contact with $S$ at $t = t_0$. While $\rho_p(t_0)$ have been defined above, we still require an expression for $\rho_S(t_0)$, the state of $S$ at $t = t_0$. We are generally interested in stationary properties at times $t \gg t_0$, at which we assume $S$ to have reached a steady state $\rho_S(t)$ which is independent of the initial state. Thus, we can choose $\rho_S(t_0)$ arbitrarily, and we select it for purposes of simplicity as

$$\rho_S(t_0) = |0\rangle\langle 0| = P_S^0, \tag{3.41}$$

i.e. the ground state projection operator acting on $S$. Accordingly, the full state $\rho_0$ of (3.40) becomes

$$\rho_0 = \frac{1}{Z} \exp\left( \sum_p -\beta_p (\hat{H}_p - \mu_p \hat{N}_p^F) \right) P_S^0. \tag{3.42}$$

### 3.2.2 Expression as a field integral

To express the model defined above as a field integral, we must calculate the total action $S$ defined in (3.31). To begin, we collect the component fields $\boldsymbol{\psi}$, $\boldsymbol{\psi}_p$, and so on in total field vectors $\tilde{\boldsymbol{\psi}}$, $\bar{\tilde{\boldsymbol{\psi}}}$ and $\tilde{\boldsymbol{\phi}}$. That is, we define $\tilde{\boldsymbol{\psi}} = [\boldsymbol{\psi}, \cdots \boldsymbol{\psi}_p \cdots]$ and similarly for the other fields. The total action of the system is then given by (3.31), with $\psi$ substituted for $\tilde{\psi}$ and $\phi$ substituted for $\tilde{\phi}$.

The evaluation of this action is a straight forward task, except for the bracket involving $\rho_0$. By (3.42) we have

$$\begin{aligned} \ln \langle \tilde{\boldsymbol{\phi}}_0^+, \bar{\tilde{\boldsymbol{\psi}}}_0^+ | \rho_0 | \tilde{\boldsymbol{\phi}}_0^-, -\tilde{\boldsymbol{\psi}}_0^- \rangle = {} & \sum_p \ln \langle \boldsymbol{\phi}_{p0}^+, \bar{\boldsymbol{\psi}}_{p0}^+ | e^{-\beta_p (\hat{H}_p - \mu_p \hat{N}_p^F)} | \boldsymbol{\phi}_{p0}^-, -\boldsymbol{\psi}_{p0}^- \rangle \\ & + \ln \langle \boldsymbol{\phi}_0^+, \bar{\boldsymbol{\psi}}_0^+ | P_S^0 | \boldsymbol{\phi}_0^-, -\boldsymbol{\psi}_0^- \rangle - \ln Z. \end{aligned} \tag{3.43}$$

From the definition (3.11) of the coherent states, it is easily realized that $\langle \boldsymbol{\phi}_0^+, \bar{\boldsymbol{\psi}}_0^+ | P_S^0 | \boldsymbol{\phi}_0^-, -\boldsymbol{\psi}_0^- \rangle = 1$. To find the remaining terms we will make use of commutation relations of exponential operators. By making use of the fundamental commutation relations of the fermionic and bosonic fields, $[\phi_i, \phi_j^\dagger] = \phi_i \phi_j^\dagger - \phi_j^\dagger \phi_i = \delta_{ij}$ and $\{\psi_i, \psi_j^\dagger\} = \psi_i \psi_j^\dagger + \psi_j^\dagger \psi_i = \delta_{ij}$, together with a Taylor expansion of the exponential, one can show that for any matrix $A$

$$e^{\boldsymbol{\phi}^\dagger \hat{\boldsymbol{\phi}}} e^{\hat{\boldsymbol{\phi}}^\dagger A \hat{\boldsymbol{\phi}}} = e^{\hat{\boldsymbol{\phi}}^\dagger A \hat{\boldsymbol{\phi}}} e^{\boldsymbol{\phi}^\dagger e^A \hat{\boldsymbol{\phi}}}, \quad \text{and} \tag{3.44}$$

$$e^{\boldsymbol{\psi}^\dagger \hat{\boldsymbol{\psi}}} e^{\hat{\boldsymbol{\psi}}^\dagger A \hat{\boldsymbol{\psi}}} = e^{\hat{\boldsymbol{\psi}}^\dagger A \hat{\boldsymbol{\psi}}} e^{\boldsymbol{\psi}^\dagger e^A \hat{\boldsymbol{\psi}}}. \tag{3.45}$$

These expressions can be used together with the definition of the coherent states (3.11) to rearrange the terms of (3.43) in such a way that the relation $\hat{\psi}_j |\boldsymbol{\phi}, \boldsymbol{\psi}\rangle = \psi_j |\boldsymbol{\phi}, \boldsymbol{\psi}\rangle$ can be used to remove the operators from the expressions. The results is

$$\ln \langle \boldsymbol{\phi}_{p0}^+, \bar{\boldsymbol{\psi}}_{p0}^+ | e^{-\beta_p (\hat{H}_p - \mu_p \hat{N}_p^F)} | \boldsymbol{\phi}_{p0}^-, -\boldsymbol{\psi}_{p0}^- \rangle = \boldsymbol{\phi}_{p0}^{+\dagger} e^{-\beta_p H_p^B} \boldsymbol{\phi}_{p0}^- - \bar{\boldsymbol{\psi}}_{p0}^+ e^{-\beta_p (H_p^F - \mu_p)} \boldsymbol{\psi}_{p0}^-. \tag{3.46}$$

Accordingly, making use of (3.43), (3.31) and (3.39), we can write the total action $S$ as

$$S[\tilde{\phi}^+, \tilde{\psi}^+, \tilde{\phi}^-, \tilde{\psi}^-] = S_S + \sum_p \left( S_p^B + S_p^F \right), \tag{3.47}$$

where

$$\frac{i}{\hbar} S_S[\phi^+, \psi^+, \phi^-, \psi^-] = \frac{i}{\hbar}\left(S^+[\phi^+, \psi^+] + S^-[\phi^-, \psi^-]\right) + \boldsymbol{\phi}_N^{-\dagger}\boldsymbol{\phi}_N^+ + \bar{\boldsymbol{\psi}}_N^-\boldsymbol{\psi}_N^+, \tag{3.48}$$

with $S^+$ and $S^-$ given respectively by (3.17) and (3.21), and where

$$\frac{i}{\hbar} S_p^B = \frac{i}{\hbar}\left(S_p^{B+} + S_p^{B-}\right) + \boldsymbol{\phi}_{pN}^{-\dagger}\boldsymbol{\phi}_{pN}^+ + \boldsymbol{\phi}_{p0}^{+\dagger} e^{-\beta_p H_p^B}\boldsymbol{\phi}_{p0}^- - \ln Z_p^B, \quad \text{and} \tag{3.49}$$

$$\frac{i}{\hbar} S_p^F = \frac{i}{\hbar}\left(S_p^{F+} + S_p^{F-}\right) + \bar{\boldsymbol{\psi}}_{pN}^-\boldsymbol{\phi}_{pN}^+ - \bar{\boldsymbol{\psi}}_{p0}^+ e^{-\beta_p(H_p^F - \mu_p)}\boldsymbol{\psi}_{p0}^- - \ln Z_p^F, \tag{3.50}$$

with

$$\begin{aligned}\frac{i}{\hbar} S_p^{B+} = & -|\boldsymbol{\phi}_{p0}^+|^2 + \frac{i}{\hbar}\sum_{n=0}^{N-1}\left\{\hbar i \boldsymbol{\phi}_{pn+1}^{+\dagger}(\boldsymbol{\phi}_{pn+1}^+ - \boldsymbol{\phi}_{pn}^+) - \boldsymbol{\phi}_{pn+1}^{+\dagger} H_p^B \boldsymbol{\phi}_{pn}^+ \Delta t\right\} \\ & - \frac{i}{\hbar}\sum_{n=0}^{N-1}\left\{\boldsymbol{\phi}_{pn+1}^{+\dagger} b_p \boldsymbol{\phi}_n^+ + \boldsymbol{\phi}_{n+1}^{+\dagger} b_p^\dagger \boldsymbol{\phi}_{pn}^+\right\}\Delta t,\end{aligned} \tag{3.51}$$

$$\begin{aligned}\frac{i}{\hbar} S_p^{F+} = & -\bar{\boldsymbol{\psi}}_{p0}^+\boldsymbol{\psi}_{p0}^+ + \frac{i}{\hbar}\sum_{n=0}^{N-1}\left\{\hbar i \bar{\boldsymbol{\psi}}_{pn+1}^+(\boldsymbol{\psi}_{pn+1}^+ - \boldsymbol{\psi}_{pn}^+) - \bar{\boldsymbol{\psi}}_{pn+1}^+ H_p^F \boldsymbol{\psi}_{pn}^+ \Delta t\right\} \\ & - \frac{i}{\hbar}\sum_{n=0}^{N-1}\left\{\bar{\boldsymbol{\psi}}_{pn+1}^+ t_p \boldsymbol{\psi}_n^+ + \bar{\boldsymbol{\psi}}_{n+1}^+ t_p^\dagger \boldsymbol{\psi}_{pn}^+\right\}\Delta t,\end{aligned} \tag{3.52}$$

$$\begin{aligned}\frac{i}{\hbar} S_p^{B-} = & -|\boldsymbol{\phi}_{pN}^-|^2 - \frac{i}{\hbar}\sum_{n=0}^{N-1}\left\{\hbar i \boldsymbol{\phi}_{pn}^{-\dagger}(\boldsymbol{\phi}_{pn+1}^- - \boldsymbol{\phi}_{pn}^-) - \boldsymbol{\phi}_{pn}^{-\dagger} H_p^B \boldsymbol{\phi}_{pn+1}^- \Delta t\right\} \\ & + \frac{i}{\hbar}\sum_{n=0}^{N-1}\left\{\boldsymbol{\phi}_{pn}^{-\dagger} b_p \boldsymbol{\phi}_{n+1}^- + \boldsymbol{\phi}_n^{-\dagger} b_p^\dagger \boldsymbol{\phi}_{pn+1}^-\right\}\Delta t, \quad \text{and}\end{aligned} \tag{3.53}$$

$$\begin{aligned}\frac{i}{\hbar} S_p^{F-} = & -\bar{\boldsymbol{\psi}}_{pN}^-\boldsymbol{\psi}_{pN}^- - \frac{i}{\hbar}\sum_{n=0}^{N-1}\left\{\hbar i \bar{\boldsymbol{\psi}}_{pn}^-(\boldsymbol{\psi}_{pn+1}^- - \boldsymbol{\psi}_{pn}^-) - \bar{\boldsymbol{\psi}}_{pn}^- H_p^F \boldsymbol{\psi}_{pn+1}^- \Delta t\right\} \\ & + \frac{i}{\hbar}\sum_{n=0}^{N-1}\left\{\bar{\boldsymbol{\psi}}_{pn}^- t_p \boldsymbol{\psi}_{n+1}^- + \bar{\boldsymbol{\psi}}_n^- t_p^\dagger \boldsymbol{\psi}_{pn+1}^-\right\}\Delta t.\end{aligned} \tag{3.54}$$

## 3.3 Integration over the leads

Combining (3.32) and (3.47), and making use of the facts that $S_S$ contains no terms from the leads, and $S_p^B$ and $S_p^F$ no terms from the other leads $q$, we can reexpress a general expectation value $\langle A[\phi, \psi]\rangle$ of fields in $S$ as

$$\langle A\rangle = \int D(+,-)A[\phi,\psi]e^{iS_S/\hbar}\prod_p \int D(\phi_p^+,\phi_p^-)e^{iS_p^B/\hbar}\int D(\psi_p^+,\psi_p^-)e^{iS_p^F/\hbar}, \tag{3.55}$$

so that the lead sections of the integral can be performed separately. By (3.49)-(3.54), the lead actions $S_p^B$ and $S_p^F$ are quadratic in the fields, so accordingly, these integrals can be performed using Gaussian integration.

Introducing a total field vector $\boldsymbol{\phi}_p = [\boldsymbol{\phi}_{p0}^+ \cdots \boldsymbol{\phi}_{pN}^+, \boldsymbol{\phi}_{p0}^- \cdots \boldsymbol{\phi}_{pN}^-]$ and similar for the other fields, the lead actions can be expressed as

$$S_p^B = \boldsymbol{\phi}_p^\dagger A_p \boldsymbol{\phi}_p + \boldsymbol{\phi}_p^\dagger B_p \boldsymbol{\phi} + \boldsymbol{\phi}^\dagger B_p^\dagger \boldsymbol{\phi}_p + \hbar i \ln Z_p^B, \quad \text{and} \tag{3.56}$$

$$S_p^F = \bar{\boldsymbol{\psi}}_p C_p \boldsymbol{\psi}_p + \bar{\boldsymbol{\psi}}_p T_p \boldsymbol{\psi} + \bar{\boldsymbol{\psi}} T_p^\dagger \boldsymbol{\psi}_p + \hbar i \ln Z_p^F, \tag{3.57}$$

where $A_p$, $B_p$, $C_p$ and $T_p$ are appropriate matrices. Making use of Gaussian integration rules found for instance in Altland and Simons[2], we find

$$\int D(\phi_p^+,\phi_p^-)e^{iS_p^B/\hbar} = \frac{\hbar i}{Z_p^B}\det A_p^{-1} e^{-i\boldsymbol{\phi}^\dagger B_p^\dagger A_p^{-1} B_p \boldsymbol{\phi}/\hbar}, \quad \text{and} \tag{3.58}$$

$$\int D(\psi_p^+,\psi_p^-)e^{iS_p^F/\hbar} = \frac{1}{\hbar i Z_p^F}\det C_p\, e^{-i\bar{\boldsymbol{\psi}} T_p^\dagger C_p^{-1} T_p \boldsymbol{\psi}/\hbar}. \tag{3.59}$$

Further, the prefactors of these expressions must in fact equal 1. To see this, consider isolated lead systems, respectively with total Hamiltonians $\hat{H} = \hat{\boldsymbol{\phi}}_p^\dagger H_p^B \hat{\boldsymbol{\phi}}_p$ and $\hat{H} = \hat{\boldsymbol{\psi}}_p^\dagger H_p^F \hat{\boldsymbol{\psi}}_p$. It should be clear that the actions of these systems are given respectively by (3.49) and (3.50), but excluding the terms involving the matrices $b_p$ and $t_p$. Thus, these actions can be written $S = \boldsymbol{\phi}_p^\dagger A_p \boldsymbol{\phi}_p$ and $S = \bar{\boldsymbol{\psi}}_p C_p \boldsymbol{\psi}_p$ respectively. Applying (3.32) to these systems, and again making use of Gaussian integration rules, we find

$$1 = \langle 1\rangle = \frac{\hbar i}{Z_p^B}\det A_p^{-1} = \frac{1}{\hbar i Z_p^F}\det C_p. \tag{3.60}$$

Thus, making use of (3.58) and (3.59), we can express (3.55) as

$$\langle A\rangle = \int D(+,-)A[\phi,\psi]e^{iS^{\text{eff}}[\phi,\psi]/\hbar}, \tag{3.61}$$

where the effective action $S^{\text{eff}}$ is given by

$$S^{\text{eff}}[\phi,\psi] = S_S + \sum_p \left(S_p^{B,\text{eff}} + S_p^{F,\text{eff}}\right), \quad \text{with} \tag{3.62}$$

$$S_p^{B,\text{eff}} = -\boldsymbol{\phi}^\dagger B_p^\dagger D_p B_p \boldsymbol{\phi}, \quad \text{and} \tag{3.63}$$

$$S_p^{F,\text{eff}} = -\bar{\boldsymbol{\psi}} T_p^\dagger G_p T_p \boldsymbol{\psi}. \tag{3.64}$$

where we have defined $D_p = A_p^{-1}$ and $G_p = C_p^{-1}$.

### 3.3.1 Green's functions

Following Altland and Simons[2], expressions for $D_p$ and $G_p$ could be obtained by direct inversion of the matrices $A_p$ and $C_p$. However, as we will see $D_p$ and $G_p$ are related to certain operator expectation values, which are in fact easier to calculate. Consider again the isolated lead systems with actions $\boldsymbol{\phi}_p^\dagger A_p \boldsymbol{\phi}_p$ and $\bar{\boldsymbol{\psi}}_p C_p \boldsymbol{\psi}_p$. Once again making use of (3.32) and laws of Gaussian integrals, we have

$$\langle \boldsymbol{\phi}_p \boldsymbol{\phi}_p^\dagger \rangle = \frac{1}{Z_p^B} \iint \frac{\mathbf{d}\boldsymbol{\phi}_p^\dagger \mathbf{d}\boldsymbol{\phi}_p}{\pi^{2Nd}} \boldsymbol{\phi}_p \boldsymbol{\phi}_p^\dagger e^{-\boldsymbol{\phi}_p^\dagger A_p \boldsymbol{\phi}_p/\hbar i} = i\hbar A_p^{-1} = i\hbar D_p, \quad \text{and} \tag{3.65}$$

$$\langle \boldsymbol{\psi}_p \bar{\boldsymbol{\psi}}_p \rangle = \frac{1}{Z_p^F} \iint \mathbf{d}\bar{\boldsymbol{\psi}}_p \mathbf{d}\boldsymbol{\psi}_p \boldsymbol{\psi}_p \bar{\boldsymbol{\psi}}_p e^{-\bar{\boldsymbol{\psi}}_p C_p \boldsymbol{\psi}_p/\hbar i} = i\hbar C_p^{-1} = i\hbar G_p. \tag{3.66}$$

Accordingly, the matrices $D_p$ and $G_p$ describe correlations between the various fields in the leads. In the literature, these objects are often referred to as Green's functions[2, 15, 7]. When dealing with more general systems, where the action need not be quadratic, we define the Green's functions directly through the relations $i\hbar D = \langle \boldsymbol{\phi}\boldsymbol{\phi}^\dagger \rangle$ and $i\hbar G = \langle \boldsymbol{\psi}\bar{\boldsymbol{\psi}} \rangle$. It is convenient to decompose these matrices into four sectors in the following manner:

$$\boldsymbol{\phi}^\dagger D \boldsymbol{\phi} = \begin{bmatrix} \boldsymbol{\phi}^{+\dagger} & \boldsymbol{\phi}^{-\dagger} \end{bmatrix} \begin{bmatrix} D^t & D^< \\ D^> & D^{\bar{t}} \end{bmatrix} \begin{bmatrix} \boldsymbol{\phi}^+ \\ \boldsymbol{\phi}^- \end{bmatrix}, \quad \text{and} \tag{3.67}$$

$$\bar{\boldsymbol{\psi}} G \boldsymbol{\psi} = \begin{bmatrix} \bar{\boldsymbol{\psi}}^+ & \bar{\boldsymbol{\psi}}^- \end{bmatrix} \begin{bmatrix} G^t & G^< \\ G^> & G^{\bar{t}} \end{bmatrix} \begin{bmatrix} \boldsymbol{\psi}^+ \\ \boldsymbol{\psi}^- \end{bmatrix}, \tag{3.68}$$

where $\boldsymbol{\phi}^+ = [\boldsymbol{\phi}_0^+ \cdots \boldsymbol{\phi}_N^+]$, $\boldsymbol{\phi}^- = [\boldsymbol{\phi}_0^- \cdots \boldsymbol{\phi}_N^-]$, and similar for the fermionic fields. Writing out the definitions in terms of the component matrices, we have

$$i\hbar D^t(t,t') = i\hbar D^t_{nm} = \langle \boldsymbol{\phi}^+(t) \boldsymbol{\phi}^{+\dagger}(t') \rangle, \tag{3.69}$$

$$i\hbar D^<(t,t') = i\hbar D^<_{nm} = \langle \boldsymbol{\phi}^+(t) \boldsymbol{\phi}^{-\dagger}(t') \rangle, \tag{3.70}$$

$$i\hbar D^>(t,t') = i\hbar D^>_{nm} = \langle \boldsymbol{\phi}^-(t) \boldsymbol{\phi}^{+\dagger}(t') \rangle, \tag{3.71}$$

$$i\hbar D^{\bar{t}}(t,t') = i\hbar D^{\bar{t}}_{nm} = \langle \boldsymbol{\phi}^-(t) \boldsymbol{\phi}^{-\dagger}(t') \rangle, \tag{3.72}$$

where $t = n\Delta t$ and $t' = m\Delta t$. Similarly, for the fermionic fields we have

$$i\hbar G^t(t,t') = i\hbar G^t_{nm} = \langle \boldsymbol{\psi}^+(t) \bar{\boldsymbol{\psi}}^+(t') \rangle, \tag{3.73}$$

$$i\hbar G^<(t,t') = i\hbar G^<_{nm} = \langle \boldsymbol{\psi}^+(t) \bar{\boldsymbol{\psi}}^-(t') \rangle, \tag{3.74}$$

$$i\hbar G^>(t,t') = i\hbar G^>_{nm} = \langle \boldsymbol{\psi}^-(t) \bar{\boldsymbol{\psi}}^+(t') \rangle, \tag{3.75}$$

$$i\hbar G^{\bar{t}}(t,t') = i\hbar G^{\bar{t}}_{nm} = \langle \boldsymbol{\psi}^-(t) \bar{\boldsymbol{\psi}}^-(t') \rangle. \tag{3.76}$$

Making use of (3.35) and (3.36), we can reexpress the bosonic Green's

functions in terms of operator expectation values as

$$D^{>}_{ij}(t,t') = -i\langle\hat{\phi}_i(t)\hat{\phi}^\dagger_j(t')\rangle/\hbar, \tag{3.77}$$
$$D^{<}_{ij}(t,t') = -i\langle\hat{\phi}^\dagger_j(t')\hat{\phi}_i(t)\rangle/\hbar, \tag{3.78}$$
$$D^{t}(t,t') = D^{>}(t,t')\theta(t-t') + D^{<}(t,t')\theta(t'-t-\Delta t), \quad \text{and} \tag{3.79}$$
$$D^{\bar{t}}(t,t') = D^{<}(t,t')\theta(t-t'-\Delta t) + D^{>}(t,t')\theta(t'-t), \tag{3.80}$$

where $\theta(t)$ is a step function such that $\theta(t) = 1$ for $t \geq 0$, and $\theta(t) = 0$ for $t < 0$. For the corresponding fermionic Green's functions, we find using (3.33) and (3.34)

$$G^{>}_{ij}(t,t') = -i\langle\hat{\psi}_i(t)\hat{\psi}^\dagger_j(t')\rangle/\hbar, \tag{3.81}$$
$$G^{<}_{ij}(t,t') = i\langle\hat{\psi}^\dagger_j(t')\hat{\psi}_i(t)\rangle/\hbar, \tag{3.82}$$
$$G^{t}(t,t') = G^{>}(t,t')\theta(t-t') + G^{<}(t,t')\theta(t'-t-\Delta t), \quad \text{and} \tag{3.83}$$
$$G^{\bar{t}}(t,t') = G^{<}(t,t')\theta(t-t'-\Delta t) + G^{>}(t,t')\theta(t'-t), \tag{3.84}$$

where the changes of sign are due to anti-commutation of the Grassmann numbers.

There is a connection between the superscript notation of the Green's functions and the relations expressed in (3.77)-(3.84). For instance, $G^<$ and $D^<$ are equal to $G^t$ and $D^t$ precisely when $t < t'$, while $G^>$ and $D^>$ are equal to $G^t$ and $D^t$ when $t > t'$. Further, examining (3.79) and (3.83), we see that the nonzero term is always the term where the operator with the largest time argument is applied last. Therefore, $G^t$ and $D^t$ are referred to as time ordered Green's functions. Similarly, examining (3.80) and (3.84), we see that see that the nonzero term is always the term where the operator with the largest time argument is applied first. $G^{\bar{t}}$ and $D^{\bar{t}}$ are therefore referred to as anti-time ordered Green's functions[2, 15].

Expressions for the Green's functions of non-interacting systems like our isolated leads, are easiest to obtain by introducing a diagonal basis of the single particle Hamiltonian. Accordingly, considering some arbitrary non-interacting system with bosonic single particle Hamiltonian $H^B$, and fermionic single particle Hamiltonian $H^F$, we introduce a basis of eigenvectors $u_i$ and $v_i$ of $H^B$ and $H^F$ respectively, and write these matrices as

$$H^B = \sum_i E^B_i u_i u^\dagger_i, \quad \text{and} \tag{3.85}$$
$$H^F = \sum_i E^F_i v_i v^\dagger_i. \tag{3.86}$$

Using these bases, we can express both the propagator $U(t-t')$ and $\rho_0$ in diagonal form, after which (3.77), (3.78), (3.81) and (3.82) are easy to evaluate.

The details can be found for instance in the book by Jacoboni[15], who derives

$$i\hbar D^<(t,t') = \sum_i n(E_i^B) e^{iE_i^B(t'-t)/\hbar} u_i u_i^\dagger, \qquad \text{and} \tag{3.87}$$

$$i\hbar D^>(t,t') = \sum_i \left[1 + n(E_i^B)\right] e^{iE_i^B(t'-t)/\hbar} u_i u_i^\dagger, \quad \text{where} \tag{3.88}$$

$$n(E) = \frac{1}{e^{\beta E} - 1}, \tag{3.89}$$

and similarly for the fermionic expressions,

$$i\hbar G^<(t,t') = -\sum_i f(E_i^F) e^{iE_i^F(t'-t)/\hbar} v_i v_i^\dagger, \qquad \text{and} \tag{3.90}$$

$$i\hbar G^>(t,t') = \sum_i \left[1 - f(E_i^F)\right] e^{iE_i^F(t'-t)/\hbar} v_i v_i^\dagger, \quad \text{where} \tag{3.91}$$

$$f(E) = \frac{1}{1 + e^{\beta(E-\mu)}}. \tag{3.92}$$

Having calculated $D^>$, $D^<$, $G^>$ and $G^<$ using these expressions, the time ordered and anti-time ordered Green's functions are easily obtained from (3.79), (3.80), (3.83) and (3.84).

### 3.3.2 Effective action

In (3.56) and (3.57), the matrices $T_p$ and $B_p$ describe the coupling of the lead fields $\boldsymbol{\phi}_p$ and $\boldsymbol{\psi}_p$ to the system fields $\boldsymbol{\phi}$ and $\boldsymbol{\psi}$. Thus, examining (3.48)-(3.54) we see that $T_p$ and $B_p$ must satisfy

$$\boldsymbol{\phi}_p^\dagger B_p \boldsymbol{\phi} = -\sum_{n=0}^{N-1} \left\{ \boldsymbol{\phi}_{pn+1}^{+\dagger} b_p \boldsymbol{\phi}_n^+ - \boldsymbol{\phi}_{pn}^{-\dagger} b_p \boldsymbol{\phi}_{n+1}^- \right\} \Delta t, \quad \text{and} \tag{3.93}$$

$$\bar{\boldsymbol{\psi}}_p T_p \boldsymbol{\psi} = -\sum_{n=0}^{N-1} \left\{ \bar{\boldsymbol{\psi}}_{pn+1}^{+} t_p \boldsymbol{\psi}_n^+ - \bar{\boldsymbol{\psi}}_{pn}^{-} t_p \boldsymbol{\psi}_{n+1}^- \right\} \Delta t, \tag{3.94}$$

for any appropriately sized vectors $\boldsymbol{\phi}_p^\dagger$, $\boldsymbol{\phi}$, $\bar{\boldsymbol{\psi}}_p$ and $\boldsymbol{\psi}$, with $\boldsymbol{\phi} = [\boldsymbol{\phi}_1^+ \cdots \boldsymbol{\phi}_N^+, \boldsymbol{\phi}_1^- \cdots \boldsymbol{\phi}_N^-]$ and so on. These expressions, together with the block decompositions (3.67) and (3.68), can be used to rewrite (3.63) and (3.64) as

$$\begin{aligned} S_p^{B,\text{eff}} = -\sum_{n=0}^{N-1}\sum_{m=0}^{N-1} \Big\{ &\boldsymbol{\phi}_n^{+\dagger} b_p^\dagger D_{pn+1,m+1}^t b_p \boldsymbol{\phi}_m^+ - \boldsymbol{\phi}_n^{+\dagger} b_p^\dagger D_{pn+1,m}^< b_p \boldsymbol{\phi}_{m+1}^- \\ &- \boldsymbol{\phi}_{n+1}^{-\dagger} b_p^\dagger D_{pn,m+1}^> b_p \boldsymbol{\phi}_m^+ + \boldsymbol{\phi}_{n+1}^{-\dagger} b_p^\dagger D_{pn,m}^{\bar{t}} b_p \boldsymbol{\phi}_{m+1}^- \Big\} \Delta t^2 \end{aligned} \tag{3.95}$$

and,

$$\begin{aligned} S_p^{F,\text{eff}} = -\sum_{n=0}^{N-1}\sum_{m=0}^{N-1} \Big\{ &\bar{\boldsymbol{\psi}}_n^{+} t_p^\dagger G_{pn+1,m+1}^t t_p \boldsymbol{\psi}_m^+ - \bar{\boldsymbol{\psi}}_n^{+} t_p^\dagger G_{pn+1,m}^< t_p \boldsymbol{\psi}_{m+1}^- \\ &- \bar{\boldsymbol{\psi}}_{n+1}^{-} t_p^\dagger G_{pn,m+1}^> t_p \boldsymbol{\psi}_m^+ + \bar{\boldsymbol{\psi}}_{n+1}^{-} t_p^\dagger G_{pn,m}^{\bar{t}} t_p \boldsymbol{\psi}_{m+1}^- \Big\} \Delta t^2. \end{aligned} \tag{3.96}$$

We first consider the terms $\bar{\psi}_n^+ t_p^\dagger G^<_{pn+1,m} t_p \psi^-_{m+1}$ from (3.96). By (3.90) we have

$$\begin{aligned}\hbar i t_p^\dagger G^<_{pnm} t_p = \hbar i t_p^\dagger G^<_p(t,t') t_p &= -\sum_i f_p(E_i^F) e^{iE_i^F(t'-t)/\hbar} t_p^\dagger v_i v_i^\dagger t_p \\ &= -\int \mathrm{d}E f_p(E) e^{iE(t'-t)/\hbar} \sum_i t_p^\dagger v_i v_i^\dagger t_p \, \delta(E - E_i^F).\end{aligned} \tag{3.97}$$

Since the leads are supposed to be macroscopic objects, we now take the limit where the lead size $L_p \to \infty$. The distance between the eigenvalues $E_i^F$ will then approach zero, so that the delta functions in the expression will be distributed with infinite density. In the limit we then obtain a continuous function which we denote

$$\Gamma_p^F(E) = 2\pi \lim_{L_p \to \infty} \sum_i t_p^\dagger v_i v_i^\dagger t_p \, \delta(E - E_i^F). \tag{3.98}$$

Thus, in the limit $L_p \to \infty$ we can write (3.97) as

$$t_p^\dagger G^<_{pnm} t_p = \frac{i}{h} \int \mathrm{d}E f_p(E) \Gamma_p^F(E) e^{iE(m-n)\Delta t/\hbar}. \tag{3.99}$$

Performing the sum over $n$ and $m$, we find

$$\sum_{n=0}^{N-1} \sum_{m=0}^{N-1} \bar{\psi}_n^+ t_p^\dagger G^<_{pn+1,m} t_p \psi^-_{m+1} = \frac{i}{h\Delta t^2} \int \mathrm{d}E f_p(E) \bar{\psi}^+(E) \Gamma_p^F(E) \psi^-(E), \tag{3.100}$$

where we have defined

$$\psi^+(E) = \sum_{n=0}^{N-1} \Delta t \psi_n^+ e^{iE(n+1)\Delta t/\hbar}, \quad \text{and} \tag{3.101}$$

$$\psi^-(E) = \sum_{n=0}^{N-1} \Delta t \psi^-_{n+1} e^{iEn\Delta t/\hbar}. \tag{3.102}$$

In a similar fashion, using (3.91), we obtain

$$t_p^\dagger G^>_{pnm} t_p = \frac{i}{h} \int \mathrm{d}E [f_p(E) - 1] \Gamma_p^F(E) e^{iE(m-n)\Delta t/\hbar}, \tag{3.103}$$

and

$$\sum_{n=0}^{N-1} \sum_{m=0}^{N-1} \bar{\psi}^-_{n+1} t_p^\dagger G^>_{pn,m+1} t_p \psi^+_m = \frac{i}{h\Delta t^2} \int \mathrm{d}E \, [f_p(E) - 1] \, \bar{\psi}^-(E) \Gamma_p^F(E) \psi^+(E). \tag{3.104}$$

For the summation of the terms involving $G_p^t$ and $G_p^{\bar{t}}$, it is convenient to make some additional definitions. We define the Fourier transform

$$\Gamma_p^F(t) = \int \mathrm{d}E \, \Gamma_p^F(E) e^{-iEt/\hbar}, \tag{3.105}$$

and the two important functions

$$\Sigma_p^r(t) = -i\tilde{\theta}(t)\Gamma_p^F(t) \quad \text{and,} \tag{3.106}$$

$$\Sigma_p^a(t) = i\tilde{\theta}(-t)\Gamma_p^F(t), \tag{3.107}$$

where the step function $\tilde{\theta}(t)$ differs from the previously defined function $\theta(t)$ only in that $\tilde{\theta}(0) = \frac{1}{2}$. We also define the inverse Fourier transforms

$$\Sigma^r(E) = \frac{1}{2\pi}\int \mathrm{d}E\, \Sigma^r(t)e^{iEt/\hbar} \quad \text{and} \tag{3.108}$$

$$\Sigma^a(E) = \frac{1}{2\pi}\int \mathrm{d}E\, \Sigma^a(t)e^{iEt/\hbar}. \tag{3.109}$$

One can show that in fact

$$\Sigma_p^r(E) = \lim_{\eta\to 0^+}\frac{1}{2\pi}\int \mathrm{d}E' \frac{\Gamma_p^F(E')}{E - E' + i\eta}, \quad \text{while} \tag{3.110}$$

$$\Sigma_p^a(E) = \Sigma_p^r(E)^\dagger = \lim_{\eta\to 0^+}\frac{1}{2\pi}\int \mathrm{d}E' \frac{\Gamma_p^F(E')}{E - E' - i\eta}. \tag{3.111}$$

Now, using (3.83), (3.99), (3.103) and (3.106), we find

$$\begin{aligned} t_p^\dagger G_{pnm}^t t_p &= t_p^\dagger G_{pnm}^< t_p + t_p^\dagger \left(G_{pnm}^> - G_{pnm}^<\right) t_p \theta(n-m) \\ &= \frac{i}{h}\int \mathrm{d}E f_p(E)\Gamma_p^F(E)e^{iE(m-n)\Delta t/\hbar} + \frac{1}{h}\Sigma_p^r(t-t') - \frac{i}{2h}\delta_{nm}t_p^\dagger t_p. \end{aligned} \tag{3.112}$$

Since the last term in this expression is proportional to $\delta_{nm}$, it will upon insertion in (3.96) result in an expression of order $\Delta t^2 \cdot N \sim 1/N$. Thus, since the approximation (3.9) already includes an error of order $1/N$, we are free to remove this term without changing the order of the approximation. Then making use of (3.108), (3.101) and (3.102), we obtain

$$\begin{aligned} &\sum_{n=0}^{N-1}\sum_{m=0}^{N-1} \bar{\boldsymbol{\psi}}_n^+ t_p^\dagger G_{pn+1,m+1}^t t_p \boldsymbol{\psi}_m^+ \\ &= \frac{1}{h\Delta t^2}\int \mathrm{d}E\, \bar{\boldsymbol{\psi}}^+(E)\left[if_p(E)\Gamma_p^F(E) + \Sigma_p^r(E)\right]\boldsymbol{\psi}^+(E). \end{aligned} \tag{3.113}$$

In a completely analogous manner, starting from (3.84), we find

$$\begin{aligned} &\sum_{n=0}^{N-1}\sum_{m=0}^{N-1} \bar{\boldsymbol{\psi}}_{n+1}^- t_p^\dagger G_{pnm}^{\bar{t}} t_p \boldsymbol{\psi}_{m+1}^- \\ &= \frac{1}{h\Delta t^2}\int \mathrm{d}E\, \bar{\boldsymbol{\psi}}^-(E)\left[if_p(E)\Gamma_p^F(E) - \Sigma_p^a(E)\right]\boldsymbol{\psi}^-(E). \end{aligned} \tag{3.114}$$

Then inserting (3.100), (3.104), (3.113) and (3.114) in (3.96), we find that the fermionic component of the effective action can be expressed as

$$\begin{aligned} \frac{i}{\hbar}S_p^{F,\mathrm{eff}} &= \frac{1}{h\hbar}\int \mathrm{d}E\,\Big\{f_p(E)\,[\bar{\boldsymbol{\psi}}^+(E) - \bar{\boldsymbol{\psi}}^-(E)]\Gamma_p^F(E)[\boldsymbol{\psi}^+(E) - \boldsymbol{\psi}^-(E)] \\ &+ \bar{\boldsymbol{\psi}}^-(E)\Gamma_p^F(E)\boldsymbol{\psi}^+(E) - i\bar{\boldsymbol{\psi}}^+(E)\Sigma_p^r(E)\boldsymbol{\psi}^+(E) + i\bar{\boldsymbol{\psi}}^-(E)\Sigma_p^a(E)\boldsymbol{\psi}^-(E)\Big\}. \end{aligned} \tag{3.115}$$

Repeating the procedure above with the corresponding bosonic expressions, we also find that the bosonic component of the effective action (3.95) can be expressed as

$$\begin{aligned} \frac{i}{\hbar}S_p^{B,\text{eff}} = \frac{1}{h\hbar}\int \mathrm{d}E \Big\{ & - n_p(E)\,[\boldsymbol{\phi}^+(E) - \boldsymbol{\phi}^-(E)]^\dagger \Gamma_p^B(E)[\boldsymbol{\phi}^+(E) - \boldsymbol{\phi}^-(E)] \\ & + \boldsymbol{\phi}^-(E)^\dagger \Gamma_p^B(E)\boldsymbol{\phi}^+(E) - i\boldsymbol{\phi}^+(E)^\dagger \Theta_p^r(E)\boldsymbol{\phi}^+(E) \\ & + i\boldsymbol{\phi}^-(E)^\dagger \Theta_p^a(E)\boldsymbol{\phi}^-(E)\Big\}, \end{aligned} \tag{3.116}$$

where we have defined

$$\Gamma_p^B(E) = \lim_{L_p\to\infty} 2\pi \sum_i b_p^\dagger u_i u_i^\dagger b_p\, \delta(E - E_i^B), \tag{3.117}$$

$$\Gamma_p^B(t) = \int \mathrm{d}E\, \Gamma_p^B(E) e^{-iEt/\hbar}, \tag{3.118}$$

$$\Theta_p^r(t) = -i\tilde{\theta}(t)\Gamma_p^B(t), \tag{3.119}$$

$$\Theta_p^a(t) = i\tilde{\theta}(-t)\Gamma_p^B(t), \tag{3.120}$$

$$\Theta^r(E) = \frac{1}{2\pi}\int \mathrm{d}E\, \Theta^r(t) e^{iEt/\hbar}, \tag{3.121}$$

$$\Theta^a(E) = \frac{1}{2\pi}\int \mathrm{d}E\, \Theta^a(t) e^{iEt/\hbar}, \tag{3.122}$$

$$\boldsymbol{\phi}^+(E) = \sum_{n=0}^{N-1} \Delta t \boldsymbol{\phi}_n^+ e^{iE(n+1)\Delta t/\hbar}, \quad \text{and} \tag{3.123}$$

$$\boldsymbol{\phi}^-(E) = \sum_{n=0}^{N-1} \Delta t \boldsymbol{\phi}_{n+1}^- e^{iEn\Delta t/\hbar}. \tag{3.124}$$

## 3.4 Transport expectation values

### 3.4.1 Operator expressions

Let $\Phi_p$ be the flux of fermions exiting the system $S$ at lead $p$, and let $N_p$ be the number of fermions in $p$. Then since $N_p$ can change only by fermions entering or exiting $S$, we must have $\dot{N}_p = \Phi_p$. Since the lead $p$ is assumed macroscopic, we can identify $N_p = \langle N_p \rangle$, and using (3.5) we get

$$\Phi_p = \frac{\mathrm{d}N_p}{\mathrm{d}t} = \frac{\mathrm{d}}{\mathrm{d}t}\mathrm{Tr}\,\rho_0 \hat{N}_p^F(t) \tag{3.125}$$

Now using (3.2), we have for any observable $A$

$$\frac{\mathrm{d}}{\mathrm{d}t}\hat{A}(t) = \frac{i}{\hbar}U^\dagger(t-t_0)[\hat{H},\hat{A}]U(t-t_0). \tag{3.126}$$

Applying this to (3.125) we get

$$\Phi_p(t) = i\mathrm{Tr}\,\rho_0 U^\dagger(t-t_0)[\hat{H},\hat{N}_p^F]U(t-t_0)/\hbar. \tag{3.127}$$

Clearly $\hat{N}_p^F$ commutes with the bosonic degrees of freedom, and since $\hat{N}_p^F$ contains only fields from lead $p$, it also commutes with $\hat{H}_S$, as well as the terms from the other leads. Thus, by (3.39) we have

$$[\hat{H}, \hat{N}_p^F] = [\hat{\boldsymbol{\psi}}_p^\dagger H_p^F \hat{\boldsymbol{\psi}}_p, \hat{N}_p^F] + [\hat{\boldsymbol{\psi}}_p^\dagger t_p \hat{\boldsymbol{\psi}}, \hat{N}_p^F] + [\hat{\boldsymbol{\psi}}^\dagger t_p^\dagger \hat{\boldsymbol{\psi}}_p, \hat{N}_p^F]. \tag{3.128}$$

Making use of some operator algebra, starting with the fundamental commutation relation $\{\psi_i, \psi_j^\dagger\} = \psi_i \psi_j^\dagger + \psi_j^\dagger \psi_i = \delta_{ij}$ and the definiton $\hat{N}_p^F = \hat{\boldsymbol{\psi}}_p^\dagger \hat{\boldsymbol{\psi}}_p$, we find that

$$[\hat{\boldsymbol{\psi}}_p^\dagger H_p^F \hat{\boldsymbol{\psi}}_p, \hat{N}_p^F] = 0, \qquad \text{while,} \tag{3.129}$$

$$[\hat{\boldsymbol{\psi}}_p^\dagger t_p \hat{\boldsymbol{\psi}}, \hat{N}_p^F] = -\hat{\boldsymbol{\psi}}_p^\dagger t_p \hat{\boldsymbol{\psi}}. \tag{3.130}$$

Taking the adjoint of the latter, we also find $[\hat{\boldsymbol{\psi}}^\dagger t_p^\dagger \hat{\boldsymbol{\psi}}_p, \hat{N}_p^F] = -[\hat{N}_p^F, \hat{\boldsymbol{\psi}}^\dagger t_p^\dagger \hat{\boldsymbol{\psi}}_p] = \hat{\boldsymbol{\psi}}^\dagger t_p^\dagger \hat{\boldsymbol{\psi}}_p$. Thus, (3.128) becomes

$$[\hat{H}, \hat{N}_p^F] = \hat{\boldsymbol{\psi}}^\dagger t_p^\dagger \hat{\boldsymbol{\psi}}_p - \hat{\boldsymbol{\psi}}_p^\dagger t_p \hat{\boldsymbol{\psi}}, \tag{3.131}$$

so that (3.127) becomes

$$\Phi_p(t) = -\frac{i}{\hbar} \mathrm{Tr}\, \rho_0 \left( \hat{\boldsymbol{\psi}}_p^\dagger(t) t_p \hat{\boldsymbol{\psi}}(t) - \hat{\boldsymbol{\psi}}^\dagger(t) t_p^\dagger \hat{\boldsymbol{\psi}}_p(t) \right) = \left\langle \hat{\Phi}_p(t) \right\rangle, \tag{3.132}$$

where clearly $\hat{\Phi}_p(t) = -i(\hat{\boldsymbol{\psi}}_p^\dagger(t) t_p \hat{\boldsymbol{\psi}}(t) - \hat{\boldsymbol{\psi}}^\dagger(t) t_p^\dagger \hat{\boldsymbol{\psi}}_p(t))/\hbar$.

Now, let $E_p^F = \hat{\boldsymbol{\psi}}_p^\dagger H_p^F \hat{\boldsymbol{\psi}}_p$ be the total energy associated with fermions in lead $p$, and let $\Phi_p^{EF}$ be the corresponding energy flux entering $p$ from $S$. Then by an argument similar to the one preceding (3.125), we have

$$\Phi_p^{EF} = \frac{\mathrm{d}E_p^F}{\mathrm{d}t} = \frac{i}{\hbar} \mathrm{Tr}\, \rho_0 U^\dagger(t - t_0) [\hat{H}, \hat{\boldsymbol{\psi}}_p^\dagger H_p^F \hat{\boldsymbol{\psi}}_p] U(t - t_0). \tag{3.133}$$

Again, the term $\hat{\boldsymbol{\psi}}_p^\dagger H_p^F \hat{\boldsymbol{\psi}}_p$ clearly commutes both with bosonic degrees of freedom, as well as with terms that are exclusive to $S$ or to the other leads. Since it also obviously commutes with itself, we obtain by again making use of (3.39) and some operator algebra,

$$[\hat{H}, \hat{\boldsymbol{\psi}}_p^\dagger H_p^F \hat{\boldsymbol{\psi}}_p] = [\hat{\boldsymbol{\psi}}_p^\dagger t_p \hat{\boldsymbol{\psi}} + \hat{\boldsymbol{\psi}}^\dagger t_p^\dagger \hat{\boldsymbol{\psi}}_p, \hat{\boldsymbol{\psi}}_p^\dagger H_p^F \hat{\boldsymbol{\psi}}_p] = -\hat{\boldsymbol{\psi}}_p^\dagger H_p^F t_p \hat{\boldsymbol{\psi}} + \hat{\boldsymbol{\psi}}^\dagger t_p^\dagger H_p^F \hat{\boldsymbol{\psi}}_p. \tag{3.134}$$

Thus, (3.133) becomes

$$\Phi_p^{EF} = -\frac{i}{\hbar} \mathrm{Tr}\, \rho_0 \left( \hat{\boldsymbol{\psi}}_p^\dagger(t) H_p^F t_p \hat{\boldsymbol{\psi}}(t) - \hat{\boldsymbol{\psi}}^\dagger(t) t_p^\dagger H_p^F \hat{\boldsymbol{\psi}}_p(t) \right) = \left\langle \hat{\Phi}_p^{EF}(t) \right\rangle, \tag{3.135}$$

with $\hat{\Phi}_p^{EF}(t) = -i(\hat{\boldsymbol{\psi}}_p^\dagger(t) H_p^F t_p \hat{\boldsymbol{\psi}}(t) - \hat{\boldsymbol{\psi}}^\dagger(t) t_p^\dagger H_p^F \hat{\boldsymbol{\psi}}_p(t))/\hbar$.

Letting $E_p^B = \hat{\boldsymbol{\phi}}_p^\dagger H_p^B \hat{\boldsymbol{\phi}}_p$ be the energy associated with bosons in $p$, and $\Phi_p^{EB}$ the corresponding energy flux, it is clear that $\Phi_p^{EB}$ can be obtained by simply repeating the steps above. The result is

$$\Phi_p^{EB} = \frac{\mathrm{d}E_p^B}{\mathrm{d}t} = \frac{i}{\hbar}\mathrm{Tr}\,\rho_0 U^\dagger(t-t_0)[\hat{H}, \hat{\boldsymbol{\phi}}_p^\dagger H_p^B \hat{\boldsymbol{\phi}}_p]U(t-t_0) \tag{3.136}$$
$$= -\frac{i}{\hbar}\mathrm{Tr}\,\rho_0 \left(\hat{\boldsymbol{\phi}}_p^\dagger(t) H_p^B b_p \hat{\boldsymbol{\phi}}(t) - \hat{\boldsymbol{\phi}}^\dagger(t) b_p^\dagger H_p^B \hat{\boldsymbol{\phi}}_p(t)\right) = \left\langle \hat{\Phi}_p^{EB}(t)\right\rangle,$$

with $\hat{\Phi}_p^{EB}(t) = -i(\hat{\boldsymbol{\phi}}_p^\dagger(t) H_p^B b_p \hat{\boldsymbol{\phi}}(t) - \hat{\boldsymbol{\phi}}^\dagger(t) b_p^\dagger H_p^B \hat{\boldsymbol{\phi}}_p(t))/\hbar$. Finally, letting $E_p = E_p^F + E_p^B$ be the total energy of lead $p$, and $\Phi_p^E$ the total energy flux entering $p$ from $S$, we have

$$\Phi_p^E = \frac{\mathrm{d}E_p}{\mathrm{d}t} = \Phi_p^{EF} + \Phi_p^{EB} = \left\langle \hat{\Phi}_p^E(t)\right\rangle, \tag{3.137}$$

with $\hat{\Phi}_p^E(t) = \hat{\Phi}_p^{EF}(t) + \hat{\Phi}_p^{EB}(t)$.

### 3.4.2 Field integral expressions

Using the relationship (3.33) between operator and field integral expectation values, we can reexpress (3.132) as

$$\Phi_p(t) = -\frac{i}{\hbar}\left\langle \bar{\boldsymbol{\psi}}_p^-(t) t_p \boldsymbol{\psi}^+(t) - \bar{\boldsymbol{\psi}}^-(t) t_p^\dagger \boldsymbol{\psi}_p^+(t)\right\rangle. \tag{3.138}$$

In order to express this in terms of fields confined within $S$ alone, we will make use of a commonly applied trick where the expectation value is reexpressed as a derivative. Thus, we define the modified action

$$S(x) = S + x(\bar{\boldsymbol{\psi}}^-(t) t_p^\dagger \boldsymbol{\psi}_p^+(t) - \bar{\boldsymbol{\psi}}_p^-(t) t_p \boldsymbol{\psi}^+(t)). \tag{3.139}$$

Making use of (3.32), it is easily verified that we then have

$$\Phi_p(t) = \frac{\mathrm{d}}{\mathrm{d}x}\int D(+,-)\prod_p \int D(\phi_p^+, \phi_p^-)\int D(\psi_p^+, \psi_p^-) e^{iS(x)/\hbar}, \tag{3.140}$$

where we are assuming evaluation at $x = 0$. The modification (3.139) changes the matrix $T$ of (3.94) and its adjoint $T^\dagger$ respectively to $T_p(x)$ and $\tilde{T}_p^\dagger(x)$, where

$$\bar{\boldsymbol{\psi}}_p T_p(x)\boldsymbol{\psi} = \bar{\boldsymbol{\psi}}_p T_p \boldsymbol{\psi} - x\bar{\boldsymbol{\psi}}_p^-(t) t_p \boldsymbol{\psi}^+(t), \quad \text{and} \tag{3.141}$$
$$\bar{\boldsymbol{\psi}}\tilde{T}_p^\dagger(x)\boldsymbol{\psi}_p = \bar{\boldsymbol{\psi}} T_p^\dagger \boldsymbol{\psi}_p + x\bar{\boldsymbol{\psi}}^-(t) t_p^\dagger \boldsymbol{\psi}_p^+(t). \tag{3.142}$$

Repeating the derivation of (3.96) with $T_p(x)$ and $\tilde{T}_p^\dagger(x)$ substituted for $T$ and $T^\dagger$, we obtain the modified effective action

$$S^{\mathrm{eff}}(x) = S^{\mathrm{eff}} + x\sum_{n=0}^{N-1}\{\bar{\boldsymbol{\psi}}_m^- t_p^\dagger G_{pm,n+1}^t t_p \boldsymbol{\psi}_n^+ - \bar{\boldsymbol{\psi}}_m^- t_p^\dagger G_{pmn}^< t_p \boldsymbol{\psi}_{n+1}^- - \tag{3.143}$$
$$\bar{\boldsymbol{\psi}}_n^+ t_p^\dagger G_{pn+1,m}^< t_p \boldsymbol{\psi}_m^+ + \bar{\boldsymbol{\psi}}_{n+1}^- t_p^\dagger G_{pnm}^{\bar{t}} t_p \boldsymbol{\psi}_m^+\}\Delta t + O(x^2),$$

where $m\Delta t = t$. Thus, by (3.140) and (3.61) we have

$$\begin{aligned}\Phi_p(t) &= \frac{\mathrm{d}}{\mathrm{d}x}\int D(+,-)e^{iS^{\mathrm{eff}}(x)/\hbar} \\ &= \frac{i}{\hbar}\sum_{n=0}^{N-1}\Big\langle \bar{\boldsymbol{\psi}}_m^- t_p^\dagger G_{pm,n+1}^t t_p \boldsymbol{\psi}_n^+ - \bar{\boldsymbol{\psi}}_m^- t_p^\dagger G_{pmn}^< t_p \boldsymbol{\psi}_{n+1}^- \\ &\qquad - \bar{\boldsymbol{\psi}}_n^+ t_p^\dagger G_{pn+1,m}^< t_p \boldsymbol{\psi}_m^+ + \bar{\boldsymbol{\psi}}_{n+1}^- t_p^\dagger G_{pnm}^{\bar{t}} t_p \boldsymbol{\psi}_m^+ \Big\rangle \Delta t.\end{aligned} \tag{3.144}$$

Now we make use of the assumption that the system will approach steady state conditions over a time scale which is short compared to $t_f - t_0$. $\Phi_p(t)$ will then essentially be independent of $t$ for all times except very close to $t_0$. Accordingly, we have

$$\begin{aligned}\Phi_p \approx \frac{1}{N}\sum_{m=1}^{N}\Phi_p(t) = \frac{i}{\hbar}\frac{1}{N}\sum_{m=1}^{N}\sum_{n=0}^{N-1}\Big\langle \bar{\boldsymbol{\psi}}_m^- t_p^\dagger G_{pm,n+1}^t t_p \boldsymbol{\psi}_n^+ - \bar{\boldsymbol{\psi}}_m^- t_p^\dagger G_{pmn}^< t_p \boldsymbol{\psi}_{n+1}^- \\ - \bar{\boldsymbol{\psi}}_n^+ t_p^\dagger G_{pn+1,m}^< t_p \boldsymbol{\psi}_m^+ + \bar{\boldsymbol{\psi}}_{n+1}^- t_p^\dagger G_{pnm}^{\bar{t}} t_p \boldsymbol{\psi}_m^+ \Big\rangle \Delta t.\end{aligned} \tag{3.145}$$

Adding (3.83) and (3.84) we find the relation

$$G_{mn}^t + G_{mn}^{\bar{t}} = G_{mn}^> + G_{mn}^< + \delta_{mn}(G_{mn}^> - G_{mn}^<). \tag{3.146}$$

Combining this with the definition $i\hbar G = \langle \boldsymbol{\psi}\bar{\boldsymbol{\psi}}\rangle$ we also obtain

$$\begin{aligned}&\Big\langle \bar{\boldsymbol{\psi}}_m^- t_p^\dagger G_{pmn}^< t_p \boldsymbol{\psi}_{n+1}^- + \bar{\boldsymbol{\psi}}_m^+ t_p^\dagger G_{pmn}^< t_p \boldsymbol{\psi}_{n+1}^+ \Big\rangle \\ &= \Big\langle \bar{\boldsymbol{\psi}}_m^- t_p^\dagger G_{pmn}^< t_p \boldsymbol{\psi}_{n+1}^+ + \bar{\boldsymbol{\psi}}_m^+ t_p^\dagger G_{pmn}^< t_p \boldsymbol{\psi}_{n+1}^- \\ &\qquad - \delta_{nm}(\bar{\boldsymbol{\psi}}_m^+ t_p^\dagger G_{pmn}^< t_p \boldsymbol{\psi}_{n+1}^- - \bar{\boldsymbol{\psi}}_m^- t_p^\dagger G_{pmn}^< t_p \boldsymbol{\psi}_{n+1}^+)\Big\rangle\end{aligned} \tag{3.147}$$

Making use of (3.146) applied to the lead Green's functions $G_p$, (3.147), (3.99) and (3.103), we can after considerable algebra rewrite (3.145) as

$$\begin{aligned}\Phi_p = &-\frac{1}{h\hbar N}\sum_{m=0}^{N-1}\sum_{n=0}^{N-1}\int \mathrm{d}E\,\Big\langle \bar{\boldsymbol{\psi}}_{m+1}^-[2f_p(E)-1]\Gamma_p^F(E)e^{iE(n-m)\Delta t/\hbar}\boldsymbol{\psi}_n^+ \\ &- \bar{\boldsymbol{\psi}}_{m+1}^- f_p(E)\Gamma_p^F(E)e^{iE(n-m-2)\Delta t/\hbar}\boldsymbol{\psi}_n^+ \\ &- \bar{\boldsymbol{\psi}}_m^+ f_p(E)\Gamma_p^F(E)e^{iE(n-m)\Delta t/\hbar}\boldsymbol{\psi}_{n+1}^-\Big\rangle\Delta t - \frac{1}{h\hbar N}\sum_{n=0}^{N-1}X_n\Delta t,\end{aligned} \tag{3.148}$$

where $X_n$ is an expression of order zero in $N$. Thus, the entire last term is of order $1/N \cdot N \cdot \Delta t \sim 1/N$, and can accordingly be dropped.

Further, as long as the energy range where $\Gamma_p^F(E)$ is nonzero is bounded, replacing the exponentials $e^{iE(n-m)\Delta t/\hbar}$ and $e^{iE(n-m-2)\Delta t/\hbar}$ with unity will also only introduce an error of order $1/N$. Doing this and making use of (3.101) and (3.102), we obtain after some rearrangement the symmetrical expression

$$\Phi_p = \int \frac{2\pi \mathrm{d}E}{h^2(t_f - t_0)} \Big\langle (1 - f_p(E))\, \bar{\boldsymbol{\psi}}^-(E)\Gamma_p^F(E)\boldsymbol{\psi}^+(E) + f_p(E)\bar{\boldsymbol{\psi}}^+(E)\Gamma_p^F(E)\boldsymbol{\psi}^-(E)\Big\rangle. \tag{3.149}$$

Field integral expressions for energy flux can be found in the same manner. Comparing (3.135) to (3.132), we see that the fermionic energy flux is obtained by simply replacing $t_p$ and $t_p^\dagger$ in (3.139) respectively with $H_p^F t_p$ and $t_p^\dagger H_p^F$. Upon making use of (3.90), (3.91) and (3.98) to rewrite the effective action as an integral over energy, the additional factor of $H_p^F$ simply becomes a factor of $E$. Thus, the fermionic energy flux can be obtained from (3.149) simply by including this additional factor of $E$. Accordingly, we have

$$\Phi_p^{EF} = \int \frac{2\pi \mathrm{d}E\, E}{h^2(t_f - t_0)} \Big\langle (1 - f_p(E))\, \bar{\boldsymbol{\psi}}^-(E)\Gamma_p^F(E)\boldsymbol{\psi}^+(E) + f_p(E)\bar{\boldsymbol{\psi}}^+(E)\Gamma_p^F(E)\boldsymbol{\psi}^-(E)\Big\rangle. \tag{3.150}$$

The bosonic energy flux is obtained in a similar manor. Repeating the steps above with the corresponding bosonic expressions, we obtain

$$\Phi_p^{EB} = \int \frac{2\pi \mathrm{d}E\, E}{h^2(t_f - t_0)} \Big\langle (n_p(E) + 1)\, \boldsymbol{\phi}^-(E)^\dagger\Gamma_p^B(E)\boldsymbol{\phi}^+(E) - n_p(E)\boldsymbol{\phi}^+(E)^\dagger\Gamma_p^B(E)\boldsymbol{\phi}^-(E)\Big\rangle. \tag{3.151}$$

### 3.4.3 Current and heat flux

In the most typical solid state applications, the model can be expressed in terms of two different types of particles: electrons which are fermions and have charge $-e$, with $e$ the elementary charge, and phonons which are bosons and have charge zero. Thus, the electrical currents at the leads are given simply by $I_p = -e\Phi_p$, or by (3.149),

$$I_p = -e \int \frac{2\pi \mathrm{d}E}{h^2(t_f - t_0)} \Big\langle (1 - f_p(E))\, \bar{\boldsymbol{\psi}}^-(E)\Gamma_p^F(E)\boldsymbol{\psi}^+(E) + f_p(E)\bar{\boldsymbol{\psi}}^+(E)\Gamma_p^F(E)\boldsymbol{\psi}^-(E)\Big\rangle. \tag{3.152}$$

To find an expression for the heat flux $q_p$ through lead $p$, we make use of the following considerations: Since we are assuming stationary conditions except very close to $t_0$, there will at $t = t_f$ have passed a total energy of $\Delta E_p = \Phi_p^E(t_f - t_0)$ into lead $p$. This energy can be decomposed as $\Delta E_p = Q + W$, where $Q$ is the transferred heat, and $W$ is electrical work having been done on that lead. If we imagine that the contact between the leads and $S$ is removed at $t = t_f$, then the lead will reequilibrate after this time. Since the lead size $L_p \to \infty$, the particle and energy content of the lead has only changed by an arbitrarily small fraction, and accordingly it must reequilibrate to the same state it was in before $t = t_0$. Thus, the chemical potential associated with the lead is still $\mu_p$, so that after equilibration the change in electrical energy can be expressed as $\Delta E_{\mathrm{el}} = W = \mu_p \Delta N_p^F = \mu_p \Phi_p(t_0 - t_f)$.

Thus, we have $\Delta E_p = \Phi_p^E(t_f - t_0) = Q + \mu_p \Phi_p(t_0 - t_f)$, and accordingly the stationary heat flux $q_p = Q/(t_f - t_0)$ can be expressed as $q_p = \Phi_p^E - \mu_p \Phi_p$. Using (3.137), (3.150), (3.151) and (3.149) we thus obtain

$$
\begin{aligned}
q_p = \int \frac{2\pi \mathrm{d}E\, E}{h^2(t_f - t_0)} \Big\langle & (n_p(E) + 1)\, \boldsymbol{\phi}^-(E)^\dagger \Gamma_p^B(E) \boldsymbol{\phi}^+(E) \\
& - n_p(E) \boldsymbol{\phi}^+(E)^\dagger \Gamma_p^B(E) \boldsymbol{\phi}^-(E) \Big\rangle \\
+ \int \frac{\mathrm{d}E\,(E - \mu)}{h\hbar(t_f - t_0)} \Big\langle & (1 - f_p(E))\, \bar{\boldsymbol{\psi}}^-(E) \Gamma_p^F(E) \boldsymbol{\psi}^+(E) \\
& + f_p(E) \bar{\boldsymbol{\psi}}^+(E) \Gamma_p^F(E) \boldsymbol{\psi}^-(E) \Big\rangle .
\end{aligned} \tag{3.153}
$$

Chapter 4

# The Linear limit and Kubo relations

(3.152) and (3.153) of the previous chapter gives expression for the currents and heat currents of a transport system under very general conditions. These currents are driven by differences between the temperatures and chemical potentials of the leads. In this chapter we will be concerned with the limit where these differences are small, so that the currents can be expressed in terms of first order approximations. In Section 4.1 we find such expressions by directly differentiating the general expressions (3.152) and (3.153). Then, in Section 4.2 we will discuss a particularly elegant set of expressions, known as the Kubo relations, which will be derived using the operator formalism. Translation into field integral expressions is considered in Section 4.3.

## 4.1 Direct linear limit of the field integral

### 4.1.1 Conductance

As discussed in Section 2.1.2, we can in the linear transport regime express the lead currents as

$$I_q = \sum_p G_{qp} V_{pq} = \sum_p G_{qp}(V_p - V_q) = \sum_p \tilde{G}_{qp} \Delta V_p, \tag{4.1}$$

where $\Delta V_p = V_p - V$ with $V$ the equilibrium potential, and where we have defined

$$\tilde{G}_{qp} = G_{qp} - \delta_{qp} \sum_r G_{qr}. \tag{4.2}$$

It is clear from (4.1) that we have

$$\tilde{G}_{qp} = \frac{\partial I_q}{\partial V_p} = -e \frac{\partial I_q}{\partial \mu_p}. \tag{4.3}$$

Thus, making use of (3.152) and (3.61), we find that for $q \neq p$

$$\tilde{G}_{qp} = e^2 \int \frac{2\pi \mathrm{d}E}{h^2(t_f - t_0)} \left\langle X_q(E) \frac{i}{\hbar} \frac{\partial S^{\text{eff}}}{\partial \mu_p} \right\rangle. \tag{4.4}$$

where

$$X_q(E) = (1 - f_q(E))\, \bar{\boldsymbol{\psi}}^-(E) \Gamma_q^F(E) \boldsymbol{\psi}^+(E) + f_q(E) \bar{\boldsymbol{\psi}}^+(E) \Gamma_q^F(E) \boldsymbol{\psi}^-(E). \tag{4.5}$$

Combining (3.62) and (3.115) we see that

$$\frac{i}{\hbar}\frac{\partial S^{\text{eff}}}{\partial \mu_p} = \frac{1}{h\hbar}\int \mathrm{d}E\, Y_p(E) + \frac{i}{\hbar}\frac{\partial S_p^{F,\text{eff}}}{\partial H_p}\frac{\partial H_p}{\partial \mu_p}, \tag{4.6}$$

where

$$Y_p(E) = \mathrm{Th}(E)\,[\bar{\boldsymbol{\psi}}^+(E) - \bar{\boldsymbol{\psi}}^-(E)]\Gamma_p^F(E)[\boldsymbol{\psi}^+(E) - \boldsymbol{\psi}^-(E)], \tag{4.7}$$

with $\mathrm{Th}(E)$ given by (2.18), and where we are allowing for the possibility that the lead Hamiltonian depends on the chemical potential. However, upon inserting (4.6) in (4.4), and making use of (3.61) again, we see that

$$e^2\int \frac{2\pi \mathrm{d}E}{h^2(t_f - t_0)}\left\langle X_q(E)\frac{i}{\hbar}\frac{\partial S_p^{F,\text{eff}}}{\partial H_p}\frac{\partial H_p}{\partial \mu_p}\right\rangle = -e\frac{\partial I_q}{\partial H_p}\frac{\partial H_p}{\partial \mu_p}. \tag{4.8}$$

But $I_q$ is the current in equilibrium, which is always zero. Thus $\partial I_q/\partial H_p = 0$, so that the term above does not contribute to (4.4), which accordingly becomes

$$\tilde{G}_{qp} = e^2\iint \frac{4\pi^2 \mathrm{d}E\,\mathrm{d}E'}{h^4(t_f - t_0)}\big\langle X_q(E)Y_p(E')\big\rangle. \tag{4.9}$$

### 4.1.2 Thermoelectric coefficients

In the presence of a temperature gradient we have by the discussion of Section 2.1.3

$$I_q = \sum_p A_{qp}\Delta T_{pq} = \sum_p A_{qp}(T_p - T_q) = \sum_p \tilde{A}_{qp}\Delta T_p, \tag{4.10}$$

where $\Delta T_p = T_p - T$ with $T$ the equilibrium temperature, and where similarly to (4.2) we have defined

$$\tilde{A}_{qp} = A_{qp} - \delta_{qp}\sum_r A_{qr}. \tag{4.11}$$

Clearly we have

$$\tilde{A}_{qp} = \frac{\partial I_q}{\partial T_p} = -\frac{1}{k_B T^2}\frac{\partial I_q}{\partial \beta_p}. \tag{4.12}$$

Thus, again assuming $q \neq p$ and using (3.152) and (3.61), we get

$$\tilde{A}_{qp} = \frac{e}{k_B T^2}\int \frac{2\pi \mathrm{d}E}{h^2(t_f - t_0)}\left\langle X_q(E)\frac{i}{\hbar}\frac{\partial S^{\text{eff}}}{\partial \beta_p}\right\rangle. \tag{4.13}$$

Using (3.62), (3.115) and (3.116) we find

$$\frac{i}{\hbar}\frac{\partial S^{\text{eff}}}{\partial \beta_p} = \frac{1}{h\hbar\beta}\int \mathrm{d}E\,\big(E\,Y_p^B(E) - (E-\mu)Y_p(E)\big) + \frac{i}{\hbar}\frac{\partial S_p^{\text{eff}}}{\partial H}\frac{\partial H}{\partial \mu_p}, \tag{4.14}$$

where

$$Y_p^B(E) = \mathrm{Tb}(E)[\boldsymbol{\phi}^+(E) - \boldsymbol{\phi}^-(E)]^\dagger \Gamma_p^B(E)[\boldsymbol{\phi}^+(E) - \boldsymbol{\phi}^-(E)], \tag{4.15}$$

with

$$\mathrm{Tb}(E) = -\frac{\partial n}{\partial E}(E) = \frac{\beta}{4\sinh^2 \beta E/2}. \tag{4.16}$$

The last term in (4.14) does not contribute to the transport coefficient for reasons similar to those given below (4.8). Thus, (4.13) becomes

$$\tilde{A}_{qp} = -\frac{e}{T}\iint \frac{4\pi^2 \mathrm{d}E\,\mathrm{d}E'}{h^4(t_f - t_0)} \left\langle X_q(E)\left[(E'-\mu)Y_p(E') - E'\,Y_p^B(E')\right]\right\rangle. \tag{4.17}$$

Similar to (4.1) and (4.10), the heat flux $q_q$ can be expressed in terms of linear transport coefficients

$$\tilde{B}_{qp} = B_{qp} - \delta_{qp}\sum_r B_{qr} = \frac{\partial q_q}{\partial V_p}, \qquad \text{and} \tag{4.18}$$

$$\tilde{C}_{qp} = C_{qp} - \delta_{qp}\sum_r C_{qr} = \frac{\partial q_q}{\partial T_p}, \tag{4.19}$$

where $B_{qp}$ and $C_{qp}$ are introduced in Section 2.1.3. Expressions for these coefficients can be found by repeating the derivations above, but starting with the heat flux expression (3.153) rather than (3.152). In the end we obtain

$$\tilde{B}_{qp} = -e\iint \frac{4\pi^2 \mathrm{d}E\,\mathrm{d}E'}{h^4(t_f - t_0)} \left\langle \left[(E-\mu)X_q(E) + E\,X_q^B(E)\right] Y_p(E')\right\rangle, \quad \text{and} \tag{4.20}$$

$$\begin{aligned}\tilde{C}_{qp} = \frac{1}{T}\iint \frac{4\pi^2 \mathrm{d}E\,\mathrm{d}E'}{h^4(t_f - t_0)} \big\langle &\left[(E-\mu)X_q(E) + E\,X_q^B(E)\right] \\ &\times \left[(E'-\mu)Y_p(E') - E'\,Y_p^B(E')\right]\big\rangle,\end{aligned} \tag{4.21}$$

where

$$X_q^B(E) = (n_q(E)+1)\,\boldsymbol{\phi}^-(E)^\dagger \Gamma_q^B(E)\boldsymbol{\phi}^+(E) - n_q(E)\boldsymbol{\phi}^+(E)^\dagger \Gamma_q^B(E)\boldsymbol{\phi}^-(E). \tag{4.22}$$

## 4.2 Kubo relations

The Kubo relations are a particularly elegant set of expressions for the transport coefficients, which express these in terms of correlation functions between the various currents. The relations were obtained by Kubo in two papers dealing with linear response, where the first[19] dealt with mechanical perturbations, and the second[20] with thermodynamic perturbations.

The treatment of mechanical perturbations, i.e. perturbations consisting of an additional term in the Hamiltonian, is a fairly straight forward procedure, where one simply makes a first order approximation to the equations of motion. However, in the system considered in Section 3.2.1, the currents are not driven by modifications to the Hamiltonian, but by deviation in the state $\rho$ from its equilibrium value. Such thermodynamic perturbations are more difficult to handle in a formal manner, and indeed Kubo based the derivations in his second paper partially on a set of heuristically justified assumptions. An overview of alternative approaches to dealing with thermodynamic perturbations is provided by Zwanzig[31]. Many of the approaches seem to handle the thermodynamic perturbations by replacing it with a mechanical perturbation which is in some sense equivalent. This is also the approach we take below.

### 4.2.1 Linear response theory of Conductance

In the operator representation, the expectation value of the particle flux is given by (3.132). Combining this with (3.42), and defining the current operator $\hat{I}_q(t) = -e\hat{\Phi}_p(t)$, we can express the electrical current at lead $q$ as

$$I_q(t) = \left\langle \hat{I}_q(t) \right\rangle = \operatorname{Tr} \hat{I}_q(t) P_S^0 \prod_r \frac{1}{Z_r} e^{-\beta_r(\hat{H}_r - \mu_r \hat{N}_r^F)}. \tag{4.23}$$

In order to find the conductance coefficients $\tilde{G}_{qp}$, we could apply (4.3) directly to this expression, taking the derivative with respect to the chemical potentials $\mu_r$. However, this approach turns out not to be so fruitful. Instead, we will make use of the fact that the conductance can also be expressed as

$$\tilde{G}_{qp} = \frac{\partial I_q}{\partial \phi_p} = -e\frac{\partial I_q}{\partial \epsilon_p}, \tag{4.24}$$

where $\phi_p$ is the electrostatic potential of lead $p$, and $\epsilon_p = -e\phi_p$. Since the electrostatic potential constitutes a mechanical perturbation, this expression can be handled using standard linear response theory.

Accordingly, we define the perturbed Hamiltonian

$$\hat{H}' = \hat{H} + \sum_p \epsilon_p N_p^F, \tag{4.25}$$

giving rise to a modified propagator $U'(t)$ satisfying $i\hbar\dot{U}' = \hat{H}'U$. Further, we assume that the system is initially in an equilibrium state of the unmodified Hamiltonian $\hat{H}$. Putting this together, we obtain by (4.24)

$$\tilde{G}_{qp} = -e\frac{\partial}{\partial \epsilon_p} \operatorname{Tr} U'(t-t_0)^\dagger \hat{I}_q U'(t-t_0) \frac{1}{Z} e^{-\beta(\hat{H} - \mu \hat{N}^F)}. \tag{4.26}$$

From here, we proceed by making a first order expansion of the propagator $U'(t)$ in the perturbing potential $\epsilon_p$. Details of the procedure are described at multiple locations in the literature[19, 2, 15]. In the end, one obtains

$$\tilde{G}_{qp} = -e\frac{i}{\hbar} \int_{t_0}^{t} \mathrm{d}t' \left\langle [\hat{N}_p^F(t'), \hat{I}_q(t)] \right\rangle. \tag{4.27}$$

Using the fact that in equilibrium correlation functions of this form depend only on time differences, together with the assumption of steady state conditions for $t \gg t_0$, we can modify the limits of the integral to obtain the expression

$$\tilde{G}_{qp} = -e\frac{i}{\hbar}\int_0^\infty \mathrm{d}t \left\langle [\hat{N}_p^F(0), \hat{I}_q(t)] \right\rangle. \tag{4.28}$$

While (4.28) is itself a valid and useful linear response formula, a more symmetric expression can be found by going through some additional steps. The details of these steps are well described by Kubo[19] and by Jacoboni[15], although in Jacoboni's argument some minor modifications must be made to convert from a canonical to a grand canonical ensemble. In the end, one obtains

$$\tilde{G}_{qp} = -\frac{1}{\hbar}\int_0^\infty \mathrm{d}t \int_0^{\hbar\beta} \mathrm{d}\tau \left\langle \hat{I}_p(-i\tau)\hat{I}_q(t) \right\rangle, \tag{4.29}$$

which is the Kubo relation for conductance.

Finally, although the equivalence of the two conductance expressions (4.3) and (4.24) is certainly intuitive, it is instructive to go through an argument showing why this is true. One way of doing that is to consider a situation where both perturbations are applied, in such a way that they are in equilibrium. Thus, we again perturb the Hamiltonian as in (4.25), but this time we assume that the leads are in an equilibrium state of the perturbed Hamiltonian $\hat{H}'$, meaning that

$$\rho(0) = P_S^0 \prod_r \frac{1}{Z_r} e^{-\beta(\hat{H}_r' - \mu\hat{N}_r^F)}. \tag{4.30}$$

Given this state, the leads are all in equilibrium with each other, but initially not with the subsystem $S$. However, it is reasonable to assume all parts of the system to equilibrate over some finite time scale, so that for large $t$, $\rho(t)$ will be an equilibrium state. In particular, since the current is always zero in equilibrium, we have for such large $t$,

$$I_q(t) = \mathrm{Tr}\, U'(t-t_0)^\dagger \hat{I}_q U'(t-t_0) P_S^0 \prod_r \frac{1}{Z_r} e^{-\beta(\hat{H}_r' - \mu\hat{N}_r^F)} = 0. \tag{4.31}$$

Taking the derivative of this expression with respect to $\epsilon_p$, we find

$$\begin{aligned} &\mathrm{Tr}\, \frac{\partial}{\partial \epsilon_p}\left(U'(t-t_0)^\dagger \hat{I}_q U'(t-t_0)\right) P_S^0 \prod_r \frac{1}{Z_r} e^{-\beta(\hat{H}_r - \mu\hat{N}_r^F)} \\ &+ \mathrm{Tr}\, U^\dagger(t-t_0)\hat{I}_q U(t-t_0) \frac{\partial}{\partial \epsilon_p} P_S^0 \prod_r \frac{1}{Z_r} e^{-\beta(\hat{H}_r' - \mu\hat{N}_r^F)} = 0. \end{aligned} \tag{4.32}$$

Next, we note that if we set $\epsilon_p = \mu - \mu_p$, then $\hat{H}_p' - \mu\hat{N}_p^F = \hat{H}_p - \mu_p\hat{N}_p^F$. Inserting this in (4.32), we obtain

$$\begin{aligned} &\mathrm{Tr}\, U^\dagger(t-t_0)\hat{I}_q U(t-t_0) \frac{\partial}{\partial \mu_p} P_S^0 \prod_r \frac{1}{Z_r} e^{-\beta(\hat{H}_r - \mu_r\hat{N}_r^F)} \\ &= \mathrm{Tr}\, \frac{\partial}{\partial \epsilon_p}\left(U'(t-t_0)^\dagger \hat{I}_q U'(t-t_0)\right) P_S^0 \prod_r \frac{1}{Z_r} e^{-\beta(\hat{H}_r - \mu\hat{N}_r^F)}. \end{aligned} \tag{4.33}$$

By (4.23) we recognize the first line of this equation as $\partial I_q/\partial\mu_p$. Further, comparing the second line to (4.26), we see that it is almost identical to $\partial I_q/\partial\epsilon_p$, except that the initial state is different. However, at large times $t$, where we have obtained steady state conditions, it is reasonable to assume that $I_q(t)$ is in fact independent of the initial state, as long as the temperature and chemical potential of the macroscopic leads are the same. Thus, under this assumption (4.33) simply states

$$\frac{\partial I_q}{\partial\mu_p} = \frac{\partial I_q}{\partial\epsilon_p}, \tag{4.34}$$

which justifies the use of (4.24) to calculate conductance.

### 4.2.2 Thermoelectric coefficients

To calculate the thermoelectric coefficient $\tilde{A}_{qp}$, we would like to repeat the procedure above, where we were able to make use of standard linear response theory since we could rephrase the definition of the transport coefficient in terms of a mechanical perturbation. However, while the electrostatic potential was a clear candidate as a replacement of the electrochemical potential, there is no obvious candidate to replace the temperature differences in (4.10). Thus, we must proceed more formally. In particular, we will want to make an argument similar to that leading from (4.31) to (4.34), which shows the equivalence of the perturbations $\Delta\mu_p$ and $\epsilon_p$.

Examining the argument, we see that the crucial step is the one below (4.32), where we obtain $\hat{H}'_p - \mu\hat{N}^F_p = \hat{H}_p - \mu_p\hat{N}^F_p$. A similar expression involving temperatures can be obtained by defining the mechanical perturbation as one given by the perturbed Hamiltonian

$$\hat{H}' = \hat{H} + \sum_p \delta_p(\hat{H}_p - \mu\hat{N}^F_p). \tag{4.35}$$

We then have $\hat{H}'_p - \mu\hat{N}^F_p = (1+\delta_p)(\hat{H}_p - \mu\hat{N}^F_p)$. Accordingly, setting $\delta_p = \beta_p/\beta - 1$, we get

$$\beta(\hat{H}'_p - \mu\hat{N}^F_p) = \beta_p(\hat{H}_p - \mu\hat{N}^F_p), \tag{4.36}$$

which is precisely on the form we seek.

With this it is a simple matter to repeat the steps from (4.31) to (4.34). We begin by differentiating (4.31) with respect to $\delta_p$, obtaining a parallel to (4.32). Then we insert (4.36), and make use of the relation $\delta_p = \beta_p/\beta - 1$ to reexpress one of the derivatives, obtaining in the end

$$\frac{\partial I_q}{\partial\beta_p} = -\frac{1}{\beta}\frac{\partial I_q}{\partial\delta_p}. \tag{4.37}$$

Then inserting this in (4.12), we see that we can calculate $\tilde{A}_{qp}$ as

$$\tilde{A}_{qp} = \frac{1}{T}\frac{\partial I_q}{\partial\delta_p}. \tag{4.38}$$

Since we have then expressed the coefficient in terms of the mechanical perturbation (4.35), we can again make use of standard linear response theory to derive the Kubo relation

$$\tilde{A}_{qp} = -\frac{1}{\hbar T}\int_0^\infty \mathrm{d}t \int_0^{\hbar\beta} \mathrm{d}\tau \left\langle \hat{q}_p(-i\tau)\hat{I}_q(t)\right\rangle , \tag{4.39}$$

where we have defined $\hat{q}_p(t) = \hat{\Phi}_p^E(t) - \mu_p\hat{\Phi}_p(t)$.

Combining the expression $q_p = \Phi_p^E - \mu_p\Phi_p$ with (3.132), (3.137) and (3.42), we can express the heat flux in lead $p$ as

$$q_q(t) = \langle\hat{q}_q(t)\rangle = \mathrm{Tr}\,\hat{q}_q(t)P_S^0\prod_r \frac{1}{Z_r}e^{-\beta_r(\hat{H}_r-\mu_r\hat{N}_r^F)} \tag{4.40}$$

Kubo relations for the remaining thermoelectric coefficients $\tilde{B}_{qp}$ and $\tilde{C}_{qp}$ can be found from this expression in a manner analogous to what we did above, by reexpressing (4.18) and (4.19) in terms of derivatives with respect to the mechanical perturbations $\epsilon_p$ and $\delta_p$. In the end we obtain

$$\tilde{B}_{qp} = -\frac{1}{\hbar}\int_0^\infty \mathrm{d}t \int_0^{\hbar\beta} \mathrm{d}\tau \left\langle \hat{I}_p(-i\tau)\hat{q}_q(t)\right\rangle , \qquad \text{and} \tag{4.41}$$

$$\tilde{C}_{qp} = -\frac{1}{\hbar T}\int_0^\infty \mathrm{d}t \int_0^{\hbar\beta} \mathrm{d}\tau \left\langle \hat{q}_p(-i\tau)\hat{q}_q(t)\right\rangle . \tag{4.42}$$

### 4.2.3 Bulk expressions

In the case of macroscopic devices one typically does not need to perform an explicit simulation of the device, since the transport coefficients $G$, $A$, $B$ etc. can be calculated relatively easily from bulk material coefficients. Thus, we will briefly consider also such bulk coefficients, starting with conductivity. In the absence of thermodynamic gradients, the conductivity $\overleftrightarrow{\sigma}$ is defined through the relation that at macroscopic size scales $\boldsymbol{j} = \overleftrightarrow{\sigma}\boldsymbol{E} = \sum_{ij}\sigma_{ij}E_j\hat{\boldsymbol{e}}_i$, $\boldsymbol{j}$ being the current density and $\boldsymbol{E}$ the electric field. Typically one also considers alternating fields, and so the relation is generalized to $\boldsymbol{j}(\omega) = \overleftrightarrow{\sigma}(\omega)\boldsymbol{E}(\omega)$, where $\boldsymbol{j}(\omega)$ and $\boldsymbol{E}(\omega)$ are respectively the Fourier transforms of $\boldsymbol{j}(t)$ and $\boldsymbol{E}(t)$.

To obtain an expression for $\overleftrightarrow{\sigma}$, we can apply the same linear response technique made use of above. Knowing that $\boldsymbol{E}$ is related to the electrostatic potential $\phi$ by $\boldsymbol{E} = -\boldsymbol{\nabla}\phi$, we introduce a perturbing potential $\epsilon(\boldsymbol{x},t) = -e\phi(\boldsymbol{x},t)$, coupled to the local number density of electrons $\hat{n}(\boldsymbol{x})$. We then simply repeat the steps leading to (4.29), with some small modifications to account for the temporal and spatial variation of $\epsilon$. Details of the derivation can again be found in the literature[19, 2, 15]. The end result is

$$\sigma_{ij}(\omega) = V\frac{1}{\hbar}\int_0^\infty \mathrm{d}t \int_0^{\hbar\beta} \mathrm{d}\tau \left\langle \hat{\bar{j}}_j(-i\tau)\hat{\bar{j}}_i(t)\right\rangle e^{-i\omega t}, \tag{4.43}$$

where $i, j \in \{x, y, z\}$, $V$ is the volume of some sufficiently large region of material, and $\bar{\hat{\boldsymbol{j}}}$ denotes the current density operator $\hat{\boldsymbol{j}}(x)$ averaged over this region. To obtain the direct current conductivity, we simply insert $\omega = 0$. In that case, (4.43) takes the same form as the Kubo relations considered above.

More generally we have $\boldsymbol{j} = -\overleftrightarrow{\sigma}\boldsymbol{\nabla}V - \overleftrightarrow{\mathcal{A}}\boldsymbol{\nabla}T$, which reduces to $\boldsymbol{j} = \overleftrightarrow{\sigma}\boldsymbol{E}$ when $\boldsymbol{\nabla}T = \boldsymbol{0} = \boldsymbol{\nabla}(V - \phi)$. The thermoelectric tensor $\overleftrightarrow{\mathcal{A}}$ is related to the coefficients $\tilde{A}_{qp}$ in the same way that $\overleftrightarrow{\sigma}$ is related to $\tilde{G}_{qp}$. Since we can reinterpret $\tilde{A}_{qp}$ as describing responses to the mechanical perturbations $\delta_p$ of (4.35), the tensor $\overleftrightarrow{\mathcal{A}}$ must also describe responses to gradients in a generalized mechanical potential $\delta(\boldsymbol{x}, t)$, coupled to the local heat density $\hat{h}(\boldsymbol{x}) - \mu\hat{n}(\boldsymbol{x})$, where $\hat{H} = \int \mathrm{d}\boldsymbol{x}\, \hat{h}(\boldsymbol{x})$. Accordingly, we may employ linear response theory in a manor completely analogous to the derivation of (4.43), obtaining

$$\mathcal{A}_{ij} = V\frac{1}{\hbar T}\int_0^\infty \mathrm{d}t \int_0^{\hbar\beta} \mathrm{d}\tau \left\langle \bar{\hat{\phi}}_{Qj}(-i\tau)\bar{\hat{j}}_i(t)\right\rangle, \tag{4.44}$$

where $\boldsymbol{\phi}_Q$ denotes heat flux, and the rest of the notation is as before.

In the linear regime the heat flux $\boldsymbol{\phi}_Q$ can also be expressed in terms of $\boldsymbol{\nabla}T$ and $\boldsymbol{\nabla}V$ as $\boldsymbol{\phi}_Q = -\overleftrightarrow{\mathcal{B}}\boldsymbol{\nabla}V - \overleftrightarrow{\mathcal{C}}\boldsymbol{\nabla}T$. The thermoelectric tensors $\overleftrightarrow{\mathcal{B}}$ and $\overleftrightarrow{\mathcal{C}}$ can be found by arguments similar to those above, and in the end we obtain

$$\mathcal{B}_{ij} = V\frac{1}{\hbar}\int_0^\infty \mathrm{d}t \int_0^{\hbar\beta} \mathrm{d}\tau \left\langle \bar{\hat{j}}_j(-i\tau)\bar{\hat{\phi}}_{Qi}(t)\right\rangle \quad \text{and,} \tag{4.45}$$

$$\mathcal{C}_{ij} = V\frac{1}{\hbar T}\int_0^\infty \mathrm{d}t \int_0^{\hbar\beta} \mathrm{d}\tau \left\langle \bar{\hat{\phi}}_{Qj}(-i\tau)\bar{\hat{\phi}}_{Qi}(t)\right\rangle. \tag{4.46}$$

Expressions similar to (4.43), (4.44), (4.45) and (4.46) were derived by Kubo in his second paper on the Kubo relations[20], as an example of how to treat thermodynamic perturbations. The equations should be compared to (4.29), (4.39), (4.41) and (4.42), which are their device analogs respectively.

In Section 2.1.3 we saw that in a simple two terminal device we could reexpress the current $I$ and the heat current $q$ in terms of new thermoelectric coefficients, as (2.30) and (2.31). In a similar manner, we can reexpress the current density $\boldsymbol{j}$ and the heat flux $\boldsymbol{\phi}_Q$ in terms of new thermoelectric tensors as

$$\boldsymbol{j} = -\overleftrightarrow{\sigma}\left(\boldsymbol{\nabla}V + \overleftrightarrow{\alpha}\boldsymbol{\nabla}T\right) \quad \text{and,} \tag{4.47}$$

$$\boldsymbol{\phi}_Q = -\overleftrightarrow{\kappa}\boldsymbol{\nabla}T + \overleftrightarrow{\pi}\boldsymbol{j}, \tag{4.48}$$

where $\overleftrightarrow{\kappa}$ is referred to as the thermal conductivity, and $\overleftrightarrow{\alpha}$ and $\overleftrightarrow{\pi}$ respectively as the Seebeck and Peltier coefficients. It is a matter of simple algebra to show that

$$\overleftrightarrow{\alpha} = \overleftrightarrow{\sigma}^{-1}\overleftrightarrow{\mathcal{A}}, \tag{4.49}$$

$$\overleftrightarrow{\pi} = \overleftrightarrow{\mathcal{B}}\overleftrightarrow{\sigma}^{-1} \quad \text{and,} \tag{4.50}$$

$$\overleftrightarrow{\kappa} = \overleftrightarrow{\mathcal{C}} - \overleftrightarrow{\mathcal{B}}\overleftrightarrow{\sigma}^{-1}\overleftrightarrow{\mathcal{A}}, \tag{4.51}$$

where the products represent matrix multiplication. These relations generalize the simple two terminal device expressions given by (2.27)-(2.29).

### 4.2.4 Correlation functions

The linear response expressions derived above all share a common format where some transport coefficient $G_{AB}$ is expressed as

$$G_{AB} = \int_0^\infty \mathrm{d}t \int_0^{\hbar\beta} \mathrm{d}\tau \left\langle \hat{A}(-i\tau)\hat{B}(t) \right\rangle = \iint_R \mathrm{d}z^\star \mathrm{d}z \, C_{AB}(z), \tag{4.52}$$

where $R = \{z : 0 \leq \mathcal{I}m\, z \leq \hbar\beta, 0 \leq \mathcal{R}e\, z < \infty\}$, and where we have defined the mixed time correlation function $C_{AB}(t + i\tau) = \langle \hat{A}(-i\tau)\hat{B}(t)\rangle$. In particular, (4.29), (4.39), (4.41), (4.42), (4.43), (4.44), (4.45) and (4.46) all have this format.

In addition to the mixed time correlation function, we define a real time correlation function $C^t_{AB}$ given by $C^t_{AB}(t) = C_{AB}(t)$ with $t$ real, and an imaginary time correlation function $C^\tau_{AB}$ given by $C^\tau_{AB}(\tau) = C_{AB}(i\tau)$ with $\tau$ real. It is easily checked that in equilibrium such correlation functions can only depend on time differences, and so we can also express the real time correlation function as $C^t_{AB}(t' - t) = \langle \hat{A}(0)\hat{B}(t' - t)\rangle = \langle \hat{A}(t)\hat{B}(t')\rangle$.

A final correlation function to be introduced, is the spectral correlation function $C^s_{AB}(E)$. To obtain this function, we make use of a tool known as the Lehmann representation[2], in which we express correlation functions in terms of exact eigenstates of the full Hamiltonian $\hat{H}$. Since $\hat{H}$ commutes with $\hat{N}^F$, these operators can be simultaneously diagonalized. Thus, introducing a basis $\{|i\rangle\}$ which diagonalizes both $\hat{H}$ and $\hat{N}^F$, we can rewrite the mixed time correlation function as

$$\begin{aligned}
\langle \hat{A}(-i\tau)\hat{B}(t)\rangle &= \mathrm{Tr}\frac{1}{Z}e^{-\beta(\hat{H}-\mu\hat{N}^F)}e^{\hat{H}\tau/\hbar}\hat{A}e^{-\hat{H}\tau/\hbar}e^{i\hat{H}t/\hbar}\hat{B}e^{-i\hat{H}t/\hbar} \\
&= \frac{1}{Z}\sum_{ij}\langle i|e^{-\beta(\hat{H}-\mu\hat{N}^F)}e^{\hat{H}\tau/\hbar}\hat{A}e^{-\hat{H}\tau/\hbar}|j\rangle\langle j|e^{i\hat{H}t/\hbar}\hat{B}e^{-i\hat{H}t/\hbar}|i\rangle \\
&= \frac{1}{Z}\sum_{ij}A_{ij}B_{ji}e^{-\beta(E_i-\mu N_i^F)+i(E_j-E_i)t/\hbar+(E_i-E_j)\tau/\hbar}.
\end{aligned} \tag{4.53}$$

where $A_{ij} = \langle i|\hat{A}|j\rangle$, $B_{ji} = \langle j|\hat{B}|i\rangle$, and $E_i$ and $N_i^F$ are respectively eigenvalues of $\hat{H}$ and $\hat{N}^F$. If we now define

$$C^s_{AB}(E) = \frac{1}{Z}\sum_{ij}A_{ij}B_{ji}e^{-\beta(E_i-\mu N_i^F)}\delta(E - E_i + E_j), \tag{4.54}$$

we can write (4.53) simply as

$$C_{AB}(t + i\tau) = \int \mathrm{d}E\, C^s_{AB}(E)e^{-iE(t+i\tau)/\hbar}. \tag{4.55}$$

In particular, the real time correlation function $C^t_{AB}(t) = C_{AB}(t)$ is then a Fourier transform of the spectral correlation function $C^s_{AB}(E)$. In the limit $L_p \to \infty$, where the leads become infinitely large, we again expect the distance between the energy levels $E_i$ to approach zero, and accordingly that $C^s_{AB}(E)$ becomes a continuous function.

In an actual calculation, one will typically obtain estimates of either the real time correlation function $C^t_{AB}$, its Fourier transform $C^s_{AB}$, or the imaginary time correlation function $C^\tau_{AB}$. Thus, the transport coefficient $G_{AB}$ can be obtained if we can reexpress (4.52) in terms of these functions. We begin by considering the relationship between $G_{AB}$ and the imaginary time correlation function. Assuming that the spectral function $C^s_{AB}(E)$ is sufficiently well behaved, we have by (4.55)

$$\frac{\partial}{\partial z^\star} C_{AB}(z) = \int \mathrm{d}E \, C_{AB}(E) \frac{\partial}{\partial z^\star} e^{-iEz/\hbar} = 0, \tag{4.56}$$

which means the function $C_{AB}(z)$ is analytic in the entire complex plane. Thus, it must equal its Taylor expansion around any point, and in particular

$$\begin{aligned} C_{AB}(z) &= \sum_{n=0}^{\infty} \frac{1}{n!} \frac{\partial^n}{\partial z^n} C_{AB}(i\hbar\beta/2)(z - i\hbar\beta/2)^n \\ &= \sum_{n=0}^{\infty} \frac{(-i)^n}{n!} \frac{\partial^n}{\partial \tau^n} C^\tau_{AB}(\hbar\beta/2)(z - i\hbar\beta/2)^n. \end{aligned} \tag{4.57}$$

Thus, given perfect knowledge of the imaginary time correlation function $C^\tau_{AB}(\tau)$ in the interval $\tau \in (0, \hbar\beta)$, we can in principle evaluate $C_{AB}(z)$ at any value of $z$, so that (4.52) can be used to calculate $G_{AB}$. However, in typical numerical calculations one will not have perfect knowledge of $C^\tau_{AB}(\tau)$, so in practice the continuation of the function to the complex plane can be a difficult task. Some approaches to solving this problem are described in the literature[30].

Next we consider the relationship between $G_{AB}$ and the spectral correlation function $C^s_{AB}(E)$. Since $C^s_{AB}$ is related to $C^t_{AB}$ by a simple Fourier transform, this will also give us the relationship between $G_{AB}$ and the real time correlation function. Making use of (4.55), we can perform the innermost integral of (4.52) to get

$$\int_0^{\hbar\beta} \mathrm{d}\tau \, C_{AB}(t + i\tau) = \hbar\beta \int \mathrm{d}E \, g(\beta E) C^s_{AB}(E) e^{-iEt/\hbar}, \tag{4.58}$$

where we have defined

$$g(x) = \int_0^1 e^{sx} \mathrm{d}s = \begin{cases} \frac{e^x - 1}{x} & \text{for } x \neq 0 \\ 1 & \text{for } x = 0 \end{cases}. \tag{4.59}$$

Defining now the function $\chi_{AB}(E) = \hbar\beta g(\beta E) C^s_{AB}(E)$, and assuming this function to be sufficiently well behaved to exchange limits and integrals, we can perform the outermost integral of (4.52) as follows:

$$G_{AB} = \int_0^\infty \mathrm{d}t \int \mathrm{d}E \lim_{\eta\to 0} \chi_{AB}(E) e^{-iEt/\hbar - \eta t} \tag{4.60}$$
$$= \lim_{\eta\to 0} \int \mathrm{d}E\, \chi_{AB}(E) \int_0^\infty \mathrm{d}t\, e^{-iEt/\hbar-\eta t} = -\hbar i \lim_{\eta\to 0} \int \mathrm{d}E\, \frac{\chi_{AB}(E)}{E - i\eta}.$$

Further, one can show that for well behaved functions $f$, we have

$$\lim_{\eta\to 0} \int \mathrm{d}x\, \frac{f(x)}{x - i\eta} = \pi i f(0) + \int \mathrm{d}x \mathbb{P}\frac{1}{x}\, f(x), \tag{4.61}$$

where $\mathbb{P}$ denotes principal value. Making use of this relation, the definition of $\chi_{AB}$, and (4.59), we can reexpress (4.60) as

$$G_{AB} = \pi\hbar^2 \beta C^s_{AB}(0) - \hbar^2 i \int \mathrm{d}E \mathbb{P}\frac{1}{E^2}\, (e^{\beta E} - 1) C^s_{AB}(E). \tag{4.62}$$

### 4.2.5 Symmetries

The correlation functions introduced above obey several important symmetry relations. Using for instance (4.62), we can translate these relations into symmetries of the transport coefficients. Consider first the real time correlation function $C^t_{AB}(t'-t) = \langle \hat{A}(t)\hat{B}(t')\rangle$. We have $C^t_{AB}(t'-t)^\star = \langle \hat{A}(t)\hat{B}(t')\rangle^\star = \langle \hat{B}(t')\hat{A}(t)\rangle = C^t_{BA}(t-t')$. Taking the inverse Fourier transform of this relation, we also find

$$C^s_{AB}(E)^\star = C^s_{BA}(E). \tag{4.63}$$

Further, exchanging $A$ and $B$ as well as the summation indices in (4.54), we get

$$C^s_{BA}(E) = \frac{1}{Z}\sum_{ij} B_{ji} A_{ij} e^{-\beta(E_j - \mu N^F_j)} \delta(E - E_j + E_i). \tag{4.64}$$

Now, in all of the Kubo relations derived above, the operators $\hat{A}$ and $\hat{B}$ were single particle operators, which in particular means they commute with $\hat{N}^F$. Accordingly, we have $A_{ij} e^{\beta\mu N^F_i} = \langle i| e^{\beta\mu \hat{N}^F} \hat{A} |j\rangle = \langle i|\hat{A} e^{\beta\mu\hat{N}^F}|j\rangle = A_{ij} e^{\beta\mu N^F_j}$. Inserting this in (4.64) we find

$$C^s_{BA}(E) = \frac{1}{Z}\sum_{ij} A_{ij} B_{ji} e^{-\beta(E_j - \mu N^F_i)} \delta(-E + E_j - E_i) = e^{-\beta E} C^s_{AB}(-E). \tag{4.65}$$

Combining this with (4.63), we also find

$$e^{\beta E} C^s_{AB}(E) = C^s_{AB}(-E)^\star. \tag{4.66}$$

Inserting (4.66) in (4.62), we obtain after some rearrangement

$$G_{AB} = \pi\hbar^2\beta C^s_{AB}(0) - 2\hbar^2 \int \mathrm{d}E\mathbb{P}\frac{1}{E^2}\,\mathcal{I}m\, C^s_{AB}(E). \tag{4.67}$$

Since by (4.66) $C_{AB}(0)$ is real, so is $G_{AB}$. This is as expected since $G_{AB}$ is defined at $\omega = 0$.

Next, we introduce the time reversal operator $\hat{\tau}$ [21]. The defining properties of this operator is that it is conjugate linear, meaning that $\hat{\tau}(c|\psi\rangle) = c^\star\hat{\tau}|\psi\rangle$, that $\hat{\tau}^2 = I$, and finally that $\hat{\tau}$ preserves the position basis, so that $\hat{\tau}\hat{\psi}_i = \hat{\psi}_i\hat{\tau}$ for any single particle state $i$ which is a real combination of position basis states. Importantly, the last fact also means we can construct a basis of many particle states $|j\rangle$ which is preserved by the time reversal operator, meaning that $\hat{\tau}|j\rangle = |j\rangle$.

So let $\{|i\rangle\}$ be such a many particle basis. We have

$$C^t_{AB}(t'-t) = \frac{1}{Z}\mathrm{Tr}\, e^{-\beta(\hat{H}-\mu\hat{N}^F)}e^{i\hat{H}t/\hbar}\hat{A}e^{-i\hat{H}t/\hbar}e^{i\hat{H}t'/\hbar}\hat{B}e^{-i\hat{H}t'/\hbar} = \frac{1}{Z}\sum_i\langle i|\hat{P}|i\rangle, \tag{4.68}$$

where we have defined $\hat{P} = e^{-\beta(\hat{H}-\mu\hat{N}^F)}e^{i\hat{H}t/\hbar}\hat{A}e^{-i\hat{H}t/\hbar}e^{i\hat{H}t'/\hbar}\hat{B}e^{-i\hat{H}t'/\hbar}$. Making use of the time reversal operator, we can rewrite this as

$$\begin{aligned} C^t_{AB}(t'-t) &= \frac{1}{Z}\sum_i\langle i|\hat{\tau}^2\hat{P}\hat{\tau}^2|i\rangle = \frac{1}{Z}\sum_{ij}\langle i|\hat{\tau}\left[|j\rangle\langle j|\hat{\tau}\hat{P}\hat{\tau}|i\rangle\right] \\ &= \frac{1}{Z}\sum_{ij}\langle j|\hat{\tau}\hat{P}\hat{\tau}|i\rangle^\star\langle i|\hat{\tau}|j\rangle = \frac{1}{Z}\sum_i\langle i|\hat{\tau}\hat{P}\hat{\tau}|i\rangle^\star = \frac{1}{Z}\left(\mathrm{Tr}\,\hat{\tau}\hat{P}\hat{\tau}\right)^\star, \end{aligned} \tag{4.69}$$

or alternatively $C^t_{AB}(t'-t)^\star = \mathrm{Tr}\,\hat{\tau}\hat{P}\hat{\tau}/Z$. Now, using the Taylor expansion of the exponential and the properties of $\hat{\tau}$, it is straight forward to show that

$$\hat{\tau}\hat{P}\hat{\tau} = e^{-\beta(\hat{\tau}\hat{H}\hat{\tau}-\mu\hat{\tau}\hat{N}^F\hat{\tau})}e^{-i\hat{\tau}\hat{H}\hat{\tau}t/\hbar}\hat{\tau}\hat{A}\hat{\tau}e^{i\hat{\tau}\hat{H}\hat{\tau}t/\hbar}e^{-i\hat{\tau}\hat{H}\hat{\tau}t'/\hbar}\hat{\tau}\hat{B}\hat{\tau}e^{i\hat{\tau}\hat{H}\hat{\tau}t'/\hbar}. \tag{4.70}$$

Accordingly we must consider how the involved operators change under the time reversal transform $\hat{X} \to \hat{\tau}\hat{X}\hat{\tau}$. We begin by considering the Hamiltonian $\hat{H}$. It can always be decomposed as $\hat{H} = \hat{H}_1 + \hat{H}_2$, where the components $\hat{H}_1$ and $\hat{H}_2$ are respectively symmetric and anti symmetric under the time reversal transform. Indeed, we can simply let $\hat{H}_1 = (\hat{H} + \hat{\tau}\hat{H}\hat{\tau})/2$ and $\hat{H}_2 = (\hat{H} - \hat{\tau}\hat{H}\hat{\tau})/2$. In solid state applications, the anti symmetric component $\hat{H}_2$ must be proportional to an external magnetic field, since an external electric field, as well as any internal force are left unchanged by time reversal.

Thus, it makes sense to generalize the Hamiltonian as $\hat{H}(b) = \hat{H}_1 + b\hat{H}_2$, where $b$ parameterizes the magnetic field strength. In fact, we will include an explicit dependence on the magnetic field in most operators, writing for instance $\hat{A}$ as $\hat{A}(b)$. In particular, we have $\hat{\tau}\hat{H}(b)\hat{\tau} = \hat{H}_1 - b\hat{H}_2 = \hat{H}(-b)$. The effect of the time reversal transform on $\hat{N}^F$ is easily found by expressing the operator in terms of a preserved single particle basis $i$. We have $\hat{\tau}\hat{N}^F\hat{\tau} = \sum_i \hat{\tau}\hat{\psi}_i^\dagger\hat{\psi}_i\hat{\tau} = \hat{N}^F$. Combining this with the effect on $\hat{H}$, we can also find the effect on the current operator $\hat{I}_p = -e\hat{\dot{\Phi}}_p$. By (3.131) and the definition of $\hat{\dot{\Phi}}_p(t)$ below (3.132), we have

$$\hat{\tau}\hat{I}_p(b)\hat{\tau} = \hat{\tau}(-ei[\hat{H}(b), \hat{N}_p^F]/\hbar)\hat{\tau} = ei[\hat{H}(-b), \hat{N}_p^F]/\hbar = -\hat{I}_p(-b). \tag{4.71}$$

Similar arguments can be applied to the other currents $\hat{q}_p$, $\hat{\boldsymbol{j}}$ and etc., and result in similar expressions. Since the operators $\hat{A}$ and $\hat{B}$ are always picked from these currents, we thus always have $\hat{\tau}\hat{A}(b)\hat{\tau} = -\hat{A}(-b)$ and $\hat{\tau}\hat{B}(b)\hat{\tau} = -\hat{B}(-b)$. Inserting all of this in (4.70), we get

$$\hat{\tau}\hat{P}(b)\hat{\tau} = e^{-\beta(\hat{H}(-b)-\mu\hat{N}^F)}e^{-i\hat{H}(-b)t/\hbar}\hat{A}(-b)e^{i\hat{H}(-b)t/\hbar}e^{-i\hat{H}(-b)t'/\hbar}\hat{B}(-b)e^{i\hat{H}(-b)t'/\hbar}. \tag{4.72}$$

Thus, by (4.69) we have

$$C_{AB}^t(b, t'-t)^\star = \frac{1}{Z}\mathrm{Tr}\,\hat{\tau}\hat{P}(b)\hat{\tau} = C_{AB}^t(-b, t-t'), \tag{4.73}$$

where we have included an explicit dependence on the magnetic field also in the correlation functions. Taking the inverse Fourier transform of this relation, we find

$$C_{AB}^s(b, E)^\star = C_{AB}^s(-b, E). \tag{4.74}$$

Evaluating (4.67) at field strength $-b$, and making use of (4.74) and the reality of $C_{AB}^s(0)$, we find

$$\begin{aligned} G_{AB}(-b) &= \pi\hbar^2\beta C_{AB}^s(-b, 0) - 2\hbar^2\int \mathrm{d}E\mathbb{P}\frac{1}{E^2}\,\mathcal{I}m\,C_{AB}^s(-b, E) \\ &= \pi\hbar^2\beta C_{AB}^s(b, 0) + 2\hbar^2\int \mathrm{d}E\mathbb{P}\frac{1}{E^2}\,\mathcal{I}m\,C_{AB}^s(b, E). \end{aligned} \tag{4.75}$$

Thus, we obtain an interpretation of the two terms in (4.67) in terms of how they behave under reversal of the magnetic field: $G_1 = \pi\hbar^2\beta C_{AB}^s(0)$ is invariant, while the term $G_2 = 2\hbar^2\int \mathrm{d}E\mathbb{P}\,\mathcal{I}m\,C_{AB}^s(E)/E^2$ changes sign. In particular this means that at zero magnetic field we must have $G_{AB} = G_1 = \pi\hbar^2\beta C_{AB}^s(0)$.

Finally, combining (4.74) and (4.63), we obtain $C_{BA}(b, E) = C_{AB}(-b, E)$. Inserting this in (4.67), it follows easily that

$$G_{BA}(b) = G_{AB}(-b). \tag{4.76}$$

These important symmetry relations were also derived by Kubo[19]. At zero magnetic field they reduce to $G_{AB} = G_{BA}$, which are the famous Onsager relations[26]. These relations have several important consequences for the thermoelectric transport coefficients introduced above. For instance they imply symmetry of the conductance and conductivity coefficients: $G_{qp} = G_{pq}$, and $\overleftrightarrow{\sigma} = \overleftrightarrow{\sigma}^T$, as well as a relation between the Peltier and Seebeck coefficients, stating $\Pi_{qp} = T\alpha_{pq}$ and $\overleftrightarrow{\pi} = T\overleftrightarrow{\alpha}^T$.

## 4.3 Four point functions

In the previous section we obtained expressions for transport coefficients in terms of real time and imaginary time correlation functions. In order to evaluate these expressions using the field integral formalism, we must translate the correlation functions into field integral expectation values. We begin by noting that the operators $\hat{A}$ and $\hat{B}$ from the previous section are always single particle operators, and accordingly are always on the form of (3.23), (3.24), or some linear combination of these. Assuming $\hat{A}$ and $\hat{B}$ to involve only fermionic fields, we have

$$\langle \hat{A}\hat{B}\rangle = \sum_{ijkl} A_{ij} B_{kl} \langle \hat{\psi}_i^\dagger \hat{\psi}_j \hat{\psi}_k^\dagger \hat{\psi}_l \rangle. \tag{4.77}$$

The generalization of this expression to include bosonic fields should be obvious. The important point is that the correlation functions of the previous section can always be expressed as a linear combination of correlation functions involving four fields.

### 4.3.1 Real time correlation functions

The real time correlation function $C^t_{AB}(t-t')$ introduced in Section 4.2.4 will be a linear combination of correlation functions on the form $\langle \hat{\psi}_i^\dagger(t)\hat{\psi}_j(t)\hat{\psi}_k^\dagger(t')\hat{\psi}_l(t')\rangle$, possibly with all or two of the fields operators replaced with boson operators. In order to express these correlation functions as field integrals, we must make a small modification, where we instead calculate $\langle \hat{\psi}_i^\dagger(t)\hat{\psi}_j(t+\Delta t)\hat{\psi}_k^\dagger(t'+\Delta t)\hat{\psi}_l(t')\rangle$. This modification will only introduce an error of order $1/N$.

The modified correlation functions are seen to be very simple generalizations of the expression to the left in (3.26), and can be expressed as a field integrals in a similar manner as that expression. In particular, we generalize the second line to the right in (3.26), expressing the correlation function in terms of two factors $U^\dagger(t-t_0)\hat{\psi}_i^\dagger U(\Delta t)\hat{\psi}_j U^\dagger(t_f - t)$ and $U(t_f - t')\hat{\psi}_k^\dagger U(\Delta t)\hat{\psi}_l U(t'-t_0)$. These can be expressed as field integrals using straight forward generalizations of (3.27) and (3.28) respectively. Putting these expressions together in a manner generalizing (3.29), we obtain

$$\langle \hat{\psi}_i^\dagger(t)\hat{\psi}_j(t)\hat{\psi}_k^\dagger(t')\hat{\psi}_l(t')\rangle \approx \langle \bar{\psi}_i^-(t)\psi_j^-(t+\Delta t)\bar{\psi}_k^+(t'+\Delta t)\psi_l^+(t')\rangle, \tag{4.78}$$

with an error of order $1/N$. Field integral expectations like the one on the right in this expression, which involves four fields, are referred to as four point functions[2].

Another source of error in (4.78) is the fact that the two expectation values are technically defined with different initial states. The operator expectation on the left is defined in the equilibrium state $e^{-\beta(\hat{H}-\mu\hat{N})}/Z$ of the full system, while the field integral expectation value on the right is defined with an initial state given by (3.42). However, as before we make use of our assumption that the system equilibrates over some finite time scale, and conclude that the different initial states are irrelevant as long as $t, t' \gg t_0$.

By combining (4.78) with (4.77) and (4.67), we can express any transport coefficient $G_{AB}$ in terms of four point functions. In particular, applying this procedure to the transport coefficients $\tilde{G}_{qp}$, $\tilde{A}_{qp}$, $\tilde{B}_{qp}$ and $\tilde{C}_{qp}$ given by (4.29), (4.39), (4.41) and (4.42), we should in principle obtain expressions equivalent to (4.9), (4.17), (4.20) and (4.21). However, this is difficult to show explicitly.

### 4.3.2 Imaginary time correlation functions

Section 4.2.4 also introduces imaginary time correlation functions, which are defined as $C^{\tau}_{AB}(\tau) = \langle \hat{A}(-i\tau)\hat{B}(0)\rangle = \mathrm{Tr} e^{-\beta(\hat{H}-\mu\hat{N}^F)} e^{\tau\hat{H}/\hbar}\hat{A}e^{-\tau\hat{H}/\hbar}\hat{B}/Z$. Again making use of the fact that $\hat{N}$ commutes with the single particle operators $\hat{A}$ and $\hat{B}$, and with the Hamiltonian, we can write this as

$$C^{\tau}_{AB}(\tau) = \frac{1}{Z}\mathrm{Tr}\, e^{-(\hbar\beta-\tau)(\hat{H}-\mu\hat{N}^F)/\hbar}\hat{A}e^{-\tau(\hat{H}-\mu\hat{N}^F)/\hbar}\hat{B} = \frac{1}{Z}\mathrm{Tr}\, Y(\hbar\beta-\tau)\hat{A}Y(\tau)\hat{B}, \tag{4.79}$$

where we have defined the imaginary time evolution operator $Y(\tau) = e^{-\tau(\hat{H}-\mu\hat{N}^F)/\hbar}$. These functions will be linear combinations of correlation function on the form $\mathrm{Tr}\, Y(\hbar\beta-\tau)\hat{\psi}^{\dagger}_i\hat{\psi}_j Y(\tau)\hat{\psi}^{\dagger}_k\hat{\psi}_l/Z$, again possibly with some field operators replaced with bosonic fields. Again, we will make a modification with an error of order $1/N$, and replace these functions with $\mathrm{Tr}\, Y(\hbar\beta-\tau)\hat{\psi}^{\dagger}_i Y(\Delta\tau)\hat{\psi}_j Y(\tau)\hat{\psi}^{\dagger}_k Y(\Delta\tau)\hat{\psi}_l/Z$, where $\Delta\tau = \hbar\beta/N$.

By introducing the factorization $Y(\tau) = Y(\Delta\tau)^N \approx (I - \tau(\hat{H}-\mu\hat{N}^F)/\hbar)^N$, we see that these expressions can be converted to field integrals by exactly the same technique that we employed in Section 3.1, and in our discussion of real time four point functions above. Going through with the procedure, we eventually end up with the expression

$$\frac{1}{Z}\mathrm{Tr}\, Y(\hbar\beta-\tau)\hat{\psi}^{\dagger}_i\hat{\psi}_j Y(\tau)\hat{\psi}^{\dagger}_k\hat{\psi}_l \approx \langle\bar{\psi}_i(\tau+\Delta\tau)\psi_j(\tau)\bar{\psi}_k(\Delta\tau)\psi_l(0)\rangle_{\tau}, \tag{4.80}$$

where the imaginary time field expectation $\langle\ \rangle_{\tau}$ is defined through

$$\langle X[\phi,\psi]\rangle_{\tau} = \frac{1}{Z}\int D(\phi,\psi)X[\phi,\psi]e^{-S^{\tau}[\phi,\psi]/\hbar} = \frac{\int D(\phi,\psi)X[\phi,\psi]e^{-S^{\tau}[\phi,\psi]/\hbar}}{\int D(\phi,\psi)\, e^{-S^{\tau}[\phi,\psi]/\hbar}}, \tag{4.81}$$

with the imaginary time action $S^{\tau}$ given by

$$-\frac{1}{\hbar}S^{\tau}[\phi,\psi] = -|\boldsymbol{\phi}_0|^2 - \bar{\boldsymbol{\psi}}_0\boldsymbol{\psi}_0 + \boldsymbol{\phi}^{\dagger}_N\boldsymbol{\phi}_0 - \bar{\boldsymbol{\psi}}_N\boldsymbol{\psi}_0 - \tag{4.82}$$

$$\frac{1}{\hbar}\sum_{n=0}^{N-1}\left\{\hbar\boldsymbol{\phi}^{\dagger}_{n+1}(\boldsymbol{\phi}_{n+1}-\boldsymbol{\phi}_n) + \hbar\bar{\boldsymbol{\psi}}_{n+1}(\boldsymbol{\psi}_{n+1}-\boldsymbol{\psi}_n) + \Omega(\boldsymbol{\phi}^{\dagger}_{n+1},\boldsymbol{\phi}_n,\bar{\boldsymbol{\psi}}_{n+1},\boldsymbol{\psi}_n)\Delta\tau\right\}$$

$$\approx -\frac{1}{\hbar}\int_0^{\hbar\beta}\mathrm{d}\tau\,\left(\hbar\boldsymbol{\phi}^{\dagger}\dot{\boldsymbol{\phi}} + \hbar\bar{\boldsymbol{\psi}}\dot{\boldsymbol{\psi}} + \Omega(\boldsymbol{\phi}^{\dagger},\boldsymbol{\phi},\bar{\boldsymbol{\psi}},\boldsymbol{\psi})\right),$$

where $\Omega(\boldsymbol{\phi}^\dagger, \boldsymbol{\phi}, \bar{\boldsymbol{\psi}}, \boldsymbol{\psi}) = H(\boldsymbol{\phi}^\dagger, \boldsymbol{\phi}, \bar{\boldsymbol{\psi}}, \boldsymbol{\psi}) - \mu\bar{\boldsymbol{\psi}}\boldsymbol{\psi}$. Again the integral expression in the final line must be considered a purely symbolic notation.

Using (4.80) and (4.77), any imaginary time correlation function can be expressed as an imaginary time field integral. By the discussion of Section 4.2.4, techniques of analytical continuation can be used to obtain transport coefficients from these correlation functions. Imaginary time field integrals is a major subject of Ref. [2].

Chapter 5

# Non-equilibrium Green's functions

In the previous two chapters, we focused on exact expressions for transport coefficients (in the limit $N \to \infty$). In order to evaluate these expressions, one must almost always apply some technique of approximation. One of the most commonly applied approximation schemes, perturbation theory, is the subject of this chapter. We will limit the discussion to the perturbative expansion of Green's functions, and leave the discussion of four point functions to the literature. See for instance the Bethe Salpeter Equation[2]. In addition, we limit the discussion to the fermionic sector.

The derivations of this chapter primarily follows Altland and Simons[2], and to some extent also Jacoboni[15]. On the other hand, much of the notational conventions are taken from Datta[7].

## 5.1 Perturbative expansion of the field integral

The effective action $S^{\text{eff}}$ is given by (3.62), where the sub-terms can be obtained from (3.115), (3.116), (3.48), (3.17) and (3.21). Examining these expressions, we observe that all terms in $S^{\text{eff}}$ are quadratic in the fields, except possibly for those stemming from the Hamiltonian $H(\boldsymbol{\phi}^\dagger, \boldsymbol{\phi}, \bar{\boldsymbol{\psi}}, \boldsymbol{\psi})$. Further, by making use of a technique known as the Hubbard Stratonovich transformation [2], we can remove all terms in the Hamiltonian that are of higher order in the fermionic fields, at the cost of introducing additional bosonic fields. Thus, assuming that this technique has been applied, we can write the effective action as

$$S^{\text{eff}}[\phi, \psi] = \boldsymbol{\phi}^\dagger A_0 \boldsymbol{\phi} + \bar{\boldsymbol{\psi}}[C_0 + C(\boldsymbol{\phi})]\boldsymbol{\psi} + S^{B,\text{int}}[\phi], \tag{5.1}$$

where $C(\boldsymbol{\phi})$ contains no constant terms, and where as indicated $S^{B,\text{int}}$ depends only on the bosonic fields, and contains all such terms in the Hamiltonian which are of order higher than quadratic.

We now define $S_0 = \boldsymbol{\phi}^\dagger A_0 \boldsymbol{\phi} + \bar{\boldsymbol{\psi}} C_0 \boldsymbol{\psi}$ and $S^{\text{int}} = S^{B,\text{int}} + \bar{\boldsymbol{\psi}} C(\boldsymbol{\phi})\boldsymbol{\psi}$, so that we can write $S^{\text{eff}} = S_0 + S^{\text{int}}$. From (3.61) we then obtain the expansion

$$\begin{aligned} \langle A \rangle &= \int D(+,-) A[\phi, \psi] e^{iS_0/\hbar} e^{iS^{\text{int}}/\hbar} \\ &= \sum_n \frac{1}{n!} \int D(+,-) A[\phi, \psi] \left(iS^{\text{int}}/\hbar\right)^n e^{iS_0/\hbar}. \end{aligned} \tag{5.2}$$

In particular, the fermionic Green's function $G$ introduced in Section 3.3.1 has the perturbative expansion

$$i\hbar G = \langle \boldsymbol{\psi} \bar{\boldsymbol{\psi}} \rangle = \sum_n \frac{1}{n!} \int D(+,-) \boldsymbol{\psi} \bar{\boldsymbol{\psi}} \left(iS^{\text{int}}/\hbar\right)^n e^{iS_0/\hbar}. \tag{5.3}$$

### 5.1.1 Unperturbed Green's functions

The first term in the expansion of the Green's function is the Green's function corresponding to the quadratic action $S_0$. It is referred to as the unperturbed Green's function, and is given by

$$i\hbar g = \langle \boldsymbol{\psi}\bar{\boldsymbol{\psi}}\rangle_0 = \int D(+,-)\boldsymbol{\psi}\bar{\boldsymbol{\psi}} e^{iS_0/\hbar} = \iint \mathbf{d}\bar{\boldsymbol{\psi}}\mathbf{d}\boldsymbol{\psi}\,\boldsymbol{\psi}\bar{\boldsymbol{\psi}} e^{-\bar{\boldsymbol{\psi}}C_0\boldsymbol{\psi}/\hbar i}, \quad \text{and} \tag{5.4}$$

$$i\hbar d = \langle \boldsymbol{\phi}\boldsymbol{\phi}^\dagger\rangle_0 = \int D(+,-)\boldsymbol{\phi}\boldsymbol{\phi}^\dagger e^{iS_0/\hbar} = \iint \mathbf{d}\boldsymbol{\phi}^\dagger\mathbf{d}\boldsymbol{\phi}\,\boldsymbol{\phi}\boldsymbol{\phi}^\dagger e^{-\boldsymbol{\phi}^\dagger A_0\boldsymbol{\phi}/\hbar i}, \tag{5.5}$$

for fermions and bosons respectively. As these expressions are Gaussian integrals, they can be evaluated in the same manner as the lead Green's functions in Section 3.3.1. Completely analogously to (3.65) and (3.66) we simply get $g = C_0^{-1}$ and $d = A_0^{-1}$.

To proceed further we write the matrix $C_0$ as

$$C_0 = C_S - \sum_p \Sigma_p, \tag{5.6}$$

where $\bar{\boldsymbol{\psi}}C_S\boldsymbol{\psi}$ contains all quadratic fermionic terms of $S_S$, and $\bar{\boldsymbol{\psi}}\Sigma_p\boldsymbol{\psi} = -S_p^{F,\text{eff}}$. Here $S_S$ and $S_p^{F,\text{eff}}$ are respectively given by (3.48) and (3.115).

Now consider a system identical to the one considered above, but isolated from the leads. By repeating the arguments above, it should be clear that the unperturbed Green's function $g_S$ of this system is given by $g_S = C_S^{-1}$. Further, since $g_S$ is the Green's function of an isolated non-interacting system, it must satisfy (3.90) and (3.91). Thus, let $\hat{\boldsymbol{\psi}}^\dagger H_S^F \boldsymbol{\psi}$ be the quadratic and fermionic sector of the Hamiltonian $\hat{H}_S$ of the system $S$. Since by (3.41) $T = 0$, we get from (3.90) and (3.91) respectively

$$i\hbar g_S^<(t,t') = 0 \qquad \text{and} \qquad i\hbar g_S^>(t,t') = e^{iH_S^F(t'-t)/\hbar}. \tag{5.7}$$

Inserting this in (3.83) and (3.84), we also find

$$i\hbar g_S^t(t,t') = e^{iH_S^F(t'-t)/\hbar}\theta(t-t') \quad \text{and,} \tag{5.8}$$

$$i\hbar g_S^{\bar{t}}(t,t') = e^{iH_S^F(t'-t)/\hbar}\theta(t'-t). \tag{5.9}$$

Using the relations $g = C_0^{-1}$ and $g_S = C_S^{-1}$, we have by (5.6)

$$g = (C_S - \Sigma_l)^{-1} = (I - g_S\Sigma_l)^{-1} g_S, \tag{5.10}$$

where $\Sigma_l = \sum_p \Sigma_p$. Like the Green's functions, the matrices $\Sigma_p$ are also naturally divided into four sectors. By (3.115), (3.101) and (3.102), we have

$$\Sigma_{p,nm} = \Delta t^2 \frac{i}{h}\int \mathrm{d}E \begin{bmatrix} f_p(E)\Gamma_p^F(E) - i\Sigma_p^r(E) & -f_p(E)\Gamma_p^F(E) \\ (1-f_p(E))\,\Gamma_p^F(E) & f_p(E)\Gamma_p^F(E) + i\Sigma_p^a(E)\end{bmatrix} e^{-i(t-t')E/\hbar}, \tag{5.11}$$

with $t = n\Delta t$ and $t' = m\Delta t$.

### 5.1.2 Interaction terms and Feynman diagrams

Since the interaction terms in $S^{\text{int}} = S^{B,\text{int}} + \bar{\psi}C(\phi)\psi$ stem from the Hamiltonian, we can by (3.48), (3.17) and (3.21) write these terms as

$$S^{B,\text{int}} = \sum_{n=0}^{N-1} \Delta t \left( U(\phi_n^{-\dagger}, \phi_{n+1}^{-}) - U(\phi_{n+1}^{+\dagger}, \phi_n^{+}) \right), \quad \text{and} \tag{5.12}$$

$$\bar{\psi}C(\phi^\dagger, \phi)\psi = \sum_{n=0}^{N-1} \Delta t \left( \bar{\psi}_n^{-} V(\phi_n^{-\dagger}, \phi_{n+1}^{-}) \psi_{n+1}^{-} - \bar{\psi}_{n+1}^{+} V(\phi_{n+1}^{+\dagger}, \phi_n^{+}) \psi_n^{+} \right), \tag{5.13}$$

where $U$ and $V$ are potentials representing the non-quadratic terms in the Hamiltonian. These can be Taylor expanded as

$$U(\phi^\dagger, \phi) = \sum_{n+m>2} \frac{\partial^n \partial^m U}{\partial^n \phi^\dagger \partial^m \phi} \phi_1^\dagger \cdots \phi_n^\dagger \phi_1 \cdots \phi_m, \quad \text{and} \tag{5.14}$$

$$V(\phi^\dagger, \phi) = \sum_{n+m>0} \frac{\partial^n \partial^m V}{\partial^n \phi^\dagger \partial^m \phi} \phi_1^\dagger \cdots \phi_n^\dagger \phi_1 \cdots \phi_m, \tag{5.15}$$

where we made use of the assumption that $S^{\text{int}}$ contains no quadratic terms. For reasons of concreteness and simplicity, we will in the following make approximations to the lowest order, and keep only a third order term in $U$ and a linear term in $V$. In particular, we let

$$U(\phi^\dagger, \phi) = \sum_{ijk} U_{ijk} (\phi_i + \phi_i^\star)(\phi_j + \phi_j^\star)(\phi_k + \phi_k^\star), \quad \text{and} \tag{5.16}$$

$$V_{ij}(\phi^\dagger, \phi) = \sum_{ijk} V_{ij}^k \left( \phi_k + \phi_k^\star \right), \tag{5.17}$$

so that

$$\begin{aligned} S^{B,\text{int}} = \sum_{n=0}^{N-1} \sum_{ijk} \Delta t U_{ijk} \Big\{ &(\phi_{n+1,i}^{-} + \phi_{ni}^{-\star})(\phi_{n+1,j}^{-} + \phi_{nj}^{-\star})(\phi_{n+1,k}^{-} + \phi_{nk}^{-\star}) \\ &- (\phi_{ni}^{+} + \phi_{n+1,i}^{+\star})(\phi_{nj}^{+} + \phi_{n+1,j}^{+\star})(\phi_{nk}^{+} + \phi_{n+1,k}^{+\star}) \Big\}, \end{aligned} \tag{5.18}$$

and

$$\begin{aligned} \bar{\psi}C(\phi^\dagger, \phi)\psi = \sum_{n=0}^{N-1} \sum_{ijk} \Delta t V_{ij}^k \times \\ \left( \bar{\psi}_{in}^{-} \left( \phi_{n+1,k}^{-} + \phi_{nk}^{-\star} \right) \psi_{j,n+1}^{-} - \bar{\psi}_{i,n+1}^{+} \left( \phi_{nk}^{+} + \phi_{n+1,k}^{+\star} \right) \psi_{jn}^{+} \right). \end{aligned} \tag{5.19}$$

Now, by (5.3) we have

$$i\hbar G = \sum_n \frac{i^n}{n!\hbar^n} \left\langle \psi \bar{\psi} \left( S^{B,\text{int}} + \bar{\psi}C(\phi^\dagger, \phi)\psi \right)^n \right\rangle_0, \tag{5.20}$$

where $\langle\,\rangle_0$ denotes the field integral expectation with respect to the unperturbed action $S_0$. Using (5.18) and (5.19) we can express the n'th order expectation as

$$\begin{aligned}&\left\langle \boldsymbol{\psi}\bar{\boldsymbol{\psi}} \left(S^{B,\text{int}} + \bar{\boldsymbol{\psi}} C(\boldsymbol{\phi}^\dagger, \boldsymbol{\phi})\boldsymbol{\psi}\right)^n \right\rangle_0 \qquad (5.21)\\ &= \sum \Delta t^n \, U_{abc} \cdots U_{def} \; V^i_{gh} \cdots V^l_{jk} \left\langle \phi^\star_a \cdots \phi^\star_b \; \phi_c \cdots \phi_d \; \bar{\psi}_e \psi_f \cdots \bar{\psi}_g \psi_h \right\rangle_0 ,\end{aligned}$$

where the sum runs over all terms in $S^{\text{int}}$ in each of the n factors. The potential factors $U_{abc}$, $V^i_{gh}$ and so on in this expression must be interpreted as having an additional negative sign if they are coupled to the forward propagating fields $\phi^+$ and $\psi^+$. The field product expectation values in (5.21) are again just Gaussian integrals, and in fact it can be shown that[2]

$$\begin{aligned}&\left\langle \phi^\star_{i_1} \cdots \phi^\star_{i_n} \; \phi_{j_1} \cdots \phi_{j_m} \; \bar{\psi}_{k_1} \psi_{l_1} \cdots \bar{\psi}_{k_p} \psi_{l_p} \right\rangle_0 \qquad (5.22)\\ &= \sum_{\sigma^B \sigma^F} (-1)^{\sigma^F} \left\langle \phi^\star_{i_1} \phi_{j_{\sigma^B_1}} \right\rangle_0 \cdots \left\langle \phi^\star_{i_n} \phi_{j_{\sigma^B_n}} \right\rangle_0 \left\langle \bar{\psi}_{k_1} \psi_{l_{\sigma^F_1}} \right\rangle_0 \cdots \left\langle \bar{\psi}_{k_p} \psi_{l_{\sigma^F_p}} \right\rangle_0\end{aligned}$$

if $n = m$, and zero otherwise. Here $\sigma^B$ and $\sigma^F$ denote permutations of respectively $n$ and $p$ elements, and $(-1)^{\sigma^F}$ denotes the sign of the permutation $\sigma^F$. Using 5.22 we can express (5.21) as

$$\begin{aligned}&\left\langle \boldsymbol{\psi}\bar{\boldsymbol{\psi}} \left(S^{B,\text{int}} + \bar{\boldsymbol{\psi}} C(\boldsymbol{\phi}^\dagger, \boldsymbol{\phi})\boldsymbol{\psi}\right)^n \right\rangle_0 \qquad (5.23)\\ &= \sum (-1)^{\sigma^F} \Delta t^n \, U_{abc} \cdots U_{def} \; V^i_{gh} \cdots V^l_{jk} \left\langle \phi^\star_a \phi_b \right\rangle_0 \cdots \left\langle \phi^\star_c \phi_d \right\rangle_0 \left\langle \bar{\psi}_e \psi_f \right\rangle_0 \cdots \left\langle \bar{\psi}_g \psi_h \right\rangle_0\\ &= \sum (-1)^{\sigma^F} \Delta t^n \, U_{abc} \cdots U_{def} \; V^i_{gh} \cdots V^l_{jk} \, (i\hbar d_{ba}) \cdots (i\hbar d_{dc}) \, (-i\hbar g_{fe}) \cdots (-i\hbar g_{hg}),\end{aligned}$$

where the sum is now also over the permutations $\sigma^B$ and $\sigma^F$. Then inserting this in (5.20), we obtain

$$\begin{aligned}i\hbar G = \sum \frac{(-1)^x \Delta t^n}{\hbar^n n!} \times \qquad (5.24)\\ iU_{abc} \cdots iU_{def} iV^i_{gh} \cdots iV^l_{jk} (i\hbar d_{ba}) \cdots (i\hbar d_{dc}) \, (i\hbar g_{fe}) \cdots (i\hbar g_{hg}).\end{aligned}$$

The terms of this sum has a useful representation in the form of diagrams/graphs. Note that each index except for two will be shared between a potential factor and a Green's function. The two remaining indices are the ones stemming from the product $\boldsymbol{\psi}\bar{\boldsymbol{\psi}}$, and accordingly these are not shared by a potential factor. Thus, each term can be represented by a diagram where the lines represent unperturbed Green's functions, and the vertices represent potential factors. The convention is then that indices of vertices are shared with the lines intersecting them. Thus, each vertex must be an intersection of either three boson lines, or two fermion lines and one boson line, and there can only be two line endpoints not intersecting a vertex. These diagrams are know as Feynman diagrams, and are extensively described in the literature[2, 15, 7, 23, 27].

Since all the potential and Green's function factors would be identical, it is easy to see that terms corresponding to the same diagram must have identical values except possibly for the sign. However, as it turns out even the sign can be determined from the diagram. It is clear that the fermion lines of the diagram must be organized in one path connecting the two free end points of the diagram, and that the remainder of the fermion lines must form a set of loops. In fact, the sign of a term corresponding to some diagram is simply $(-1)^l$, where $l$ is the number of fermion loops in the diagram[15, 23, 27].

Thus, two terms in (5.23) have the same value if their Feynman diagrams are topologically identical, i.e. if they can be made identical by a continuous transformation. Further, as shown for instance by Jacoboni[15], all terms in (5.24) which correspond to Feynman diagrams that are not connected, end up canceling, and can accordingly be omitted from the sum. Finally, it is also shown in the literature[2, 15] that for any connected Feynman diagram with $n$ vertices, there is exactly $n!$ terms in (5.24) which corresponds to that diagram.

Thus, by the previous two paragraphs, we can omit the terms having disconnected diagrams, perform the sums over the $n!$ terms corresponding to the same diagrams, and finally write (5.24) as

$$
\begin{aligned}
i\hbar G = \sum (-1)^l \frac{iU_{abc}\Delta t}{\hbar} \cdots \frac{iU_{def}\Delta t}{\hbar} \frac{iV^i_{gh}\Delta t}{\hbar} \cdots \frac{iV^l_{jk}\Delta t}{\hbar} \\
\times (i\hbar d_{ba}) \cdots (i\hbar d_{dc})\, (i\hbar g_{fe}) \cdots (i\hbar g_{hg}),
\end{aligned}
\tag{5.25}
$$

where the sum runs over all topologically distinct connected Feynman diagrams with two end points, $l$ is the number of fermion loops in each particular diagram, and the potential factors and Green's functions are those represented by the vertices and lines respectively.

### 5.1.3 Self energy and the Dyson equation

We now define some terminology. A connected diagram is a diagram containing no isolated groups. A free index is an index on a vertex which is not connected to a Green's function line. A free index diagram is a connected diagram which has no free endpoints, but which has two free indices. A reducible diagram is a free index diagram that can be split into two components by removing a single Green's function line, in such a way that there is one of the original free indices on each new component. An irreducible diagram is a free index diagram that is not reducible.

Any diagram in the expansion (5.25) can be turned into a free index diagram by removing the two free endpoint lines. If the result is then a reducible diagram, a single Green's function line can be removed to obtain two components which are both free index diagrams. If any of these components are reducible, a single Green's function can again be removed to split also this component into two new free index diagrams. The process can be continued until one is left with a finite sequence of irreducible diagrams. Thus, the value of any diagram in (5.25) can be written as

$$
D = (i\hbar g_{ij}) A_{jk} (i\hbar g_{kl}) B_{lm} \cdots (i\hbar g_{pq}) C_{qr} (i\hbar g_{rs}),
\tag{5.26}
$$

where $A_{jk}$, $B_{lm}$, $C_{qr} \cdots$ represent irreducible diagrams, $j, k, l, m, q$ and $r$ being the free indices. Thus, (5.25) can be written as

$$i\hbar G_{is} = \sum_{n=1}^{\infty} \sum_{A^1 A^2 \cdots A^n} (i\hbar g_{ij}) A^1_{jk} (i\hbar g_{kl}) A^2_{lm} \cdots (i\hbar g_{pq}) A^n_{qr} (i\hbar g_{rs}), \tag{5.27}$$

where the factors $A^n_{ij}$ correspond to irreducible diagrams, and the second sum is over all such. We now define

$$\Sigma_{s,ij} = i\hbar \sum_A A_{ij}, \tag{5.28}$$

where the sum is over all irreducible diagrams. We refer to the matrix $\Sigma_s$ as the scattering self energy, in order to distinguish it from the lead self energies $\Sigma_p$. Inserting (5.28) in (5.27), we get

$$i\hbar G_{is} = \sum_{n=1}^{\infty} \sum_{jklm \cdots pqr} i\hbar g_{ij} \Sigma_{s,jk} g_{kl} \Sigma_{s,lm} \cdots g_{pq} \Sigma_{s,qr} g_{rs} = i\hbar \sum_{n=1}^{\infty} \left[ g(\Sigma_s g)^{n-1} \right]_{is}, \tag{5.29}$$

which can be reexpressed as

$$G = \sum_{n=0}^{\infty} g(\Sigma_s g)^n = g + g\Sigma_s \sum_{n=0}^{\infty} g(\Sigma_s g)^n = g + g\Sigma_s G. \tag{5.30}$$

This equation is known as the Dyson equation[2, 15]. With some simple rearrangement it can also be expressed as

$$G = (g^{-1} - \Sigma_s)^{-1} \tag{5.31}$$

Defining the total self energy $\Sigma = \Sigma_s + \Sigma_l$, we get by (5.10)

$$G = (C_S - \Sigma_l - \Sigma_s)^{-1} = (g_S^{-1} - \Sigma)^{-1} = (I - g_S \Sigma)^{-1} g_S. \tag{5.32}$$

## 5.2 The NEGF equations

### 5.2.1 Dyson equation as a difference equation

Making use of the last expression in (5.32), we can rewrite the Dyson equation as $(I - g_S \Sigma) G = g_S$. If we explicitly include the block structure corresponding to time arguments, this becomes

$$G_{nm} - \sum_{l=0}^{N-1} \sum_{p=0}^{N-1} g_{,Snl} \Sigma_{lp} G_{pm} = g_{S,nm}. \tag{5.33}$$

Making use of (5.7)-(5.9) it is easy to show that $g^{<}_{S,nm} = 0$, $g^{>}_{S,nm} = e^{-iH^F_S \Delta t/\hbar} g^{>}_{S,n-1,m}$, $g^{t}_{S,nm} = e^{-iH^F_S \Delta t/\hbar} g^{t}_{S,n-1,m} - \frac{i}{\hbar}\delta_{nm}$ and $g^{\bar{t}}_{S,nm} = e^{-iH^F_S \Delta t/\hbar} g^{\bar{t}}_{S,n-1,m} + \frac{i}{\hbar}\delta_{n-1,m}$. If we define matrices

$$M = \begin{bmatrix} e^{-iH^F_S \Delta t/\hbar} & 0 \\ 0 & e^{-iH^F_S \Delta t/\hbar} \end{bmatrix}, \quad P = \begin{bmatrix} I & 0 \\ 0 & 0 \end{bmatrix}, \quad \text{and } Q = \begin{bmatrix} 0 & 0 \\ 0 & I \end{bmatrix}, \tag{5.34}$$

these relations can be combined in a single matrix expression as

$$M g_{S,n-1,m} = g_{S,nm} + \frac{i}{\hbar}(P\delta_{nm} - Q\delta_{n-1,m}). \tag{5.35}$$

Now making use of (5.35) and (5.33) to evaluate $MG_{n-1,m} - G_{nm}$, we obtain the difference equation

$$MG_{n-1,m} - G_{nm} - \frac{i}{\hbar}\sum_{l=0}^{N-1}(P\Sigma_{nl} - Q\Sigma_{n-1,l})G_{lm} = \frac{i}{\hbar}(P\delta_{nm} - Q\delta_{n-1,m}). \tag{5.36}$$

Since we are interested in the limit $N \to \infty$, we can expand $M$ to the first order in $\Delta t$, to get

$$M \approx I - iH^{FF}_S \Delta t/\hbar, \quad \text{where } H^{FF}_S = \begin{bmatrix} H^F_S & 0 \\ 0 & H^F_S \end{bmatrix}. \tag{5.37}$$

Making use of this approximation, we can rewrite (5.36) as

$$\begin{aligned} &\hbar i(G_{nm} - G_{n-1,m}) - \Delta t H^{FF}_S G_{n-1,m} \\ &\qquad - \sum_{l=0}^{N-1}(P\Sigma_{nl} - Q\Sigma_{n-1,l})G_{lm} = P\delta_{nm} - Q\delta_{n-1,m}. \end{aligned} \tag{5.38}$$

### 5.2.2 Contributions to the self energy

Now consider the irreducible diagrams contributing to (5.28). These can be divided into two classes depending on whether or not the free indices sit on the same vertex. We will refer to diagrams where the free indices sit on the same vertex as single vertex diagrams, and those where they sit on different vertices as double vertex diagrams. In any single vertex diagram, the remaining index on the free vertex must be attached to a boson line. Analogously to (5.25), one can show that the sum of all connected diagrams with a single free boson end point is $\langle\phi_i\rangle$ or $\langle\phi^\star_i\rangle$ depending on the type of endpoint. Thus, summing over all single vertex diagrams we in fact obtain

$$\frac{i\Delta t}{\hbar}\sum_k V^k_{ij}\langle\phi_k + \phi^\star_k\rangle. \tag{5.39}$$

Examining (5.19) we see that the vertices always couple fields with the same superscript. Also keeping in mind the comment below (5.21), we express (5.39) in matrix form as

$$\frac{i\Delta t}{\hbar}\sum_k \begin{bmatrix} -V^k_{ij}\langle\phi^+_k+\phi^{+\star}_k\rangle & 0 \\ 0 & V^k_{ij}\langle\phi^-_k+\phi^{-\star}_k\rangle \end{bmatrix}. \tag{5.40}$$

Further, essentially repeating the arguments leading to (3.35), we see that $\langle\phi^+_k\rangle = \langle\hat{\phi}_k\rangle = \langle\phi^-_k\rangle$. Accordingly, we write the sum of all single vertex diagrams as

$$-\frac{i\Delta t}{\hbar}\begin{bmatrix} \delta H^F(t) & 0 \\ 0 & -\delta H^F(t) \end{bmatrix} = -\frac{i\Delta t}{\hbar}(P-Q)\delta H^{FF}(t), \tag{5.41}$$

where

$$\delta H^F_{ij}(t) = \sum_k V^k_{ij}\langle\phi^+_k(t)+\phi^{+\star}_k(t)\rangle = \sum_k V^k_{ij}\langle\phi^-_k(t)+\phi^{-\star}_k(t)\rangle. \tag{5.42}$$

A diagram in the expansion of $\langle\phi_k\rangle$ must contain a free boson line connected to a vertex, connected to some diagram with two end points. Summing over all such diagrams, it should be clear that

$$\langle\phi_k\rangle = \sum_{lmn}(i\hbar d_{kl})U_{lmn}(i\hbar D_{mn}) + \sum_{lmn}(i\hbar d_{kl})V^l_{mn}(i\hbar G_{mn}). \tag{5.43}$$

Now we turn to the double vertex diagrams. Each one of the two free vertices in the diagram is connected to one free fermion line, and one free boson line. Thus, removing the two free vertices we are left with an irreducible diagram containing two free boson end points, and two free fermion end points. Since an analogue of (5.25) applies also to four point functions, we see that these diagrams are precisely the irreducible diagrams occurring in the expansion of the four point function

$$\left\langle\psi_i\bar{\psi}_j(\phi_k+\phi^\star_k)(\phi_l+\phi^\star_l)\right\rangle. \tag{5.44}$$

Accordingly, we denote the sum of all such diagrams $\left\langle\psi_i\bar{\psi}_j(\phi_k+\phi^\star_k)(\phi_l+\phi^\star_l)\right\rangle_{\mathrm{Irr}}$, so that summing over all double vertex diagrams we get

$$\frac{(i\Delta t)^2}{\hbar^2}\sum_{jklm} V^k_{ij}V^l_{mn}\left\langle\psi_j\bar{\psi}_m(\phi_k+\phi^\star_k)(\phi_l+\phi^\star_l)\right\rangle_{\mathrm{Irr}}. \tag{5.45}$$

Again making use of the comment below (5.21), as well as the fact that (5.19) only couples fields with the same superscript, we can express this in matrix form as

$$\frac{(i\Delta t)^2}{\hbar^2}\sum_{jklm} V^k_{ij}V^l_{mn}\begin{bmatrix} \left\langle\psi^+_j\bar{\psi}^+_m x^+_k x^+_l\right\rangle_{\mathrm{Irr}} & -\left\langle\psi^+_j\bar{\psi}^-_m x^+_k x^-_l\right\rangle_{\mathrm{Irr}} \\ -\left\langle\psi^-_j\bar{\psi}^+_m x^-_k x^+_l\right\rangle_{\mathrm{Irr}} & \left\langle\psi^-_j\bar{\psi}^-_m x^-_k x^-_l\right\rangle_{\mathrm{Irr}} \end{bmatrix}, \tag{5.46}$$

where we have defined $x_i(t) = \phi_i(t) + \phi_i^\star(t)$. Accordingly, we write the sum of all double vertex diagrams as

$$-\frac{i\Delta t^2}{\hbar}\Sigma_s(t,t') = -\frac{i\Delta t^2}{\hbar}\begin{bmatrix} \Sigma_s^t(t,t') & -\Sigma_s^<(t,t') \\ -\Sigma_s^>(t,t') & \Sigma_s^{\bar{t}}(t,t') \end{bmatrix}, \quad \text{where} \tag{5.47}$$

$$\Sigma_{s,in}^t(t,t') = -\frac{i}{\hbar}\sum_{jklm} V_{ij}^k V_{mn}^l \left\langle \psi_j^+(t)\bar{\psi}_m^+(t')x_k^+(t)x_l^+(t')\right\rangle_{\text{Irr}}, \tag{5.48}$$

$$\Sigma_{s,in}^<(t,t') = -\frac{i}{\hbar}\sum_{jklm} V_{ij}^k V_{mn}^l \left\langle \psi_j^+(t)\bar{\psi}_m^-(t')x_k^+(t)x_l^-(t')\right\rangle_{\text{Irr}}, \tag{5.49}$$

$$\Sigma_{s,in}^>(t,t') = -\frac{i}{\hbar}\sum_{jklm} V_{ij}^k V_{mn}^l \left\langle \psi_j^-(t)\bar{\psi}_m^+(t')x_k^-(t)x_l^+(t')\right\rangle_{\text{Irr}}, \quad \text{and} \tag{5.50}$$

$$\Sigma_{s,in}^{\bar{t}}(t,t') = -\frac{i}{\hbar}\sum_{jklm} V_{ij}^k V_{mn}^l \left\langle \psi_j^-(t)\bar{\psi}_m^-(t')x_k^-(t)x_l^-(t')\right\rangle_{\text{Irr}}. \tag{5.51}$$

An important approximation to these expressions is the self consistent Born approximation[2], where one makes the approximation

$$\left\langle \psi_j\bar{\psi}_m x_k x_l\right\rangle_{\text{Irr}} \approx \left\langle \psi_j\bar{\psi}_m\right\rangle \left\langle x_k x_l\right\rangle = -\hbar^2 G_{jm}(D_{kl} - D_{lk}^\star). \tag{5.52}$$

This expression can be further approximated by replacing the full Green's functions with the unperturbed ones, so that $\left\langle \psi_j\bar{\psi}_m x_k x_l\right\rangle_{\text{Irr}} \approx -\hbar^2 g_{jm}(d_{kl} - d_{kl}^\star)$. The latter is referred to simply as the Born approximation.

### 5.2.3 Continuum limit

Now, inserting the sum of all single vertex diagrams (5.41), and the sum of all double vertex diagrams (5.47) in (5.28), we get

$$\Sigma_{s,nm} = \Delta t\,(P-Q)\delta H^{FF}(t)\delta_{nm} + \Delta t^2\Sigma_s(t,t'), \tag{5.53}$$

where $t = n\Delta t$ and $t' = m\Delta t$. Combining this with the definition of the total self energy $\Sigma$ above (5.32), we have

$$\Sigma_{nm} = \Delta t\,(P-Q)\delta H^{FF}(t)\delta_{nm} + \Delta t^2\Sigma(t,t'), \tag{5.54}$$

where

$$\Sigma(t,t') = \Sigma_s(t,t') + \sum_p \frac{\Sigma_{p,nm}}{\Delta t^2}, \tag{5.55}$$

with $\Sigma_{pnm}$ given by (5.11). Inserting (5.54) in (5.38), we obtain

$$\begin{aligned} &\hbar i(G_{nm} - G_{n-1,m}) - \Delta t H_S^{FF} G_{n-1,m} \qquad (5.56)\\ &-\Delta t P\delta H^{FF}(n\Delta t)G_{nm} - \Delta t Q\delta H^{FF}(n\Delta t - \Delta t)G_{n-1,m} \\ &-\Delta t^2 \sum_{l=0}^{N-1}(P\Sigma(n\Delta t, l\Delta t) - Q\Sigma(n\Delta t - \Delta t, l\Delta t))G_{lm} = P\delta_{nm} - Q\delta_{n-1,m}. \end{aligned}$$

Note that the order of this equation is $1/N$, except near $n = m$, where it is of order $N^0$. So consider the expression $\Delta t P(\delta H^{FF}(n\Delta t)G_{nm} - \delta H^{FF}(n\Delta t - \Delta t)$ $G_{n-1,m})$. Since $\delta H$ is differentiable as a function of time, and $G$ is differentiable everywhere except at $t = t'$, the expression must be of order $1/N^2$ everywhere except at $n = m$, where it is of order $1/N$. Thus, the expression is always of higher order than (5.56), and so it can be added to the left side of the equation without changing the order of the approximation. Similarly, the expression

$$\Delta t^2 \sum_{l=0}^{N-1} Q(\Sigma(n\Delta t, l\Delta t) - \Sigma(n\Delta t - \Delta t, l\Delta t))G_{lm} \tag{5.57}$$

is also of order $1/N^2$, since $\Sigma$ is differentiable everywhere except at $t = t'$. Thus, this expression can also be added to the left side of (5.56). Making both of these modifications, we obtain

$$\begin{aligned}\hbar i(G_{nm} - G_{n-1,m}) - \Delta t H_C^{FF}(n\Delta t - \Delta t)G_{n-1,m} \\ -\Delta t^2 \sum_{l=0}^{N-1} (P - Q)\Sigma(n\Delta t, l\Delta t)G_{lm} = P\delta_{nm} - Q\delta_{n-1,m},\end{aligned} \tag{5.58}$$

where we have defined $H_C^{FF}(t) = H_S^{FF} + \delta H^{FF}(t)$.

Multiplying (5.58) by $(P - Q)/\Delta t$, and then finally taking the continuum limit $\Delta t \to 0$, we obtain the integro-differential equation

$$(P - Q)\left(\hbar i \frac{\partial G}{\partial t}(t, t') - H_C^{FF}(t)G(t, t')\right) - \int_{t_0}^{t_f} \mathrm{d}t''\, \Sigma(t, t'')G(t'', t') = \delta(t - t'), \tag{5.59}$$

If one is to obtain the strictly correct continuum limit of (5.58), one must take care to make the appropriate interpretation of the delta function in this expression. Indeed, adopting the standard interpretation of the delta function will result in slightly modified Green's functions where the step functions $\theta$ in (3.83) and (3.84) are replaced with step functions $\tilde{\theta}$, with $\tilde{\theta}(0) = 1/2$ rather than 1. However, since this modification only affects the time ordered and anti time ordered Green's functions at $t = t'$, and since $G^t(t, t)$ and $G^{\bar{t}}(t, t)$ are never used in the calculation of expectation values, this consideration is not so important in practice. Thus, we will in fact make the modification $\theta \to \tilde{\theta}$ to the Green's functions discussed in the remainder of the chapter.

We now assume the existence of a finite correlation time $\tau$ such that $\Sigma(t, t') \approx 0$ for $|t - t'| \gg \tau$. This allows us to replace the limits on the integral in (5.59) with $\pm\infty$, thereby obtaining

$$\begin{bmatrix} \hbar i\partial_t - H_C^F(t) & 0 \\ 0 & -\hbar i\partial_t + H_C^F(t) \end{bmatrix} G(t, t') - \int_{-\infty}^{\infty} \mathrm{d}t''\, \Sigma(t, t'')G(t'', t') = \delta(t - t'), \tag{5.60}$$

where we have defined the corrected Hamiltonian $H_C^F(t) = H_S^F + \delta H^F(t)$, and introduced the short notation $\partial_t = \partial/\partial t$. Similar equations can be found in the literature[2, 15].

### 5.2.4 Energy representation

Now we make use of the assumption that the system approaches a steady state, so that for $t \gg t_0$ all functions depend only on the argument difference $t - t'$. Thus, $H_C^F(t)$ becomes a constant $H_C^F$, while $G$ and $\Sigma$ become functions $G(t-t')$ and $\Sigma(t-t')$. We define energy representations of these functions as Fourier transforms, i.e. we let

$$G(E) = \int_{-\infty}^{\infty} \mathrm{d}t\, G(t-t') e^{iE(t-t')/\hbar}, \tag{5.61}$$

while analogous expressions define $\Sigma(E)$, as well as the component functions $\Sigma_p(E)$, $\Sigma^<(E)$, $G^t(E)$ and so on. Making use of the inverse Fourier transform, we can invert these relationships, obtaining

$$G(t-t') = \frac{1}{h} \int_{-\infty}^{\infty} \mathrm{d}E\, G(E) e^{-iE(t-t')/\hbar}, \tag{5.62}$$

and similar for the other functions.

Taking the Fourier transform of (5.55) and making use of the block decompositions of (5.47) and (5.11), we obtain the relationship

$$\Sigma(E) = \begin{bmatrix} \Sigma^t(t,t') & -\Sigma^<(t,t') \\ -\Sigma^>(t,t') & \Sigma^{\bar{t}}(t,t') \end{bmatrix} \tag{5.63}$$

where

$$\Sigma^t(E) = \Sigma_s^t(E) + \sum_p \left( i f_p(E) \Gamma_p^F(E) + \Sigma_p^r(E) \right), \tag{5.64}$$

$$\Sigma^<(E) = \Sigma_s^<(E) + i \sum_p f_p(E) \Gamma_p^F(E) \tag{5.65}$$

$$\Sigma^>(E) = \Sigma_s^>(E) + i \sum_p \left( f_p(E) - 1 \right) \Gamma_p^F(E), \quad \text{and} \tag{5.66}$$

$$\Sigma^{\bar{t}}(E) = \Sigma_s^{\bar{t}}(E) + \sum_p \left( i f_p(E) \Gamma_p^F(E) - \Sigma_p^a(E) \right). \tag{5.67}$$

Finally, inserting the inverse Fourier transform of $\Sigma(E)$ in (5.60), and then Fourier transforming the entire equation, we obtain

$$\begin{bmatrix} E - H_C^F & 0 \\ 0 & -E + H_C^F \end{bmatrix} G(E) - \Sigma(E) G(E) = I, \tag{5.68}$$

the energy representation of the Dyson equation.

### 5.2.5 Component Green's functions

Making use of (3.81)-(3.84), one easily shows that $G^<(t,t')^\dagger = -G^<(t',t)$, $G^>(t,t')^\dagger = -G^>(t',t)$ and $G^t(t,t')^\dagger = -G^{\bar{t}}(t',t)$. Fourier transforming these relations, we obtain $G^<(E)^\dagger = -G^<(E)$, $G^>(E)^\dagger = -G^>(E)$, and $G^t(E)^\dagger = -G^{\bar{t}}(E)$. Further, adding (3.83) and (3.84), and keeping in mind the modification $\theta \to \tilde{\theta}$, we get $G^t(t,t') + G^{\bar{t}}(t,t') = G^<(t,t') + G^>(t,t')$. Fourier transforming this we obtain

$$G^t(E) + G^{\bar{t}}(E) = G^<(E) + G^>(E). \tag{5.69}$$

Now we define new Green's functions

$$G^r(E) = G^t(E) - G^<(E), \quad \text{and} \tag{5.70}$$

$$G^a(E) = G^t(E) - G^>(E), \tag{5.71}$$

which we refer to respectively as the retarded and advanced Green's functions. Making use of (5.69) and the other relations of the previous paragraph, we can also express these functions as

$$G^r(E) = G^>(E) - G^{\bar{t}}(E), \qquad \text{and} \tag{5.72}$$

$$G^a(E) = G^<(E) - G^{\bar{t}}(E) = G^r(E)^\dagger. \tag{5.73}$$

Analogously to (5.70) and (5.71), we also define retarded and advanced self energies respectively as

$$\Sigma^r(E) = \Sigma^t(E) - \Sigma^<(E), \quad \text{and} \tag{5.74}$$

$$\Sigma^a(E) = \Sigma^t(E) - \Sigma^>(E). \tag{5.75}$$

Similar expressions define the scattering self energies $\Sigma_s^r(E)$ and $\Sigma_s^a(E)$.

Inserting the block decompositions of $G$ and $\Sigma$ given by (3.68) and (5.63) in (5.68), we get

$$\begin{bmatrix} E - H_C^F - \Sigma^t(E) & \Sigma^<(E) \\ \Sigma^>(E) & -E + H_C^F - \Sigma^{\bar{t}}(E) \end{bmatrix} \begin{bmatrix} G^t(E) & G^<(E) \\ G^>(E) & G^{\bar{t}}(E) \end{bmatrix} = \begin{bmatrix} I & 0 \\ 0 & I \end{bmatrix}, \tag{5.76}$$

or in terms of the component functions,

$$\left(E - H_C^F - \Sigma^t(E)\right) G^t(E) + \Sigma^<(E) G^>(E) = I, \tag{5.77}$$

$$\left(E - H_C^F - \Sigma^t(E)\right) G^<(E) + \Sigma^<(E) G^{\bar{t}}(E) = 0, \tag{5.78}$$

$$\Sigma^>(E) G^t(E) + \left(-E + H_C^F - \Sigma^{\bar{t}}(E)\right) G^>(E) = 0, \quad \text{and} \tag{5.79}$$

$$\Sigma^>(E) G^<(E) + \left(-E + H_C^F - \Sigma^{\bar{t}}(E)\right) G^{\bar{t}}(E) = I. \tag{5.80}$$

Subtracting the first two of these equations, and then making use of (5.70), (5.72) and (5.74), we obtain $\left(E - H_C^F - \Sigma^r(E)\right) G^r(E) = I$, and upon inverting

$$G^r(E) = \left(E - H_C^F - \Sigma^r(E)\right)^{-1}. \tag{5.81}$$

Similarly, subtracting (5.79) and (5.80), we find $\left(E - H_C^F + \Sigma^{\bar{t}}(E) - \Sigma^>(E)\right) G^r(E) = I$. Then multiplying on the right by $E - H_C^F - \Sigma^r(E)$ and making use of (5.81) we get

$$\Sigma^r(E) = \Sigma^>(E) - \Sigma^{\bar{t}}(E), \tag{5.82}$$

which in combination with (5.74) yields

$$\Sigma^t(E) + \Sigma^{\bar{t}}(E) = \Sigma^<(E) + \Sigma^>(E). \tag{5.83}$$

Then combining this with (5.75), we also find

$$\Sigma^a(E) = \Sigma^<(E) - \Sigma^{\bar{t}}(E). \tag{5.84}$$

Note that (5.83), (5.82) and (5.84) respectively form analogs of (5.69), (5.72) and (5.73).

Now adding (5.77) and (5.79) and making use of (5.75), (5.84) and (5.71), we get $\left(E - H_C^F - \Sigma^a(E)\right) G^a(E) = I$, and upon inverting

$$G^a(E) = \left(E - H_C^F - \Sigma^a(E)\right)^{-1}. \tag{5.85}$$

By (5.73) we can equate this to the adjoint of (5.81), and upon inverting we get $\Sigma^a(E) = \Sigma^r(E)^\dagger$. Making use of (5.74), (5.84), (5.64)-(5.67) and (3.111) we then also find $\Sigma_s^a(E) = \Sigma_s^r(E)^\dagger$.

Solving (5.74) and (5.73) for $\Sigma^t(E)$ and $G^{\bar{t}}(E)$ respectively, and then inserting the resulting expressions in (5.78), we get $\left(E - H_C^F - \Sigma^r(E)\right) G^<(E) - \Sigma^<(E)G^a(E) = 0$. Then finally multiplying this expression on the left by $G^r(E)$, and making use of (5.81) we obtain

$$G^<(E) = G^r(E)\Sigma^<(E)G^a(E). \tag{5.86}$$

Together with an expression for the self energies, (5.86) and (5.81) are the only equations one needs to solve to find all Green's functions, since the remaining ones are then easily obtained using the relations (5.70)-(5.73). (5.86) and (5.81) are identified as the main NEGF equations by Datta[7].

## 5.3 The spectral density

Following Datta[7], we define the spectral density $A(E)$ in terms of the Green's functions introduced above as

$$A(E) = i\left(G^r(E) - G^a(E)\right). \tag{5.87}$$

It follows immediately from this definition and (5.73) that $A(E)$ is Hermitian. Further, making use of the definitions (5.70) and (5.71), we can also express the spectral density as

$$A(E) = i\left(G^>(E) - G^<(E)\right). \tag{5.88}$$

Defining

$$\Gamma(E) = i\left(\Sigma^r(E) - \Sigma^a(E)\right), \tag{5.89}$$

we also have the relations

$$A(E) = G^a(E)\Gamma(E)G^r(E) = G^r(E)\Gamma(E)G^a(E), \tag{5.90}$$

derivations of which can be found in the book by Datta[7].

Inserting the Fourier representation (5.61) of the Green's functions in (5.88), we obtain

$$A(E) = \int_{-\infty}^{\infty} A(t-t')e^{iE(t-t')/\hbar}, \tag{5.91}$$

where in general we define

$$A(t,t') = i\left(G^>(t,t') - G^<(t,t')\right). \tag{5.92}$$

Making use of (3.81) and (3.82), we see that

$$A_{ij}(t,t') = \frac{1}{\hbar}\left(\langle\psi_i(t)\bar{\psi}_j(t')\rangle + \langle\psi_j(t')\bar{\psi}_i(t)\rangle\right) = \frac{1}{\hbar}\left\langle\left\{\psi_i(t),\bar{\psi}_j(t')\right\}\right\rangle, \tag{5.93}$$

meaning that $A(t,t')$ is in fact the anti-commutator correlation function of the field operators.

Taking the inverse Fourier transform of (5.91), and making use of the fundamental anti-commutation relations of the field operators, we find

$$\int_{-\infty}^{\infty} \frac{\operatorname{Tr} A(E)}{2\pi}\mathrm{d}E = \sum_i \hbar A_{ii}(t,t) = \sum_i 1 = s^F, \tag{5.94}$$

where $s^F$ is the total number of fermion single particle states. Thus, it is reasonable to identify the quantity $\operatorname{Tr} A(E)/2\pi$ with the contribution to $s^F$ from different energies, i.e. the Density of states. In the literature[15, 7] it is in fact shown that in the absence of many particle interactions

$$\frac{\operatorname{Tr} A(E)}{2\pi} = \sum_i \delta(E - E_i), \tag{5.95}$$

where the sum is over all fermionic single particle states. The quantity on the right is of course the single particle density of states $D(E)$. Accordingly, we regard $\operatorname{Tr} A(E)/2\pi$ as a meaningful generalization of the density of states outside of the single particle approximation, and make the general identification[2] $D(E) = \operatorname{Tr} A(E)/2\pi$. Naturally, the diagonal elements $A_{ii}(E)/2\pi$ are then identified with the projected density of states $D_i(E)$.

Finally, there are some important relations between the various Green's functions in equilibrium. These can be derived by once again making use of the Lehmann representation, as in Section 4.2.4, to obtain equilibrium expressions for $G^<$ and $G^>$. These representations can then be used to find a relationship between the two Green's functions, which can be reexpressed in terms of the spectral density using (5.88). The details can be found for instance in the book by Jacoboni[15], who derives the equilibrium relation

$$G^<(E) = if(E)A(E), \tag{5.96}$$

where $f(E)$ is the Fermi function defined in (3.92). Multiplying (5.96) by $(E - H_C^F - \Sigma^r(E))$ and $(E - H_C^F - \Sigma^a(E))$ on the left and right respectively, and making use of (5.86) and (5.90), we also obtain the equilibrium relation

$$\Sigma^<(E) = if(E)\Gamma(E). \tag{5.97}$$

## 5.4 Transport expressions

From (3.152) we see that we can calculate the electrical current from the expectation values $\langle \bar{\psi}_i^-(E)\psi_j^+(E)\rangle$ and $\langle \bar{\psi}_i^+(E)\psi_j^-(E)\rangle$ as

$$\begin{aligned} I_p = -e\int \frac{2\pi \mathrm{d}E}{h^2(t_f - t_0)} \sum_{ij} \Big( &(1 - f_p(E))\,\Gamma_{pij}^F(E)\left\langle \bar{\psi}_i^-(E)\psi_j^+(E)\right\rangle \\ &+ f_p(E)\Gamma_{pij}^F(E)\left\langle \bar{\psi}_i^+(E)\psi_j^-(E)\right\rangle \Big). \end{aligned} \tag{5.98}$$

By (3.101) and (3.102) we have

$$\left\langle \bar{\psi}_i^-(E)\psi_j^+(E)\right\rangle = \sum_{nm} \Delta t^2 \left\langle \bar{\psi}_{n+1,i}^-\psi_{mj}^+\right\rangle e^{iE(m-n+1)\Delta t/\hbar}. \tag{5.99}$$

Making use of (3.82) and taking the limit $\Delta t \to 0$ this becomes

$$\begin{aligned} \left\langle \bar{\psi}_i^-(E)\psi_j^+(E)\right\rangle &= -i\hbar \int_{t_0}^{t_f} \mathrm{d}t \int_{t_0}^{t_f} \mathrm{d}t'\, G_{ji}^<(t', t) e^{iE(t'-t)/\hbar} \\ &\approx -i\hbar \int_{t_0}^{t_f} \mathrm{d}t \int_{-\infty}^{\infty} \mathrm{d}t'\, G_{ji}^<(t' - t) e^{iE(t'-t)/\hbar} = -i\hbar(t_f - t_0)G_{ji}^<(E), \end{aligned} \tag{5.100}$$

where we have made use of the assumption that $G^<$ has a finite correlation time to replace the innermost integration limits. In a completely analogous manor, from (3.81) we also find

$$\left\langle \bar{\psi}_i^+(E)\psi_j^-(E)\right\rangle = -i\hbar(t_f - t_0)G_{ji}^>(E). \tag{5.101}$$

Inserting (5.100) and (5.101) back in (5.98), and then making use of (5.88), we obtain

$$I_p = \frac{ei}{h}\int \mathrm{d}E\, \mathrm{Tr}\, \Gamma_p^F(E)\big(G^<(E) - if_p(E)A(E)\big). \tag{5.102}$$

By (3.153), the heat current $q_p$ can be decomposed into a bosonic (phonon) component $q_p^l$ and a fermionic (electron) component $q_p^e$ such that $q_p = q_p^l + q_p^e$.

Since we are limiting our discussion to the fermionic sector, we will only consider the fermionic component, which can be expressed as

$$q_p^e = \int \frac{2\pi \mathrm{d}E\,(E-\mu)}{h^2(t_f - t_0)} \sum_{ij} \Big( (1 - f_p(E))\, \Gamma^F_{pij}(E) \left\langle \bar{\psi}_i^-(E) \psi_j^+(E) \right\rangle \tag{5.103}$$
$$+ f_p(E) \Gamma^F_{pij}(E) \left\langle \bar{\psi}_i^+(E) \psi_j^-(E) \right\rangle \Big).$$

Inserting (5.100) and (5.101) in this expression, we obtain in a manor analogous to the derivation of (5.102)

$$q_p^e = -\frac{i}{h} \int \mathrm{d}E\,(E-\mu) \mathrm{Tr}\, \Gamma^F_p(E) \big( G^<(E) - i f_p(E) A(E) \big). \tag{5.104}$$

Comparing (5.102) and (5.104) to (2.11) and (2.12), we see that the expressions for $I_p$ and $q_p^e$ are respectively on the form of (2.11) and (2.12), with the energy resolved particle flux $i_p(E)$ given by

$$i_p(E) = -\frac{i}{h} \mathrm{Tr}\, \Gamma^F_p(E) \big( G^<(E) - i f_p(E) A(E) \big). \tag{5.105}$$

This is also the expression presented by Datta[7].

# Chapter 6

# Markov and semiclassical approximations

In this chapter, we will discuss two remaining important transport formalisms, which are master equations, and the Boltzmann equation. Master equations apply when the dynamics of the model can be regarded Markovian, i.e. when the future development depends only on the current state. Typically a Markovian approximation can be made when the system develops more slowly than some characteristic non-Markovian time scale. The Boltzmann equation on the other hand, will apply when in addition to such slow temporal variations, there are also slow spatial variations, i.e when the system is homogeneous over long length scales.

In Section 6.1 we will make a Markovian approximation to the NEGF equations introduced in the previous chapter, and derive a general nonlinear master equation. The three remaining sections will be devoted to the Boltzmann equation. For this discussion, we will make use of two important tools known as the Wigner transform and the Moyal expansion, which are introduced in Section 6.2. Then, in Section 6.3 we will discuss how these techniques can be generalized, in such a way that they are applicable to a solid state system with multiple bands. Finally, in Section 6.4 we discuss how the Boltzmann equation arises within this context.

## 6.1 The Markov approximation and the general Master equation

Combining (3.82) with (3.25), we have

$$\langle A\rangle = -i\hbar\sum_{ij} A_{ij}G^<_{ij}(t,t) = -i\hbar\mathrm{Tr}\,AG^<(t,t), \tag{6.1}$$

for any single particle observable $A$. Thus, the matrix $-i\hbar G^<(t,t)$ plays the role of a single particle density operator $\rho_e(t)$, since we then have $\langle A\rangle = \mathrm{Tr}\,A\rho_e(t)$. An important note however, is that while the trace of the density operator is usually set to one, in this case we instead have $\mathrm{Tr}\,\rho_e = N_f$, the total number of fermions. One could argue that the latter is in any case a more appropriate convention when one is dealing with multiple identical particles.

Thus, we make the identification $\rho_e(t) = -i\hbar G^<(t,t)$, and obtain the derivative

$$\frac{\mathrm{d}\rho_e}{\mathrm{d}t}(t) = -i\hbar\frac{\mathrm{d}}{\mathrm{d}t}\left(G^<(t,t')|_{t=t'}\right) = -i\hbar\left(\frac{\partial}{\partial t}+\frac{\partial}{\partial t'}\right)G^<(t,t')|_{t=t'}. \tag{6.2}$$

Writing out the upper right component of (5.60), we get

$$\left(\hbar i \frac{\partial}{\partial t} - H_C^F(t)\right) G^<(t,t') = \int_{-\infty}^{\infty} dt'' \left(\Sigma^r(t,t'')G^<(t'',t') + \Sigma^<(t,t'')G^a(t'',t')\right), \tag{6.3}$$

where we have made use of the definition (5.74) and the relation (5.73). Taking the adjoint of (6.3), and then interchanging the time arguments, we obtain also an equation involving the other time derivative:

$$\hbar i \frac{\partial}{\partial t'} G^<(t,t') + G^<(t,t')H_C^F(t') = \tag{6.4}$$
$$- \int_{-\infty}^{\infty} dt'' \left(G^r(t,t'')\Sigma^<(t'',t') + G^<(t,t'')\Sigma^a(t'',t')\right).$$

Here we have made use some relations concerning the adjoint of Green's functions, all of which either appear in the beginning of Section 5.2.5, or are easily obtained from relations appearing there.

Finally adding (6.3) and (6.4), we get by (6.2)

$$-\frac{\mathrm{d}\rho_e}{\mathrm{d}t}(t) - \frac{i}{\hbar}[H_C^F(t), \rho_e(t)] = \int_{-\infty}^{\infty} dt'' \left(\Sigma^r(t,t'')G^<(t'',t) - G^<(t,t'')\Sigma^a(t'',t)\right)$$
$$- \int_{-\infty}^{\infty} dt'' \left(G^r(t,t'')\Sigma^<(t'',t) - \Sigma^<(t,t'')G^a(t'',t)\right). \tag{6.5}$$

To proceed further, we will once more make use of the assumption that the various Green's functions have finite correlation times $\tau$, and that they are essentially zero whenever the time arguments are further apart than this characteristic time scale, i.e. when $|t - t'| \gg \tau$. We also assume that these correlation times are typically different for the different Green's functions, so that we have separate correlation times $\tau_F^<$, $\tau_F^r$, $\tau_B^<$, $\tau_B^r$ and so on, corresponding to the Green's functions $G^<$, $G^r$, $D^<$, $D^r$ and so on. The Markov approximation is based upon the assumption that the correlation time $\tau_F^<$ of the lesser Green's function $G^<$ is much larger than all other significant time scales. In particular, we then assume $\tau_F^< \gg \tau_B^r$ and $\tau_B^<$. These conditions essentially amount to the assumption that scattering interactions are sufficiently weak.

The important aspect of this assumption, is that $\Sigma^r$ and $\Sigma^a$ will only be significantly different from zero in a range where $G^<$ is approximately constant. Thus, on the right hand side of (6.5), we may approximate $G^<(t,t'') \approx G^<(t'',t) \approx G^<(t,t)$. Doing this, we find

$$\int_{-\infty}^{\infty} dt'' \left(\Sigma^r(t,t'')G^<(t'',t) - G^<(t,t'')\Sigma^a(t'',t)\right) \approx R(t)\rho_e(t) + \rho_e(t)R^\dagger(t), \tag{6.6}$$

where we have defined

$$R(t) = -\frac{i}{\hbar}\int_{-\infty}^{\infty} dt' \Sigma^r(t,t'). \tag{6.7}$$

By the discussion in Section 5.1.2 and 5.1.3, the self energies $\Sigma^r$ and $\Sigma^<$ can be expressed as integrals involving products of the various Green's functions. In these expression also, we can apply the Markov approximation to exchange factors of $G^<$ with a factor of $\rho_e(t)$. Thus, if $\rho_e$ is known at $t$, the self energy expressions together with (5.81) forms a self contained set of equations which can be solved for $\Sigma^r(t,t')$, $\Sigma^<(t,t')$ and $G^r(t,t')$ without needing any more information about $G^<$. Accordingly, we can express the solutions as functions of $\rho_e$, so that $\Sigma^r = \Sigma^r(\rho_e, t-t')$, $\Sigma^< = \Sigma^<(\rho_e, t-t')$ and $G^r = G^r(\rho_e, t-t')$. We are here assuming time independent dynamics, so that the solutions must be invariant under time translation. Inserting for instance the expression for $\Sigma^r$ in (6.7), we have

$$R(t) = -\frac{i}{\hbar}\int_{-\infty}^{\infty} dt' \Sigma^r(\rho_e(t), t-t') = R(\rho_e(t)), \tag{6.8}$$

so that the matrix $R$ is also some function of $\rho_e$. Likewise, for the last integral in (6.5), we have

$$\int_{-\infty}^{\infty} dt'' \left(G^r(t,t'')\Sigma^<(t'',t) - \Sigma^<(t,t'')G^a(t'',t)\right) \tag{6.9}$$

$$= \int_{-\infty}^{\infty} dt'' \left(G^r(\rho_e, t-t'')\Sigma^<(\rho_e, t''-t) - \Sigma^<(\rho_e, t-t'')G^a(\rho_e, t''-t)\right) = S(\rho_e(t)).$$

Finally, combining (5.42) and (5.43), we see that we do not need to apply the Markov approximation to the expression for $\delta H^F$, since it already only involves $G^<(t,t)$. Accordingly, $H_C^F(t) = H_S^F + \delta H^F(\rho_e) = H_C^F(\rho_e)$, and we can write (6.5) simply as

$$-\frac{\mathrm{d}\rho_e}{\mathrm{d}t} - \frac{i}{\hbar}[H_C^F(\rho_e), \rho_e] = R(\rho_e)\rho_e + \rho_e R^\dagger(\rho_e) - S(\rho_e). \tag{6.10}$$

It makes sense to rewrite this expression by separating out the Hermitian and anti-Hermitian components of the Matrix $R$. Thus, we write

$$R = \frac{1}{2}\Gamma + \frac{i}{\hbar}\delta H_M^F, \tag{6.11}$$

where $\Gamma$ and $\delta H_M^F$ are both Hermitian matrices. Inserting this in (6.10), we finally obtain the general Markovian nonlinear master equation

$$\frac{\mathrm{d}\rho_e}{\mathrm{d}t} = -\frac{i}{\hbar}[H_e(\rho_e), \rho_e] - \frac{1}{2}\left\{\Gamma(\rho_e), \rho_e\right\} + S(\rho_e). \tag{6.12}$$

Here curly brackets denote anti commutation, and

$$H_e(\rho_e) = H_C^F(\rho_e) + \delta H_M^F(\rho_e) = H_S^F + \delta H^F(\rho_e) + \delta H_M^F(\rho_e). \tag{6.13}$$

From (6.7) and (6.11) it should be clear that

$$\delta H_M(\rho_e) = -\frac{1}{2}\int_{-\infty}^{\infty} dt' \left(\Sigma^r(\rho_e, t-t') + \Sigma^a(\rho_e, t'-t)\right), \quad \text{while} \tag{6.14}$$

$$\Gamma(\rho_e) = -\frac{i}{\hbar}\int_{-\infty}^{\infty} dt' \left(\Sigma^r(\rho_e, t-t') - \Sigma^r(\rho_e, t'-t)\right). \tag{6.15}$$

## 6.2 The Weyl-Wigner transform

### 6.2.1 Weyl quantization

The subject of this section is the Weyl and Wigner transforms. The introduction of the Weyl transform in particular, can be motivated from the problem of quantization, which concerns how one may transform a function $g(\boldsymbol{x}, \boldsymbol{p})$ defined in classical phase space, into a corresponding Hilbert space operator $\hat{g} = g(\hat{\boldsymbol{x}}, \hat{\boldsymbol{p}})$. Quantization is a simple procedure when the involved function is a simple polynomial where all terms contain only one of the variables $\boldsymbol{x}$ or $\boldsymbol{p}$. One may then simply replace these variables with the corresponding operators, replacing for instance the term $\boldsymbol{p}^2$ with $\hat{\boldsymbol{p}}^2$ and so on. However, if we generalize the situation simply by including cross terms in $\boldsymbol{x}$ and $\boldsymbol{p}$, the problem becomes more complex. Since the operators $\hat{\boldsymbol{x}}$ and $\hat{\boldsymbol{p}}$ do not commute, there will in general be several inequivalent ways to represent such cross terms as operators[14]. And if the function $g$ is not even analytic, the problem becomes even more difficult.

The procedure known as Weyl quantization seeks to resolve these problems by employing a Fourier transform[13]. Taking the Fourier transform of the function $g$, and then the inverse Fourier transform of the result, we obtain

$$g(\boldsymbol{x}', \boldsymbol{p}') = \frac{1}{(2\pi)^6} \int \mathrm{d}^3\boldsymbol{k} \int \mathrm{d}^3\boldsymbol{r} \int \mathrm{d}^3\boldsymbol{x} \int \mathrm{d}^3\boldsymbol{p}\, g(\boldsymbol{x}, \boldsymbol{p}) e^{i\boldsymbol{k}\cdot(\boldsymbol{x}'-\boldsymbol{x}) - i(\boldsymbol{p}'-\boldsymbol{p})\cdot\boldsymbol{r}}. \quad (6.16)$$

From this expression it is a simple matter to obtain a Hilbert space operator $g(\hat{\boldsymbol{x}}, \hat{\boldsymbol{p}})$ by simply substituting $\boldsymbol{x}'$ and $\boldsymbol{p}'$ with $\hat{\boldsymbol{x}}$ and $\hat{\boldsymbol{p}}$. Thus, we obtain the expression

$$\hat{g} = g(\hat{\boldsymbol{x}}, \hat{\boldsymbol{p}}) = \frac{1}{(2\pi)^6} \int \mathrm{d}^3\boldsymbol{k} \int \mathrm{d}^3\boldsymbol{r} \int \mathrm{d}^3\boldsymbol{x} \int \mathrm{d}^3\boldsymbol{p}\, g(\boldsymbol{x}, \boldsymbol{p}) e^{i\boldsymbol{k}\cdot(\hat{\boldsymbol{x}}-\boldsymbol{x}) - i(\hat{\boldsymbol{p}}-\boldsymbol{p})\cdot\boldsymbol{r}}, \quad (6.17)$$

which is well defined for any Fourier transformable function $g$. The operator $\hat{g}$ is referred to as the Weyl transform of the function $g$.

It will also be convenient to have more direct expressions for the Weyl transform, in terms of the position and momentum bases. To obtain these, we make use of the Baker-Campbell-Hausdorff theorem[15, 24], which states

$$e^{i\boldsymbol{k}\cdot\hat{\boldsymbol{x}}} e^{-i\hat{\boldsymbol{p}}\cdot\boldsymbol{r}} = e^{i\boldsymbol{k}\cdot\hat{\boldsymbol{x}} - i\hat{\boldsymbol{p}}\cdot\boldsymbol{r} + \hbar i\boldsymbol{k}\cdot\boldsymbol{r}/2}, \quad (6.18)$$

to reexpress (6.17) as

$$\hat{g} = \frac{1}{(2\pi)^6} \int \mathrm{d}^3\boldsymbol{k} \int \mathrm{d}^3\boldsymbol{r} \int \mathrm{d}^3\boldsymbol{x} \int \mathrm{d}^3\boldsymbol{p}\, g(\boldsymbol{x}, \boldsymbol{p}) e^{-i\boldsymbol{k}\cdot\hbar\boldsymbol{r}/2} e^{i\boldsymbol{k}\cdot(\hat{\boldsymbol{x}}-\boldsymbol{x})} e^{-i(\hat{\boldsymbol{p}}-\boldsymbol{p})\cdot\boldsymbol{r}}. \quad (6.19)$$

Making use of the position and momentum bases respectively, we have

$$e^{i\boldsymbol{k}\cdot\hat{\boldsymbol{x}}} = \int \mathrm{d}^3\boldsymbol{y}\, e^{i\boldsymbol{k}\cdot\boldsymbol{y}} |\boldsymbol{y}\rangle\langle\boldsymbol{y}|, \qquad \text{and} \quad (6.20)$$

$$e^{-i\hat{\boldsymbol{p}}\cdot\boldsymbol{r}} = \int \mathrm{d}^3\boldsymbol{q}\, e^{-i\boldsymbol{q}\cdot\boldsymbol{r}} |\boldsymbol{q}\rangle\langle\boldsymbol{q}|. \quad (6.21)$$

Inserting (6.20) in (6.19) we obtain after making use of the Fourier representation of the delta function to integrate over $\boldsymbol{k}$ and $\boldsymbol{y}$,

$$\hat{g} = \frac{1}{h^3} \int \mathrm{d}^3\boldsymbol{x}' \int \mathrm{d}^3\boldsymbol{x} \int \mathrm{d}^3\boldsymbol{p}\, g(\boldsymbol{x},\boldsymbol{p}) |\boldsymbol{x}+\boldsymbol{x}'/2\rangle\langle \boldsymbol{x}+\boldsymbol{x}'/2| e^{-i(\hat{\boldsymbol{p}}-\boldsymbol{p})\cdot\boldsymbol{x}'/\hbar}, \quad (6.22)$$

where $\boldsymbol{x}' = \boldsymbol{r}/\hbar$. Making use of the translation formula $e^{i\hat{\boldsymbol{p}}\cdot\boldsymbol{x}'/\hbar}|\boldsymbol{x}\rangle = |\boldsymbol{x}-\boldsymbol{x}'\rangle$, we find in particular $\langle \boldsymbol{x}+\boldsymbol{x}'/2|e^{-i(\hat{\boldsymbol{p}}-\boldsymbol{p})\cdot\boldsymbol{x}'/\hbar} = \langle \boldsymbol{x}-\boldsymbol{x}'/2|e^{i\boldsymbol{p}\cdot\boldsymbol{x}'/\hbar}$. Inserting this back in (6.22), we finally obtain

$$\hat{g} = \frac{1}{h^3} \int \mathrm{d}^3\boldsymbol{x}' \int \mathrm{d}^3\boldsymbol{x} \int \mathrm{d}^3\boldsymbol{p}\, e^{i\boldsymbol{p}\cdot\boldsymbol{x}'/\hbar} g(\boldsymbol{x},\boldsymbol{p}) |\boldsymbol{x}+\boldsymbol{x}'/2\rangle\langle \boldsymbol{x}-\boldsymbol{x}'/2|, \quad (6.23)$$

where the Weyl transform is expressed in terms of the position basis. Instead inserting (6.21) in (6.19), and essentially repeating the steps above, we can also express the Weyl transform in the momentum basis, as

$$\hat{g} = \frac{1}{h^3} \int \mathrm{d}^3\boldsymbol{x} \int \mathrm{d}^3\boldsymbol{p} \int \mathrm{d}^3\boldsymbol{p}'\, e^{-i\boldsymbol{p}'\cdot\boldsymbol{x}/\hbar} g(\boldsymbol{x},\boldsymbol{p}) |\boldsymbol{p}+\boldsymbol{p}'/2\rangle\langle \boldsymbol{p}-\boldsymbol{p}'/2|. \quad (6.24)$$

### 6.2.2 The Wigner transform, and the Wigner distribution

Starting from (6.23), and making use of the delta function normalization $\langle \boldsymbol{x}|\boldsymbol{x}'\rangle = \delta(\boldsymbol{x}-\boldsymbol{x}')$, as well as the Fourier representation of the delta function, one can show that

$$g(\boldsymbol{x},\boldsymbol{p}) = \int \mathrm{d}^3\boldsymbol{x}'\, e^{-i\boldsymbol{p}\cdot\boldsymbol{x}'/\hbar} \langle \boldsymbol{x}+\boldsymbol{x}'/2|\hat{g}|\boldsymbol{x}-\boldsymbol{x}'/2\rangle. \quad (6.25)$$

Thus, (6.25) gives the inverse of the Weyl transform, which is referred to as the Wigner transform[15]. The Wigner transform can also be expressed in the momentum basis. Indeed, starting from (6.24) we find in a similar manner

$$g(\boldsymbol{x},\boldsymbol{p}) = \int \mathrm{d}^3\boldsymbol{p}'\, e^{i\boldsymbol{p}'\cdot\boldsymbol{x}/\hbar} \langle \boldsymbol{p}+\boldsymbol{p}'/2|\hat{g}|\boldsymbol{p}-\boldsymbol{p}'/2\rangle. \quad (6.26)$$

Clearly, the Wigner transform simplifies significantly if $\hat{g}$ is diagonal in either the position or momentum bases. That is, if $\langle \boldsymbol{x}|\hat{g}|\boldsymbol{x}'\rangle = g(\boldsymbol{x})\delta(\boldsymbol{x}-\boldsymbol{x}')$ or $\langle \boldsymbol{p}|\hat{g}|\boldsymbol{p}'\rangle = g(\boldsymbol{p})\delta(\boldsymbol{p}-\boldsymbol{p}')$. By (6.25) and (6.26), we have respectively in these two cases simply $g(\boldsymbol{x},\boldsymbol{p}) = g(\boldsymbol{x})$ and $g(\boldsymbol{x},\boldsymbol{p}) = g(\boldsymbol{p})$. In particular, since $\langle \boldsymbol{x}|I|\boldsymbol{x}'\rangle = \delta(\boldsymbol{x}-\boldsymbol{x}')$, the Wigner transform of the identity operator is 1.

Assume now that we are given two operators $\hat{A}$ and $\hat{B}$, whose Wigner transforms are respectively $A^W(\boldsymbol{x},\boldsymbol{p})$ and $B^W(\boldsymbol{x},\boldsymbol{p})$. Then making use of (6.23) and (6.25), we can write the trace of the product of the two operators as

$$\begin{aligned}\operatorname{Tr}\hat{A}\hat{B} &= \operatorname{Tr}\frac{1}{h^3}\int \mathrm{d}^3\boldsymbol{x}'\int \mathrm{d}^3\boldsymbol{x}\int \mathrm{d}^3\boldsymbol{p}\, e^{i\boldsymbol{p}\cdot\boldsymbol{x}'/\hbar}A^W(\boldsymbol{x},\boldsymbol{p})|\boldsymbol{x}+\boldsymbol{x}'/2\rangle\langle\boldsymbol{x}-\boldsymbol{x}'/2|\hat{B}\\
&= \frac{1}{h^3}\int \mathrm{d}^3\boldsymbol{x}\int \mathrm{d}^3\boldsymbol{p}\, A^W(\boldsymbol{x},\boldsymbol{p})\int \mathrm{d}^3\boldsymbol{x}'\, e^{i\boldsymbol{p}\cdot\boldsymbol{x}'/\hbar}\langle\boldsymbol{x}-\boldsymbol{x}'/2|\hat{B}|\boldsymbol{x}+\boldsymbol{x}'/2\rangle\\
&= \frac{1}{h^3}\int \mathrm{d}^3\boldsymbol{x}\int \mathrm{d}^3\boldsymbol{p}\, A^W(\boldsymbol{x},\boldsymbol{p})\int \mathrm{d}^3\boldsymbol{x}'\, e^{-i\boldsymbol{p}\cdot\boldsymbol{x}'/\hbar}\langle\boldsymbol{x}+\boldsymbol{x}'/2|\hat{B}|\boldsymbol{x}-\boldsymbol{x}'/2\rangle\\
&= \frac{1}{h^3}\int \mathrm{d}^3\boldsymbol{x}\int \mathrm{d}^3\boldsymbol{p}\, A^W(\boldsymbol{x},\boldsymbol{p})B^W(\boldsymbol{x},\boldsymbol{p}).\end{aligned} \tag{6.27}$$

In particular, since the Wigner transform of $I$ is 1, we get the important special case

$$\operatorname{Tr}\hat{A} = \frac{1}{h^3}\int \mathrm{d}^3\boldsymbol{x}\int \mathrm{d}^3\boldsymbol{p}\, A^W(\boldsymbol{x},\boldsymbol{p}). \tag{6.28}$$

It should also be noted that the Wigner transform preserves the adjoint operation. Indeed, we have

$$\begin{aligned}A^W(\boldsymbol{x},\boldsymbol{p})^\star &= \int \mathrm{d}^3\boldsymbol{x}'\, e^{i\boldsymbol{p}\cdot\boldsymbol{x}'/\hbar}\langle\boldsymbol{x}-\boldsymbol{x}'/2|\hat{A}^\dagger|\boldsymbol{x}+\boldsymbol{x}'/2\rangle \qquad (6.29)\\
&= \int \mathrm{d}^3\boldsymbol{x}'\, e^{-i\boldsymbol{p}\cdot\boldsymbol{x}'/\hbar}\langle\boldsymbol{x}+\boldsymbol{x}'/2|\hat{A}^\dagger|\boldsymbol{x}-\boldsymbol{x}'/2\rangle,\end{aligned}$$

which is the Wigner transform of $\hat{A}^\dagger$.

The Wigner transform of the density operator is referred to as the Wigner distribution, and denoted $f^W$. That is, we have[15, 29, 25]

$$f^W(\boldsymbol{x},\boldsymbol{p}) = \int \mathrm{d}^3\boldsymbol{x}'\, e^{-i\boldsymbol{p}\cdot\boldsymbol{x}'/\hbar}\langle\boldsymbol{x}+\boldsymbol{x}'/2|\hat{\rho}|\boldsymbol{x}-\boldsymbol{x}'/2\rangle. \tag{6.30}$$

Since the Wigner transform preserves the adjoint operation, it follows from the Hermiticity of $\hat{\rho}$ that $f^W(\boldsymbol{x},\boldsymbol{p})$ is a real function. However, it does not follow from the positive definiteness of $\hat{\rho}$ that the Wigner distribution is necessarily positive, and in general this is not the case. But in the special cases where $\hat{\rho}$ is diagonal in the position or momentum basis, $f^W(\boldsymbol{x},\boldsymbol{p})$ will be positive, since in those cases we respectively have $f^W(\boldsymbol{x},\boldsymbol{p}) \sim \langle\boldsymbol{x}|\hat{\rho}|\boldsymbol{x}\rangle \geq 0$ and $f^W(\boldsymbol{x},\boldsymbol{p}) \sim \langle\boldsymbol{p}|\hat{\rho}|\boldsymbol{p}\rangle \geq 0$.

From (6.28) together with our choice of normalization $\operatorname{Tr}\hat{\rho} = N_f$, we find

$$\frac{1}{h^3}\int \mathrm{d}^3\boldsymbol{x}\int \mathrm{d}^3\boldsymbol{p}\, f^W(\boldsymbol{x},\boldsymbol{p}) = N_f. \tag{6.31}$$

Further, by (6.27) we have for any observable $A$

$$\langle A\rangle = \operatorname{Tr}\hat{\rho}\hat{A} = \frac{1}{h^3}\int \mathrm{d}^3\boldsymbol{x}\int \mathrm{d}^3\boldsymbol{p}\, f^W(\boldsymbol{x},\boldsymbol{p})A^W(\boldsymbol{x},\boldsymbol{p}). \tag{6.32}$$

Thus, the Wigner function plays the role of a probability distribution in phase space. However, $f^W$ is not fully interpretable as a probability distribution, since as mentioned it does not generally have to be positive.

### 6.2.3 The Moyal product and the Moyal expansion

In order to express equations of motion for the Wigner distribution, we will need to find the Wigner transform of a product of Weyl transforms. That is, if $\hat{A}$ and $\hat{B}$ are respectively the Weyl transforms of $A(\boldsymbol{x}, \boldsymbol{p})$ and $B(\boldsymbol{x}, \boldsymbol{p})$, we need an expression for the Wigner transform of $\hat{A}\hat{B}$. This object is referred to as the Moyal product of the functions $A$ and $B$, and denoted $A \star B$[2, 10, 25]. Making use of (6.23) and (6.25), and then repeated use of the delta function normalization of the position basis and the Fourier representation of the delta function to integrate away the dummy variables, we eventually end up with the integral representation

$$\begin{aligned}&(A \star B)(\boldsymbol{x}, \boldsymbol{p}) \\ &= \frac{16}{h^6} \iint \mathrm{d}^3\boldsymbol{y}\mathrm{d}^3\boldsymbol{q}\mathrm{d}^3\boldsymbol{r}\mathrm{d}^3\boldsymbol{q}'\, e^{2i(\boldsymbol{q}'-\boldsymbol{p})\cdot(\boldsymbol{y}-\boldsymbol{x})/\hbar - 2i(\boldsymbol{q}-\boldsymbol{p})\cdot(\boldsymbol{r}-\boldsymbol{x})/\hbar} A(\boldsymbol{y}, \boldsymbol{q}) B(\boldsymbol{r}, \boldsymbol{q}')\end{aligned} \tag{6.33}$$

of the Moyal product. If either both $A$ and $B$ are independent of $\boldsymbol{x}$, or both $A$ and $B$ are independent of $\boldsymbol{p}$, then the Moyal product simplifies to an ordinary product, so that $(A \star B)(\boldsymbol{x}, \boldsymbol{p}) = A(\boldsymbol{x}, \boldsymbol{p})B(\boldsymbol{x}, \boldsymbol{p})$.

In a closed system the density matrix obeys the Von Neumann equation

$$\frac{\mathrm{d}\hat{\rho}}{\mathrm{d}t} = -\frac{i}{\hbar}[\hat{H}, \hat{\rho}]. \tag{6.34}$$

Taking the Wigner transform of both sides of this equation, we obtain the equation

$$\frac{\partial f^W}{\partial t} = -\frac{i}{\hbar}(H \star f^W - f^W \star H) = \{\{H, f^W\}\}, \tag{6.35}$$

where $H$ is the Wigner transform of the Hamiltonian $\hat{H}$, and we have introduced the Moyal bracket, defined by the expression

$$\{\{A, B\}\} = \frac{A \star B - B \star A}{i\hbar}. \tag{6.36}$$

If the system instead obeys the master equation (6.12), we obtain the modified equation

$$\frac{\partial f^W}{\partial t} = \{\{H, f^W\}\} - \frac{1}{2}\left(\Gamma^W \star f^W + f^W \star \Gamma^W\right) + S[f^W], \tag{6.37}$$

where $\Gamma^W$ and $S[f^W]$ are respectively the Wigner transforms of the operators $\Gamma$ and $S(\rho)$ appearing in (6.12).

Systematic approximations to these equations can be obtained by expressing the Moyal product in terms of a differential expansion. To find this expansion, we start by Taylor expanding the functions $A$ and $B$ in $\boldsymbol{x}$:

$$A(\boldsymbol{x} + \boldsymbol{y}, \boldsymbol{q}) = \sum_{n=0}^{\infty} \frac{1}{n!} \sum_{i_1 \cdots i_n} \frac{\partial^n A}{\partial x_{i_1} \cdots \partial x_{i_n}}(\boldsymbol{x}, \boldsymbol{q}) y_{i_1} y_{i_2} \cdots y_{i_n}, \tag{6.38}$$

and similarly for $B$

$$B(\boldsymbol{x}+\boldsymbol{r},\boldsymbol{q}') = \sum_{m=0}^{\infty} \frac{1}{m!} \sum_{i_1\cdots i_m} \frac{\partial^m B}{\partial x_{i_1}\cdots \partial x_{i_m}}(\boldsymbol{x},\boldsymbol{q}') r_{i_1} r_{i_2}\cdots r_{i_m}. \tag{6.39}$$

At this point it becomes highly simplifying to introduce left acting differential operators. Thus, we define operators $\frac{\overleftarrow{\partial}}{\partial x}$ by the relation

$$f(x)\frac{\overleftarrow{\partial}}{\partial x} = \frac{\partial}{\partial x} f(x). \tag{6.40}$$

To make the distinction as clear as possible, we will in this context also write the ordinary right acting differential operator as $\frac{\overrightarrow{\partial}}{\partial x}$. Thus, for functions $f(x)$ and $g(x)$, we have

$$f(x)\frac{\overleftarrow{\partial}}{\partial x} g(x) = \frac{\partial f}{\partial x}(x) g(x), \qquad \text{while} \tag{6.41}$$

$$f(x)\frac{\overrightarrow{\partial}}{\partial x} g(x) = f(x)\frac{\partial g}{\partial x}(x). \tag{6.42}$$

With these definitions, one can start with (6.38) and verify that

$$e^{2i(\boldsymbol{q}'-\boldsymbol{p})\cdot\boldsymbol{y}/\hbar} A(\boldsymbol{x}+\boldsymbol{y},\boldsymbol{q}) = \sum_{n=0}^{\infty} \frac{1}{n!} A(\boldsymbol{x},\boldsymbol{q}) \left(\frac{i\hbar}{2}\frac{\overleftarrow{\partial}}{\partial \boldsymbol{x}}\cdot\frac{\overrightarrow{\partial}}{\partial \boldsymbol{p}}\right)^n e^{2i(\boldsymbol{q}'-\boldsymbol{p})\cdot\boldsymbol{y}/\hbar}, \tag{6.43}$$

while similarly, one can use (6.39) to verify that

$$e^{-2i(\boldsymbol{q}-\boldsymbol{p})\cdot\boldsymbol{r}/\hbar} B(\boldsymbol{x}+\boldsymbol{r},\boldsymbol{q}') = \sum_{m=0}^{\infty} \frac{1}{m!} e^{-2i(\boldsymbol{q}-\boldsymbol{p})\cdot\boldsymbol{r}/\hbar} \left(-\frac{i\hbar}{2}\frac{\overleftarrow{\partial}}{\partial \boldsymbol{p}}\cdot\frac{\overrightarrow{\partial}}{\partial \boldsymbol{x}}\right)^m B(\boldsymbol{x},\boldsymbol{q}'). \tag{6.44}$$

Inserting both these expressions in (6.33), and then again making use of the Fourier representation of the delta function to perform the remaining integrals, we obtain the differential expansion

$$\begin{aligned} &(A\star B)(\boldsymbol{x},\boldsymbol{p}) \qquad (6.45) \\ &= \sum_{n=0}^{\infty}\sum_{m=0}^{\infty} \frac{1}{n!m!} A(\boldsymbol{x},\boldsymbol{p}) \left(\frac{i\hbar}{2}\frac{\overleftarrow{\partial}}{\partial \boldsymbol{x}}\cdot\frac{\overrightarrow{\partial}}{\partial \boldsymbol{p}}\right)^n \left(-\frac{i\hbar}{2}\frac{\overleftarrow{\partial}}{\partial \boldsymbol{p}}\cdot\frac{\overrightarrow{\partial}}{\partial \boldsymbol{x}}\right)^m B(\boldsymbol{x},\boldsymbol{p}). \end{aligned}$$

Reindexing the sum, and using the binomial theorem, we can simplify the expression further as

$$\begin{aligned} &(A\star B)(\boldsymbol{x},\boldsymbol{p}) \qquad (6.46) \\ &= \sum_{N=0}^{\infty} \frac{1}{N!} A(\boldsymbol{x},\boldsymbol{p}) \sum_{m=0}^{N} \binom{N}{m} \left(\frac{i\hbar}{2}\frac{\overleftarrow{\partial}}{\partial \boldsymbol{x}}\cdot\frac{\overrightarrow{\partial}}{\partial \boldsymbol{p}}\right)^{N-m} \left(-\frac{i\hbar}{2}\frac{\overleftarrow{\partial}}{\partial \boldsymbol{p}}\cdot\frac{\overrightarrow{\partial}}{\partial \boldsymbol{x}}\right)^m B(\boldsymbol{x},\boldsymbol{p}) \\ &= \sum_{N=0}^{\infty} \frac{1}{N!} A(\boldsymbol{x},\boldsymbol{p}) \left(\frac{i\hbar}{2}\right)^N \left(\frac{\overleftarrow{\partial}}{\partial \boldsymbol{x}}\cdot\frac{\overrightarrow{\partial}}{\partial \boldsymbol{p}} - \frac{\overleftarrow{\partial}}{\partial \boldsymbol{p}}\cdot\frac{\overrightarrow{\partial}}{\partial \boldsymbol{x}}\right)^N B(\boldsymbol{x},\boldsymbol{p}). \end{aligned}$$

Thus, interpreting the exponential function as its Taylor expansion, we have the formal identity

$$A \star B = A \left[ \exp \frac{i\hbar}{2} \left( \frac{\overleftarrow{\partial}}{\partial \boldsymbol{x}} \cdot \frac{\overrightarrow{\partial}}{\partial \boldsymbol{p}} - \frac{\overleftarrow{\partial}}{\partial \boldsymbol{p}} \cdot \frac{\overrightarrow{\partial}}{\partial \boldsymbol{x}} \right) \right] B, \tag{6.47}$$

in agreement with typical literature expressions[2]. To save some space we now introduce the short notation $\overleftarrow{\partial}_{\boldsymbol{q}} \overrightarrow{\partial}_{\boldsymbol{r}} = \overrightarrow{\partial}_{\boldsymbol{r}} \overleftarrow{\partial}_{\boldsymbol{q}} = \frac{\overleftarrow{\partial}}{\partial \boldsymbol{q}} \cdot \frac{\overrightarrow{\partial}}{\partial \boldsymbol{r}} = \frac{\overrightarrow{\partial}}{\partial \boldsymbol{r}} \cdot \frac{\overleftarrow{\partial}}{\partial \boldsymbol{q}}$. (6.47) can then be written as $A \star B = A\, e^{\frac{i\hbar}{2}(\overleftarrow{\partial}_{\boldsymbol{x}} \overrightarrow{\partial}_{\boldsymbol{p}} - \overleftarrow{\partial}_{\boldsymbol{p}} \overrightarrow{\partial}_{\boldsymbol{x}})} B$. The reverse Moyal product $B \star A$ can then be evaluated as

$$\begin{aligned} B \star A &= B\, e^{\frac{i\hbar}{2}(\overleftarrow{\partial}_{\boldsymbol{x}} \overrightarrow{\partial}_{\boldsymbol{p}} - \overleftarrow{\partial}_{\boldsymbol{p}} \overrightarrow{\partial}_{\boldsymbol{x}})} A = A\, e^{\frac{i\hbar}{2}(\overrightarrow{\partial}_{\boldsymbol{x}} \overleftarrow{\partial}_{\boldsymbol{p}} - \overrightarrow{\partial}_{\boldsymbol{p}} \overleftarrow{\partial}_{\boldsymbol{x}})} B \\ &= A\, e^{\frac{i\hbar}{2}(\overleftarrow{\partial}_{\boldsymbol{p}} \overrightarrow{\partial}_{\boldsymbol{x}} - \overleftarrow{\partial}_{\boldsymbol{x}} \overrightarrow{\partial}_{\boldsymbol{p}})} B = A\, e^{-\frac{i\hbar}{2}(\overleftarrow{\partial}_{\boldsymbol{x}} \overrightarrow{\partial}_{\boldsymbol{p}} - \overleftarrow{\partial}_{\boldsymbol{p}} \overrightarrow{\partial}_{\boldsymbol{x}})} B. \end{aligned} \tag{6.48}$$

Accordingly, the Moyal bracket of (6.36) can be written as

$$\begin{aligned} \{\{A, B\}\} &= A \frac{e^{\frac{i\hbar}{2}(\overleftarrow{\partial}_{\boldsymbol{x}} \overrightarrow{\partial}_{\boldsymbol{p}} - \overleftarrow{\partial}_{\boldsymbol{p}} \overrightarrow{\partial}_{\boldsymbol{x}})} - e^{-\frac{i\hbar}{2}(\overleftarrow{\partial}_{\boldsymbol{x}} \overrightarrow{\partial}_{\boldsymbol{p}} - \overleftarrow{\partial}_{\boldsymbol{p}} \overrightarrow{\partial}_{\boldsymbol{x}})}}{i\hbar} B \\ &= \frac{2}{\hbar} A \left( \sin \frac{\hbar}{2} (\overleftarrow{\partial}_{\boldsymbol{x}} \overrightarrow{\partial}_{\boldsymbol{p}} - \overleftarrow{\partial}_{\boldsymbol{p}} \overrightarrow{\partial}_{\boldsymbol{x}}) \right) B, \end{aligned} \tag{6.49}$$

where the sine function is also to be interpreted as its Taylor expansion. In particular, the Wigner transformed Von Neumann equation (6.35) then becomes

$$\frac{\partial f^W}{\partial t} = \frac{2}{\hbar} H \left( \sin \frac{\hbar}{2} (\overleftarrow{\partial}_{\boldsymbol{x}} \overrightarrow{\partial}_{\boldsymbol{p}} - \overleftarrow{\partial}_{\boldsymbol{p}} \overrightarrow{\partial}_{\boldsymbol{x}}) \right) f^W. \tag{6.50}$$

The Moyal anti-commutator $A \star B + B \star A$ can be evaluated in a similar manor, yielding

$$\frac{A \star B + B \star A}{2} = A \left( \cos \frac{\hbar}{2} (\overleftarrow{\partial}_{\boldsymbol{x}} \overrightarrow{\partial}_{\boldsymbol{p}} - \overleftarrow{\partial}_{\boldsymbol{p}} \overrightarrow{\partial}_{\boldsymbol{x}}) \right) B, \tag{6.51}$$

and accordingly we can write the Wigner transformed master equation (6.37) as

$$\begin{aligned} \frac{\partial f^W}{\partial t} &= \frac{2}{\hbar} H \Big( \sin \frac{\hbar}{2} (\overleftarrow{\partial}_{\boldsymbol{x}} \overrightarrow{\partial}_{\boldsymbol{p}} - \overleftarrow{\partial}_{\boldsymbol{p}} \overrightarrow{\partial}_{\boldsymbol{x}}) \Big) f^W \\ &\quad - \Gamma^W \Big( \cos \frac{\hbar}{2} (\overleftarrow{\partial}_{\boldsymbol{x}} \overrightarrow{\partial}_{\boldsymbol{p}} - \overleftarrow{\partial}_{\boldsymbol{p}} \overrightarrow{\partial}_{\boldsymbol{x}}) \Big) f^W + S[f^W]. \end{aligned} \tag{6.52}$$

## 6.3 Multi-band systems

### 6.3.1 Bloch states, Wannier states and Envelope transformations

The purpose of this section is to generalize the Wigner transform to multi-band systems. This will be done by utilizing a technique known as envelope transformations or envelope functions, which will seamlessly merge the concept of band structures with the formalism introduced above. Since band structure arises in systems with a periodic potential, we begin by discussing such systems, and some mathematics which naturally arises there.

Consider a single particle in tree dimensions, having the Hamiltonian

$$\hat{H} = \frac{\hat{\boldsymbol{p}}^2}{2m} + V(\hat{\boldsymbol{x}}), \tag{6.53}$$

where $V(\boldsymbol{x})$ is a potential satisfying the periodic conditions

$$V(\boldsymbol{x}) = V(\boldsymbol{x} + \boldsymbol{b}_i), \qquad i \in \{1,2,3\}. \tag{6.54}$$

One can then show that the eigenstates $|\psi\rangle$ of the system must themselves satisfy[11]

$$\langle \boldsymbol{x} + \boldsymbol{b}_i|\psi\rangle = e^{i\boldsymbol{k}\cdot\boldsymbol{b}_i}\langle \boldsymbol{x}|\psi\rangle, \qquad i \in \{1,2,3\}. \tag{6.55}$$

Such states are known as Bloch states, and the vector $\boldsymbol{k}$ as the vector of Bloch indices, or alternatively, as the crystal momentum. The Bloch states are denoted $|n, \boldsymbol{k}\rangle$, where $n$ is the band index, needed to distinguish different Bloch states with the same crystal momentum.

Starting with a basis of Bloch states, one can define so called Wannier states[2] as

$$|n, \boldsymbol{j}\rangle = \int_B \frac{\mathrm{d}^3\boldsymbol{k}}{\sqrt{V_B}} e^{-i\boldsymbol{k}\cdot\boldsymbol{x}_{\boldsymbol{j}} + i\theta(\boldsymbol{k})}|n, \boldsymbol{k}\rangle, \tag{6.56}$$

where $\boldsymbol{j}$ is a vector of integers, $\boldsymbol{x}_{\boldsymbol{j}} = j_1\boldsymbol{b}_1 + j_2\boldsymbol{b}_2 + j_3\boldsymbol{b}_3$, $B = \{\boldsymbol{k} : |\boldsymbol{k}\cdot\boldsymbol{b}_i| \le \pi\}$, $V_B$ is the volume of $B$, and $\theta$ is some function of $\boldsymbol{k}$. The definition can be inverted to obtain the relation

$$\frac{1}{\sqrt{V_B}}\sum_{\boldsymbol{j}} e^{i\boldsymbol{k}'\cdot\boldsymbol{x}_{\boldsymbol{j}} - i\theta(\boldsymbol{k}')}|n, \boldsymbol{j}\rangle = \sum_{\boldsymbol{j}}\int_B \frac{\mathrm{d}^3\boldsymbol{k}}{V_B} e^{i(\boldsymbol{k}'-\boldsymbol{k})\cdot\boldsymbol{x}_{\boldsymbol{j}} - i\theta(\boldsymbol{k}') + i\theta(\boldsymbol{k})}|n, \boldsymbol{k}\rangle = |n, \boldsymbol{k}'\rangle. \tag{6.57}$$

Using the defining property of the Bloch states, we find that the Wannier states satisfy the relation

$$\langle \boldsymbol{x}|n, \boldsymbol{j}\rangle = \int_B \frac{\mathrm{d}^3\boldsymbol{k}}{\sqrt{V_B}} e^{-i\boldsymbol{k}\cdot\boldsymbol{x}_{\boldsymbol{j}} + i\theta(\boldsymbol{k}) - i\boldsymbol{k}\cdot\boldsymbol{x}_{\boldsymbol{j}'-\boldsymbol{j}}}\langle \boldsymbol{x} + \boldsymbol{x}_{\boldsymbol{j}'-\boldsymbol{j}}|n, \boldsymbol{k}\rangle = \langle \boldsymbol{x} + \boldsymbol{x}_{\boldsymbol{j}'-\boldsymbol{j}}|n, \boldsymbol{j}'\rangle, \tag{6.58}$$

meaning that different Wannier states of any single band are related by translation.

If we assume the Bloch states to be delta function normalized, so that $\langle n, \boldsymbol{k}|m, \boldsymbol{k}'\rangle = \delta_{nm}\delta(\boldsymbol{k} - \boldsymbol{k}')$ and $I = \sum_n \int_B \mathrm{d}^3\boldsymbol{k}\, |n, \boldsymbol{k}\rangle\langle n, \boldsymbol{k}|$, then making use of (6.56) and its inverse (6.57), it is straight forward to show that

$$I = \sum_{n\boldsymbol{j}} |n, \boldsymbol{j}\rangle\langle n, \boldsymbol{j}|, \tag{6.59}$$

while $\langle n, \boldsymbol{j}' | m, \boldsymbol{j} \rangle = \delta_{nm}\delta_{\boldsymbol{j}\boldsymbol{j}'}$. Thus, the Wannier states form an orthonormal basis. In particular, the fact that they are orthonormal means they must be localized, and accordingly the Wannier state basis is a basis of localized states, which is invariant under translations that respect the periodic symmetry.

Finally, we define envelope operators $\hat{E}_n$ as

$$\hat{E}_n = \hbar^{3/2} \int_B \mathrm{d}^3\boldsymbol{k}\, e^{-i\theta(\boldsymbol{k})} |\hbar\boldsymbol{k}\rangle\langle n, \boldsymbol{k}|, \tag{6.60}$$

where $|\hbar\boldsymbol{k}\rangle$ is the momentum state with momentum $\boldsymbol{p} = \hbar\boldsymbol{k}$. Given a state $|\psi\rangle$, we refer to $\hat{E}_n|\psi\rangle$ as an envelope transformation of $|\psi\rangle$. The functions $\psi_n^E(\boldsymbol{x}) = \langle \boldsymbol{x}|\hat{E}_n|\psi\rangle$ are referred to as envelope functions[4]. These satisfy

$$\psi_n^E(\boldsymbol{x}) = \hbar^{3/2} \int_B \mathrm{d}^3\boldsymbol{k}\, e^{-i\theta(\boldsymbol{k})} \langle \boldsymbol{x}|\hbar\boldsymbol{k}\rangle\langle n, \boldsymbol{k}|\psi\rangle = \int_B \frac{\mathrm{d}^3\boldsymbol{k}}{(2\pi)^{\frac{3}{2}}}\, e^{i\boldsymbol{k}\cdot\boldsymbol{x} - i\theta(\boldsymbol{k})} \langle n, \boldsymbol{k}|\psi\rangle. \tag{6.61}$$

Thus, the envelope functions are in fact Fourier transforms of functions which are zero for $\boldsymbol{k} \notin B$. Crucially, this means that the envelope functions are almost always more smooth than the wave function $\psi(\boldsymbol{x}) = \langle \boldsymbol{x}|\psi\rangle$. Further, we have in particular

$$\psi_n^E(\boldsymbol{x_j}) = \int_B \frac{\mathrm{d}^3\boldsymbol{k}}{(2\pi)^{\frac{3}{2}}}\, e^{i\boldsymbol{k}\cdot\boldsymbol{x_j} - i\theta(\boldsymbol{k})} \langle n, \boldsymbol{k}|\psi\rangle = \sqrt{\frac{V_B}{2\pi}} \langle n, \boldsymbol{j}|\psi\rangle. \tag{6.62}$$

This tells us both that there is an important relationship between envelope transformations and Wannier states, and by (6.59) also that all information about the wave function is contained in the value of the envelope functions at the discrete set of points $\boldsymbol{x_j}$.

It is easily seen that the envelope operators satisfy $\hat{E}_n^\dagger \hat{E}_n = \hat{P}_n$ and $\hat{E}_n \hat{E}_m^\dagger = \delta_{nm}\hat{P}_B$, where $\hat{P}_n$ and $\hat{P}_B$ are projection operators such that $\hat{P}_n|m, \boldsymbol{k}\rangle = \delta_{nm}|m, \boldsymbol{k}\rangle$, while $\hat{P}_B|\boldsymbol{p}\rangle = |\boldsymbol{p}\rangle$ if $\boldsymbol{p}/\hbar \in B$ and zero otherwise. We then also have

$$\sum_n \hat{E}_n^\dagger \hat{E}_n = I. \tag{6.63}$$

Given an operator $\hat{A}$, we refer to $\hat{E}_n \hat{A} \hat{E}_m^\dagger$ as an envelope transformation of $\hat{A}$. Using (6.63) we have

$$\mathrm{Tr}\,\hat{A} = \sum_n \mathrm{Tr}\,\hat{E}_n^\dagger \hat{E}_n \hat{A} = \sum_n \mathrm{Tr}\,\hat{E}_n \hat{A} \hat{E}_n^\dagger. \tag{6.64}$$

Further, we also have

$$\hat{E}_n \hat{A} \hat{B} \hat{E}_m^\dagger = \sum_l \hat{E}_n \hat{A} \hat{E}_l^\dagger \hat{E}_l \hat{B} \hat{E}_m^\dagger. \tag{6.65}$$

Combining (6.64) and (6.65) we get

$$\operatorname{Tr} \hat{A}\hat{B} = \sum_{nm} \operatorname{Tr} \hat{E}_n \hat{A} \hat{E}_m^\dagger \hat{E}_m \hat{B} \hat{E}_n^\dagger \tag{6.66}$$

so that in particular

$$\langle A \rangle = \operatorname{Tr} \hat{A}\hat{\rho} = \sum_{nm} \operatorname{Tr} \hat{E}_n \hat{\rho} \hat{E}_m^\dagger \hat{E}_m \hat{A} \hat{E}_n^\dagger. \tag{6.67}$$

### 6.3.2 The multi-band Wigner transformation

We now define a multi-band version of the Wigner transform, as the composition of the standard Wigner transform with an envelope transformation. That is, given an operator $\hat{A}$, we combine (6.60) and (6.26), defining

$$\begin{aligned} A^w_{nm}(\boldsymbol{x},\boldsymbol{k}) &= (\hat{E}_n \hat{A} \hat{E}_m^\dagger)^W(\boldsymbol{x}, \hbar\boldsymbol{k}) = \int \mathrm{d}^3\boldsymbol{p}'\, e^{i\boldsymbol{p}'\cdot\boldsymbol{x}/\hbar} \langle \hbar\boldsymbol{k} + \boldsymbol{p}'/2 | \hat{E}_n \hat{A} \hat{E}_m^\dagger | \hbar\boldsymbol{k} - \boldsymbol{p}'/2 \rangle \\ &= 8 \iint_{B^2} \mathrm{d}^3\boldsymbol{k}' \mathrm{d}^3\boldsymbol{k}''\, e^{i(\boldsymbol{k}-\boldsymbol{k}'')\cdot\boldsymbol{x}/\hbar} \langle n, \boldsymbol{k}' | \hat{A} | m, \boldsymbol{k}'' \rangle \delta(\boldsymbol{k}' + \boldsymbol{k}'' - 2\boldsymbol{k}). \end{aligned} \tag{6.68}$$

We also define $A^w(\boldsymbol{x},\boldsymbol{k})$ as the matrix with elements $A^w_{nm}(\boldsymbol{x},\boldsymbol{k})$. Similar definitions, as well as analogs to the derivations below, can be found in the literature[3].

Like the standard Wigner transform, the multi-band Wigner transform respects the adjoint operation. Indeed, $(A^\dagger)^w_{nm} = (\hat{E}_n \hat{A}^\dagger \hat{E}_m^\dagger)^W = ((\hat{E}_m \hat{A} \hat{E}_n^\dagger)^\dagger)^W = (\hat{E}_m \hat{A} \hat{E}_n^\dagger)^{W\star} = A^{w\star}_{mn} = (A^{w\dagger})_{nm}$. Further, for $\boldsymbol{k}', \boldsymbol{k}'' \in B$ we have $|(\boldsymbol{k}'+\boldsymbol{k}'')\cdot\boldsymbol{b}_i| \le |\boldsymbol{k}'\cdot\boldsymbol{b}_i| + |\boldsymbol{k}'\cdot\boldsymbol{b}_i| \le 2\pi$, so that also $(\boldsymbol{k}'+\boldsymbol{k}'')/2 \in B$. Thus, because of the delta function in (6.68), we must have $A^w(\boldsymbol{x},\boldsymbol{k}) = 0$ whenever $\boldsymbol{k} \notin B$. Making use of this together with (6.64) and (6.28), we get

$$\operatorname{Tr} \hat{A} = \sum_n \int \mathrm{d}^3\boldsymbol{x} \int \frac{\mathrm{d}^3\boldsymbol{k}}{(2\pi)^3} A^w_{nn}(\boldsymbol{x},\boldsymbol{k}) = \int \mathrm{d}^3\boldsymbol{x} \int_B \frac{\mathrm{d}^3\boldsymbol{k}}{(2\pi)^3} \operatorname{Tr} A^w(\boldsymbol{x},\boldsymbol{k}). \tag{6.69}$$

Similarly, we can combine (6.66) and (6.27) to get

$$\operatorname{Tr} \hat{A}\hat{B} = \int \mathrm{d}^3\boldsymbol{x} \int_B \frac{\mathrm{d}^3\boldsymbol{k}}{(2\pi)^3} \operatorname{Tr} A^w(\boldsymbol{x},\boldsymbol{k}) B^w(\boldsymbol{x},\boldsymbol{k}). \tag{6.70}$$

Finally, using (6.65) we have

$$\begin{aligned} (\hat{A}\hat{B})^w_{nm} = (\hat{E}_n \hat{A}\hat{B} \hat{E}_m)^W &= \sum_l (\hat{E}_n \hat{A} \hat{E}_l^\dagger \hat{E}_l \hat{B} \hat{E}_m^\dagger)^W \\ &= \sum_l (\hat{E}_n \hat{A} \hat{E}_l^\dagger)^W \star (\hat{E}_l \hat{B} \hat{E}_m^\dagger)^W = \sum_l A^w_{nl} \star B^w_{lm}. \end{aligned} \tag{6.71}$$

Thus, defining a matrix Moyal product by $(A \star B)_{nm} = \sum_l A_{nl} \star B_{lm}$, we have

$$(\hat{A}\hat{B})^w = A^w \star B^w. \tag{6.72}$$

Making use of the differential expansion (6.47), we can write the matrix Moyal product as

$$(A \star B)_{nm} = \sum_l A_{nl} \star B_{lm} = \sum_l A_{nl} e^{\frac{i}{2}(\overleftarrow{\partial}_{\boldsymbol{x}} \overrightarrow{\partial}_{\boldsymbol{k}} - \overleftarrow{\partial}_{\boldsymbol{k}} \overrightarrow{\partial}_{\boldsymbol{x}})} B_{lm} \tag{6.73}$$
$$= (A\, e^{\frac{i}{2}(\overleftarrow{\partial}_{\boldsymbol{x}} \overrightarrow{\partial}_{\boldsymbol{k}} - \overleftarrow{\partial}_{\boldsymbol{k}} \overrightarrow{\partial}_{\boldsymbol{x}})} B)_{nm}.$$

However, since the ordinary matrix product is not commutative, the matrix Moyal bracket does not take the simple form of (6.49). But under the assumption that both $A$ and $B$ are Hermitian, we can instead write it as

$$\{\{A, B\}\} = \frac{A e^{\frac{i}{2}(\overleftarrow{\partial}_{\boldsymbol{x}} \overrightarrow{\partial}_{\boldsymbol{k}} - \overleftarrow{\partial}_{\boldsymbol{k}} \overrightarrow{\partial}_{\boldsymbol{x}})} B - B e^{\frac{i}{2}(\overleftarrow{\partial}_{\boldsymbol{x}} \overrightarrow{\partial}_{\boldsymbol{k}} - \overleftarrow{\partial}_{\boldsymbol{k}} \overrightarrow{\partial}_{\boldsymbol{x}})} A}{i\hbar} \tag{6.74}$$
$$= \frac{A e^{\frac{i}{2}(\overleftarrow{\partial}_{\boldsymbol{x}} \overrightarrow{\partial}_{\boldsymbol{k}} - \overleftarrow{\partial}_{\boldsymbol{k}} \overrightarrow{\partial}_{\boldsymbol{x}})} B - (A e^{\frac{i}{2}(\overleftarrow{\partial}_{\boldsymbol{x}} \overrightarrow{\partial}_{\boldsymbol{k}} - \overleftarrow{\partial}_{\boldsymbol{k}} \overrightarrow{\partial}_{\boldsymbol{x}})} B)^\dagger}{i\hbar} = \frac{2}{\hbar} \mathcal{I}m\, A e^{\frac{i}{2}(\overleftarrow{\partial}_{\boldsymbol{x}} \overrightarrow{\partial}_{\boldsymbol{k}} - \overleftarrow{\partial}_{\boldsymbol{k}} \overrightarrow{\partial}_{\boldsymbol{x}})} B,$$

where for any matrix $X$ we define $\mathcal{I}m\, X$ as the anti-Hermitian component $(X - X^\dagger)/2i$.

Generalizing the proceedings of Section 6.2.2 in the obvious manor, we define the multi-band Wigner distribution as the multi-band Wigner transform of the density operator:

$$\rho^w_{nm}(\boldsymbol{x}, \boldsymbol{k}) = \int \mathrm{d}^3\boldsymbol{p}'\, e^{i\boldsymbol{p}'\cdot\boldsymbol{x}/\hbar} \langle \hbar\boldsymbol{k} + \boldsymbol{p}'/2 | \hat{E}_n \hat{\rho} \hat{E}^\dagger_m | \hbar\boldsymbol{k} - \boldsymbol{p}'/2 \rangle. \tag{6.75}$$

Since the multi-band Wigner transform respects the adjoint, $\rho^w$ is a Hermitian matrix. Further, by (6.69) and (6.70) we have

$$N_f = \int \mathrm{d}^3\boldsymbol{x} \int_B \frac{\mathrm{d}^3\boldsymbol{k}}{(2\pi)^3} \operatorname{Tr} \rho^w(\boldsymbol{x}, \boldsymbol{k}) \qquad \text{and} \tag{6.76}$$
$$\langle A \rangle = \int \mathrm{d}^3\boldsymbol{x} \int_B \frac{\mathrm{d}^3\boldsymbol{k}}{(2\pi)^3} \operatorname{Tr} \rho^w(\boldsymbol{x}, \boldsymbol{k}) A^w(\boldsymbol{x}, \boldsymbol{k}). \tag{6.77}$$

Taking the multi-band Wigner transform of the Von Neumann equation 6.34, and making use of the fact that both $H^w$ and $\rho^w$ are Hermitian, we find

$$\frac{\partial \rho^w}{\partial t} = \{\{H^w, \rho^w\}\} = \frac{2}{\hbar} \mathcal{I}m\, H^w e^{\frac{i}{2}(\overleftarrow{\partial}_{\boldsymbol{x}} \overrightarrow{\partial}_{\boldsymbol{k}} - \overleftarrow{\partial}_{\boldsymbol{k}} \overrightarrow{\partial}_{\boldsymbol{x}})} \rho^w. \tag{6.78}$$

Applying the same procedure to the Master equation (6.12), we find

$$\frac{\partial \rho^w}{\partial t} = \{\{H^w, \rho^w\}\} - \frac{1}{2}\left(\Gamma^w \star \rho^w + \rho^w \star \Gamma^w\right) + S^w[\rho^w] \tag{6.79}$$
$$= \frac{2}{\hbar} \mathcal{I}m\, H^w e^{\frac{i}{2}(\overleftarrow{\partial}_{\boldsymbol{x}} \overrightarrow{\partial}_{\boldsymbol{k}} - \overleftarrow{\partial}_{\boldsymbol{k}} \overrightarrow{\partial}_{\boldsymbol{x}})} \rho^w - \mathcal{R}e\, \Gamma^w e^{\frac{i}{2}(\overleftarrow{\partial}_{\boldsymbol{x}} \overrightarrow{\partial}_{\boldsymbol{k}} - \overleftarrow{\partial}_{\boldsymbol{k}} \overrightarrow{\partial}_{\boldsymbol{x}})} \rho^w + S^w[\rho^w],$$

where we define $\mathcal{R}e\, X$ as the Hermitian component $(X + X^\dagger)/2$.

## 6.4 Classical limit and the Boltzmann equation

In the classical limit $\hbar \to 0$, we identify the Wigner distribution $f^W(\boldsymbol{x}, \boldsymbol{p})$ with the classical distribution function $f(\boldsymbol{x}, \boldsymbol{p})$. Taking the limit $\hbar \to 0$ of the Wigner transformed Von Neumann equation (6.50), amounts to keeping only the terms of the differential expansion that are of order zero in $\hbar$. Thus, we obtain the classical Liouville equation

$$\frac{\partial f}{\partial t} = H(\overleftarrow{\partial}_{\boldsymbol{x}} \overrightarrow{\partial}_{\boldsymbol{p}} - \overleftarrow{\partial}_{\boldsymbol{p}} \overrightarrow{\partial}_{\boldsymbol{x}}) f = \frac{\partial H}{\partial \boldsymbol{x}} \cdot \frac{\partial f}{\partial \boldsymbol{p}} - \frac{\partial H}{\partial \boldsymbol{p}} \cdot \frac{\partial f}{\partial \boldsymbol{x}}. \tag{6.80}$$

Applying the same procedure to (6.52), we obtain the Boltzmann equation

$$\frac{\partial f}{\partial t} = \frac{\partial H}{\partial \boldsymbol{x}} \cdot \frac{\partial f}{\partial \boldsymbol{p}} - \frac{\partial H}{\partial \boldsymbol{p}} \cdot \frac{\partial f}{\partial \boldsymbol{x}} - \Gamma f + S[f], \tag{6.81}$$

where $\Gamma^W$ has also been identified with a classical function $\Gamma$.

Since the terms of order zero in $\hbar$ are also the lowest order terms in the differential expansion, the classical limit amounts to an assumption that all functions are slowly varying with $\boldsymbol{x}$ or $\boldsymbol{p}$. Of these, the easiest to justify is slow variations with $\boldsymbol{x}$. However, in a solid state system this assumption is in fact not justified, since the potential has large variations at an atomic size scale. However, we may get around this by first applying a set of envelope transformations, since these will normally smoothen the state, and if appropriately selected, also the potential.

Applying the Wigner transform after such envelope transformations, we are of course actually applying the multi-band Wigner transformation. Thus, when taking the limit of slow variation with $\boldsymbol{x}$, it is more appropriate to start out with (6.79). Keeping only the terms of differential order zero and two, we obtain

$$\begin{aligned} \frac{\partial \rho^w}{\partial t} &= -\frac{i}{\hbar}[H^w, \rho^w] - \frac{1}{2}\{\Gamma^w, \rho^w\} + \frac{1}{\hbar}\mathcal{R}e\, H^w(\overleftarrow{\partial}_{\boldsymbol{x}} \overrightarrow{\partial}_{\boldsymbol{k}} - \overleftarrow{\partial}_{\boldsymbol{k}} \overrightarrow{\partial}_{\boldsymbol{x}})\rho^w \\ &\quad + \frac{1}{\hbar}\mathcal{I}m\, \Gamma^w(\overleftarrow{\partial}_{\boldsymbol{x}} \overrightarrow{\partial}_{\boldsymbol{k}} - \overleftarrow{\partial}_{\boldsymbol{k}} \overrightarrow{\partial}_{\boldsymbol{x}})\rho^w + S^w[\rho^w]. \end{aligned} \tag{6.82}$$

In order to obtain a multi-band version of the Boltzmann equation, we must transform this equation into a corresponding diagonal format. We do this by making use of a first order perturbation argument. Thus, we begin by making the decompositions $H^w = H_0 + \Delta H$, $\Gamma^w = \Gamma_0 + \Delta\Gamma$ and $\rho^w = \rho_0 + \Delta\rho$, where in all cases the matrix with subscript 0 is the diagonal component of the matrix on the left. We also define $S_0[g]$ as the diagonal component of the functional $S^w[g]$. Then defining the functional

$$\begin{aligned} \Lambda[g] = &-\frac{i}{\hbar}[\Delta H, g] - \frac{1}{2}\{\Delta\Gamma, g\} + S^w[g] - S_0[\rho_0] \\ &+ \frac{1}{\hbar}\mathcal{R}e\, H^w(\overleftarrow{\partial}_{\boldsymbol{x}} \overrightarrow{\partial}_{\boldsymbol{k}} - \overleftarrow{\partial}_{\boldsymbol{k}} \overrightarrow{\partial}_{\boldsymbol{x}})g + \frac{1}{\hbar}\mathcal{I}m\, \Gamma^w(\overleftarrow{\partial}_{\boldsymbol{x}} \overrightarrow{\partial}_{\boldsymbol{k}} - \overleftarrow{\partial}_{\boldsymbol{k}} \overrightarrow{\partial}_{\boldsymbol{x}})g, \end{aligned} \tag{6.83}$$

we can write (6.82) as

$$\frac{\partial \rho^w}{\partial t} = -\frac{i}{\hbar}[H_0, \rho^w] - \frac{1}{2}\{\Gamma_0, \rho^w\} + \Lambda[\rho^w] + S_0[\rho_0]. \tag{6.84}$$

Now we will make the assumption that the magnitude of the functional $\Lambda$ is in some sense small, and the ansatz that this causes the non-diagonal component $\Delta\rho$ to be small as well. Thus, we can make a linear expansion of $S^w[\rho^w]$ in $\Delta\rho$. Further, since both $\Lambda$ and $\Delta\rho$ are small, any term in $\Lambda[\rho^w]$ containing $\Delta\rho$ will in fact be second order, and so to the first order we can make the substitution $\Lambda[\rho^w] = \Lambda[\rho_0]$ in (6.84). In particular, for the non-diagonal element $\rho_{nm}$ of $\rho^w$ we then get

$$\frac{\partial \rho_{nm}}{\partial t} = -\frac{i}{\hbar}(E_n - E_m)\rho_{nm} - \frac{1}{2}(\Gamma_n + \Gamma_m)\rho_{nm} + \Lambda[\rho_0]_{nm}, \tag{6.85}$$

where $E_n$ and $\Gamma_n$ are respectively the elements of the diagonal matrices $H_0$ and $\Gamma_0$. Noting that $\Gamma_n < 0$ would cause an unphysical exponential growth of some $\rho_{nm}$, we will assume $\Gamma_n > 0$. (6.85) then implies that effects of the initial value of $\rho_{nm}$ will quickly die out, and that we will be left with the long term solution

$$\rho_{nm} = \int_{-\infty}^{t} \mathrm{d}t'\, e^{-\frac{i}{\hbar}(E_n - E_m)(t-t') - \frac{1}{2}(\Gamma_n + \Gamma_m)(t-t')}\Lambda(t')_{nm}, \tag{6.86}$$

where we have defined $\Lambda(t) = \Lambda[\rho_0(t)]$. Assuming that the frequency $(E_n - E_m)/\hbar$ is much faster than the temporal variations of $\Lambda$, there will be a significant cancellation between different regions of the integral, and the leading contribution will come from the region $t' \approx t$, where the exponential is decreasing most rapidly. Thus, we may substitute $\Lambda(t') \approx \Lambda(t)$, obtaining

$$\rho_{nm}(t) = \frac{\Lambda(t)_{nm}}{i(E_n - E_m)/\hbar + (\Gamma_n + \Gamma_m)/2}. \tag{6.87}$$

Accordingly, we have verified the ansatz that small values of $\Lambda$ will cause the non-diagonal component of $\rho^w$ to be small, and we have identified the precise condition as $\hbar\Lambda(t)_{nm} \ll |E_n - E_m|$.

Finally, we turn to the diagonal elements of (6.84). Denoting the diagonal elements of $\rho^w$ as $f_n$, the diagonal elements of $S[\rho_0]$ as $S_n[\boldsymbol{f}]$, and keeping in mind that we are still making the substitution $\Lambda[\rho^w] = \Lambda[\rho_0]$, we obtain

$$\frac{\partial f_n}{\partial t} = \frac{1}{\hbar}\frac{\partial E_n}{\partial \boldsymbol{x}} \cdot \frac{\partial f_n}{\partial \boldsymbol{k}} - \frac{1}{\hbar}\frac{\partial E_n}{\partial \boldsymbol{k}} \cdot \frac{\partial f_n}{\partial \boldsymbol{x}} - \Gamma_n f_n + S_n[\boldsymbol{f}], \tag{6.88}$$

which is a multi-band generalization of the Boltzmann equation (6.81). It is common to approximate the scattering term $S_n[\boldsymbol{f}]$ as a functional which is linear and local in $\boldsymbol{x}$. Thus, introducing the integral kernel $\Gamma_{nm}(\boldsymbol{k}, \boldsymbol{k}')$, we have[15]

$$S_n[\boldsymbol{f}](\boldsymbol{x}, \boldsymbol{k}) = \sum_m \int \mathrm{d}^3\boldsymbol{k}'\, \Gamma_{nm}(\boldsymbol{k}, \boldsymbol{k}') f_m(\boldsymbol{x}, \boldsymbol{k}'), \tag{6.89}$$

Inserting this in (6.88) and making use of the fact that $\sum_n \iint \mathrm{d}^3\boldsymbol{k}\mathrm{d}^3\boldsymbol{x} f_n(\boldsymbol{x},\boldsymbol{k})$ is constant, and that the integral of the differential terms cancel by Gauss theorem, we find

$$\sum_n \iint \mathrm{d}^3\boldsymbol{k}\mathrm{d}^3\boldsymbol{x} \left(\Gamma_n(\boldsymbol{x},\boldsymbol{k}) - \sum_m \int \mathrm{d}^3\boldsymbol{k}'\, \Gamma_{mn}(\boldsymbol{k}',\boldsymbol{k})\right) f_n(\boldsymbol{x},\boldsymbol{k}) = 0, \qquad (6.90)$$

and since this must hold for all distributions $f_n$, we must have

$$\Gamma_n(\boldsymbol{x},\boldsymbol{k}) = \sum_m \int \mathrm{d}^3\boldsymbol{k}'\, \Gamma_{mn}(\boldsymbol{k}',\boldsymbol{k}). \qquad (6.91)$$

Inserting this and (6.89) in (6.88), we obtain

$$\begin{aligned} \frac{\partial f_n}{\partial t} &= \frac{1}{\hbar}\frac{\partial E_n}{\partial \boldsymbol{x}} \cdot \frac{\partial f_n}{\partial \boldsymbol{k}} - \frac{1}{\hbar}\frac{\partial E_n}{\partial \boldsymbol{k}} \cdot \frac{\partial f_n}{\partial \boldsymbol{x}} \\ &\quad + \sum_m \int \mathrm{d}^3\boldsymbol{k}' \left(\Gamma_{nm}(\boldsymbol{k},\boldsymbol{k}') f_m(\boldsymbol{x},\boldsymbol{k}') - \Gamma_{mn}(\boldsymbol{k}',\boldsymbol{k}) f_n(\boldsymbol{x},\boldsymbol{k})\right), \end{aligned} \qquad (6.92)$$

which is the standard form of the multi-band Boltzmann equation[15, 11, 22].

Part II

# Progress towards the implementation of a general purpose thermoelectric transport solver for heterostructures

# Chapter 7

# Context of published and submitted works

## 7.1 Motivation

My work has been motivated by the problem of calculating thermoelectric transport coefficients in nanoscale heterostructures. For most thermoelectric applications, the Boltzmann equation (6.92) is relied upon as the appropriate transport formalism. By the discussion of Section 6.4 this is certainly justified in many bulk systems, since one can then find a single envelope transform which transforms the Hamiltonian into a function that is both diagonal and independent of $\boldsymbol{x}$. Indeed, in the bulk case it would usually be the failure of the Markov approximation which would make the Boltzmann equation inappropriate. In cases where the field strength is very high, there could also be problems with the assumption of slow spatial variations, but this is usually not the case when calculating thermoelectric coefficients, since we are typically interested in the linear regime.

In nanoscale heterostructures however, the situation is different. In the case of a non-periodic heterostructure, there will not exist an envelope transform which makes the Hamiltonian independent of position, even when no field is applied. And since the envelope transformed Hamiltonian will have variations on the nanoscale, the step from (6.79) to (6.82) is not justified, since it assumes slow spatial variations. In the case of a periodic heterostructure such as a superlattice, one will be able to find an envelope transform which makes the Hamiltonian independent of position, so (6.82) should in fact apply. However, if the superlattice period is large, the bands of the structure will be very close in energy, so that the requirement $\Lambda(t)_{nm} \ll |E_n - E_m|$ from below (6.87) may fail. Accordingly, in this case the step from (6.82) to (6.88) is not justified.

Thus, the applicability of the Boltzmann equation to the problem at hand is not guarantied. This can also be argued from the discussion in Section 2.3, by which it seems clear that if the size scale of the heterostructure is gradually increased, there will be a transition from the coherent to the incoherent regime. Clearly, the Boltzmann equation will not be able to pick up this transition, since it is not at all concerned with the coherence of the particles it describes. However, the fact that the Boltzmann equation is not strictly justified, does not by itself guarantee that calculations will deviate significantly from other methods.

Because of this, one of the first questions we wanted to answer, was how big of a difference one can expect between a semiclassical model like the Boltzmann equation, and a more general method. Work by Wacker[28] has shown that there can be a substantial difference between the Boltzmann equation and NEGF in superlattice transport. However, his work was mostly concerned with high field applications, and not with the linear regime. In terms of the functional $\Lambda$, Wacker's work shows that the Boltzmann equation can be expected to fail when the first term of the second line of (6.83) is to large. For thermoelectric applications, where we are typically in the linear regime, this term can be assumed to be arbitrarily small, and so we are more concerned with the terms from the first line of (6.83), particularly those involving $\Gamma$ and $S$, which describe scattering.

In order to directly observe whether Wacker's results generalize to the linear regime, we set out to construct a relatively simple transport model, which could be compared to the Boltzmann equation in a straight forward manor. As material system superlattices of Mercury-Cadmium-Telluride was chosen, since this was the system on which we were likely to do the later calculations. As a scattering model, we made use of the Büttiker approximation. This was mostly because this approximation can be implemented in ballistic transport frameworks.

## 7.2 The Büttiker approximation

Within the NEGF framework, the Büttiker approximation can be considered as an approximation to the scattering self energy $\Sigma_s$ introduced in (5.47). The approximation is expressed in terms of the Fourier transformed self energy $\Sigma_s(E)$ as

$$\Sigma_s^r(E) = \sum_\phi \Sigma_{s\phi}^r(E), \quad \text{and} \tag{7.1}$$

$$\Sigma_s^<(E) = \sum_\phi i\Gamma_{s\phi}(E) g_\phi(E), \tag{7.2}$$

where $\Gamma_{s\phi} = i(\Sigma_{s\phi}^r - \Sigma_{s\phi}^{r\dagger})$ is positively definite, and $g_\phi(E) \geq 0$. As we discuss to some length in our third paper (Paper III), it seems highly plausible that $\Gamma_s(E) = i(\Sigma_s^r(E) - \Sigma_s^a(E))$ is positively definite, in which case (7.1) is satisfied with $\{\Sigma_{s\phi}^r\}_\phi = \{\Sigma_s^r\}$. Thus, the approximation consists of the the assumption that the lesser scattering self energy can be expressed as (7.2) with the same set of matrices $\Sigma_{s\phi}^r$. In our third paper we consider an important special case of this, namely the one where the self energies $\Sigma_s^r(E)$ and $\Sigma_s^<(E)$ commute at all energies. In that case we may choose $\Sigma_{s\phi}^r = \theta_\phi |\phi\rangle\langle\phi|$, where the vectors $|\phi\rangle$ make up an orthonormal basis of eigenvectors, and $\theta_\phi$ are the corresponding eigenvalues. Since $\Sigma_s^<(E)$ has the same eigenvector expansion, (7.2) follows immediately.

Combining the Büttiker approximation with (5.63)-(5.67), and also making use of (5.74), we see that the total self energies $\Sigma^r(E)$ and $\Sigma^<(E)$ can be written

as

$$\Sigma^r(E) = \sum_\alpha \Sigma^r_\alpha(E), \quad \text{and} \tag{7.3}$$

$$\Sigma^<(E) = \sum_\alpha i\Gamma_\alpha(E)h_\alpha(E), \tag{7.4}$$

where the summation index $\alpha$ runs both over the leads $p$ and the indices $\phi$, and where $h_p(E) = f_p(E)$ and $h_\phi(E) = g_\phi(E)$. Thus, we can reinterpret the equations as describing a transport problem without scattering, but with an additional set of "fictitious leads" indexed by $\phi$, in which the electrons have arbitrary distributions $g_\phi$ rather than equilibrium distributions $f_\phi$. These fictitious leads are sometimes referred to as Büttiker probes.

Since there is no scattering in the reinterpreted problem, electron motion within the system $S$ can be treated in a coherent framework, while the effects of both the real and fictitious leads can be dealt with by Landauer-Büttiker theory, as discussed in Chapter 2. In fact, it was in precisely that context Büttiker first introduced fictitious leads in order to study the effects of coherence loss in electronic devices[5]. In our case, the fact that we could reinterpret the model as a ballistic problem in this way, meant it could be implemented in the ballistic transport framework Kwant[12, 1].

In order to define a scattering model within the Büttiker approximation, one must specify the self energy operators $\Sigma^r_{s\phi}$, as well as a model of the distribution functions $g_\phi$. In our calculations we set $\Sigma^r_{s\phi}(E) = -i\hbar/2\tau \cdot P_\phi$, where $\tau$ is a constant relaxation time, and $P_\phi$ is a projection operator projecting onto a small group of atoms. Further, we let the scatting mechanism be elastic and local, meaning that at each fictitious lead, the energy resolved current $i_\phi(E) = 0$. This latter condition implicitly defines the distribution functions $g_\phi(E)$.

## 7.3 Discussion of the first two publications

Our first publication (Paper I) is mainly a documentation of calculations we made in order to test our implementation of the Büttiker approximation within Kwant. These tests were done by comparing the calculations to the Boltzmann equation, which we expected to be in agreement with the model in those cases, since the systems were all either bulk or short period super lattices. And indeed, the deviations between the two methods are quite small, and seem to have origins that are understandable. Partially, the deviation originates in the finite value of the scattering time $\tau$, which even in Bulk system will result in small errors. Another major source of discrepancy, is that in order to make the calculations computationally tractable, we had to modify the scattering mechanism, so that the probe operators $P_\phi$ did not cover every atom. However, we believe that both of these discrepancies are under control, as the Büttiker results seem to converge toward the Boltzmann results when the relevant parameters are adjusted.

In our second publication (Paper II), we compare the Büttiker approximation to the Boltzmann equation in superlattices of longer period, and we also study the effect of increasing the scattering time. By the discussion above, these are precisely the systems in which we would expect to observe major discrepancies between these two methods. And indeed this is precisely what we observe. In particular, upon increasing the super lattice period, we do observe a clear transition which is not picked up by the Boltzmann equation. In addition to the Boltzmann equation, we also compare these results to a semiclassical model based upon the incoherent transport expressions of Section 2.3.1, and indeed this is in much better agreement with our results at long periods. Accordingly, the observed transition seems to be a coherent-incoherent transition.

Concluding from this, it seems that if our results are to be reliable, we can not base our calculations on the Boltzmann equation, but must make use of a different formalism. It could be argued, as some of our reviewers did, that the scattering mechanism employed in our calculations is somewhat to simple and specialized to make very broad conclusions about this. However, if nothing else, our results at least show that applicability of the Boltzmann equation is not guarantied, and that considerable care must be taken.

## 7.4 Choice of formalism and method for the general transport framework

Having determined that the Boltzmann equation is not sufficiently general for our purposes, the next step was to find a formalism which is. Considering the discussions of part I, we see that there are at least five different approaches that should be considered. First, one may consider a direct evaluation of the path integral expressions of Chapter 3, with the time discretization size $N$ set to a finite value. Secondly, we could make use of the perturbative NEGF framework, covered in Chapter 5. As a third option, we could calculate the thermoelectric coefficients by using the Kubo relations, or other linear response expressions, as described in Chapter 4. These expressions could also be evaluated directly through path integration, or we could make use of a perturbative expansion of the four point functions. The final option would be to make use of a Markovian master equation, as discussed in Section 6.1.

Let us first consider the question of path integration versus perturbative calculations. Both of these approaches involve approximations. In the path integral approach, the approximation consists of setting a finite value for $N$, while in the perturbative approach the approximation lies in the fact that we can only include a subset of the infinitely many Feynman diagrams. In the path integral approach, the computational burden increases very fast both with $N$ and with the number of included single particle states. Thus, we are limited by the need to keep these parameters at a small value, which will again severely limit the quantitative accuracy of the method. In addition, it is commonly agreed that such non-perturbative techniques are required mostly in cases involving strong correlations, such as super conductors and fractional quantum hall systems.

Next, let us consider the Markovian master equation approach. This would in a sense be the most natural generalization of the Boltzmann equation, since we are only modifying the assumptions of coherence between states, while keeping the assumption of Markovian dynamics. However, in many solid state systems the Markovian approximation is not formally justified. Consider the discussion below (6.5). In the case of phonon scattering, the time scale $\tau_B^r$ can be estimated from the optical phonon frequency, which typically lies in the range $10^{13}$-$10^{14}$ Hz. Thus, we estimate $\tau_B^r \sim 10^{-14}$-$10^{-13}$ s = 10-100 ps. On the other hand, the time scale $\tau_F^<$ satisfies $\tau_F^< \lesssim \tau_p$, the momentum relaxation time. Since $\tau_p$ may also often extend into the range 10-100 ps, the condition $\tau_F^< \gg \tau_B^r$ for Markovian dynamics will often not hold. In addition to this, the computational requirements of the master equation approach is very similar to those of the non-Markovian NEGF method, the only difference being that NEGF requires an additional integral over energy.

Considering the discussion of the previous two paragraphs, it seems clear that the most appropriate formalisms are the two perturbative approaches, NEGF together with the perturbative evaluation of the Kubo relations. In choosing between these two methods, we originally concluded that the approach based on the Kubo relations would be best suited, since this is a linear formalism, and we are interested in linear thermoelectric coefficients. However, before being able to make much progress on this approach, we changed our mind and decided to make use of the NEGF formalism. The reason for this was the increased generality of the NEGF method, the fact that it is formulated in terms of Green's functions rather than four point functions, and thus seems easier to handle, and finally that we figured we could in any case calculate the linear transport coefficients by taking a numerical derivative of the currents. However, in retrospect this choice seems to have been a mistake, as will be discussed further in the concluding chapter (Chapter 9).

Having decided to make use of the NEGF formalism, the next step was to select a method by which to solve the NEGF equations (5.81) and (5.86). We quickly discovered that the RGF method[6] is particularly well suited to solve (5.81) in quasi-one-dimensional systems. However, in the presence of scattering, this equation must be solved self consistently with (5.86), and with expressions for the scattering self energies $\Sigma_s^r$ and $\Sigma_s^<$. In the literature, the approach most commonly applied for this purpose, is to make use of an iterative procedure.

In the general case, the computational scaling of each iteration is $\sim (MN_g)^2$, where $M$ is the dimension of the matrices $G(E)$, and $N_g$ is the size of the integration grid. In a quasi-one-dimensional system $M \sim N_b Z$, where $N_b$ is the number of bands included, and $Z$ is the extent of the system in the $z$ direction. Further, $N_g = N_E N_{kx} N_{ky}$, where $N_E$, $N_{kx}$ and $N_{ky}$ are respectively the sizes of integration grids over energy, and over the Bloch indices $k_x$ and $k_y$. Thus, the full scaling of the method is $\sim N_{itt} N_b^2 Z^2 N_E^2 N_{kx}^2 N_{ky}^2$, with $N_{itt}$ being the number of iterations required for convergence. Accordingly, without the introduction of drastic approximations, the solution of (5.81) and (5.86) is a daunting computational task.

We spent some time experimenting with the benchmarked nanoscale transport software NEMO5[8], which contains routines for the iterative solution of (5.81) and (5.86) within the self consistent Born approximation. However, while this software is probably very well suited for the nanoscale electronic devices for which it is written, it turned out to be difficult to repurpose it for our somewhat different application. In addition, the fact that we required both a large extent in the $z$ direction, as well as a dense integration grid over $k_x$ and $k_y$, makes the scaling of the iterative approach a big problem.

## 7.5 Discussion of the third paper

Accordingly, we decided to abandon both our work on NEMO5, and in fact the iterative approach all together. Instead, we decided to focus on Monte Carlo methods, motivated somewhat by the success of such methods in solving the Boltzmann equation (see Section 8.1 below, and Refs. [15, 22]).

Our third paper (Paper III) is concerned with this subject. The paper describes a Monte Carlo algorithm for evaluating the NEGF lead current given by (5.102). We prove formally that under certain conditions, the expectation value of the estimator equals the lead current. In particular, this is under a particular positivity condition which we show to be satisfied in the self consistent Born approximation. Further, the method is exclusively concerned with the solution of (5.86), which must be assumed to satisfy a certain condition of linearity. In order for this linearity condition to be satisfied, $\Sigma_s^<$ must be a linear function of $G^<$, while $G^r$ can not be a function of $G^<$, so that (5.81) and (5.86) can not be coupled. Accordingly, the linearity requirement is a major limitation of the method. We discuss some options for getting around this. In particular, we argue that with appropriate linearization the method is in any case applicable to linear transport, and we show this explicitly in the case of elastic scattering.

On the other hand, the positivity requirement seems to be an absolute requirement of the method. It is presently unclear under what conditions this is fulfilled beyond the self consistent Born approximation, but we do find some heuristic justification for the assumption in the general case, as long as we specialize to stationary transport.

We also perform some numerical tests of the method. In particular, we calculate conductances of a few short nanowires, and use this to estimate the conductivity in the wires. We compare the results to standard methods like the iterative approach, and find the Monte Carlo results to be in agreement with the alternative methods within a few percent. The relative standard deviation of the Monte Carlo results is also about one percent, so this is as expected.

When it comes to the performance of the method, the nanowire system is not the best suited model for testing this. This is because the Monte Carlo method must be expected to be most competitive in systems with a dependence on a crystal momentum $\boldsymbol{k}$, of which there is none in short nanowires. Indeed, part of the reason for employing the Monte Carlo method in the first place, is to avoid the numerically costly integration grid over $\boldsymbol{k}$, which must be employed by iterative methods. We nevertheless decided to do the first calculations on nanowires, since the introduction of a $\boldsymbol{k}$-dependence complicates the implementation, and we felt that the initial tests should be performed with as simple of an implementation as possible. For the same reason we again made use of Büttikers approximation as the scattering model.

Chapter 8

# Results not submitted for publication

In this chapter I present some calculations done during the course of my work, which are either not sufficiently new or have not matured sufficiently far for publication. Nevertheless, they serve as useful illustrations, and may form important elements of a future general transport framework.

In Section 8.1 I describe some experiments with the Boltzmann Monte Carlo method, which was the first method for transport calculations I pursued. Section 8.1.1 includes results obtained using a direct simulation approach, where currents are estimated directly from the flow of particles through a simulated system. Section 8.1.2 includes a few simple tests of a method where Boltzmann Monte Carlo is combined with the Green-Kubo relations, by which linear transport coefficients can be calculated from correlation functions at equilibrium. While the direct method is applied to superlattices, the Green-Kubo method is only tested in a simple bulk system.

In Section 8.3 I discuss some experiments with a method for making the calculation of the retarded Green's function $G^r$ more efficient. This was work performed before we landed on the RGF method as the best suited method for this purpose in quasi-one-dimensional systems. Put shortly, the method treats exactly only those states which have an energy close to the range of interest, and handles the remaining states using a perturbative expansion.

Finally, in Section 8.2 I present some further results of the NEGF Monte Carlo method, which were omitted from our third paper (Paper III) for reasons of brevity. While the paper illustrates the method on a set of thin nanowires, Section 8.2 includes results obtained from simulations of a small quantum dot, and of a thin film.

## 8.1 Boltzmann Monte Carlo

As discussed above, the Boltzmann equation is only applicable in a fairly limited range when quantitative accuracy is required. However, most of the important physical effects are still at work, so it serves well for illustrative purposes. In fact, two such effects, namely inelastic scattering and a self consistent electrostatic potential, have in this work only been investigated using the Boltzmann Monte Carlo method. In all other calculations presented in this thesis, scattering is assumed elastic, and electrostatic interactions between the charge carriers are ignored.

This is not because these effects are unimportant. As will be shown below, they are in fact very important. Instead, in the case of the NEGF Monte Carlo calculations, it is simply because there are more fundamental complications and problems that must be solved first. In the case of our calculations using Kwant, these were made almost exclusively for the purpose of investigating the coherent-incoherent transition, and accordingly the introduction of further complications seemed unnecessary.

The Boltzmann Monte Carlo method is well described in the literature[22, 15]. Put shortly, it involves the explicit simulation of particles as they move through a device or material according to the classical equations of motion

$$\frac{\mathrm{d}\boldsymbol{x}}{\mathrm{d}t} = \boldsymbol{v} = \frac{1}{\hbar}\frac{\partial E}{\partial \boldsymbol{k}}, \tag{8.1}$$

$$\frac{\mathrm{d}\boldsymbol{k}}{\mathrm{d}t} = \frac{1}{\hbar}\boldsymbol{F} = -\frac{1}{\hbar}\frac{\partial E}{\partial \boldsymbol{x}}, \tag{8.2}$$

interrupted by random discontinuous transformations of the momentum $\boldsymbol{k}$, occurring with a rate given by the scattering rate $\Gamma(\boldsymbol{k}', \boldsymbol{k})$ from (6.92). In a multi-band simulation, these scattering events may also change the band index. It can be shown that this process results in the simulated particles being distributed according to a probability proportional to the distribution function $f(\boldsymbol{x}, \boldsymbol{k})$ which solves the Boltzmann equation[22, 15]. Thus, the process can be used to estimate any measurable quantity that can be expressed in terms of this distribution function.

### 8.1.1 Direct approach

In the direct simulation approach, we directly simulate a charge carrier, or a set of charge carriers, moving according to the procedure described above in the presence of an applied field, or other non-equilibrium perturbation. We may either consider a situation like that described in chapters 2 and 3.2, where some electronic device is connected to a set of leads, or we may simulate some bulk/periodic region of material. In the latter case, one will typically implement periodic boundary conditions, where only a single small region of material is simulated, and the charge carriers are instantly transported to the opposite side whenever they hit a boundary of this region.

Currents can be estimated from the rates by which the carriers exits different leads, or from the average velocity of particles at different locations. The spatial distribution of the particles can be used to calculate variations in charge carrier concentration. These variations are used to evaluate an electrostatic potential, which will act back on the charge carriers through (8.2), resulting in a self consistent potential.

The calculations shown below are performed with highly simplifying approximations. The charge carrier bands are assumed parabolic, according to the relation $E = E_0 + U(\boldsymbol{x}) + \frac{\hbar^2}{2m^\star}\boldsymbol{k}^2$, where $U(\boldsymbol{x})$ is the electrostatic potential, $E_0$ is a material specific band minimum, and $m^\star$ is a material specific effective mass. A single simulation may contain different materials in different regions, and so there may be a spatial dependency in the parameters $E_0$ and $m^\star$ as well.

Further, the scattering mechanism is also based on a fairly unphysical phenomenological approximation, where the charge carriers are affected by a continuous thermal noise. Thus, rather than introducing discontinuous scattering events, we modify (8.2) to

$$\frac{\mathrm{d}\boldsymbol{k}}{\mathrm{d}t} = \frac{1}{\hbar}\boldsymbol{F} - \gamma\boldsymbol{k} + \frac{1}{\hbar}\sqrt{2\gamma m^\star k_B T}\boldsymbol{\eta}(t), \tag{8.3}$$

where $T$ is the local lattice temperature, and $\boldsymbol{\eta}$ is a stochastic noise term satisfying $\langle \eta_i(t)\eta_j(t')\rangle = \delta_{ij}\delta(t-t')$. It can be shown that in the absence of a force term $\boldsymbol{F}$, the mean square $\langle \boldsymbol{k}^2\rangle$ will satisfy the equation

$$\frac{\mathrm{d}}{\mathrm{d}t}\langle \boldsymbol{k}^2\rangle = \frac{2\gamma}{\hbar^2}\left(3m^\star k_B T - \langle \hbar^2\boldsymbol{k}^2\rangle\right), \tag{8.4}$$

which has the correct equilibrium solution $\langle \frac{\hbar^2\boldsymbol{k}^2}{2m^\star}\rangle = \frac{3}{2}k_B T$.

Finally, the geometries of the simulations are quasi-one-dimensional. Thus, the material composition, electrostatic potential, and temperature varies only along a single direction. As mentioned, the electrostatic potential is calculated self consistently. The temperature on the other hand, is either constant, or varies linearly.

#### 8.1.1.1 IV- and qV-characteristics

Among the most common things to calculate using Boltzmann Monte Carlo simulations, are IV-characteristics, which show how the current $I$ varies with an applied voltage $V$. The IV-characteristic is an important characteristic of several different electronic devices. In Figure 8.1a we show the IV-characteristic of a superlattice, calculated by the method described above. The simulations were performed with a well thickness of 20, a barrier thickness of 2, and we imposed periodic boundary conditions over a single superlattice period. The effective mass in both layers was set to $m^\star = 1$, the barrier height to $b = E_{0b} - E_{0w} = 2$, and the temperature was set to $k_B T = 1$. The scattering rate parameter $\gamma$ was set to 0.1. Finally, the solution of the Poisson equation to obtain the self consistent potential makes use of a permittivity of $\epsilon = 1$. The units of these parameters are arbitrary.

Note that the characteristic does not contain a region of negative differential conductivity, which one expects to see in certain high quality superlattices[28, 15]. This is because this classical model is not able to capture such resonant effects. Accordingly, the model is more appropriate when dealing with superlattices of poor quality, or with broad wells, as discussed in our second publication (Paper II).

Instead the superlattice current grows rapidly with the applied field. Starting out an order of magnitude lower than the bulk current, it grows exponentially with increasing $V$, and eventually approaches the bulk value. This can be understood from the fact that at high fields, the field will excite the particles into a higher state of energy, and so a larger proportion will have energies above the barrier height, where they are able to contribute to the current. We also see that the superlattice currents lie closer to the bulk values at higher charge carrier concentrations. This is because charge realignment of the superlattice cell will tend to reduce the barrier height. These effects are illustrated in Figure 8.1, which shows more details from the simulations.

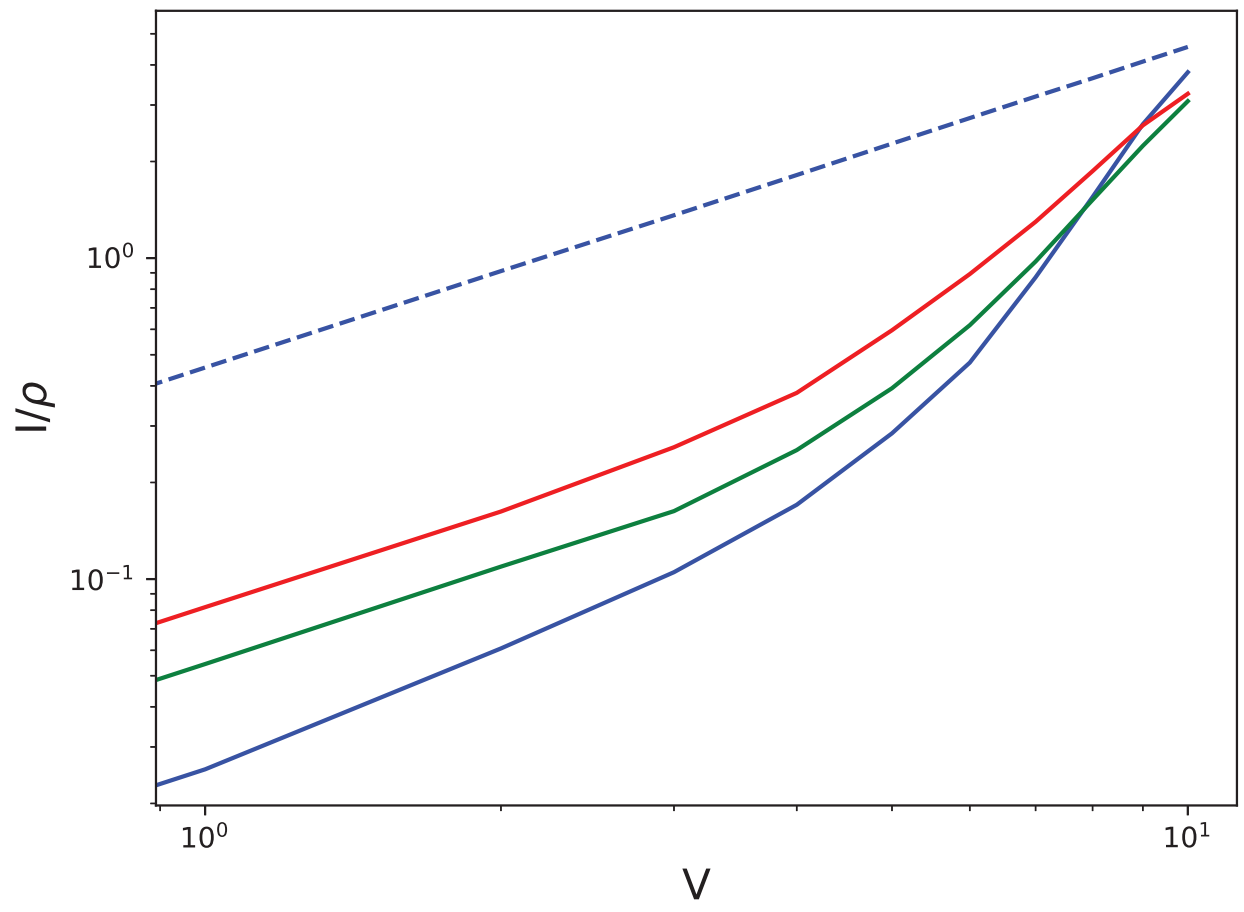

(a) IV-characteristic.

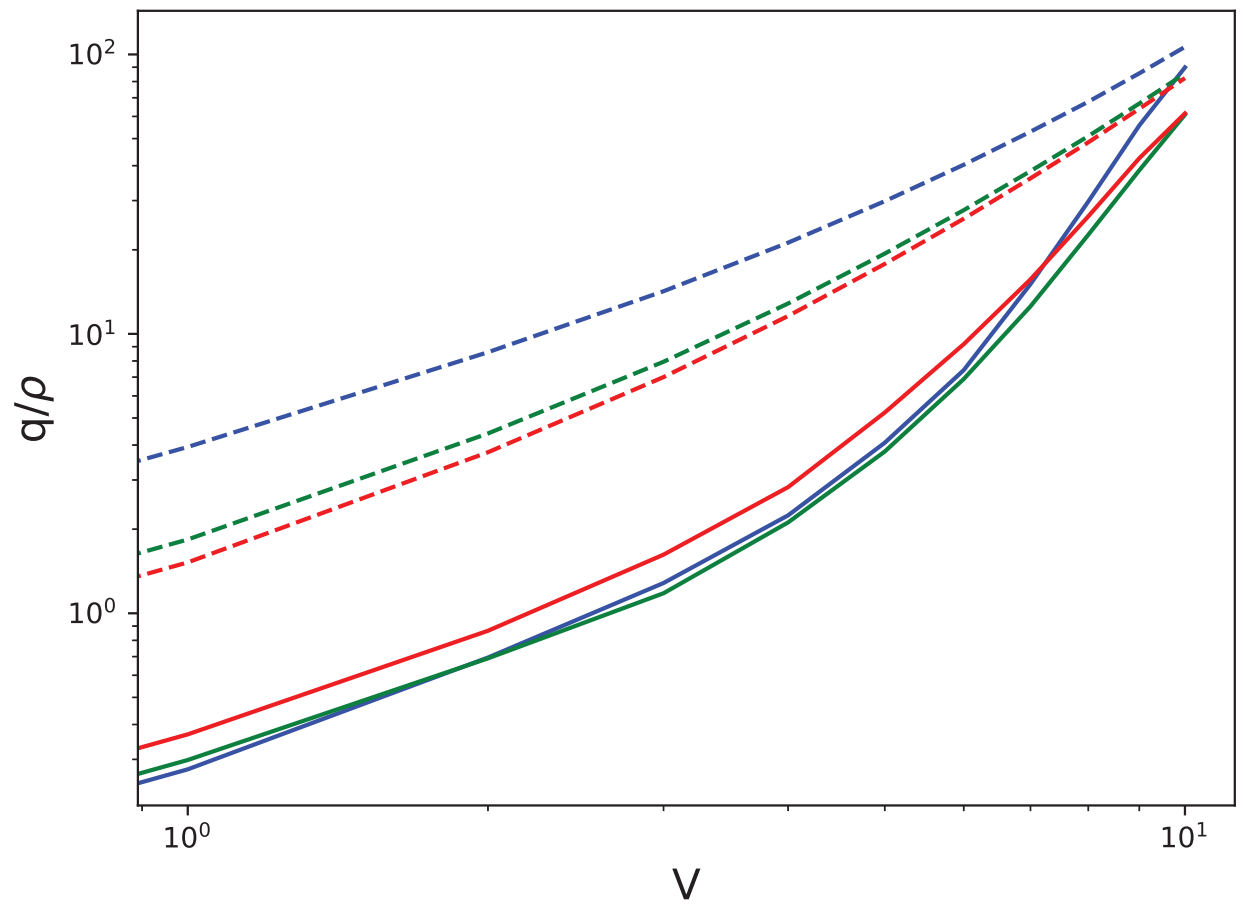

(b) qV-characteristic.

Figure 8.1: IV and qV characteristics of a superlattice heterostructure, according to an incoherent classical transport model. The IV-characteristic in (a) shows how the current $I$ varies with the potential $V$ over a single period. The current is normalized by the charge carrier density $\rho$. The qV-characteristic in (b) shows instead how the heat current $q$ varies with $V$. The heat current is also normalized by $\rho$. Results shown in blue, green and red are respectively calculated with $\rho = 0.01$, $\rho = 1.0$ and $\rho = 2.0$. For comparison, corresponding bulk currents are included as dashed lines. In bulk, the normalized current $I/\rho$ is independent of $\rho$, so only a single dashed line is shown in (a). All units are arbitrary.

In Figure 8.1b we show the qV-characteristic of the same superlattice. Unlike the normalized current, the normalized heat current $q/\rho$ has a dependency on $\rho$ also in bulk. This is because heat is a measure of entropy, and the average entropy per particle decreases with increasing density. In fact, the heat current is calculated as $q = J_E - \mu J$, where $J_E$ is the energy flux, $J$ the particle flux, and $\mu \sim \ln \rho$ is the chemical potential. However, in the superlattice the dependency of $q/\rho$ on $\rho$ is reversed, with $q/\rho$ increasing with the carrier concentration. This is again because charge realignment reduces the barrier height, so that more particles are able to participate in the current. We also observe that unlike $I/\rho$, $q/\rho$ is a nonlinear function of $V$ also in the bulk case. This is because the field excites the particles to higher energies, so that they carry more heat.

#### 8.1.1.2 Linear transport coefficients

In our work, we are interested in the linear conductivity, and in linear thermoelectric transport coefficients. To calculate such coefficients using the direct

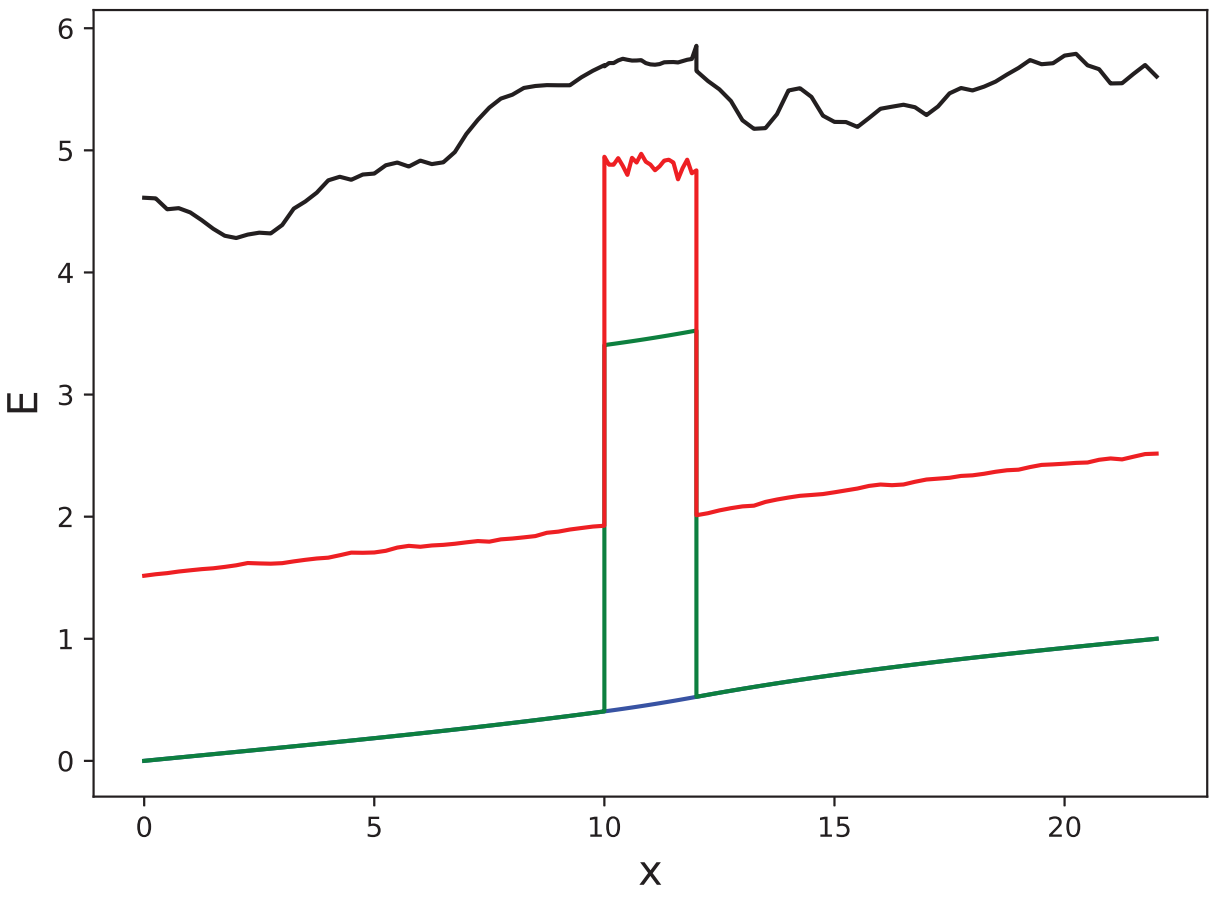

(a) $\rho = 0.01$, $V = 1$

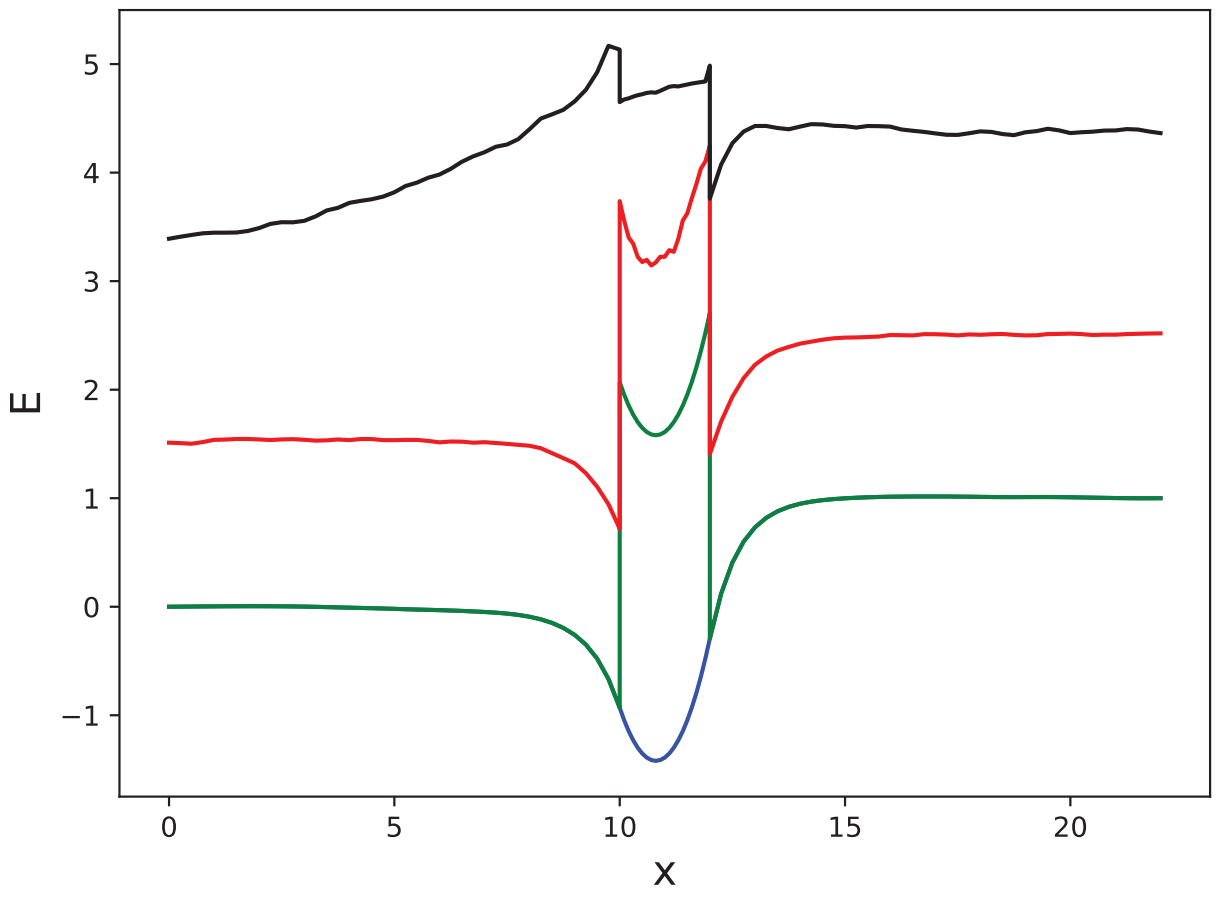

(b) $\rho = 2$, $V = 1$

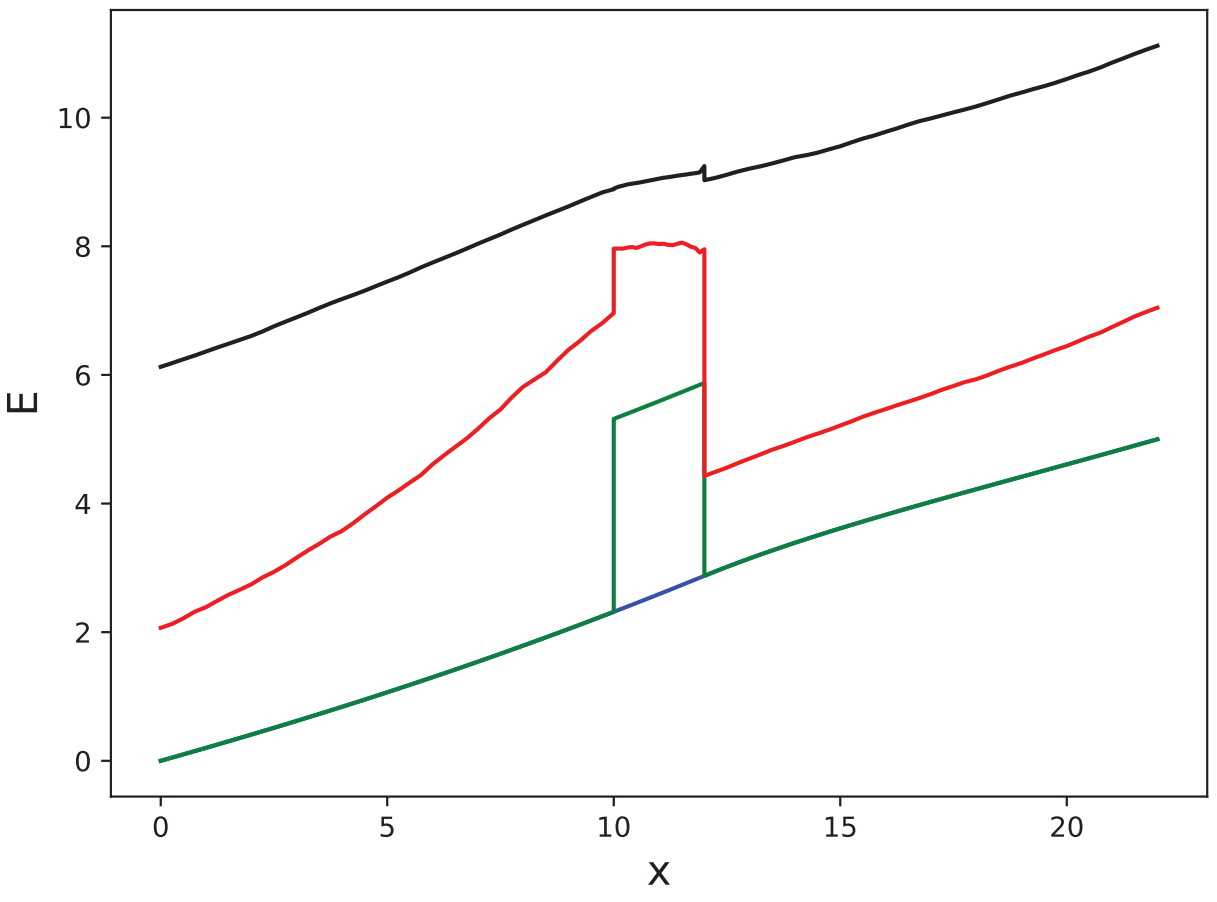

(c) $\rho = 0.01$, $V = 5$

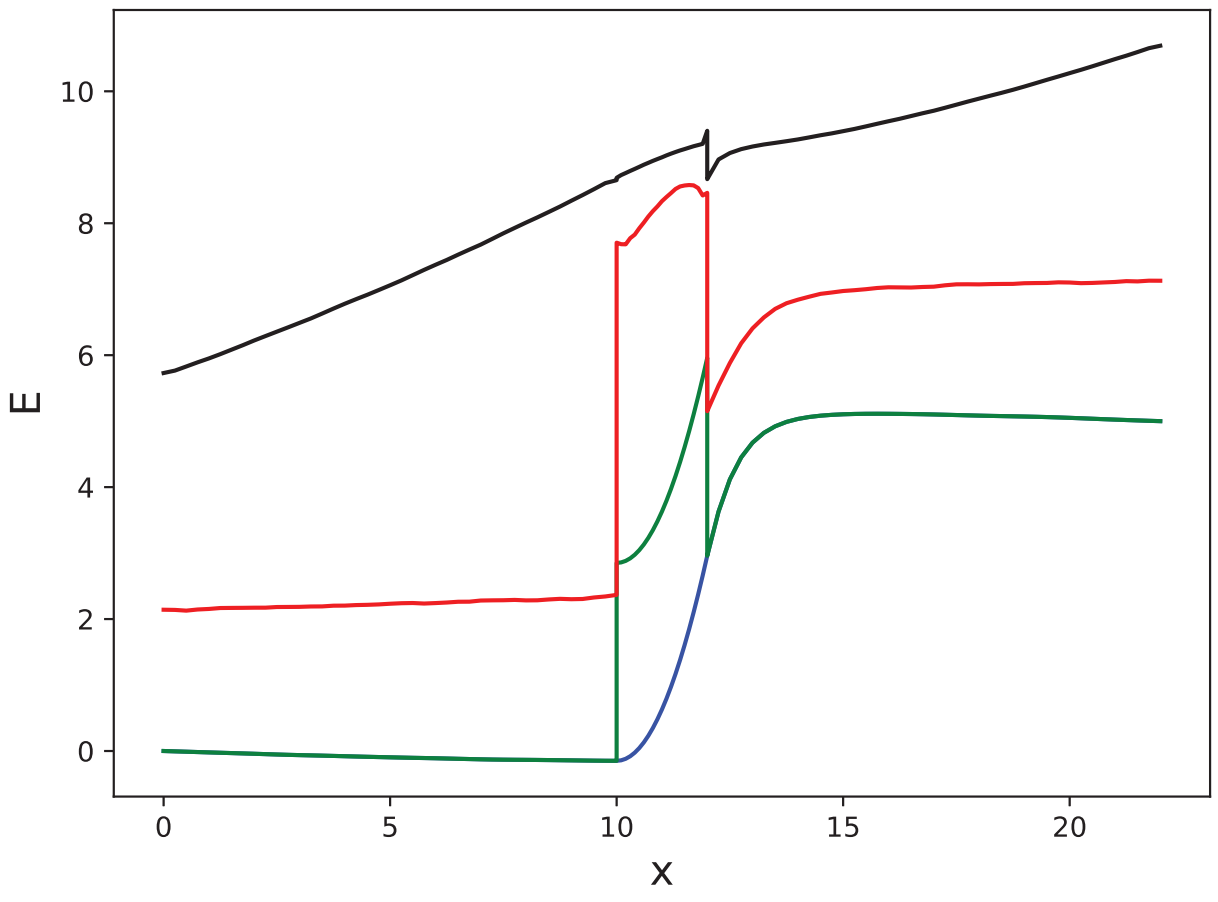

(d) $\rho = 2$, $V = 5$

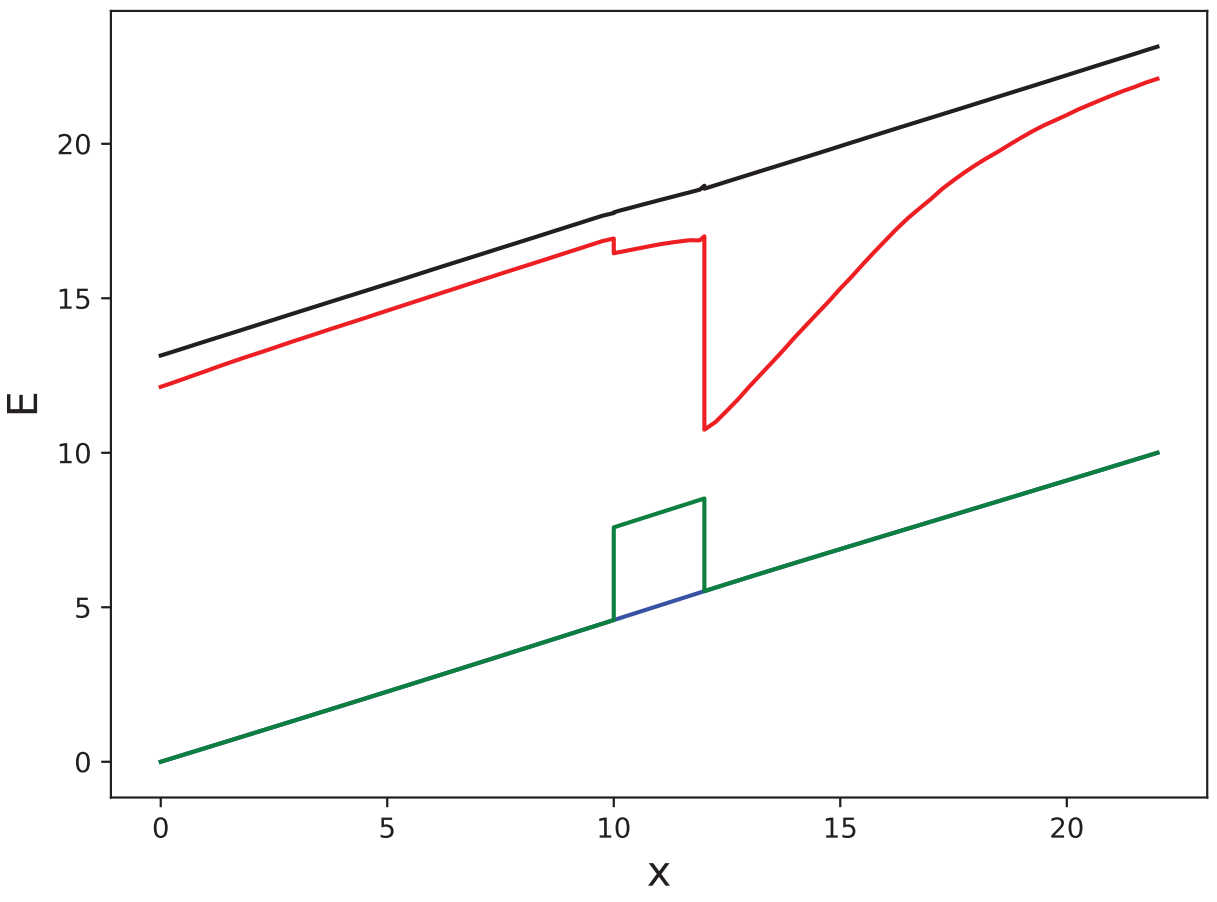

(e) $\rho = 0.01$, $V = 10$

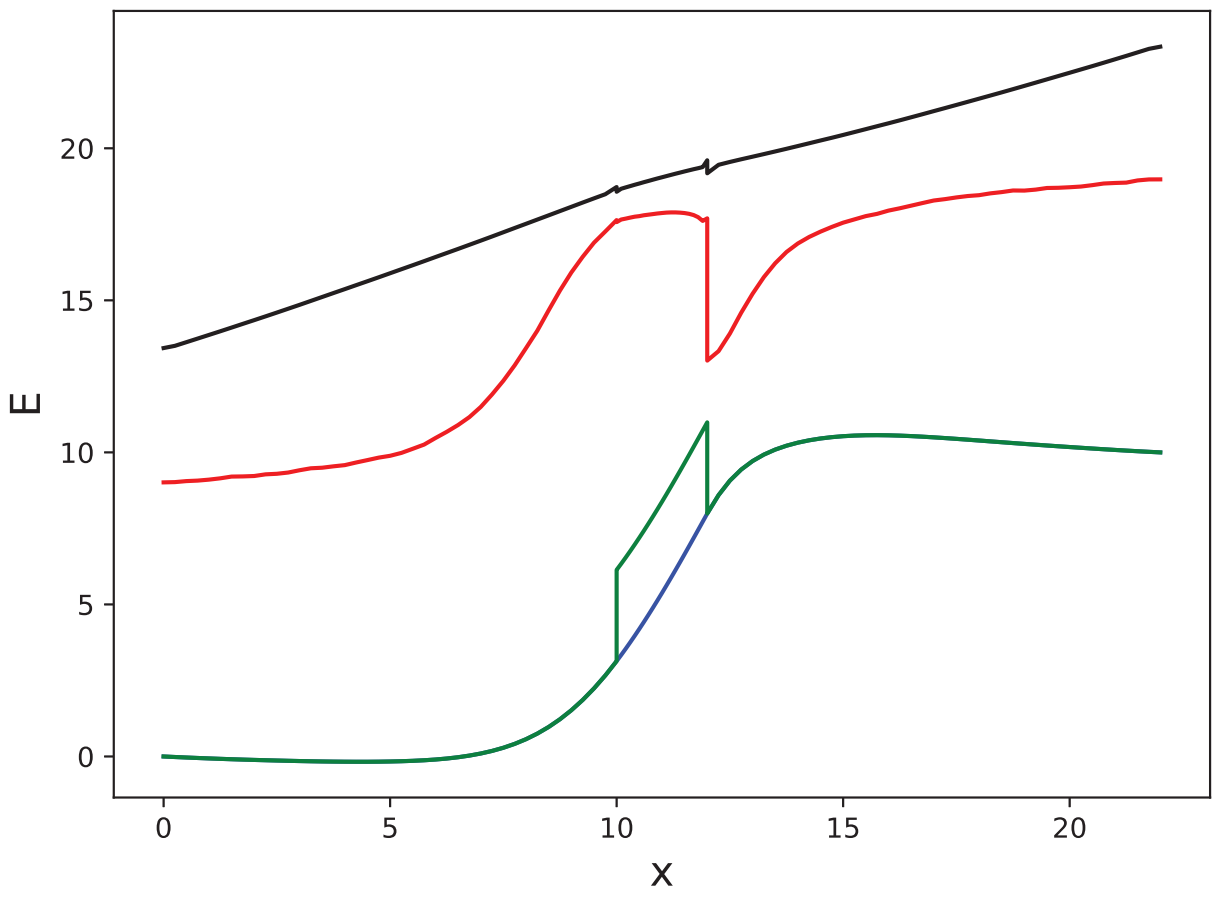

(f) $\rho = 2$, $V = 10$

Figure 8.2: Transport through superlattice. $x$ denotes position along direction of transport, $\rho$ is charge carrier concentration, and $V$ is the potential applied over a single period. The electrostatic potential energy is shown in blue, and the conduction band minimum in green. The average particle energy $\langle E \rangle$ is shown in red, while the black curve shows $\langle Ev \rangle / \langle v \rangle$, where $v$ is particle velocity. The irregularities in the black curve at low fields is due to numerical noise. All units are arbitrary.

simulation approach, we apply a small field $V$, so that we are in the linear regime of the IV- and qV-characteristics of Figure 8.1, and then calculate the linear transport coefficient as response/stimuli. In Figure 8.3, we have calculated the mobility $\mu = \sigma/\rho = I/\rho V$ and the Seebeck coefficient $\alpha$ of a superlattice using this approach. The Seebeck coefficient is found using the relationship $\alpha = \Pi/T$ mentioned at the end of Section 4.2.5, where the Peltier coefficient is calculated as $\Pi = q/I$. The voltage difference $V$ over a single superlattice period was set to $V = 0.5$. Observing Figure 8.1, this should be small enough to be more or less in the linear regime.

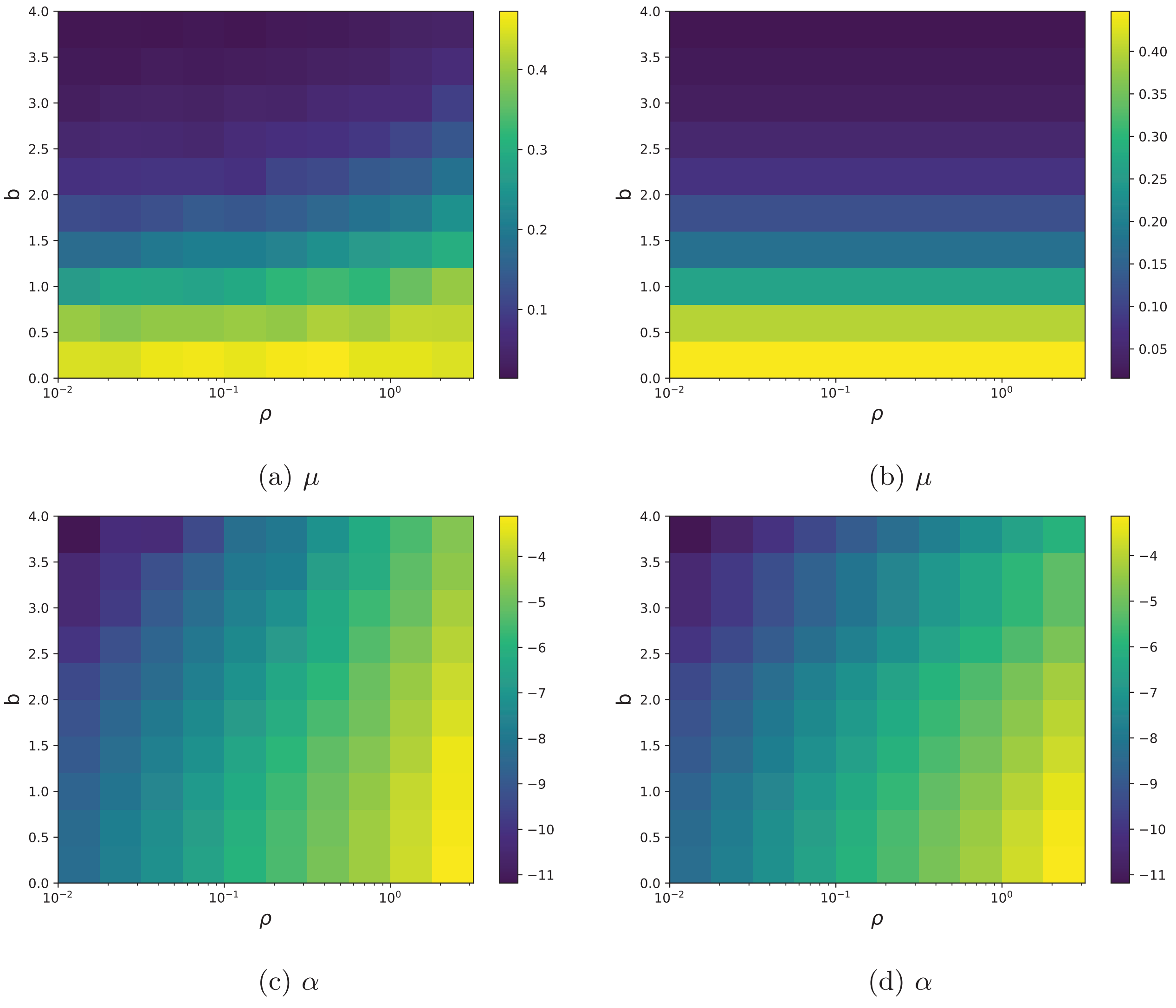

Figure 8.3: Mobility $\mu$ and Seebeck coefficient $\alpha$ of a superlattice structure, as a function of barrier height $b$ and charge carrier concentration $\rho$. The figures to the left are calculated with a self consistent Poisson potential, while in the figures to the right no such potential is included. All units are arbitrary.

Figure 8.3 illustrates how the transport coefficients vary with the barrier height $b$, and with the charge carrier concentration $\rho$. The other parameters are set identically as in the previous calculations. We have also included calculations both with and without a self consistent Poisson potential, in order to examine the importance of calculating this field. Observe first that, as expected, the mobility $\mu$ decreases rapidly when the barrier height is increased. The largest difference between the calculations with and without self consistent Poisson potential is also observed in the mobility. As expected, in the absence of a self consistent potential, the mobility is independent of $\rho$, but when such a potential is included, $\mu$ tends to fall of more slowly with $b$ at high values of $\rho$. Again, this is because charge realignment tends to reduce the effective barrier height, as illustrated in Figure 8.2b.

The Seebeck coefficient $\alpha$ tends to increase in absolute value both as the barrier height $b$ is increased, and as the charge carrier density $\rho$ is decreased. The reason why $|\alpha|$ increases with $b$ can again be understood from Figure 8.2. Since $q \sim \langle Ev \rangle$ and $I \sim \langle v \rangle$, we have $\alpha \sim q/I \sim \langle Ev \rangle / \langle v \rangle$, which is included as a black curve in Figure 8.2. As illustrated by the figure, this curve is forced to pass above the barrier.

Some difference can be observed between the calculations with and without self consistent Poisson potential also in the Seebeck coefficient, although the effect is less visible than it is for the mobility. In particular, at the highest values of $\rho$ and $b$, $\alpha$ is increased from about $-6$ to about $-5$ when a self consistent potential is included. In the absence of a self consistent potential, the dependency of $\alpha$ on $\rho$ is due only to the reduction of the chemical potential.

It deserves mentioning that the effects of charge realignment both on the mobility and on the Seebeck coefficient is underestimated to some extent in these calculations. This is because this particular semiclassical approach is not able to model tunneling through the barriers. Again observing Figure 8.2b, we see that also the shape of the barrier is different at high $\rho$, in such a way that if electrons were able to tunnel through thin sections of the barrier, then the effective barrier height would be reduced even further. At very high charge carrier concentrations, this could potentially remove almost the entire effect of the barrier.

#### 8.1.1.3 Currents induced by temperature gradients

In order to estimate the electronic component $k_e$ of the thermal conductivity using the direct simulation approach, it is necessary to impose a temperature gradient over the region of simulation. This poses a major problem, since we can then no longer impose periodic boundary conditions. A potential difference can be formulated in terms of an electrical field, which can be periodic even at $V > 0$. Thus, periodic boundary conditions can be imposed, as we have done above. An explicit temperature gradient however, will always explicitly break the periodic translation symmetry, so that this is no longer possible.

Thus, we must instead perform the simulation in a finite region, without periodic boundary conditions, but instead including two leads between which the currents can flow. This poses a problem, since the presence of the leads will introduce contact effects which must be dealt with in some way. As a test of the approach we have calculated the Seebeck coefficient again, using a more direct approach where we estimate it from the electrical current induced by a temperature gradient. In particular, we have performed a simulation where the charge carrier density as well as the electrostatic potential is equal between the leads, but where the temperature differs by $\Delta T = 0.4$. This will induce a current of $I = \sigma(V + \alpha\Delta T) = \sigma(\alpha\Delta T - e\Delta\mu)$, so that we can estimate the Seebeck coefficient as

$$\alpha = \frac{I/\sigma + e(\mu_1 - \mu_2)}{\Delta T}, \tag{8.5}$$

where the chemical potential is estimated according to the non-degenerate model

$$\mu = T\ln\rho - \frac{3}{2}T\ln T - \frac{1}{2}T\ln 2\pi. \tag{8.6}$$

The result is shown in Figure 8.4a. Comparing this to Figure 8.3c, we see that the results are very similar, but that the result in Figure 8.4a is slightly smaller in absolute magnitude. This can be understood from the fact that the mobility, and thus indirectly other transport coefficients, will be underestimated in a simulation of finite size, due to an additional contact resistance.

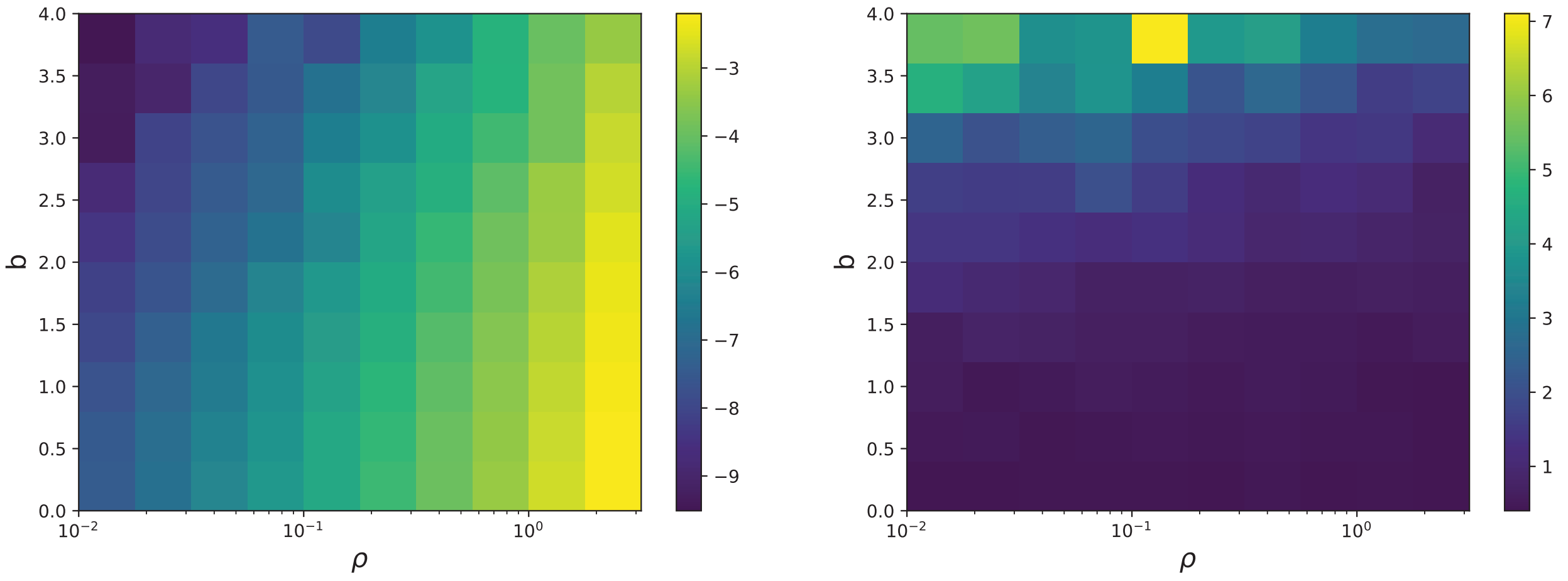

(a) $\alpha$

(b) $L$

Figure 8.4: Seebeck coefficient $\alpha$ and Lorenz coefficient $L$ of a superlattice structure, as a function of barrier height $b$ and charge carrier concentration $\rho$. The coefficients are calculated from currents induced by temperature gradients.

In Figure 8.4b we show the Lorenz coefficient, which is estimated as $L = k_e/\sigma T = (q - \Pi I)/\Delta T/\sigma T$. This result appears more noisy than the others, which is probably because more Monte Carlo samples are required to obtain an accurate average. In addition, we observe that the Lorenz coefficient becomes very high for large barriers. To understand the reason for this, we examine Figure 8.5. We observe that at the location of the barrier, the energy average $\langle Ev\rangle/\langle v\rangle$ increases with the barrier height, so that we always have $\langle Ev\rangle/\langle v\rangle > b$. However, in fact it increases even more rapidly in the regions outside of the barrier.

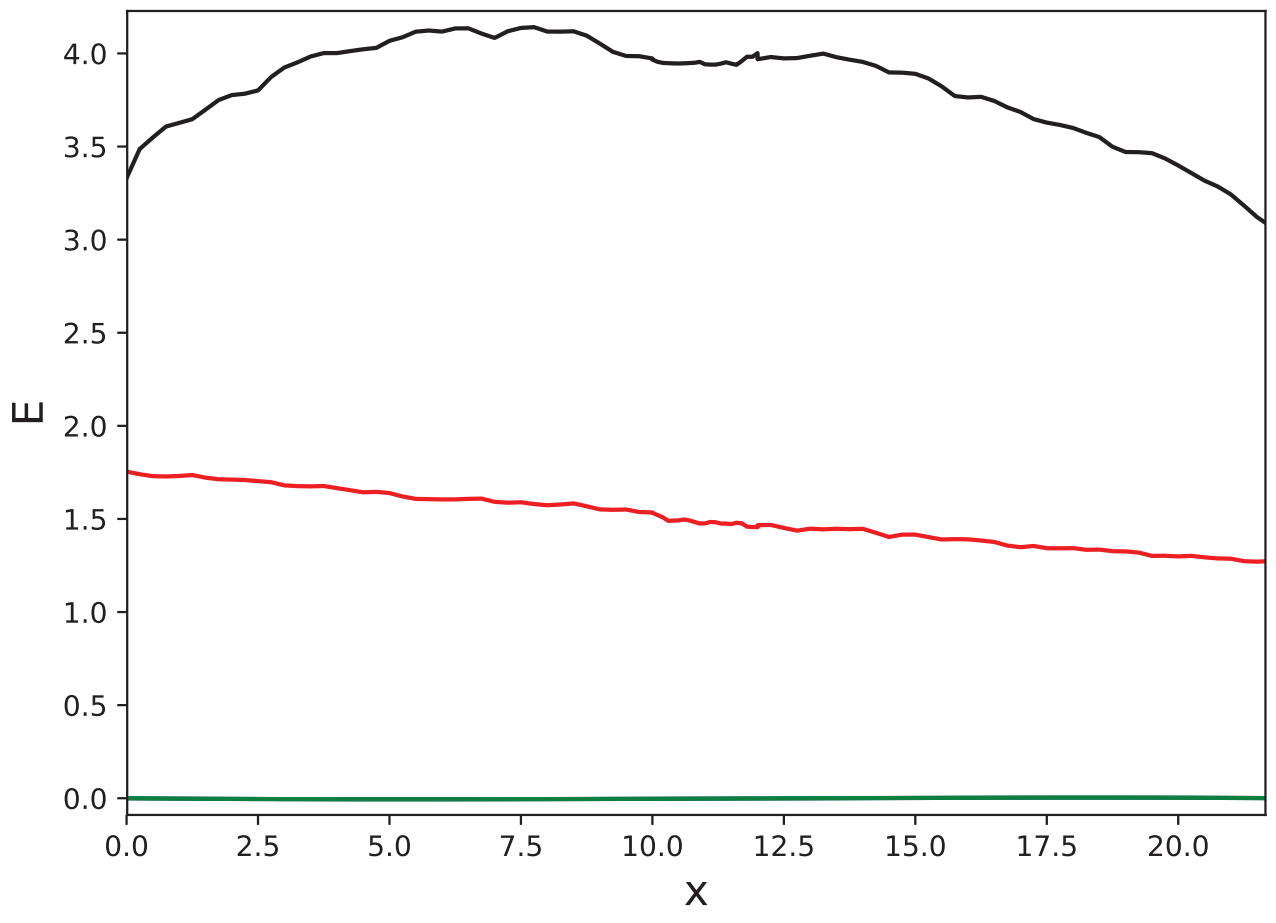

(a) $b = 0$

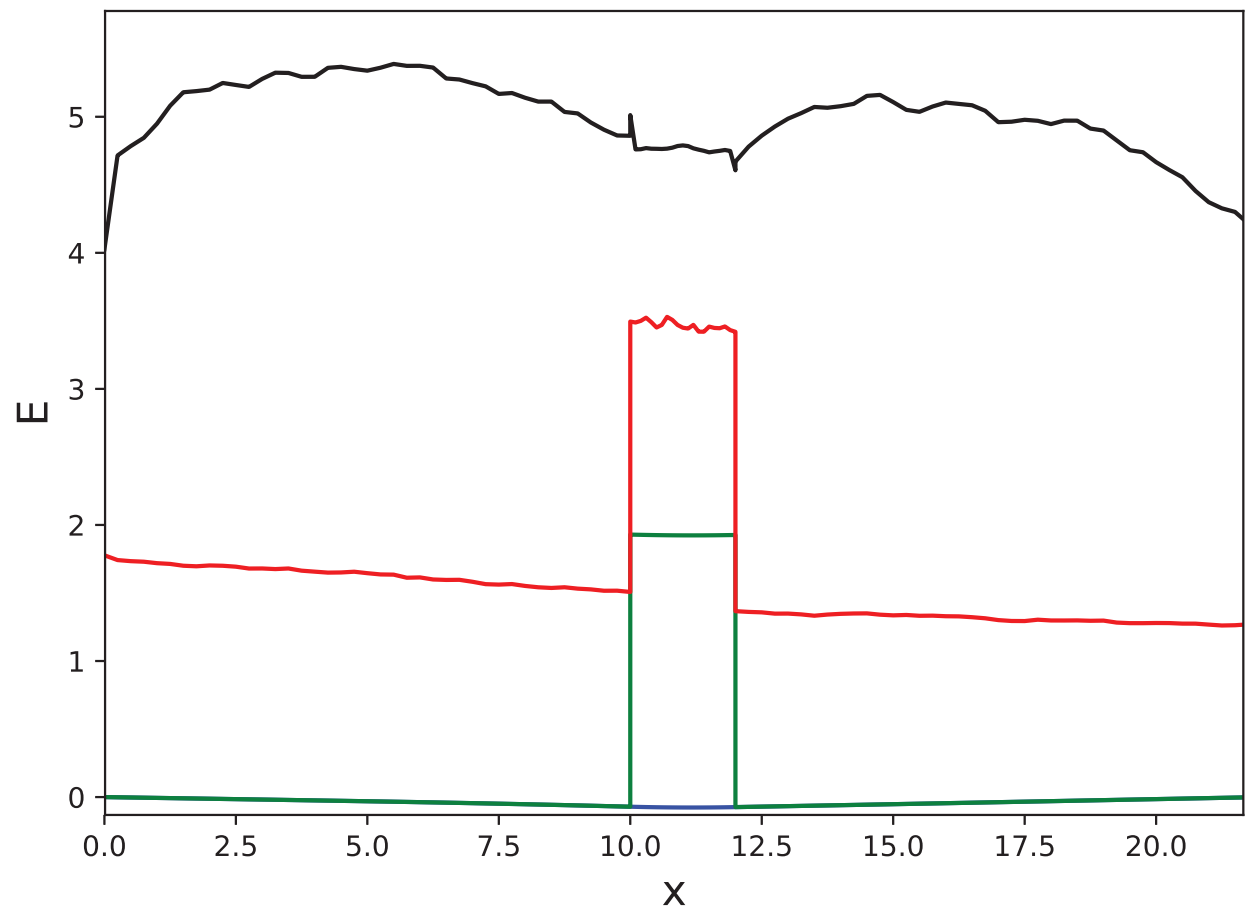

(b) $b = 2$

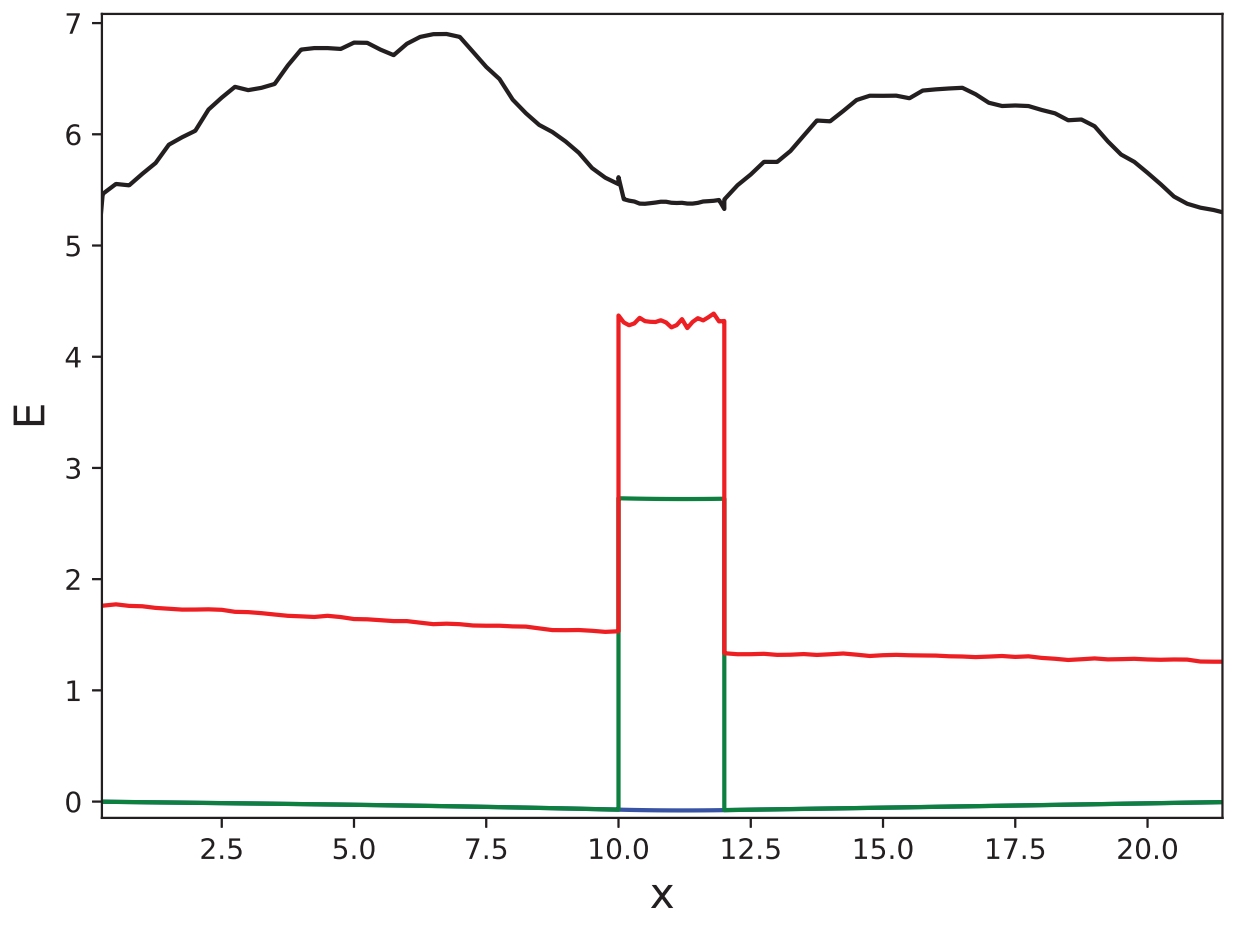

(c) $b = 2.8$

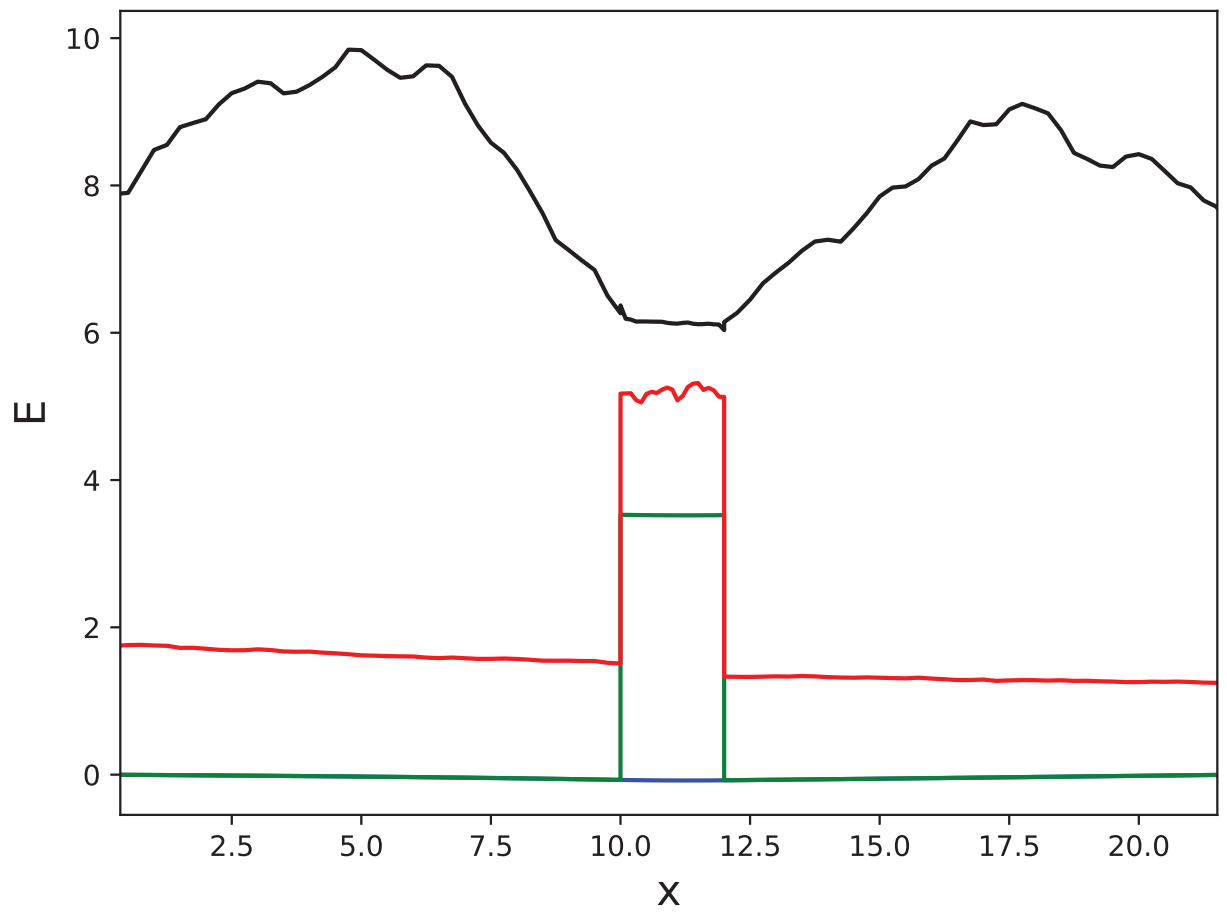

(d) $b = 3.6$

Figure 8.5: Transport through superlattice, induced by temperature gradients. $b$ denotes the height of the barrier. The electrostatic potential energy is shown in blue, and the conduction band minimum in green. All four figures are obtained with the charge carrier concentration set to $\rho = 0.01$. The average particle energy $\langle E \rangle$ is shown in red, while the black curve shows $\langle Ev \rangle / \langle v \rangle$, where $v$ is particle velocity. All units are arbitrary.

This is somewhat counter intuitive, and indeed reversed relative to Figure 8.2. However, it can be understood in the following way: The average $\langle Ev \rangle / \langle v \rangle$ is effectively a measure of the electronic component $q_e$ of the heat flow. There is also a lattice contribution $q_l$ to the heat flow, and in stationary state the total heat flow $q = q_e + q_l$ must be conserved throughout the material. The fraction of the total heat flow which is due to electrons will be $q_e/q = k_e/k = k_e/(k_e + k_l)$, where $k_e$ and $k_l$ are respectively the electronic and lattice contributions to the heat conductivity $k$. Now, since only a small fraction of the electrons is able to pass over the barrier, the electronic heat conductivity $k_e$ is significantly reduced at the location of the barrier. Thus, let us give all quantities a superscript $b$ to signify its value in the barrier region. We then have

$$\frac{q_e}{q_e^b} = \frac{q_e/q}{q_e^b/q} = \frac{k_e}{k_e^b} \cdot \frac{k_e^b + k_l^b}{k_e + k_l}. \tag{8.7}$$

We can in general expect also the lattice contribution to $k$ to be somewhat reduced in the barrier region, but not exponentially like the electronic contribution. Thus, we expect the quantity on right of the equation above to be larger than one, and accordingly $q_e$ and also $\langle Ev \rangle / \langle v \rangle$ will be larger in regions outside of the barriers.

Thus, the discussed effect seen in Figure 8.5 is in fact a real effect. However, the effect is probably overestimated in these calculations, since we are assuming a constant temperature gradient. This implies we are implicitly assuming $k_e^b + k_l^b = k_e + k_l$, so that by the equation above $q_e/q_e^b = k_e/k_e^b \gg 1$. In addition to this, the heat flow is overestimated somewhat due to contact effects in the simulation. This is seen from Figure 8.5a, in which there is no barrier, so that the heat current should be homogeneous in a true bulk system. Accordingly, since the heat flow is being overestimated by two distinct effects, it is unclear whether the behavior of the Lorenz coefficient seen in Figure 8.4b will be seen in real systems. This is particularly the case in superlattices with thin wells, where electrons will not have time to interchange energy with the lattice before reaching the next barrier.

#### 8.1.1.4 Variation of the well thickness

In order to optimize the superlattice for thermoelectric applications, we must study not only how the relevant transport coefficients vary with material properties like charge carrier concentration and barrier height, but also with parameters describing the geometry of the superlattice. In a simple superlattice, like the one above, there are only two such parameters, namely the thicknesses of the two distinct layer types. In the absence of tunneling, it is quite clear that the barrier layer should be as thin as possible, so that it contributes a minimal resistance. However, when tunneling is considered, this conclusion does not hold, since having a to thin barrier could potentially remove the entire effect. Since the transport model considered here is not able to account for tunneling effects, we can thus not correctly optimize the well thickness, and we have accordingly set it arbitrarily to 2.

In the case of the well layer however, significant effects of varying the thickness will be present even in the classical model, so that we can at least illustrate its optimization. Thus, in Figure 8.6 we study how a few thermoelectric transport coefficients depend on the well thickness, which we denote $d$. In figures 8.6a and 8.6b we show respectively how the mobility and the Seebeck coefficient varies with $d$, and in 8.6c we show the power factor $\sigma\alpha^2$. The simulations are performed with a charge carrier concentration of $\rho = 0.01$ and a barrier height of $b = 2$. The other parameters are as above. As expected, the mobility increases when the distance between the barriers is increased. However, the absolute value of the Seebeck coefficient tends to drop. The reason for this is illustrated in Figure 8.7. There we see that the energy average $\langle Ev\rangle/\langle v\rangle$ is only raised in a moderately sized region around the barrier. Thus, as the thickness of the wells is increased, the fraction of the material in which $\langle Ev\rangle/\langle v\rangle$ is enlarged drops. Thus, the ratio $\pi = q/I \sim \langle Ev\rangle/\langle v\rangle$ will also drop, as will the Seebeck coefficient $\alpha = \pi/T$.

Examining Figure 8.6c, we see that even though the Seebeck coefficient is largest in the limit $d = 0$, the power factor is actually largest in the bulk limit $d \to \infty$, indicating that the increasing mobility is more important. However, for optimal thermoelectric properties, one must consider not only the power factor, but also the thermal conductivity $\kappa$ of the material. In fact, if we define $z$ as the ratio between the power factor and $\kappa$, one can show that unconstrained by other concerns, the material which is best suited for any thermoelectric application is the material in which $zT$ is higher, where $T$ is absolute temperature. Thus, as was briefly mentioned in the introduction, our goal is not to optimize the power factor, but the dimensionless figure of merit

$$zT = \frac{\sigma\alpha^2}{\kappa}T = \frac{\sigma\alpha^2 T}{\kappa_e + \kappa_l} = \frac{\alpha^2}{L}\left(1 + \frac{\kappa_l}{\kappa_e}\right)^{-1}. \tag{8.8}$$

To optimize this quantity, we will need a model also of the two contributions to the thermal conductivity, $\kappa_e$ and $\kappa_l$. The electronic component may be calculated as $\kappa_e = \rho\mu LT$. As mentioned, it is questionable whether the Monte Carlo simulations discussed here result in values for the Lorenz coefficient that can be trusted. Because of this, we will arbitrarily make use of a constant model with $L = 1$. Further, inspired by the Matthiessen rule[15], we model the lattice contribution to $\kappa$ as

$$\frac{1}{\kappa_l} = \frac{1}{\kappa_{l0}} + \frac{1}{\kappa_{lB}}, \tag{8.9}$$

where $\kappa_{l0}$ is the bulk value of $\kappa_l$, and $\kappa_{lB}$ describes a reduction from scattering on the barriers. We expect the barrier contribution $1/\kappa_{lB}$ to be inversely proportional to the well thickness $d$. Thus, defining the length scale $\lambda_l = d\kappa_{l0}/\kappa_{lB}$, we may write the model as

$$\kappa_l = \kappa_{l0}\left(1 + \frac{\lambda_l}{d}\right)^{-1}. \tag{8.10}$$

In Figure 8.6d we show the figure of merit as a function of $d$, assuming a lattice contribution to $\kappa$ given by (8.10), with $\kappa_{l0} = 1$ and $\lambda_l = 100$. Unlike the

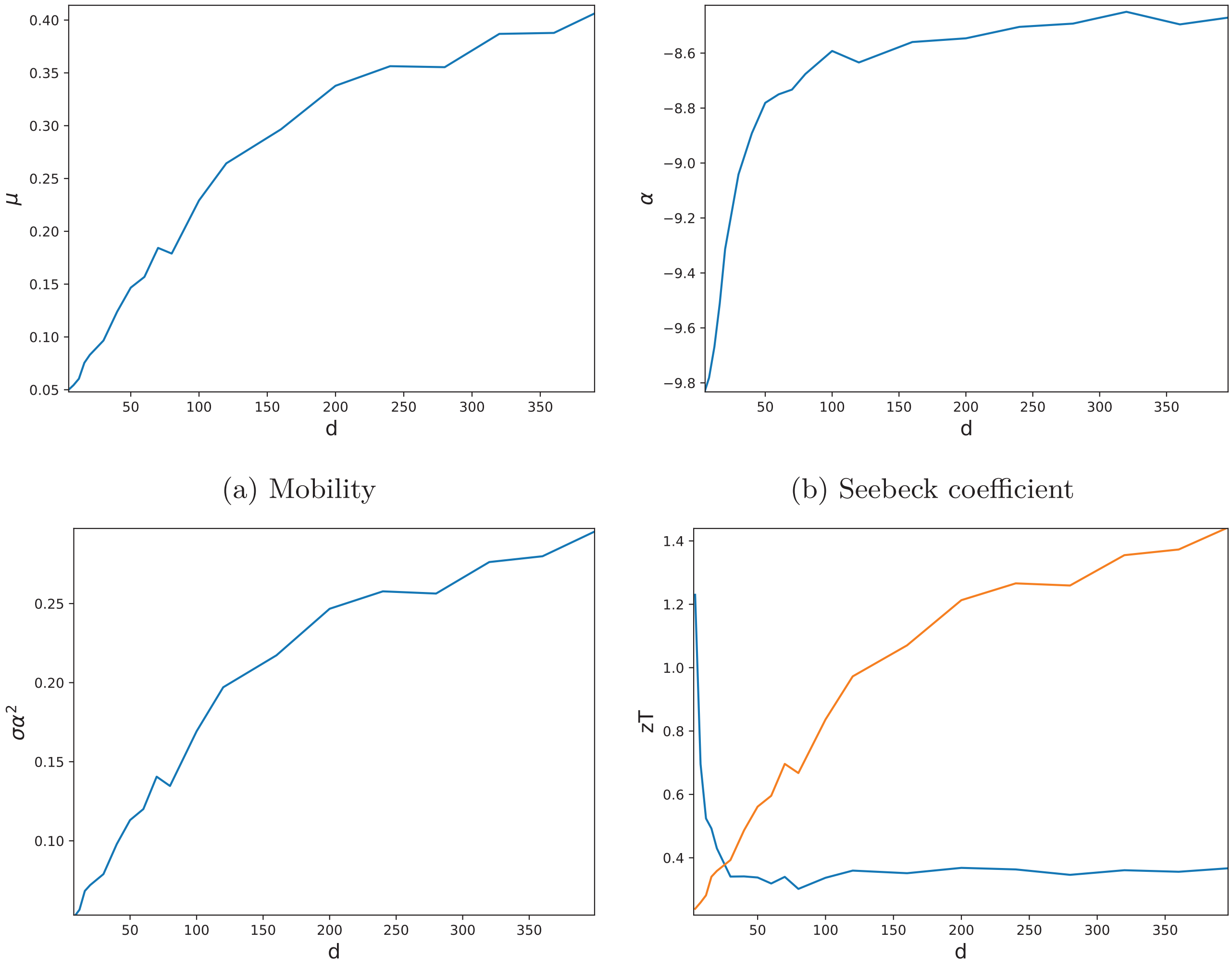

(a) Mobility

(b) Seebeck coefficient

(c) Power factor

(d) Figure of merit

Figure 8.6: Mobility $\mu$, Seebeck coefficient $\alpha$, power factor $\sigma\alpha^2$, and figure of merit $zT$ of a superlattice structure, as a function of well thickness $d$. In (d), two different models have been used for the lattice component of the thermal conductivity: The curve in orange assume a constant value $\kappa_l = 0.2$, while the curve in blue assumes $\kappa_l = (1 + 100/d)^{-1}$. All units are arbitrary.

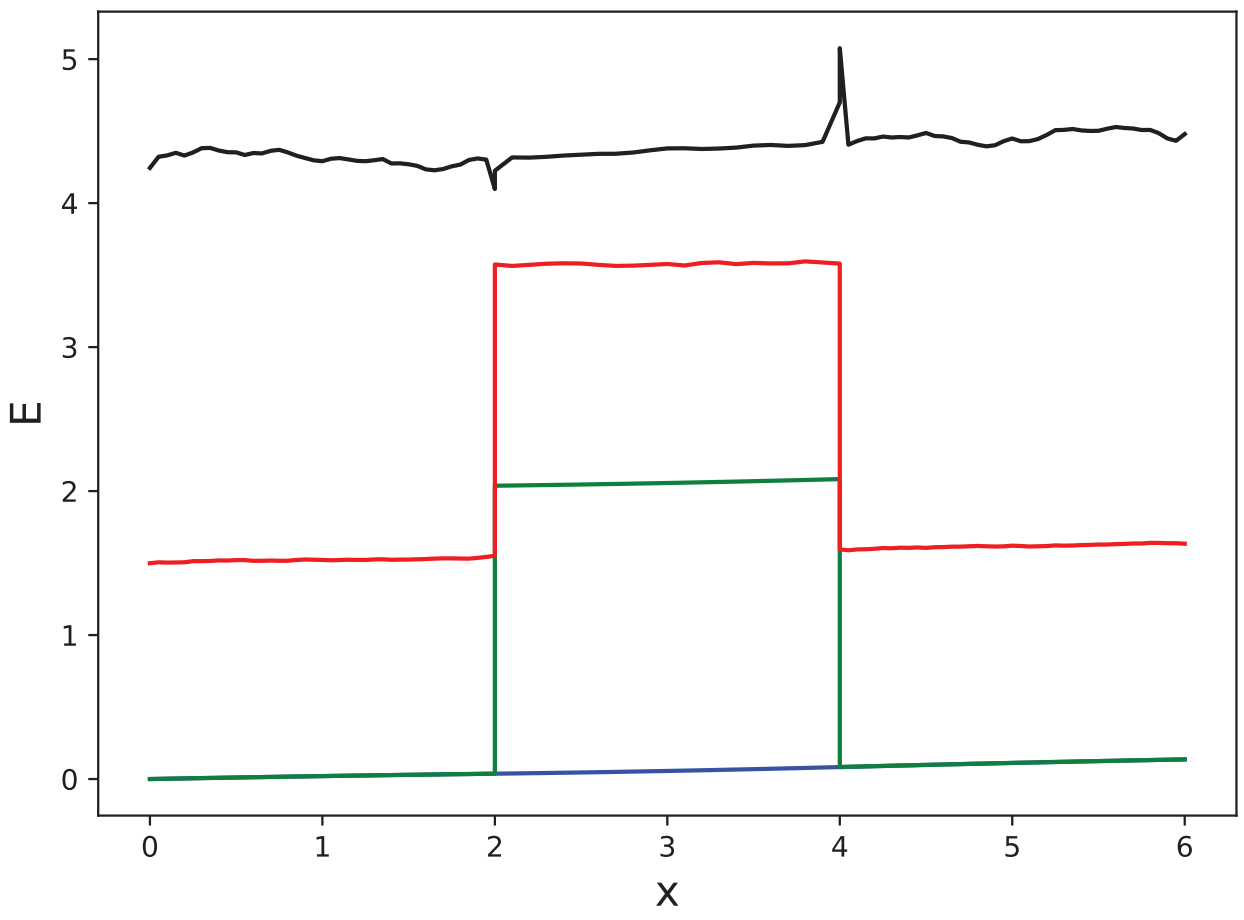

(a) $d = 4$

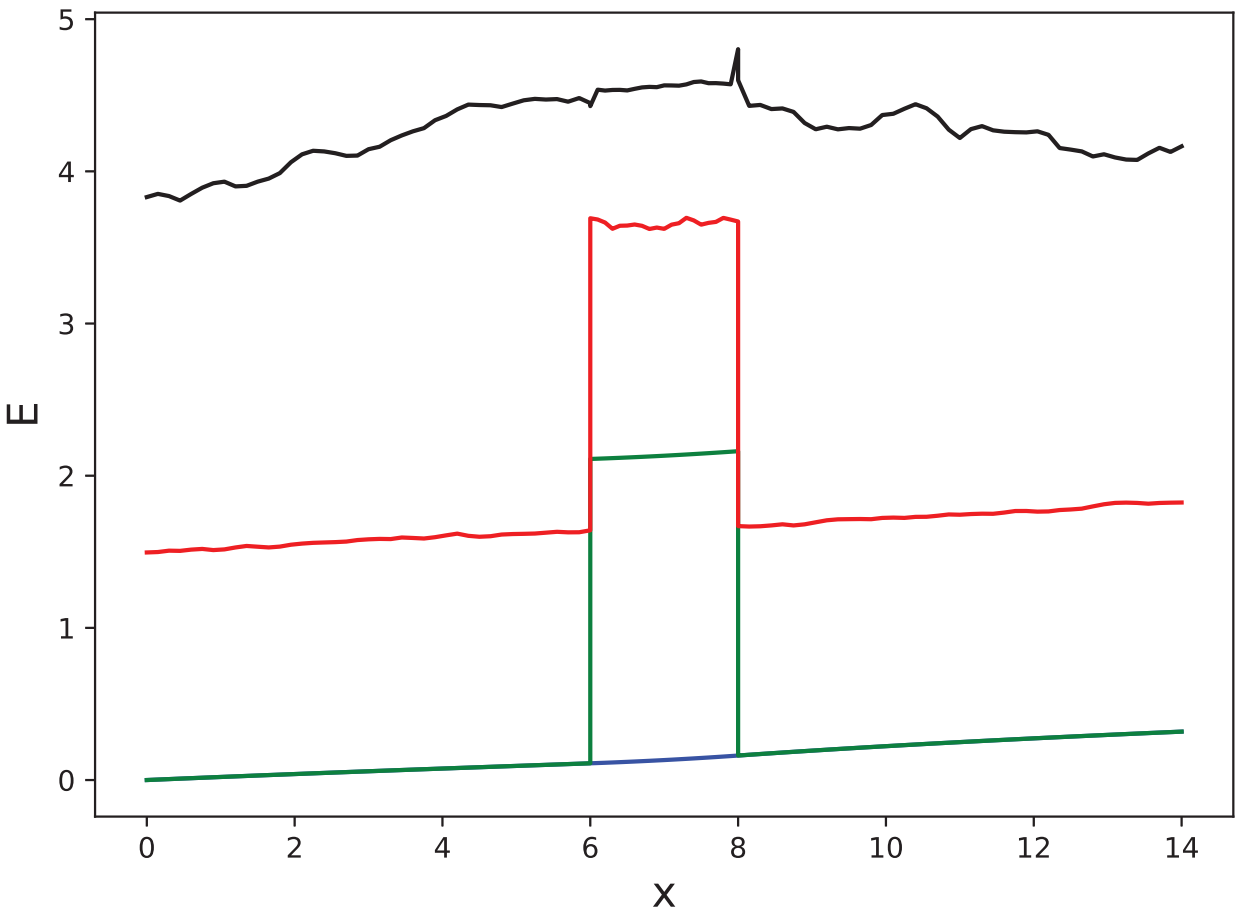

(b) $d = 12$

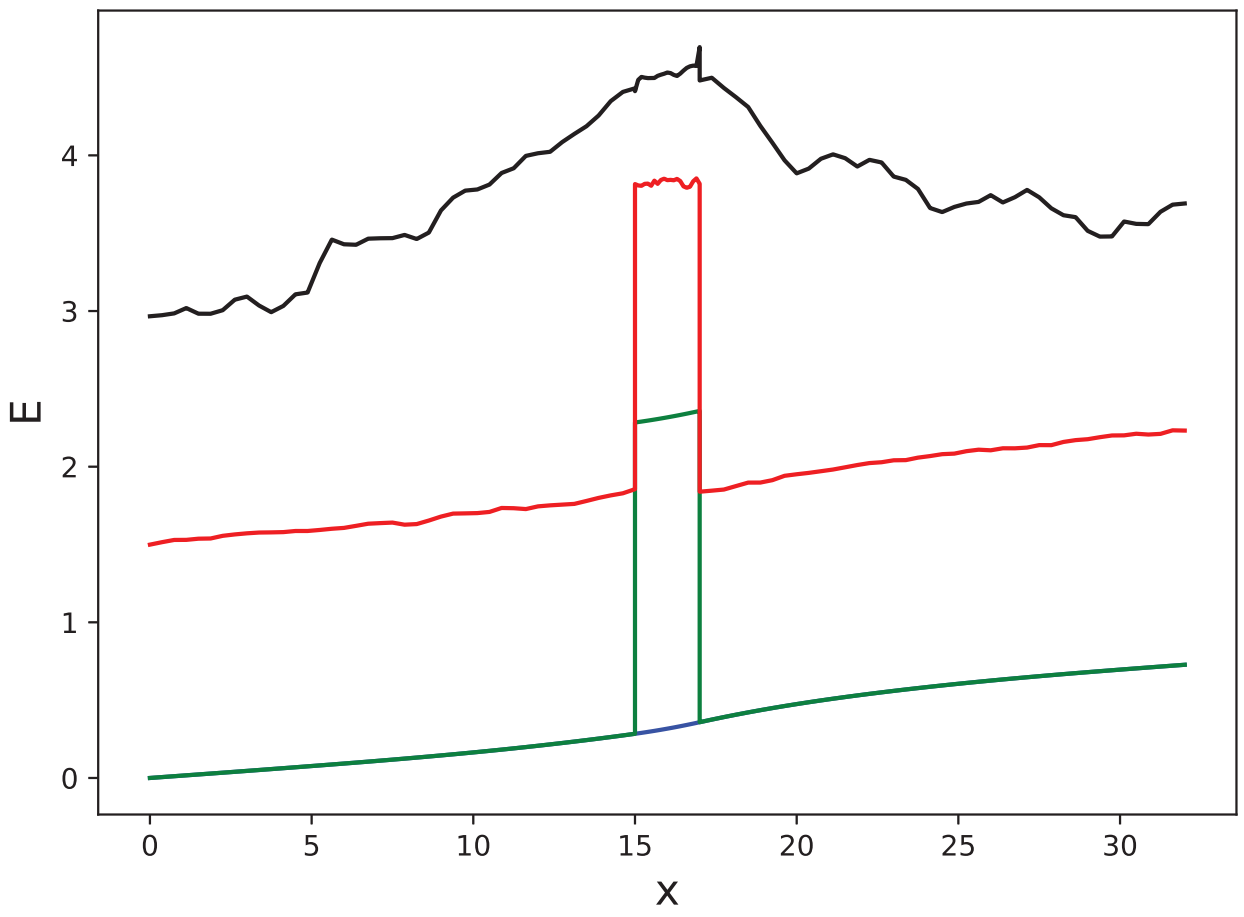

(c) $d = 30$

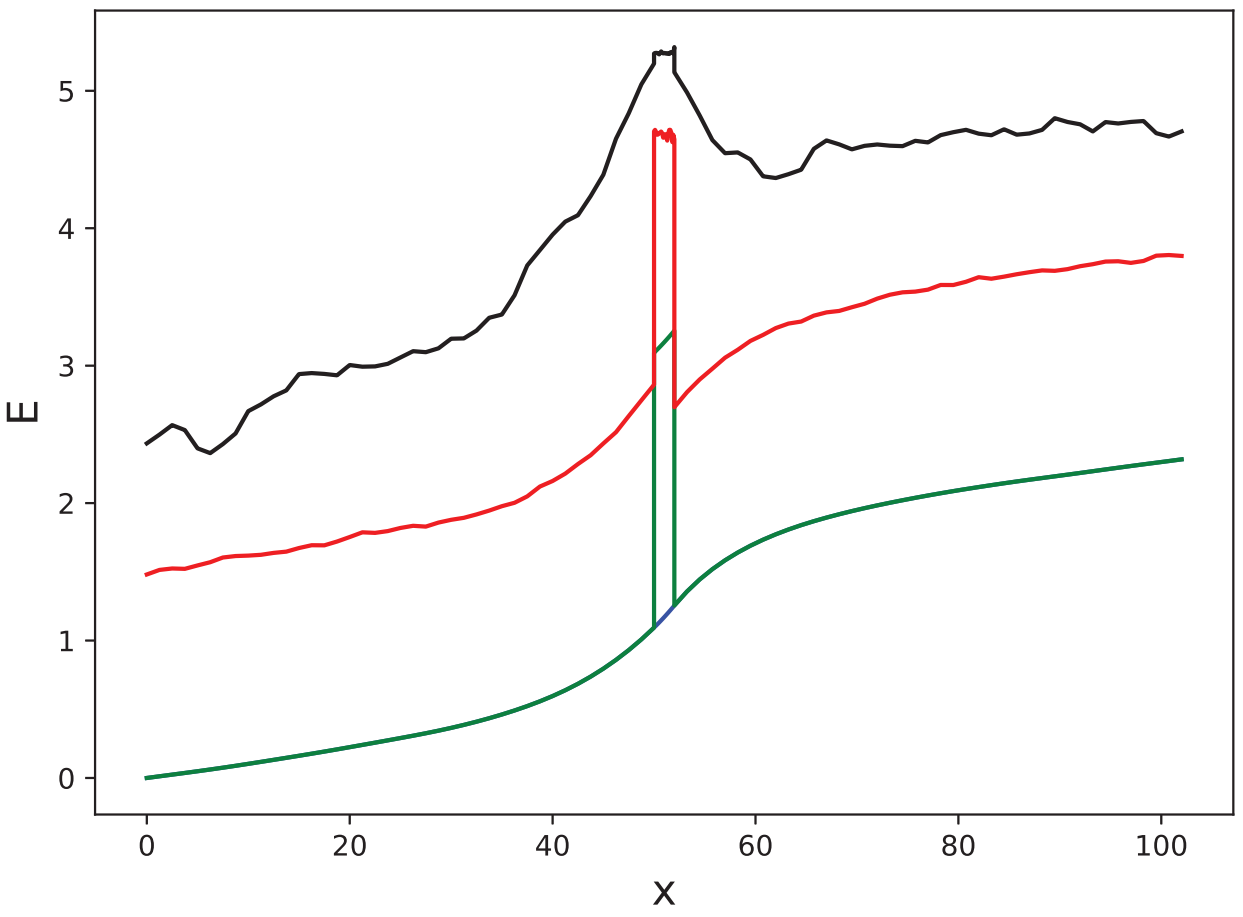

(d) $d = 100$

Figure 8.7: Transport through superlattice. $d$ denotes the thickness of the wells, or equivalently the distance between the barriers. The electrostatic potential energy is shown in blue, and the conduction band minimum in green. All four figures are obtained with the charge carrier concentration set to $\rho = 0.01$, and a barrier height of $b = 2$. The average particle energy $\langle E \rangle$ is shown in red, while the black curve shows $\langle Ev \rangle / \langle v \rangle$, where $v$ is particle velocity. All units are arbitrary.

power factor, this particular model for $zT$ is maximal in the limit $d = 0$, and in fact seems to get a significant boost in the region where $|\alpha|$ is enlarged. However, in reality (8.10) is unlikely to apply all the way down to $d = 0$, since for thin wells we will probably have a transition to coherent transport, as discussed in Section 2.3. Thus, to illustrate the importance of the model for $\kappa_l$, Figure 8.6d also include a model for $zT$ where we assume a constant lattice thermal conductivity of $\kappa_l = 0.2$. In this model, the improvement near $d = 0$ is completely gone, and instead the dependence of $zT$ on $d$ resembles that of the power factor, illustrating that the optimization of thermoelectric materials is a nontrivial task.

#### 8.1.1.5 Discussion

The model employed in the simulations above, is too simplified to allow for making conclusions about real materials. Indeed, knowing that this would in any case not be possible, we have not made use of parameters from existing materials, and have made all calculations in arbitrary units. Nevertheless, when it comes to the importance of different physical effects, a few conclusions can still be drawn from the above discussions. Awareness of these effects is important to understand which elements of the model must be generalized to obtain reliable results.

The first important conclusion is drawn from the equation (8.8). We see that in order to obtain a high value of $zT$, we will typically want the ratio

$$\frac{\kappa_e}{\kappa_l} = \frac{\mu L}{\kappa_l}\rho T \tag{8.11}$$

to be quite high. Thus, we expect that the optimal thermoelectric performance is usually obtained at quite high charge carrier concentrations $\rho$, so accordingly it is important to be able to model such high charge carrier concentrations. In particular, this means it will be necessary to model a self consistent Poisson potential, since as seen from Figure 8.2 there will be a significant effect of charge realignment at high values of $\rho$. Figure 8.3 also shows that the Poisson potential can have a significant effect on the transport coefficients.

While a self consistent Poisson potential has been included in the simulations above, there are two important effects of high charge carrier concentrations which are not accounted for. First, the simulations assume non-degenerate statistics, where electrons are distributed according to a Boltzmann distribution. At high $\rho$, this is not a correct description, and we must instead use degenerate statistics and a Fermi distribution. Secondly, at particularly high values of $\rho$, we must expect the chemical potential to lie far above the band minimum. In this region, the employed assumption of parabolic bands will not be realistic, so that we must make use of a more accurate description of the band structure.

Next, two important conclusions can be drawn from Figure 8.6d. First of all, in thermoelectric applications, it is impossible to optimize the electronic properties of the material alone, since the model used to describe lattice heat transport will significantly affect the behavior of $zT$. Thus, a complete model must also include a model of phonons. Secondly, due the effect of $\kappa_l$, one can not apriori know at which value of the well width $d$ one will obtain the optimal $zT$. Indeed, the two models employed in Figure 8.6d yield completely opposite results, with one having maximal $zT$ at $d = 0$, and the other in the bulk limit $d \to \infty$. Given a more realistic phonon transport model, which correctly accounts for coherent transport at small $d$, it is entirely possible that the optimal value of $d$ will lie in some intermediate range. Thus, it is important to be able to model the entire range of superlattice geometries, limited neither to small or to large periods.

Another important effect related to the geometry of the superlattice, is of course tunneling through the barrier. As mentioned, it is important to be able model tunneling in order to find the optimal barrier thickness. In addition, we have also mentioned that if tunneling is not accounted for, one may end up underestimating the effects of charge realignment on the effective barrier height.

Finally, the discussion of figures 8.4 and 8.5 illustrates two important points concerning the modeling of heat flows. Firstly, in a wide well super lattice, there will usually be an interchange of energy between electron and lattice contributions to the heat flow, so that these two contributions can not be considered separately, but must actually be modeled self consistently. Secondly, since the direct simulation approach is not able do model currents induced by temperature gradients without breaking the periodic symmetry, one is inevitably stuck with some kind of contact effects, which can be quite severe. In order to remove these effects, one will have to simulate a larger region, containing more than a single superlattice period, which will inevitably increase the computational requirements. This is the approach we have followed in our first two publications, Papers I-II. The alternative is to make use of an indirect approach, such as the Kubo relations.

### 8.1.2 Green-Kubo relations

As discussed in Section 6.4, the Boltzmann equation is obtained from quantum transport in the limit $\hbar \to 0$. Accordingly, this limit can also be applied to the Kubo relations (4.43)-(4.46). In this limit the integrand of these expressions will be approximately constant in the innermost integration range, so that the innermost integral can be evaluated explicitly, yielding a factor of $\hbar\beta$. Further, in the classical limit $\hbar \to 0$ we can also replace operators such as $\hat{\boldsymbol{j}}(t)$ with classical functions. Thus, (4.43)-(4.46) become

$$\sigma_{ij} = \frac{V}{k_B T} \int_0^\infty \mathrm{d}t \left\langle \bar{j}_j(0) \bar{j}_i(t) \right\rangle , \tag{8.12}$$

$$\mathcal{A}_{ij} = \frac{V}{k_B T^2} \int_0^\infty \mathrm{d}t \left\langle \bar{\phi}_{Qj}(0) \bar{j}_i(t) \right\rangle , \tag{8.13}$$

$$\mathcal{B}_{ij} = \frac{V}{k_B T} \int_0^\infty \mathrm{d}t \left\langle \bar{j}_j(0) \bar{\phi}_{Qi}(t) \right\rangle \quad \text{and,} \tag{8.14}$$

$$\mathcal{C}_{ij} = \frac{V}{k_B T^2} \int_0^\infty \mathrm{d}t \left\langle \bar{\phi}_{Qj}(0) \bar{\phi}_{Qi}(t) \right\rangle . \tag{8.15}$$

These equations represent the classical limit of the Kubo relations (4.43)-(4.46). Relations of this type are often referred to as Green-Kubo relations, since they were independently discovered by Green. The correlation functions in the expressions can be estimated from Monte Carlo simulations at equilibrium, and accordingly these relations can be used to calculate thermoelectric transport coefficients without the need of imposing explicit fields or temperature gradients. As discussed in the previous section, this is a particularly important advantage when modeling the effects of temperature gradients.

In this section, some results are presented which were obtained using the relations (8.12)-(8.15). As in the previous section, we assume parabolic bands and non-degenerate statistics. However, a more realistic scattering mechanism is utilized, compared to the phenomenological model of (8.3), which was employed in the previous section. In particular, the calculations of this section make use of a model of acoustic phonon scattering described by Jacoboni[15]. On the other hand, our implementation has not progressed far enough to model heterostructures like those of the previous section, so it is limited to simple bulk simulations.

The fact that we are assuming non-degenerate statistics allows us to simplify the relations (8.12)-(8.15) even further. This is because under this assumption, and an additional assumption of weak interactions, the particles in the system are uncorrelated. This significantly simplifies the evaluation of the involved correlation functions. Consider for example (8.12). The average current density in the system can be expressed as

$$\bar{\boldsymbol{j}}(t) = \frac{-e}{V} \sum_{n=1}^{N} v_n, \tag{8.16}$$

where we assume that there are $N$ particles within the volume $V$, and where $v_n$ is the velocity of the $n$'th particle. Thus, since the particles are uncorrelated, we find

$$\left\langle \bar{j}_j(0) \bar{j}_i(t) \right\rangle = \frac{e^2}{V^2} \sum_{n=1}^{N} \langle v_{nj}(0) v_{ni}(t) \rangle = \frac{Ne^2}{V^2} \langle v_j(0) v_i(t) \rangle \,, \tag{8.17}$$

where we have used that the correlation function is independent of $n$, since the particles are indistinguishable. Inserting this in (8.12), and using $\rho = N/V$, we obtain

$$\sigma_{ij} = \frac{e^2 \rho}{k_B T} \int_0^{\infty} \mathrm{d}t \, \langle v_j(0) v_i(t) \rangle \,. \tag{8.18}$$

Similarly, the average heat flux in the system can be expressed as

$$\bar{\boldsymbol{j}}(t) = \frac{1}{V} \sum_{n=1}^{N} (E_n - \mu) v_n, \tag{8.19}$$

where $\mu$ is the chemical potential, and $E_n$ is the energy of the $n$'th particle. Repeating the arguments above, we see that (8.13)-(8.15) simplify to

$$\mathcal{A}_{ij} = -\frac{e\rho}{k_B T^2}\int_0^\infty \mathrm{d}t \left\langle (E(0)-\mu)v_j(0)v_i(t)\right\rangle, \tag{8.20}$$

$$\mathcal{B}_{ij} = -\frac{e\rho}{k_B T}\int_0^\infty \mathrm{d}t \left\langle v_j(0)(E(t)-\mu)v_i(t)\right\rangle \quad \text{and,} \tag{8.21}$$

$$\mathcal{C}_{ij} = \frac{1}{k_B T^2}\int_0^\infty \mathrm{d}t \left\langle (E(0)-\mu)v_j(0)(E(t)-\mu)v_i(t)\right\rangle. \tag{8.22}$$

The correlation functions on the right are easy to estimate in a Monte Carlo simulation. The Seebeck coefficient, Peltier coefficient and the electronic component of the the thermal conductivity, can then be found from $\overleftrightarrow{\mathcal{A}}$, $\overleftrightarrow{\mathcal{B}}$ and $\overleftrightarrow{\mathcal{C}}$ using (4.49)-(4.51).

The scattering mechanism is taken from Ref. [15]. In particular, we make use of model C of chapter 9.3, which describes inelastic scattering on acoustic phonons, assuming spherical and parabolic electron bands. The model is formulated in terms of the integral kernel $\Gamma_{nm}(\boldsymbol{k},\boldsymbol{k}')$ introduced above (6.89). Since we only have a single band, the band indices $n$ and $m$ are subsumed. The model states

$$\begin{aligned}&\Gamma(\boldsymbol{k},\boldsymbol{k}') \\ &= \frac{qE_l^2}{8\pi^2\rho_m c_l}\big(n(q)\delta(E(k')-E(k)+\hbar q c_l)+(n(q)+1)\delta(E(k')-E(k)-\hbar q c_l)\big),\end{aligned} \tag{8.23}$$

where $E(k) = k^2/2m^\star$ ,$q = |\boldsymbol{k}-\boldsymbol{k}'|$, $E_l$ is a deformation potential constant, $c_l$ is the speed of sound for longitudinal modes, $\rho_m$ is the mass density of the material, and

$$n(q) = \frac{1}{e^{\beta\hbar q c_l}-1}. \tag{8.24}$$

The implementation of this model as a Monte Carlo process follows Ref. [16], which describes how to draw new Bloch-vectors $\boldsymbol{k}$ with a probability distribution proportional to (8.23). The simulations were performed with $c_l = 0.1$ and $E_l^2/\rho_m c_l = 100$.

Since (8.23) is a fairly complicated model of scattering, it is important to test the implementation. A particularly important test, is to check whether the particles end up with the correct distribution in $\boldsymbol{k}$-space. Since the simulation coccus at equilibrium, they should be distributed according to the Boltzmann distribution

$$p(\boldsymbol{k}) = \frac{1}{\sqrt{2\pi k_B T m^\star}} e^{-\frac{\boldsymbol{k}^2}{2k_B T m^\star}}. \tag{8.25}$$

As a very simple test, we can simply simulate a few electrons, plot their position in k-space, and check if the resulting distribution seems reasonable. Figure 8.8 shows a scatter plot of the position in $\boldsymbol{k}$-space of 1000 electrons, after an equilibration time of $t = 50$. The result is consistent with a Gaussian distribution centered at the origin.

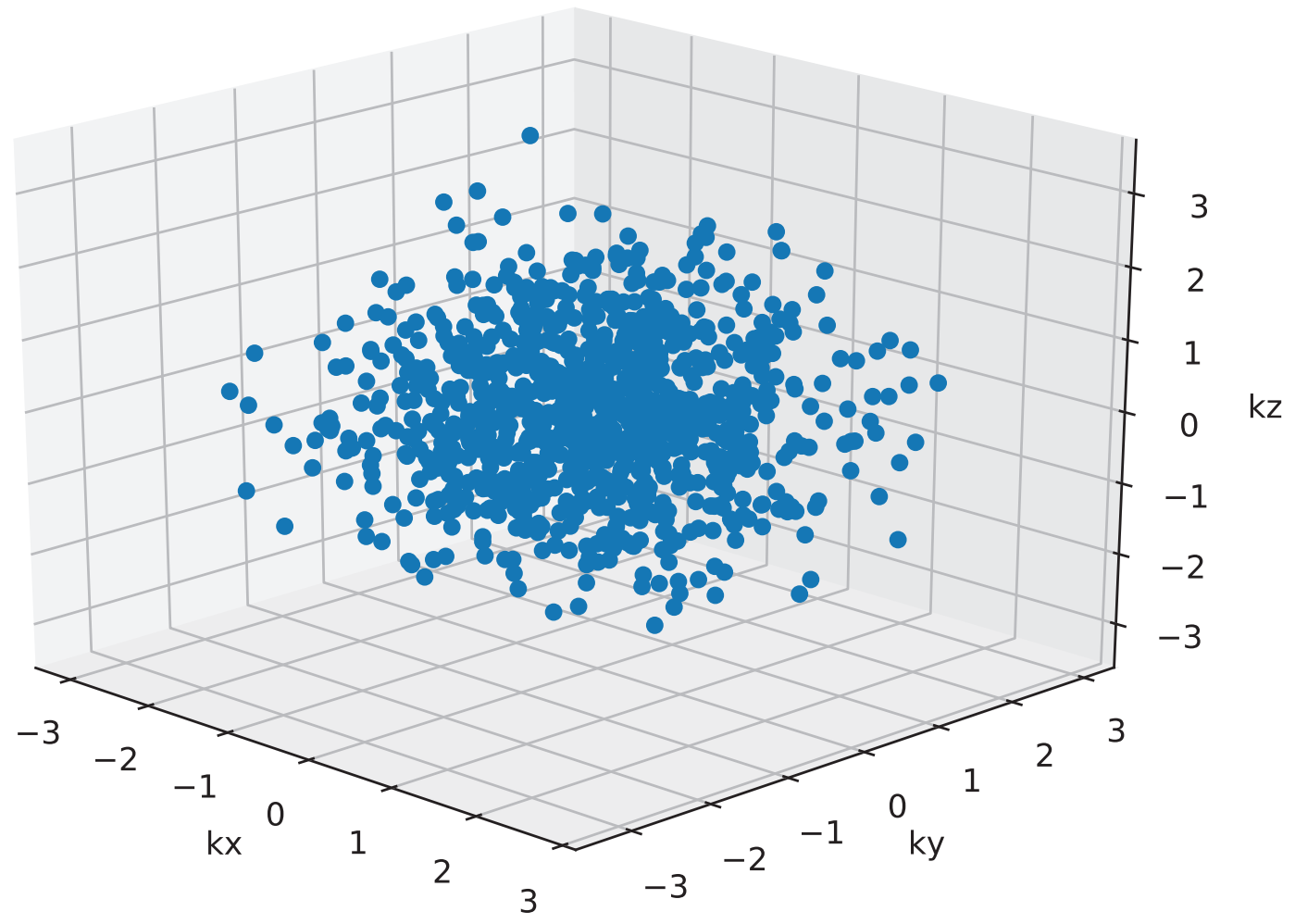

Figure 8.8: $\boldsymbol{k}$-space scatter plot of 1000 electrons after equilibration.

As a more quantitative test, we study the distribution of energies of the electrons. It is easy to show that the energies $E = k^2/2m^\star$ will have a distribution $\sim \sqrt{E}e^{-\beta E}$, and that the average energy is $\frac{3}{2}k_B T$. Figure 8.9a shows how the energy average of 10 000 electrons develops during a Monte Carlo simulation. The electrons are initialized with $E = 0$, and are equilibrated over a time scale of about $t = 20$. After this the energy average is approximately constant and $\approx 1.5$. Since the simulations were performed with $k_B T = 1$, this fits perfectly with the theory. In Figure 8.9b we show the distribution of energies at $t = 50$, where the equilibration should be complete. Again the Monte Carlo result fits perfectly with the analytical result.

Having tested that the simulations yield the correct equilibrium distribution, we turn to time dependent phenomena. First, we estimate the energy- and momentum relaxation times, which are defined respectively as

$$\frac{1}{\tau_E} = \left\langle \frac{E(\boldsymbol{k}') - E(\boldsymbol{k})}{E(\boldsymbol{k}')t} \right\rangle \quad \text{and,} \tag{8.26}$$

$$\frac{1}{\tau_m} = \left\langle \frac{(\boldsymbol{k}' - \boldsymbol{k}) \cdot \boldsymbol{k}'}{|\boldsymbol{k}'|^2 t} \right\rangle, \tag{8.27}$$

where $t$ is the time between scattering events, and $\boldsymbol{k}'$ and $\boldsymbol{k}$ are respectively the momentums before and after a scattering event. These quantities are estimated by averaging over multiple scattering events in a simulation. The averaging starts only after equilibration. We obtain $\tau_E = 4.7$, and $\tau_m = 0.095$. In Figure 8.9a we saw that the electrons are equilibrated over a time scale of about $t = 20 \approx 4\tau_E$. Thus, the energy equilibration time is some small number times $\tau_E$, which seems very reasonable.

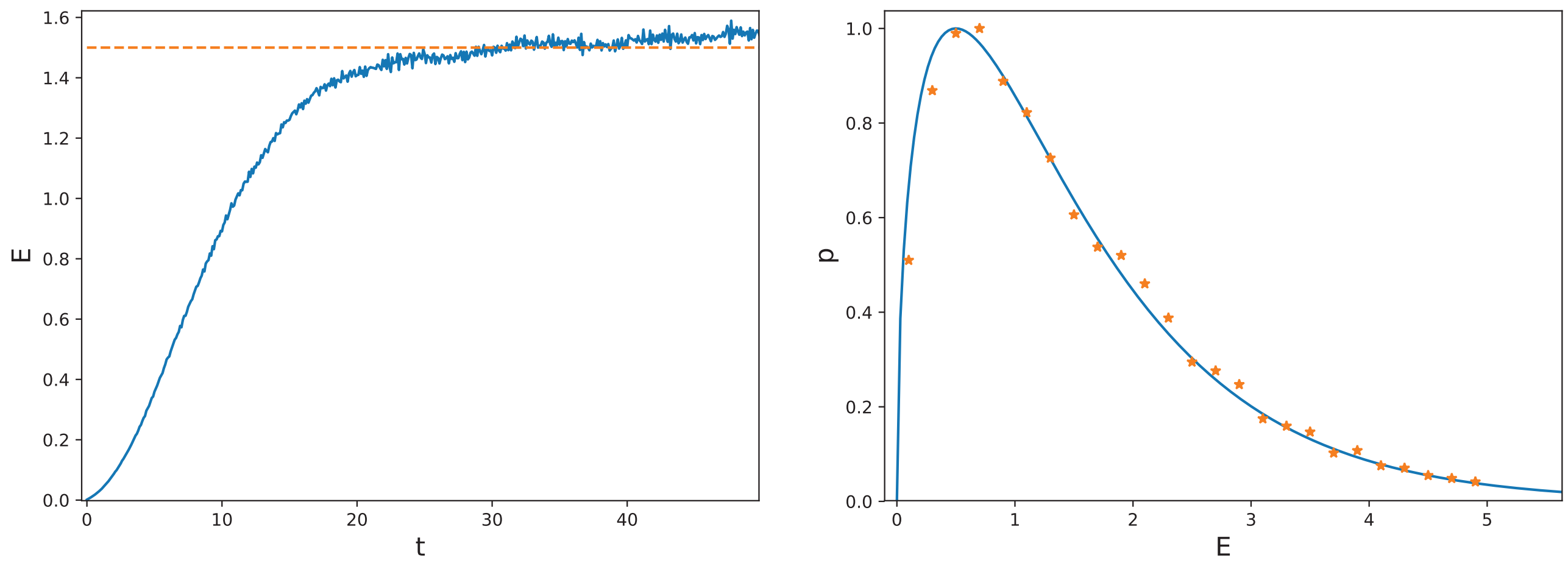

(a) Average energy

(b) Distribution at $t = 50$

Figure 8.9: Electron energy. (a) shows the average energy as a function of time, and (b) shows the distribution of energies at $t = 50$. The orange stars represent a histogram obtained from the Monte Carlo simulation, while the curve in blue represents the analytical Boltzmann distribution (8.25). The distributions have been normalized so that $\max p(E) = 1$.

Next, we turn to the correlation functions of (8.18)-(8.22). These functions are calculated by first equilibrating the simulated particles, and then approximating the expectation values as mixed ensemble and time averages. For instance, the correlation function $\langle v_j(0)v_i(t)\rangle$ is estimated as

$$\langle v_j(0)v_i(t)\rangle \approx \frac{1}{NM}\sum_{n=1}^{N}\sum_{m=1}^{M} v_{nj}(t_0 + m\Delta t)v_{ni}(t_0 + m\Delta t + t), \tag{8.28}$$

where $t_0$ is larger than the equilibration time, $N$ is the number of simulated particles, and $M\Delta t$ is a sufficiently large range of time. The correlation functions $\langle E(0)v_j(0)v_i(t)\rangle$, $\langle v_j(0)E(t)v_i(t)\rangle$ and $\langle E(0)v_j(0)E(t)v_i(t)\rangle$ are estimated in a similar manor. The results are shown in Figure 8.10. Note that many of these correlation functions are related by exchange of time arguments, and that in those cases they are identical due to time reversal symmetry. For instance, $\langle v_j(0)v_i(t)\rangle = \langle v_i(0)v_j(t)\rangle$, and $\langle v_j(0)E(t)v_i(t)\rangle = \langle E(0)v_i(0)v_j(t)\rangle$. Whenever this is the case, only one of the functions are shown in Figure 8.10.

Examining Figure 8.10, we first observe that all of the non-diagonal results, with $i \neq j$, are essentially zero within the expected accuracy of the calculation. Further, the diagonal results with $i = j$ are identical for all values of $i$. Both of these facts follow as a consequence of the spherical symmetry of the model. The fact that these symmetries are realized in the simulations is another important consistency check of the implementation. We also observe that the diagonal correlation functions seem to fall of over a time scale $\sim 0.1$, which fits well with the calculated value of $\tau_m = 0.095$.

As a final consistency check, we may consider the well known fact that the conductivity $\sigma$ is related to the diffusion constant $D$ by $\sigma = e^2\rho D/k_B T$, which

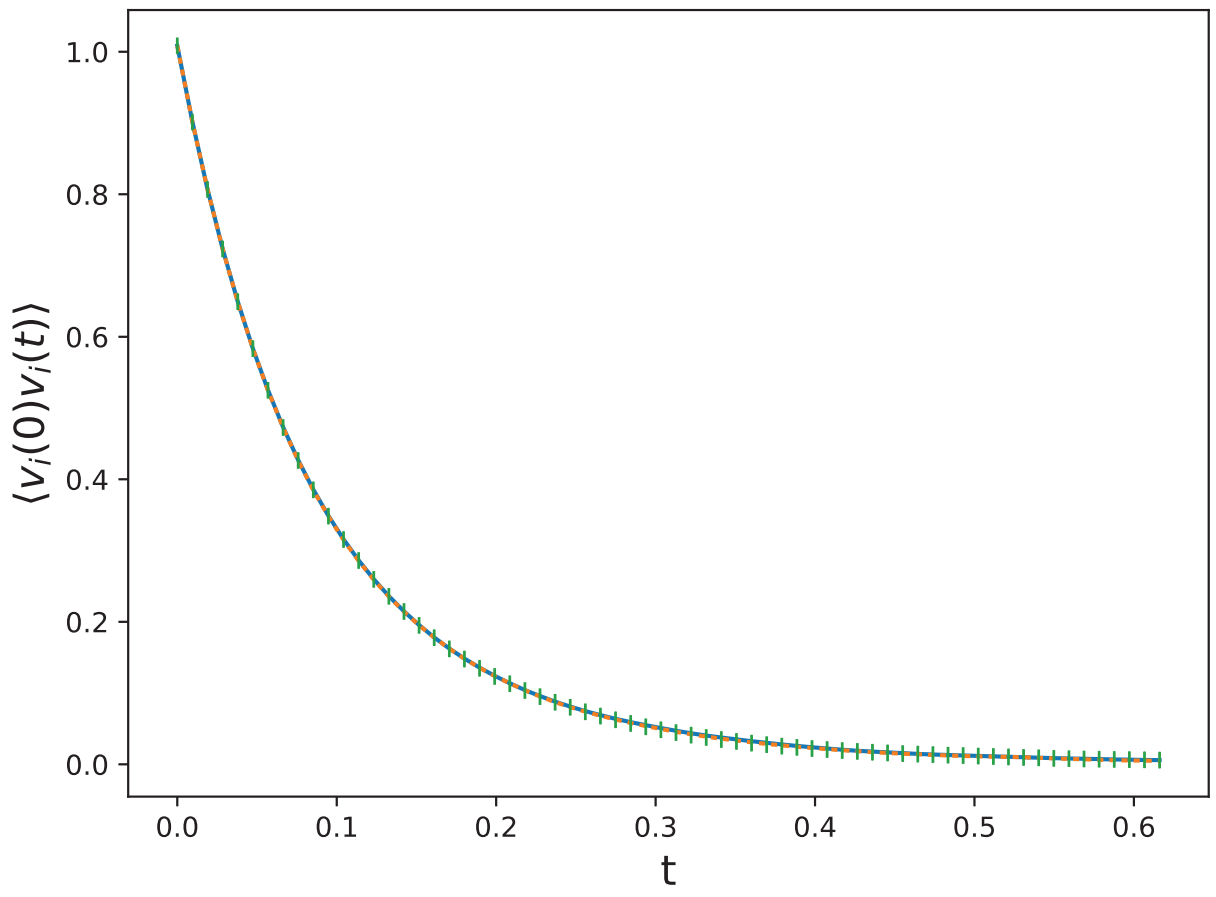

(a)

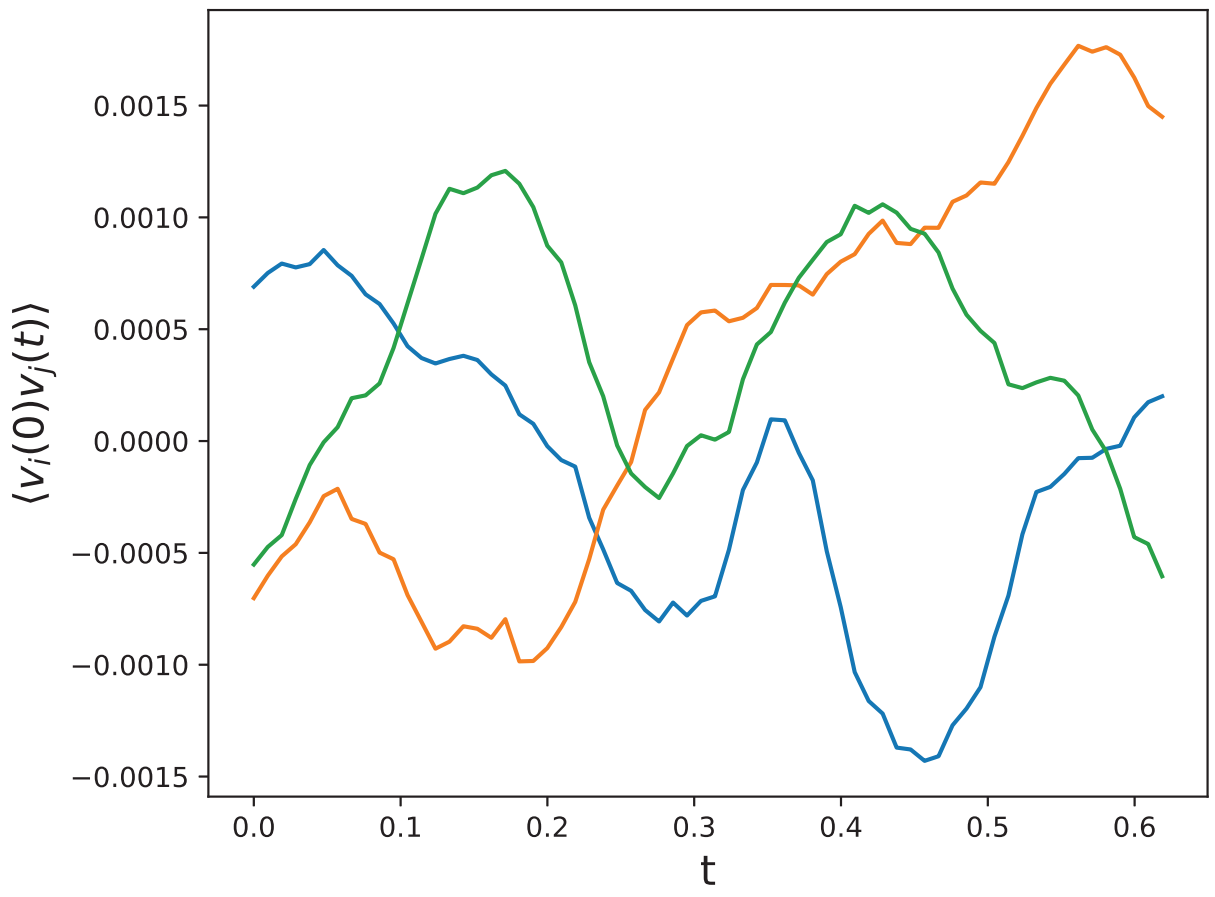

(b)

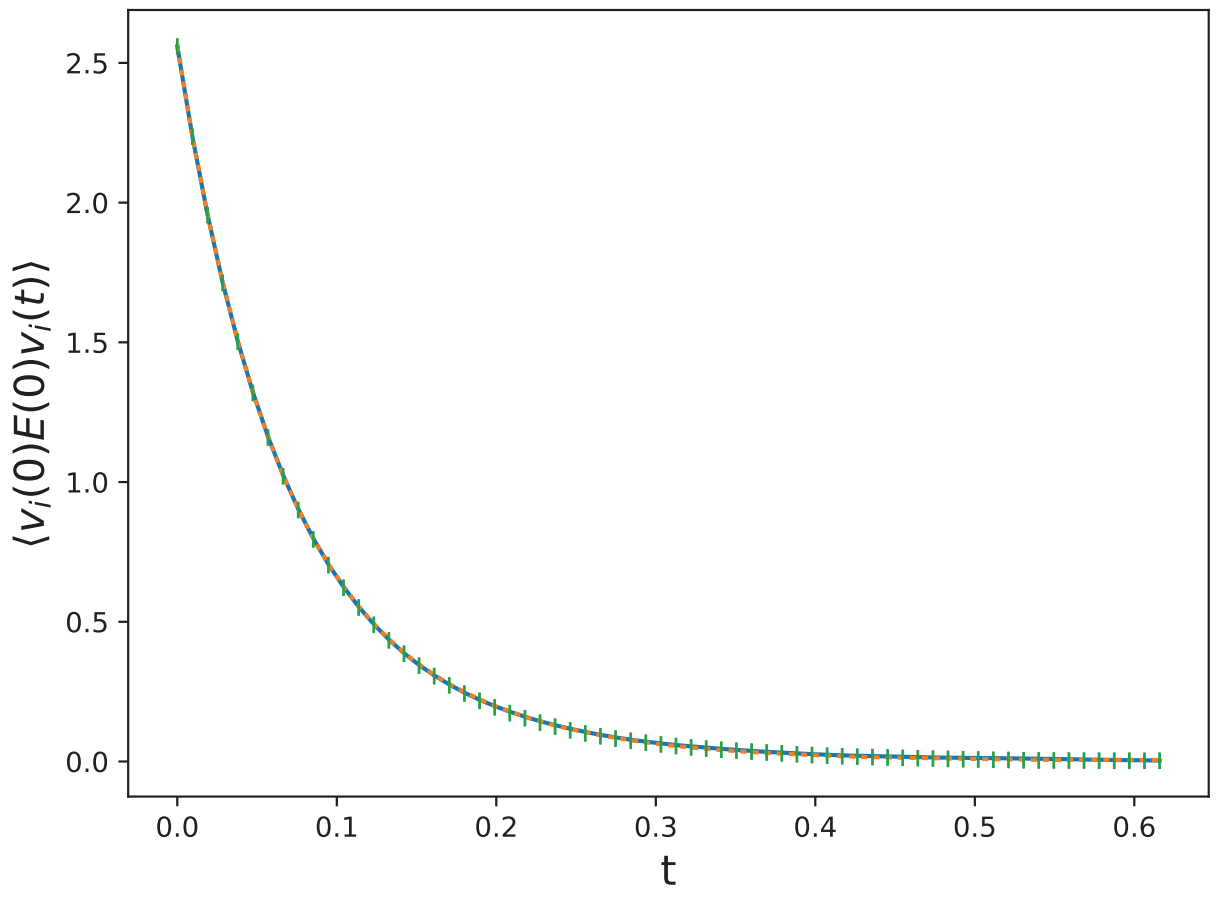

(c)

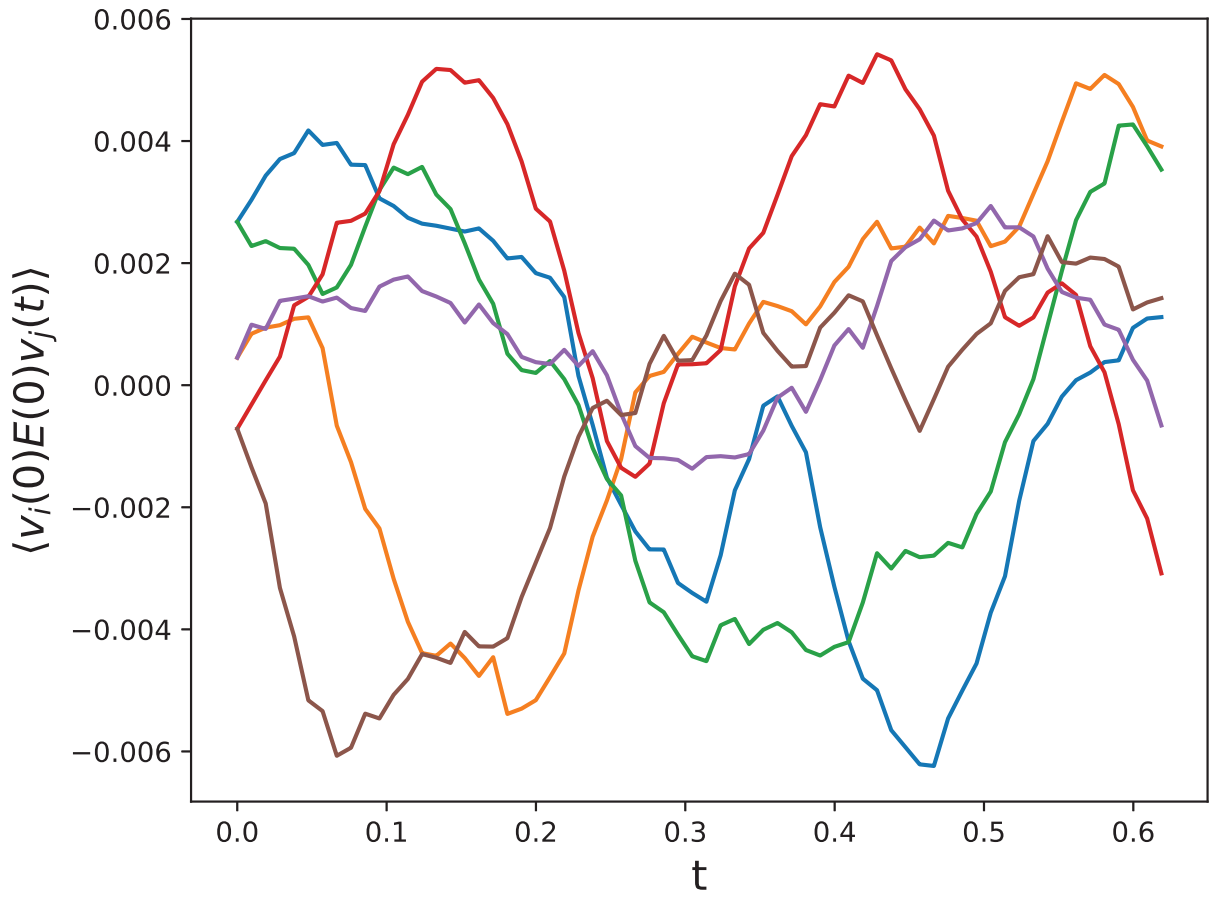

(d)

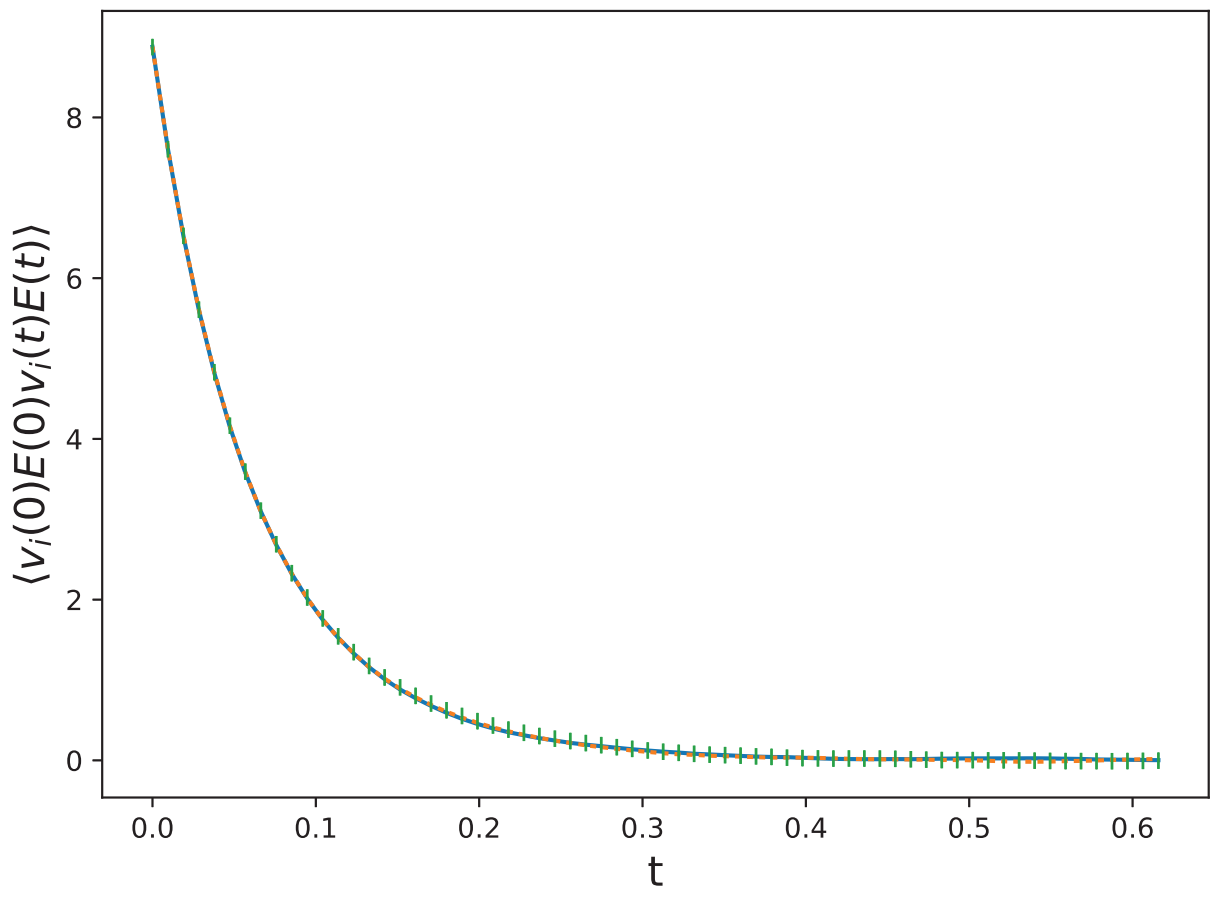

(e)

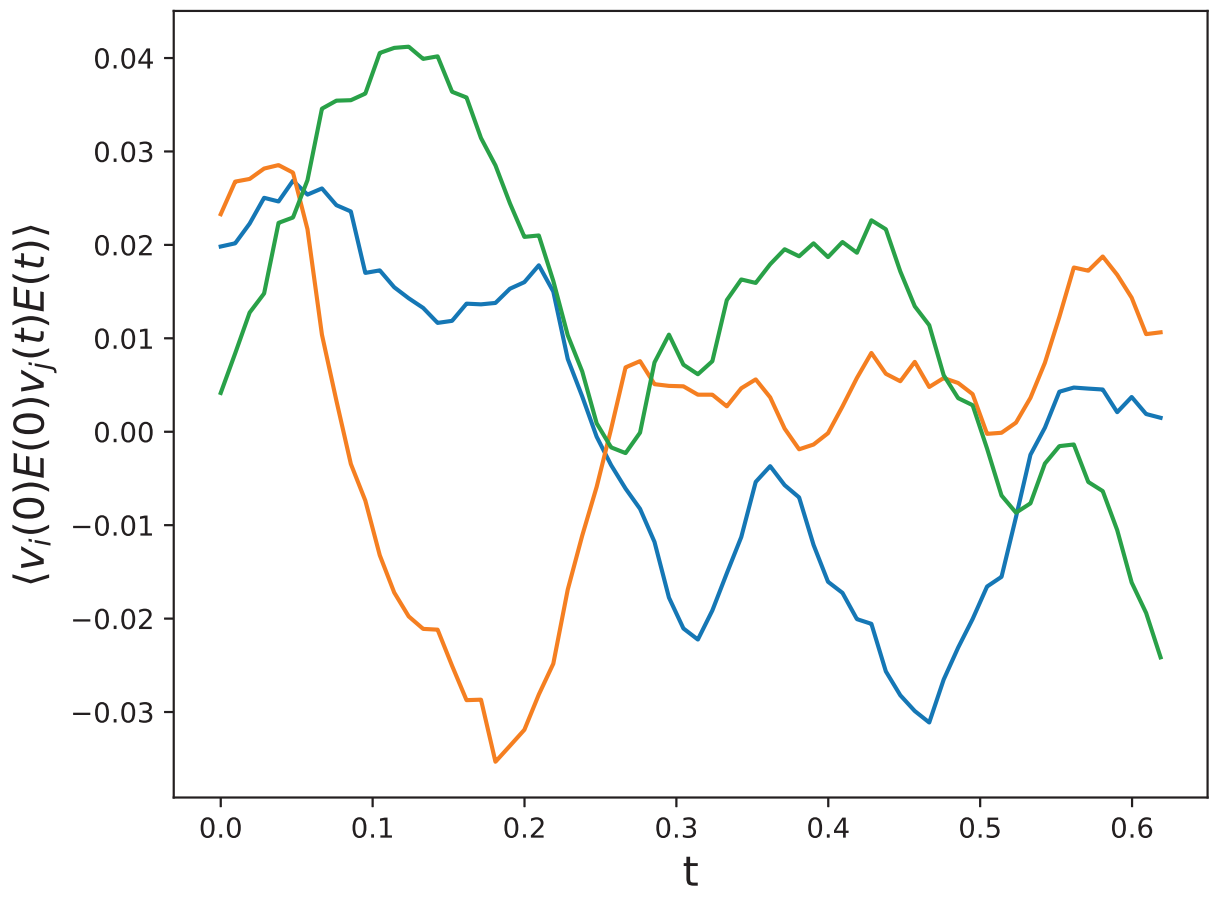

(f)

Figure 8.10: Correlation functions. The top, middle and bottom rows respectively show $\langle v_j(0)v_i(t)\rangle$, $\langle v_j(0)E(0)v_i(t)\rangle$, and $\langle v_j(0)E(0)v_i(t)E(t)\rangle$. The figures on the left show results with $i = j$, while those where $i \neq j$ are shown on the right. On the left, results shown in blue, orange and green respectively represent the cases $ii = xx$, $yy$ and $zz$. On the right, the cases $ij = xy$, $xz$, $yz$, $yx$, $zx$ and $zy$ are respectively shown in blue, orange, green, red purple and brown.

follows from the Einstein relation[7, 15]. Comparing this to (8.18), we see that we must have

$$D = \int_0^\infty \mathrm{d}t \, \langle v_i(0) v_i(t) \rangle \,, \tag{8.29}$$

where $i$ is any of $x, y$ or $z$. This can be used as a consistency check, since the diffusion constant can also be estimated more directly by considering how the variance of the particles position varies with time. In fact, it is not hard to see that in three dimensions we should have $\sigma_{\boldsymbol{x}}^2 = \langle x^2 \rangle + \langle y^2 \rangle + \langle z^2 \rangle = 6Dt$. Thus, in Figure 8.11 we show Monte Carlo estimates of $\sigma_{\boldsymbol{x}}^2$ as a function of the simulation time $t$. The dependency of $\sigma_{\boldsymbol{x}}^2$ on $t$ takes on a linear character after $\sim t = 20$. This fits well with the fact that the electrons are equilibrated at about this time, as seen in Figure 8.9a.

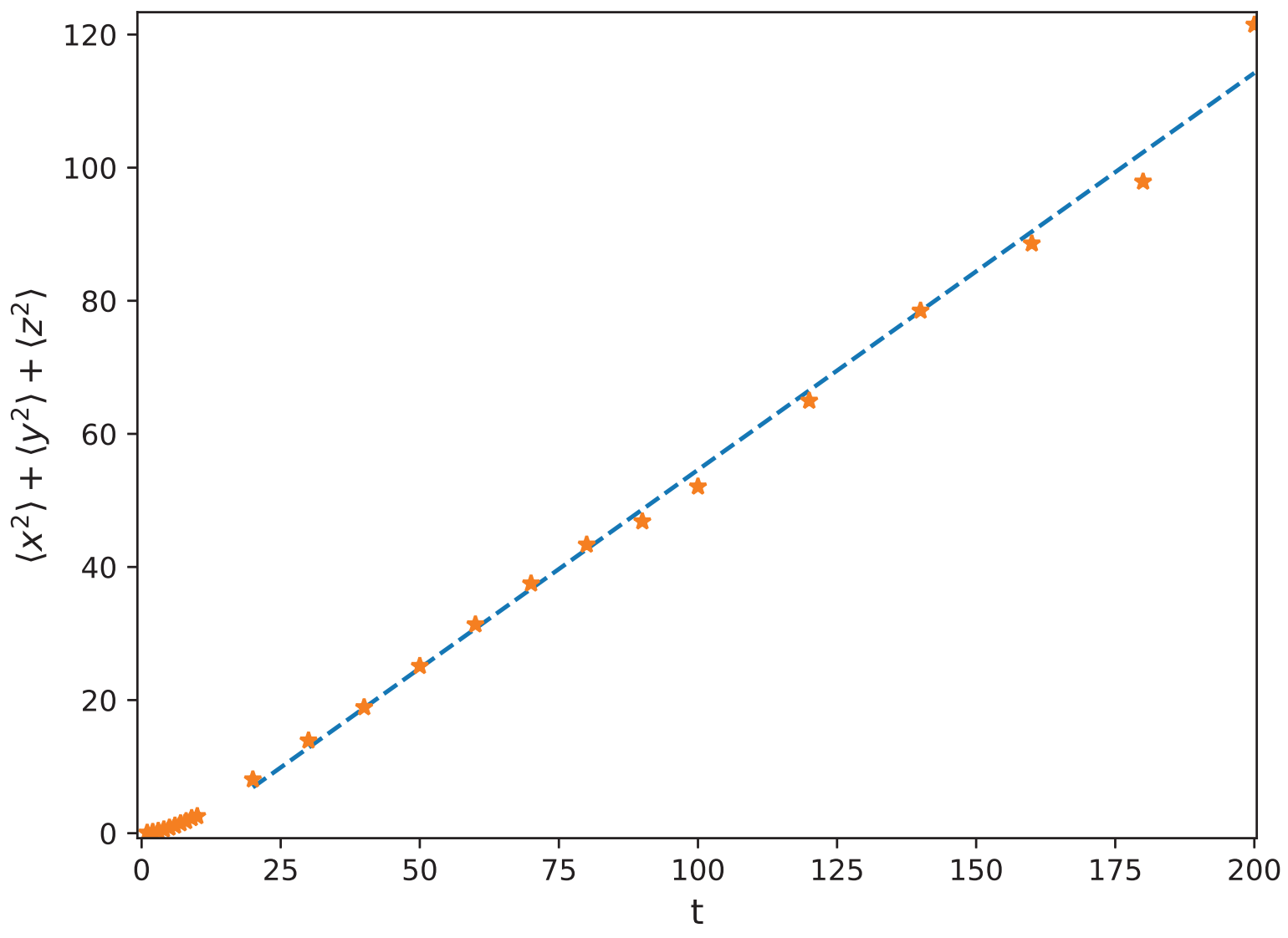

Figure 8.11: Estimation of diffusion constant. The orange stars show estimates of $\sigma_{\boldsymbol{x}}^2 = \langle x^2 \rangle + \langle y^2 \rangle + \langle z^2 \rangle$ after various simulation times $t$. The dashed line in blue shows a linear fit, fitted only against the points with $t \geq 20$.

Figure 8.11 also includes a linear model, which is fitted against the Monte Carlo results where $t \geq 20$. This model is $-5.04 + 0.596t$. Thus, we estimate the diffusion constant as $D = 0.596/6 = 0.099$. Numerically integrating the function $\langle v_i(0) v_i(t) \rangle$ shown in Figure 8.10a, we obtain $D = 0.095$ by (8.29). Within the expected accuracy, these results are equal, so the implementation passes also this consistency check.

Finally, we calculate the thermoelectric transport tensors $\overleftrightarrow{\sigma}$, $\overleftrightarrow{\alpha}$, $\overleftrightarrow{\pi}$ and $\overleftrightarrow{\kappa_e}$, using (8.18)-(8.22) and (4.49)-(4.51). The simulations are performed with $k_B T = 1$ and $\mu = 0$, which yields a charge carrier density of $\rho = 0.785$. We also assume units where $k_B = e = 1$. The results are shown in table 8.1.

Again we note that the off diagonal results are essentially zero, and that the diagonal elements of each tensor are practically identical. Again, this is a consequence of the spherical symmetry of the system, and indeed follows from the fact that the correlation functions also have these properties. We also note that the conductivity has the expected value given the formula $\sigma = e^2\rho D/k_BT = 0.785 \cdot 0.95 = 0.746$. The Peltier and Seebeck coefficients are identical since $T = 1$, and have a value of $\alpha \approx -2$. We also calculate the Lorenz coefficient to be $L = \kappa_e/\sigma T \approx 0.15/0.075 = 2$.

| | $\sigma$ | | | $\alpha$ | | |
|---|---|---|---|---|---|---|
| | x | y | z | x | y | z |
| x | 0.07451 | 1.074e-4 | 8.425e-5 | -2.021 | -0.002883 | -7.182e-04 |
| y | 1.074e-4 | 0.07453 | 5.231e-5 | -3.348e-4 | -2.017 | 0.01105 |
| z | 8.425e-5 | 5.231e-5 | 0.07474 | 0.02104 | 0.008486 | -2.024 |
| | $\pi$ | | | $\kappa_e$ | | |
| | x | y | z | x | y | z |
| x | -2.021 | -3.348e-4 | 0.02104 | 0.1505 | 1.002e-5 | 0.004471 |
| y | -0.002883 | -2.017 | 0.008486 | 1.002e-5 | 0.1491 | -0.003897 |
| z | -7.182e-4 | 0.01105 | -2.024 | 0.004471 | -0.003897 | 0.1504 |

Table 8.1: Thermoelectric transport tensors.

In the literature[22] we find analytical estimates of $\alpha$ and $L$ within certain approximations. These are $\alpha = -k_B/e(\beta\mu + s + 5/2)$ and $L = k_B^2/e^2(s + 5/2)$, where $s$ is an exponent describing the energy dependence of the scattering rates. For acoustic phonon scattering with parabolic bands[22], $s = -1/2$. Accordingly. we should have $\alpha = -(0 - 1/2 + 5/2) = -2$ and $L = -1/2 + 5/2 = 2$, which fits perfectly with the simulation results.

#### 8.1.2.1 Discussion

Clearly, the application of this method to the simple bulk system above an overkill, since the simple analytical estimates yield the same results. But the application to the bulk system serves as a test of the implementation. Indeed, the agreement with the analytical estimates, together with the other test results, indicates that we can put some trust in the implementation. Thus, the next step would be to apply the method to superlattices, after which we could repeat the studies of the previous section.

The indirect Green-Kubo method described here would be much better suited for that purpose, compared to the direct method made use of there. The fact that we are making use of a more realistic scattering mechanism is only one of the reasons for this. In addition, it is a major advantage that we can perform the simulation at equilibrium, since this entails that we will not have to solve the Poisson equation self consistently with the transport problem, and that we can always apply periodic boundary conditions. Further, when applying the direct method, we always have to make sure the perturbations we apply are small enough for the transport to be within the linear regime. On the other hand, if we choose the perturbations to small, the signal to noise ratio will be very low, which can also cause erroneous results. This trade-of is something we do not have to consider when using the indirect method presented here.

However, while the implementation is better suited than that of the previous section, it is still lacking important aspects. First of all, like the direct implementation, it is still limited to parabolic bands. This could however be easily remedied, for instance by making use of a tight binding model like the ones we have used in our publications (papers I-II). A larger problem is the limitation to non-degenerate statistics. The derivation of (8.18)-(8.22) assumed that the particles are uncorrelated, which will not be the case if the statistics is degenerate. In addition, the introduction of degenerate corrections to the scattering rates seem to cause a drastic increase in the variance of the correlation functions. However, both of these issues could probably be resolved by more careful consideration.

A more serious problem is the fact that the method is still limited to classical transport, which means we are not able to model quantum transport phenomena such as tunneling and resonance between barriers. Tunneling can probably be dealt with in terms of pre-calculated transmission functions, like we try in our second publication (Paper II). Resonances however, can only be dealt with in a classical transport framework if we assume coherence throughout the super cell, so that the miniband dispersion relation of the superlattice can be utilized. The extent to which that assumption affects the results is the subject in our second publication.

## 8.2 Perturbative approximations to $G^r$

This section concerns the evaluation of the retarded Green's function $G^r$ through the solution of (5.81). This is an important step in the evaluation of currents through the NEGF formalism, and indeed in any perturbative transport framework formulated in terms of Green's functions. A major problem when dealing with large systems like those considered in our publications, is the time it takes to solve this equation. Solving the equation through direct inversion would require an amount of time proportional to $N^3$, where $N$ is the side of the involved matrices. In an atomistic model $N = n_o n_a$, where $n_a$ is the number of atoms in the system, and $n_o$ is the number of states included per atom. Accordingly, direct inversion is fairly intractable in large systems.

In our two first publications (Papers I-II), the solution of (5.81) is done by the external transport framework Kwant[12, 1], which is able to use a variety of different methods. In our third paper (Paper III) we make use of the RGF-algorithm, which is described in detail in the literature[6]. The RGF-algorithm assumes that the system can be divided into a set of $n_Z$ slices, in such a way that there is only nearest neighbor coupling between the slices. It then scales as $n_X^3 n_Z(1 + n_C)$, where $n_X$ is the number of states per slice, and $n_C$ is the number of columns required in the block decomposition of $G^r(E)$. In particular, if only the diagonal of $G^r(E)$ is required, then the RGF-algorithm scales as $n_X^3 n_Z$, while if the full matrix is required it scales as $n_X^3 n_Z^2$. In any case, this is a significant improvement compared to direct inversion, as long as one is dealing with a quasi-one-dimensional system, where $n_X$ is smaller than $n_Z$.

Before deciding to use the RGF algorithm, I spent some time considering alternative ways to speed up the solution of (5.81). This section deals in particular with one such approach, which was quite successful. It involves treating states with energies far away from the range of consideration using perturbation theory, rather than the exact expressions. The computational requirements are still cubically scaling with system size, so the approach is not able to compete with the RGF-algorithm in large quasi-one-dimensional systems. However, it may have an advantage over RGF in more general systems.

We begin by considering (5.81) in the absence of scattering, so that it can be written as

$$G^r(E) = \left( E - H - \sum_p \Sigma_p^r(E) \right)^{-1}, \tag{8.30}$$

where $\Sigma_p^r$ is the retarded lead self energy associated with lead $p$. Combining (5.105) with (5.86), (5.65) and (5.90) it is easy to see that in the absence of scattering

$$i_p(E) = \frac{1}{h} \sum_q \mathrm{Tr}\, \Gamma_p^F(E) G^r(E) \Gamma_p^F(E) G^a(E)(f_q(E) - f_p(E)), \tag{8.31}$$

where we have also used that

$$\Gamma(E) = i\left(\Sigma^r(E) - \Sigma^a(E)\right) = \sum_q \Gamma_p^F(E), \tag{8.32}$$

which follows from (5.74), (5.75), (5.65) and (5.66). Comparing (8.31) to (2.16), we see that we must have

$$\bar{\mathcal{T}}_{qp}(E) = \mathrm{Tr}\, \Gamma_p^F(E) G^r(E) \Gamma_p^F(E) G^a(E), \tag{8.33}$$

an equation which is also derived by Datta[7]. The accuracy of the following approximations will be judged in terms of their ability to reproduce the correct transmission function $\bar{\mathcal{T}}(E)$, where the subscripts $qp$ have been omitted, since we are considering a system with only two leads.

The perturbative procedure requires that the states of our system is divided into two groups based on their energies. One of these groups will be treated exactly, and one perturbatively. However, the states in terms of which a model is formulated, does usually not have a wide range of different energies. For instance, the tight binding model of CdTe employed in our publications (Papers I-II), only has two different types of states: s-orbitals with energy 4.95 eV, and p-orbitals with energy 0.52 eV. This is too coarse grid of energies for us to expect the method to work particularly well.

To introduce a larger variety of different energies, we proceed as follows: First we construct small groups of states localized close to each other. In our case these groups consist of four neighboring unit cells of CdTe. Indexing these groups by $i$, the total Hamiltonian of the system can be written as $H = \sum_{ij} H_{ij}$, where $H_{ii}$ represents the Hamiltonian of group $i$ by itself, and the terms with $i \neq j$ represent interactions between the groups. We then diagonalize each of the operators $H_{ii}$ to get a new set of states $|il\rangle$, with energies equal to the corresponding eigenvalues $\epsilon_{il}$. The resulting grid of energies is much denser than that of the original model, as illustrated in Figure 8.12.

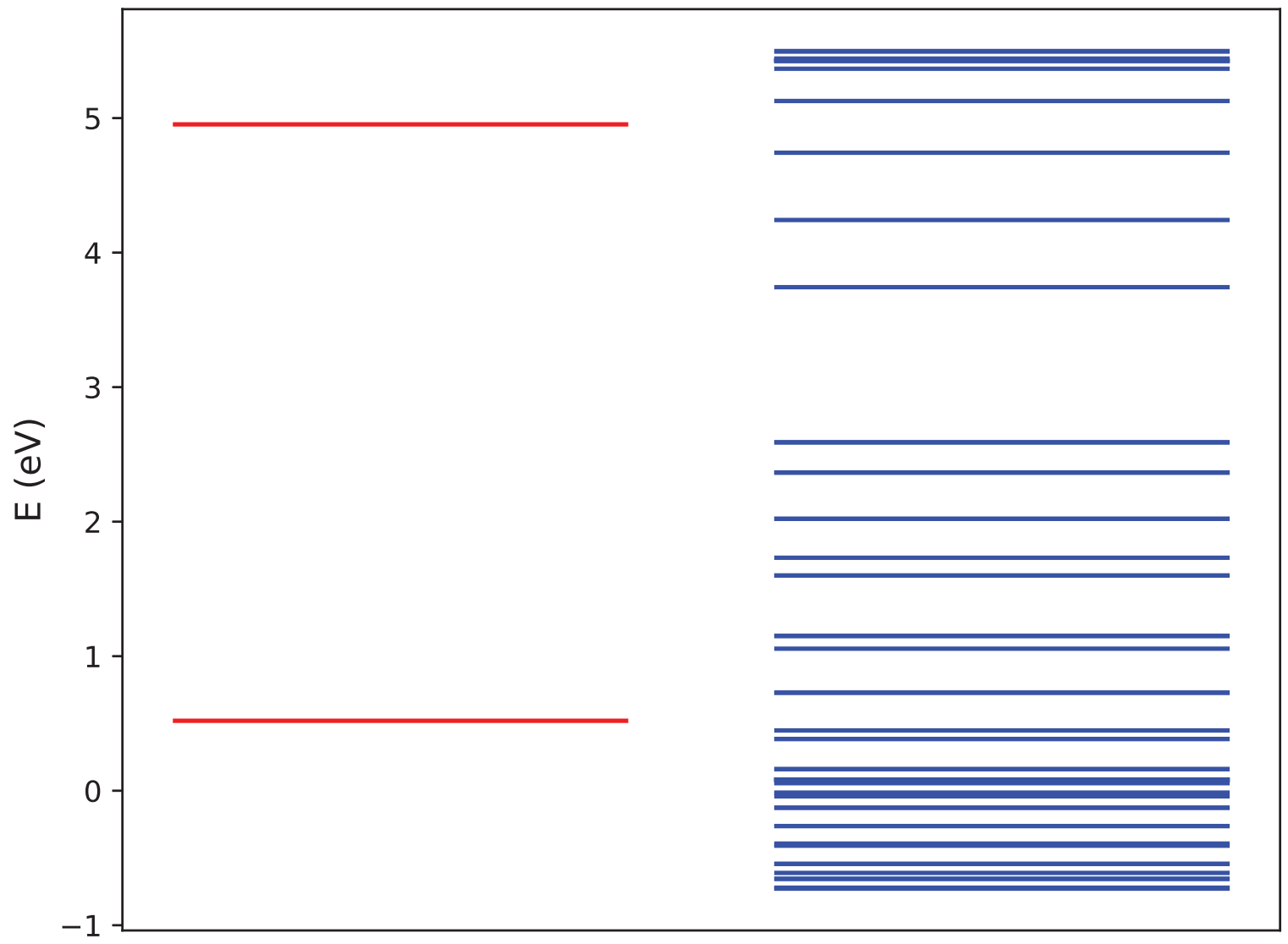

Figure 8.12: Energy levels in a CdTe system. The red lines on the left show the two energy levels of the original tight binding model, while those on the right show the energy levels after applying the procedure described in the text, where four neighboring unit cells of CdTe are grouped together, and the groups isolated Hamiltonian is diagonalized, resulting in a new set of states.

Next, we define an energy range $[E_1, E_2]$, where states $|il\rangle$ such that $E_1 \leq \epsilon_{il} \leq E_2$ will be treated exactly, and the others perturbatively. In the calculations below, we have unless otherwise specified set $E_1 = 0.4$ eV and $E_2 = 5.2$ eV, which results in roughly a third of the energies $\epsilon_{il}$ being in the range $[E_1, E_2]$. Following a fairly standard procedure[27], we define a projection operator $P$ such that $P|il\rangle = |il\rangle$ if $E_1 \leq \epsilon_{il} \leq E_2$, and $P|il\rangle = 0$ otherwise. We also define $Q = I - P$. The total Hamiltonian can then be written

$$H = (Q + P)H(Q + P) = H_0 + H^\star + V + V^\dagger, \qquad (8.34)$$

where we have defined $H_0 = PHP$, $H^\star = QHQ$, and $V = PHQ$.

The first approximation we will consider involves interaction with the leads. These interactions are described by the matrices $\Gamma_p^F$ from (8.35), with $p \in \{1, 2\}$ since there are only two leads. The approximation consists of replacing these operators with $P\Gamma_p^F P$, which means we are allowing only states with energy within the range $[E_1, E_2]$ to interact with the leads. The consequences of this approximation is shown in Figure 8.13. There we show the transmission function of a system consisting of 32 unit cells of CdTe, arranged along the z-axis, and employing the tight bind model of our first publication. Periodic boundary conditions are imposed in the x- and y-directions, and for the lead self energies we use a simple diagonal model $\Sigma_p^r(E) = -iP_p/2\tau$, with $\tau = 2$ eV$^{-1}$, and $P_p$ a projection operator projecting upon the atoms nearest lead $p$.

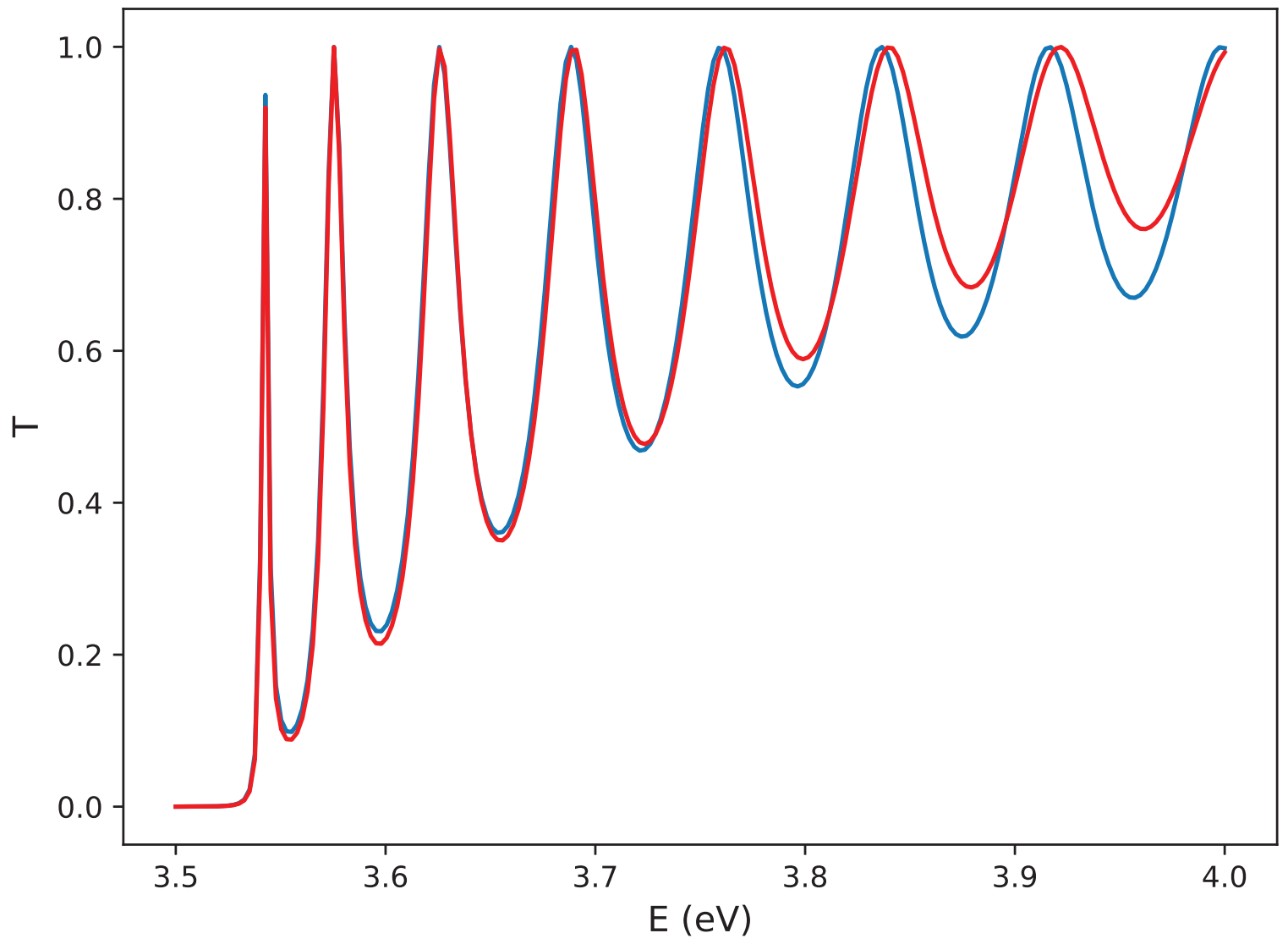

Figure 8.13: Transmission function of a CdTe system. The curve in blue shows the unapproximated transmission function, while the curve in red shows the result after the approximation $\Gamma_p^F = P\Gamma_p^F P$.

Figure 8.13 contains both the exact result, and the result of the approximation $\Gamma_p^F = P\Gamma_p^F P$. We observe that the exact and the approximate results are quite similar in the low lying range $E \sim 3.6$ eV, but that deviations start to become significant at higher energies, closer to the upper limit $E_2 = 5.2$ eV. In the long run we could consider making corrections also to the self energies, which might reduce the deviations seen in Figure 8.13. For now however, such corrections have not been implemented. The model with $\Gamma_p^F = P\Gamma_p^F P$ is thus to be regarded as the target result, to which further approximations will be compared.

Within the approximation $\Gamma_p^F = P\Gamma_p^F P$, we can use the fact that $P^2 = P$ to write (8.31) as

$$\bar{\mathcal{T}}(E) = \mathrm{Tr}\, P\Gamma_1^F(E)PPG^r(E)PP\Gamma_2^F(E)PPG^a(E)P. \tag{8.35}$$

In order to calculate the transmission function, we only need the projected Green's function $PG^r(E)P$. Combining (8.30) with (8.34) we obtain

$$\left(E - H_0 - H^\star - V - V^\dagger - \Sigma_l^r(E)\right) G^r(E) = I, \tag{8.36}$$

where we have defined $\Sigma_l^r(E) = \sum_p \Sigma_p^r(E)$. Multiplying (8.36) on both sides by $P$, and using that $P^2 = P$, $Q^2 = Q$, and $PQ = QP = 0$, we obtain

$$\left(E - H_0 - \Sigma_l^r(E)\right) PG^r(E)P - VQG^r(E)P = P. \tag{8.37}$$

Similarly, multiplying (8.36) on the left by $Q$ and on the right by $P$, we get

$$(E - H^\star)\, QG^r(E)P - V^\dagger PG^r(E)P = 0. \tag{8.38}$$

Solving (8.38) for $QG^r(E)P$, we find $QG^r(E)P = (E - H^\star)^{-1} V^\dagger PG^r(E)P$, and upon inserting this in (8.37) we obtain the equation

$$\left(E - H_0 - \Sigma_l^r(E) - V\,(E - H^\star)^{-1} V^\dagger\right) PG^r(E)P = P. \tag{8.39}$$

Thus, if we consider inversion only in the subspace spanned by the states $|il\rangle$ with energy in the range $[E_1, E_2]$, we obtain the expression

$$PG^r(E)P = (E - H_0 - \Sigma^r(E))^{-1}, \tag{8.40}$$

which is on the same form as (8.30), but where we are using a corrected self energy $\Sigma^r(E) = \Sigma_l^r(E) + \Sigma_R^r(E)$, with

$$\Sigma_R^r(E) = V\,(E - H^\star)^{-1} V^\dagger. \tag{8.41}$$

The perturbative approximation consists of approximations to (8.41). We begin by writing $H^\star$ as

$$H^\star = \sum_{il} \epsilon_{il} |il\rangle\langle il| + \sum_{i \neq j} H_{ij}^\star, \tag{8.42}$$

where the first sum is only over $l$ such that $\epsilon_{il} \notin [E_1, E_2]$, and where $H_{ij}^\star = QH_{ij}Q$. We then define $D = \sum_{il} \epsilon_{il} |il\rangle\langle il|$, and $U = \sum_{i \neq j} H_{ij}^\star$, and we write

$$\begin{aligned}(E - H^\star)^{-1} &= (E - D - U)^{-1} = (E - D)^{-1} \left(I - U(E - D)^{-1}\right)^{-1} \\ &= (E - D)^{-1} \sum_{n=0}^{\infty} \left[U(E - D)^{-1}\right]^n,\end{aligned} \tag{8.43}$$

assuming converge of the sum. Finally, we insert (8.43) in (8.41), and obtain

$$\Sigma_R^r(E) = V\,(E - D)^{-1} \sum_{n=0}^{\infty} \left[U(E - D)^{-1}\right]^n V^\dagger. \tag{8.44}$$

Perturbative approximations of various orders are now obtained simply by replacing $\infty$ by some finite number $N$. Note that the order of the expansion refers to the total number of interaction factors $U$ or $V$. Thus, the term with $n = 0$ is a second order term, the term with $n = 1$ a third order term and so on.

Finite order termination of (8.44) are easy to implement using a sparse matrix library, as we have done using Python. In Figure 8.14 we show results of such an implementation, applied to the same system studied in Figure 8.13. The reference result is the result shown in red in Figure 8.13, and is shown in black in Figure 8.14a. The perturbative result of order zero, obtained by setting $\Sigma_R^r(E) = 0$, lies quite far from the reference. However, already the second order result, with $\Sigma_R^r(E) = V\,(E - D)^{-1}\,V^\dagger$, is almost indistinguishably from the reference result. So are also the higher order results. Thus, in Figure 8.14b we show the errors $|\mathcal{T}_n(E) - \mathcal{T}(E)|$ in a log plot, in order to more clearly illustrate the improvement in accuracy with increasing order.

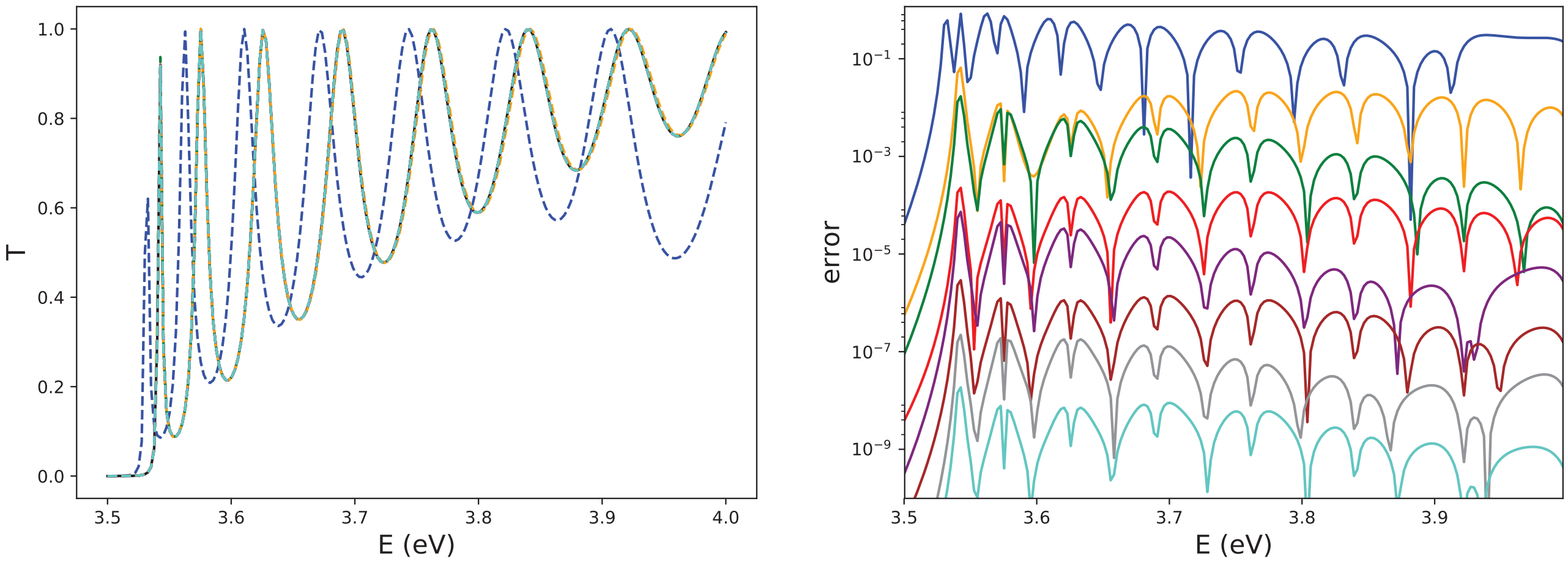

(a) Transmission function

(b) Error of perturbative expansion

Figure 8.14: Transmission function of a CdTe system. (a) shows $\mathcal{T}(E)$ itself, while (b) shows the error $|\mathcal{T}_n(E) - \mathcal{T}(E)|$, where $\mathcal{T}_n(E)$ is a perturbative approximation of order $n$. In (a) $\mathcal{T}(E)$ is shown as a solid black line, while the perturbative results are shown as dashed lines. In both figures, the blue, orange, green, red, purple, brown, grey and turquoise lines respectively represent perturbative results of order zero, two, three, four, five, six, seven and eight.

As one would expect, the error varies quite a bit with energy. To remove this variation, we can consider the maximal error in some range of interest, which we here set arbitrarily to $[3.5, 3.7]$. In Figure 8.15 we show how the maximal error in this range varies both with the perturbation order $n$, and with the two energy limits $E_1$ and $E_2$. Note that the most accurate results are limited by the numerical accuracy, and flatten out at $\sim 10^{-13}$. The dependency of the error on $n$ is seen to fit well with a simple exponential model $ar^n$. The most important parameter from this fit is the base $r$, which tells us how fast the perturbative approximation converges towards the correct result. Figure 8.16 shows how this parameter depends on the choice of limits $E_1$ and $E_2$. Other than the fact that $r$ decreases when the energy range $[E_1, E_2]$ is increased, the most important thing to note is that we always have $r < 1$, which indicates that we at least in this case, always have convergence in the limit $n \to \infty$.

In addition to the accuracy of the perturbative approximations, we must also compare their computation time requirements to that of the exact solution. Figure 8.17a illustrates how the computation time depends on the perturbation order. It increases close to linearly, which makes sense since for each new order a new term in the expansion (8.44) must be calculated. We note however that the growth rate goes down slightly after $n = 8$. This is probably related to the fact that we have divided the system into 8 groups. Since we have nearest neighbor interaction between these groups, the sparse matrices $[U(D-E)^{-1}]^n$ from (8.44) will be maximally dense after $\sim 8$ orders.

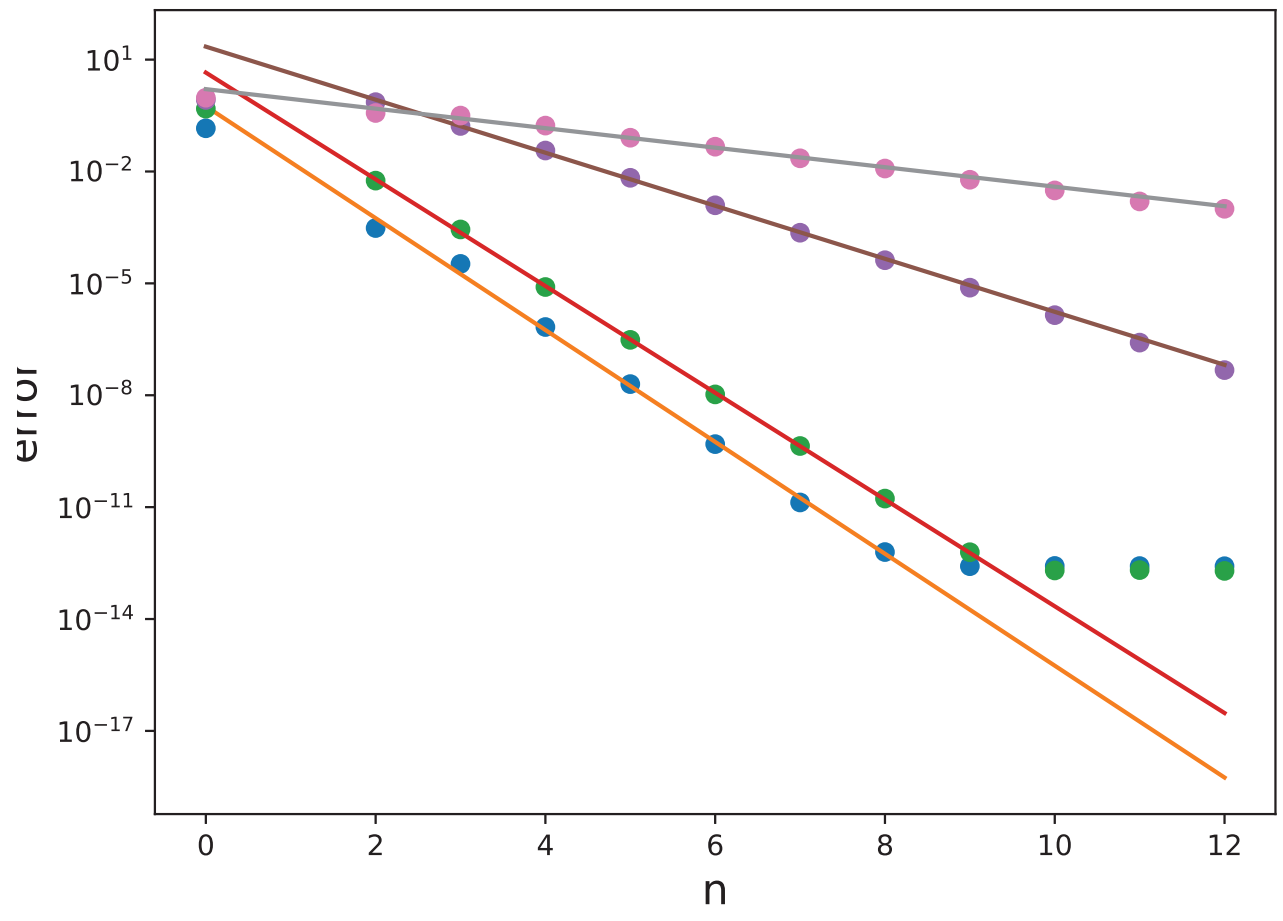

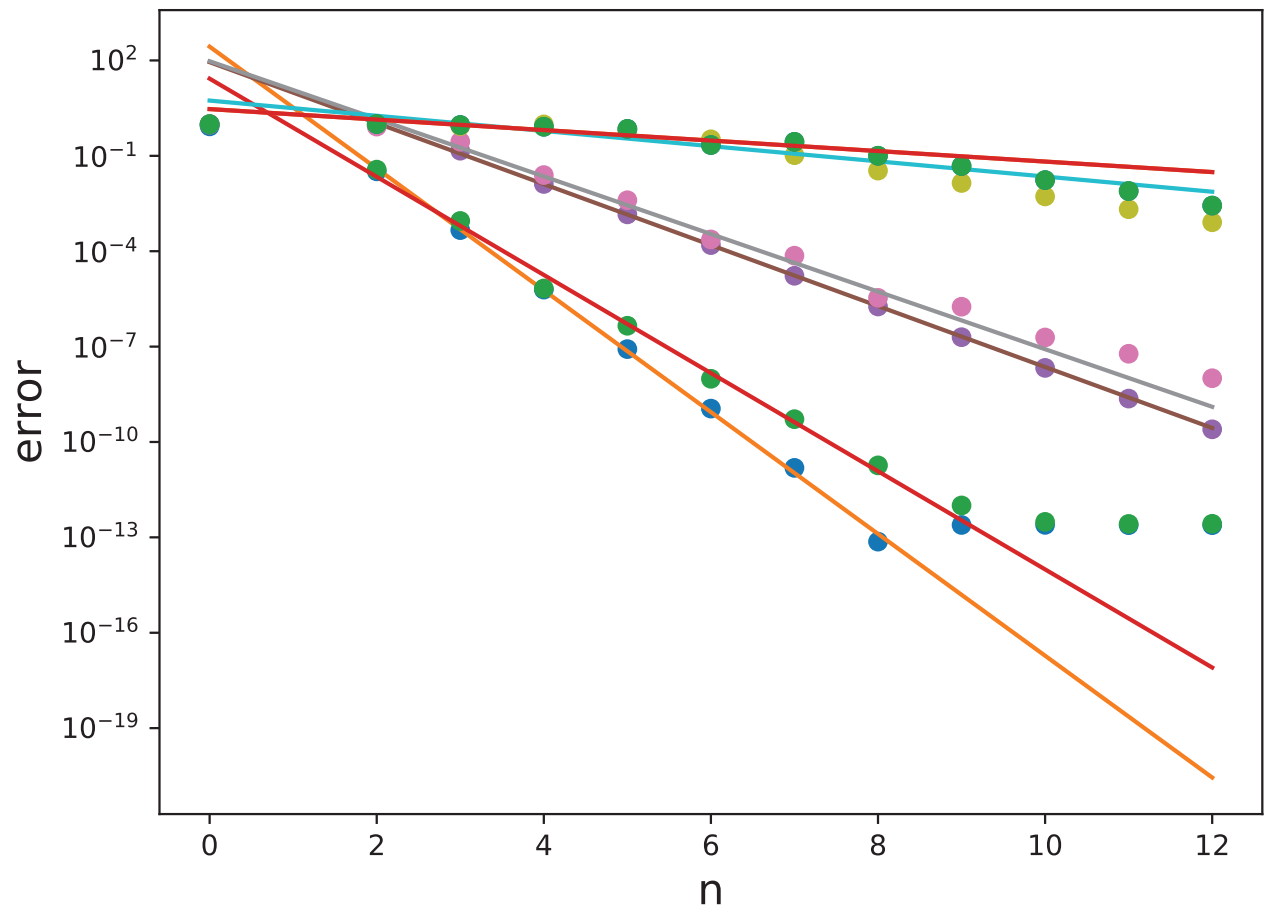

Figure 8.15: Maximal error $|\mathcal{T}_n(E) - \mathcal{T}(E)|$ of the perturbative approximation in the energy range $[3.5, 3.7]$, as a function of the perturbation order $n$. In (a) $E_1 = -\infty$, and the results shown in blue, green, purple and pink are obtained respectively with $E_2 = 5.12$, 4.74, 4.24 and 3.74 eV. In (b) $E_2 = \infty$, and results obtained with $E_1 = -0.13$, 0.38, 1.06, 1.60, 2.37, 2.59, and 3.74 eV are shown respectively in blue, green, purple, pink, yellow, blue and green. The solid lines represent exponential models $ar^n$ fitted against the results where $2 \leq n \leq 8$.

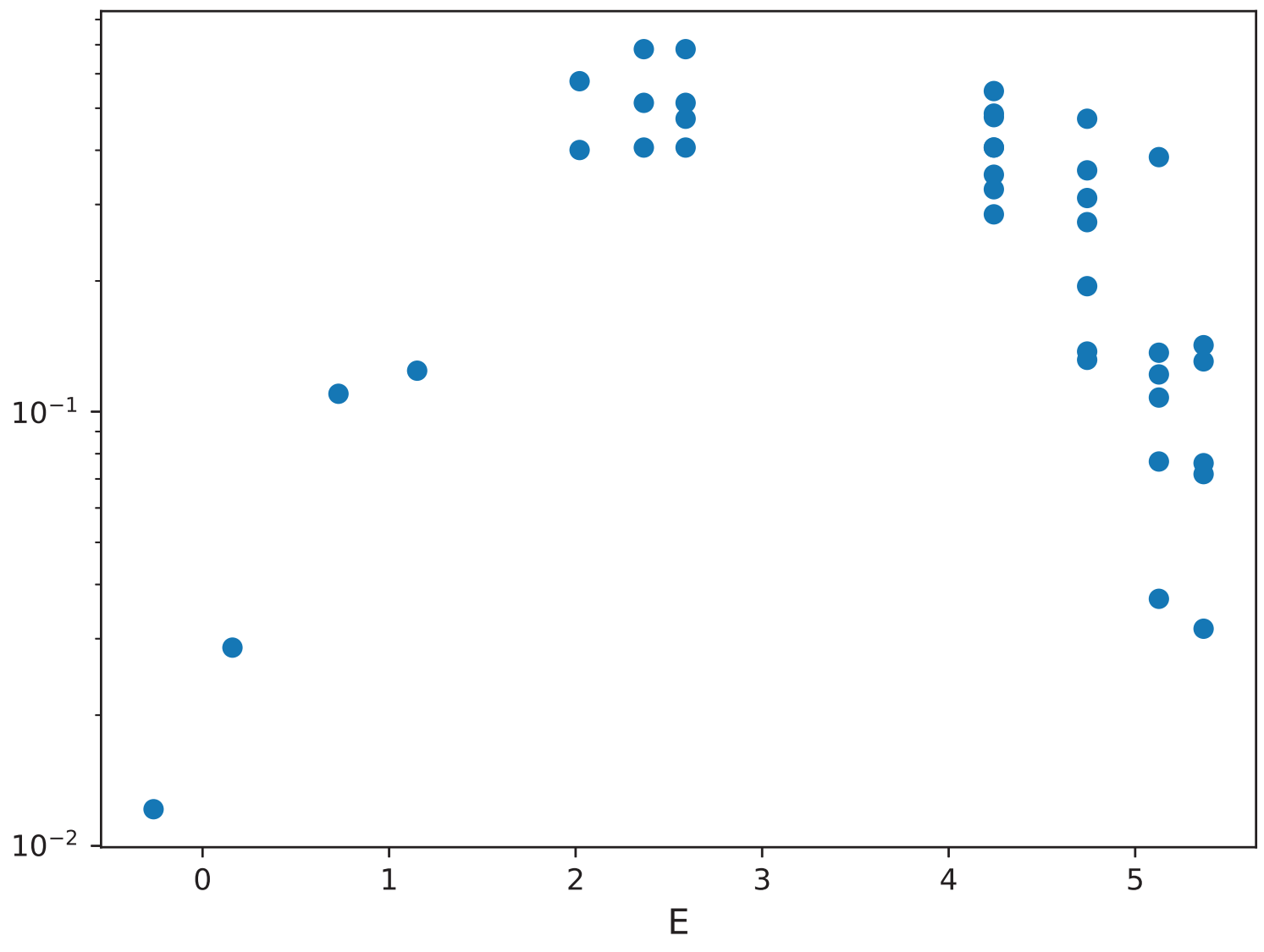

Figure 8.16: Dependency of convergence rate on the energy range $[E_1, E_2]$. We show all possible combinations of $E_2 \in \{5.12, 4.74, 4.24, 3.74, \infty\}$ and $E_1 \in \{-\infty, -0.13, 0.38, 1.06, 1.60, 2.37, 2.59, 3.74\}$. The y-axis shows the parameter $r$ from the exponential fits illustrated in Figure 8.15, and the x-axis shows the energy $\epsilon_{il} \notin [E_1, E_2]$ which is closest to the range $[4.5, 4.7]$, i.e. the closest perturbatively treated energy.

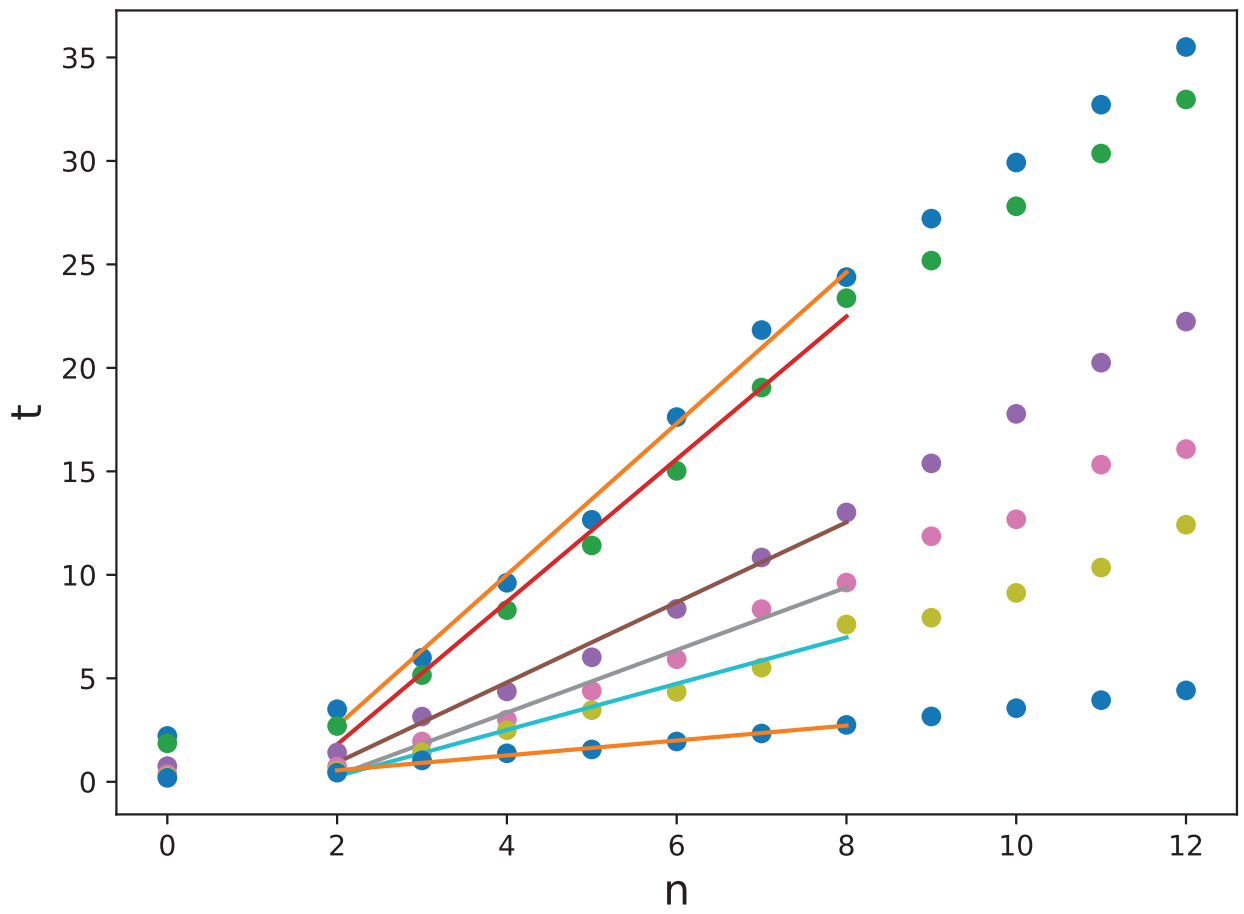

(a) Computation time vs perturbation order

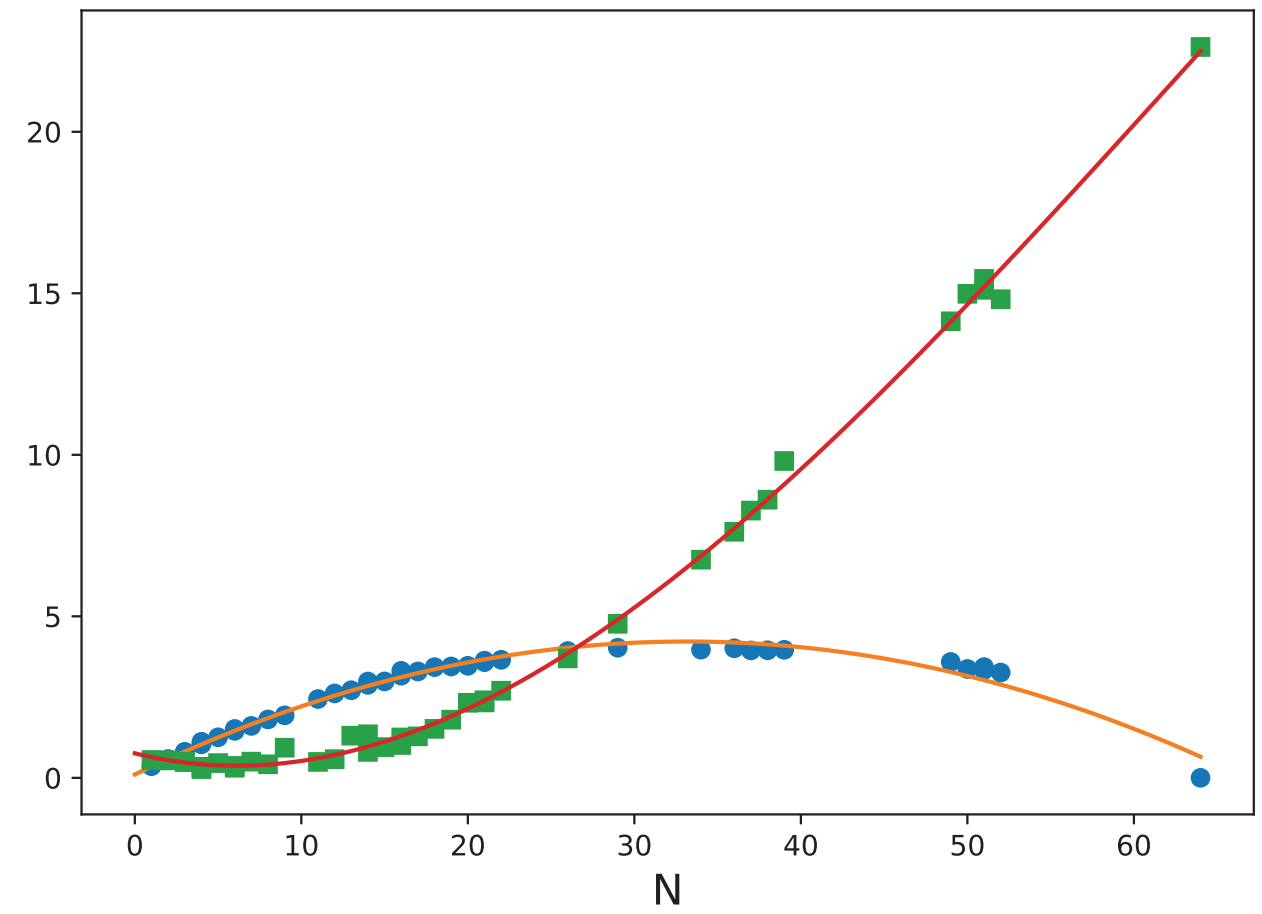

(b) Fit coefficients $a$ and $b$

Figure 8.17: Computation time of the perturbative method. (a) Shows how the computation time depends on the perturbation order $n$, for six different arbitrarily chosen energy ranges $[E_1, E_2]$. We also show linear fits $an + b$, fitted in the range $n \in [2, 8]$. (b) shows how the fit coefficients $a$ and $b$ vary with $N$, the number of states $|0l\rangle$ with energy in $[E_1, E_2]$, i.e. the number of exactly treated states. The constant contribution $b$ is shown in green, while $a$ is shown in blue. We also show a second order polynomial fitted to $a(N)$, and a third order polynomial fitted to $b(N)$.

Figure 8.17a also shows linear fits, obtained in the range $n \in [2, 8]$. Figure 8.17b shows how the fit coefficients depend on the chosen energy range $[E_1, E_2]$. The relevant parameter here is the number of exactly treated states $N = \sum_{E_1 \leq \epsilon_{0l} \leq E_2} 1$, since this determines the size of all dense and sparse matrices involved in the calculation. The exact functional dependency of the computation time on $N$ is determined by internal details of the the sparse matrix library, and is beyond our scope here. We do however note that the coefficient $a$, describing the order-proportional contribution to the computation time, is roughly proportional to $N(64-N)$. This is probably related to the fact that the matrices $[U(D-E)^{-1}]^n V^\dagger$ from (8.44) have dimension $8(64-N) \times 8N$.

Finally, in Figure 8.18 we compare the computation time requirements to the obtained accuracy. As long as we do not require an accuracy better than $10^{-4}$, the fastest calculation is always that shown as grey diagonal crosses. These represent the perturbative approximation where only a single type of state with energy 3.74 eV is treated exactly. This is the least accurate method in terms of convergence with $n$, but still converges faster in terms of the required computation time, due to the speed of the operations. According to projections based on our fitted models, this trend will continue beyond the twelfth order, so that the approximation with $N = 1$ will be the fastest method at all levels of accuracy, until it catches up with the exact solution at an accuracy of $\sim 10^{-24}$.

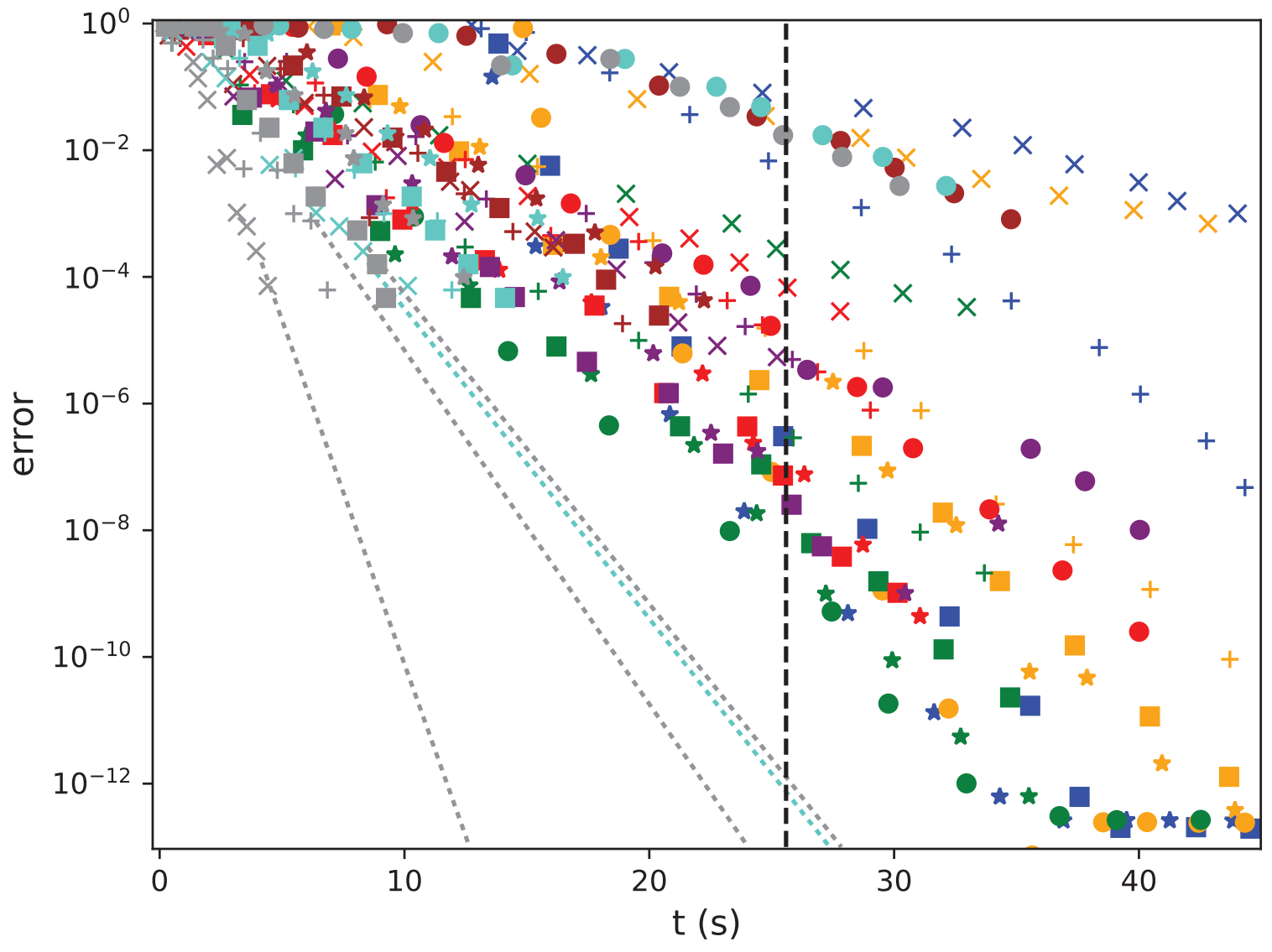

Figure 8.18: Perturbative approximation accuracy as a function of computation time. Results in blue, orange, green, red, purple, brown, turquoise and grey are respectively obtained with $E_1 = -\infty$, -0.13, 0.38, 1.06, 1.60, 2.37, 2.59, and 3.74 eV. Results where $E_2 = \infty$, 5.12, 4.74, 4.24 and 3.74 eV are shown respectively as circles, stars, squares, and horizontal and diagonal crosses. The order of the result always increases towards the right, and goes up to 12 in all cases. For some of the results we also include projections based on the fitted models. These are shown as dotted lines. The vertical dashed line in black represents the average computation time of the exact solution.

## 8.3 NEGF Monte Carlo

The NEGF Monte Carlo method is described in our third paper (Paper III). Here we just describe some additional results, which were omitted from that work for reasons of brevity. In Section 8.3.1 we use the method to calculate transport coefficients of a small quantum dot. We did this calculation originally as a proof of concept in a small system where the computational requirements are not very large. The quantum dot results also illustrates that the method can be used to calculate thermoelectric coefficients, such as the Seebeck coefficient and the electron contribution to the thermal conductance. In Section 8.3.2 we test the method on a thin film. This was done in order to test the method in a system with k-point dependency.

### 8.3.1 Quantum dot

In this section we show results of transport calculations in a simple two by two by two unit cells cubic quantum dot of CdTe, employing the tight binding model from our first publication (Paper I). Scattering is implemented within Büttikers approximation, with $\Sigma^r(E) = -i\Gamma I/2$, where $\Gamma = 0.06$ eV. The scattering is assumed elastic and local. As in the previous section, we set $\Sigma^r_p = -i\Gamma_l P_p/2$, but this time with $\Gamma_l = 0.2$ eV. The transmission function is shown in Figure 8.19, which also illustrates that both the error estimates and the spread of the Monte Carlo results converge as $1/\sqrt{n}$, in agreement with theory. The Monte Carlo results of Figure 8.19 are compared to a direct solution approach. This approach involves solving equation (5.86) together with an equation for $\Sigma^<_s(E)$ derived from Büttikers approximation and the elastic criterion. This is described further in Paper III. The direct solution is obtained with and without scattering, in order to show the magnitude of scattering effects and thereby demonstrate that the Monte Carlo approach estimates these effects correctly.

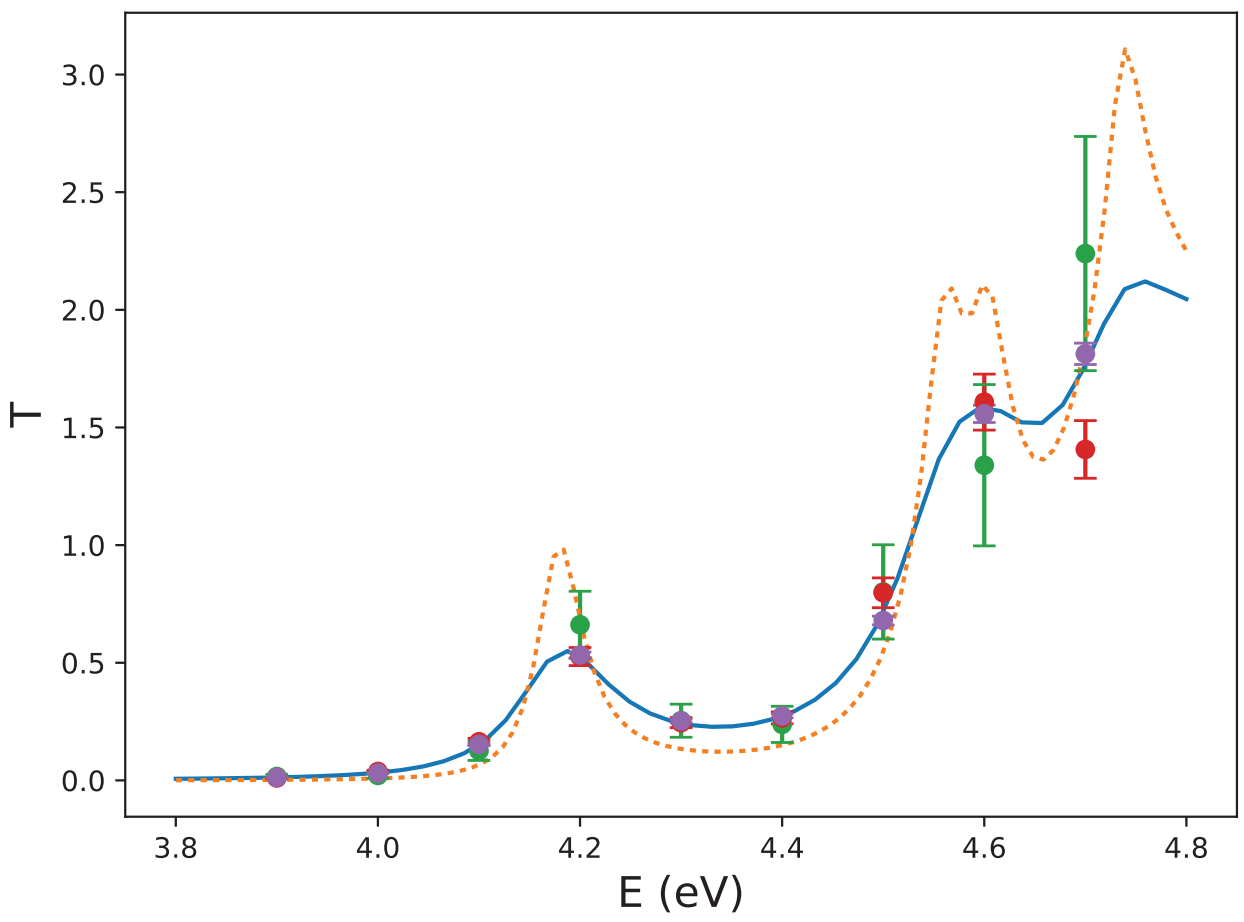

(a) Transmission function

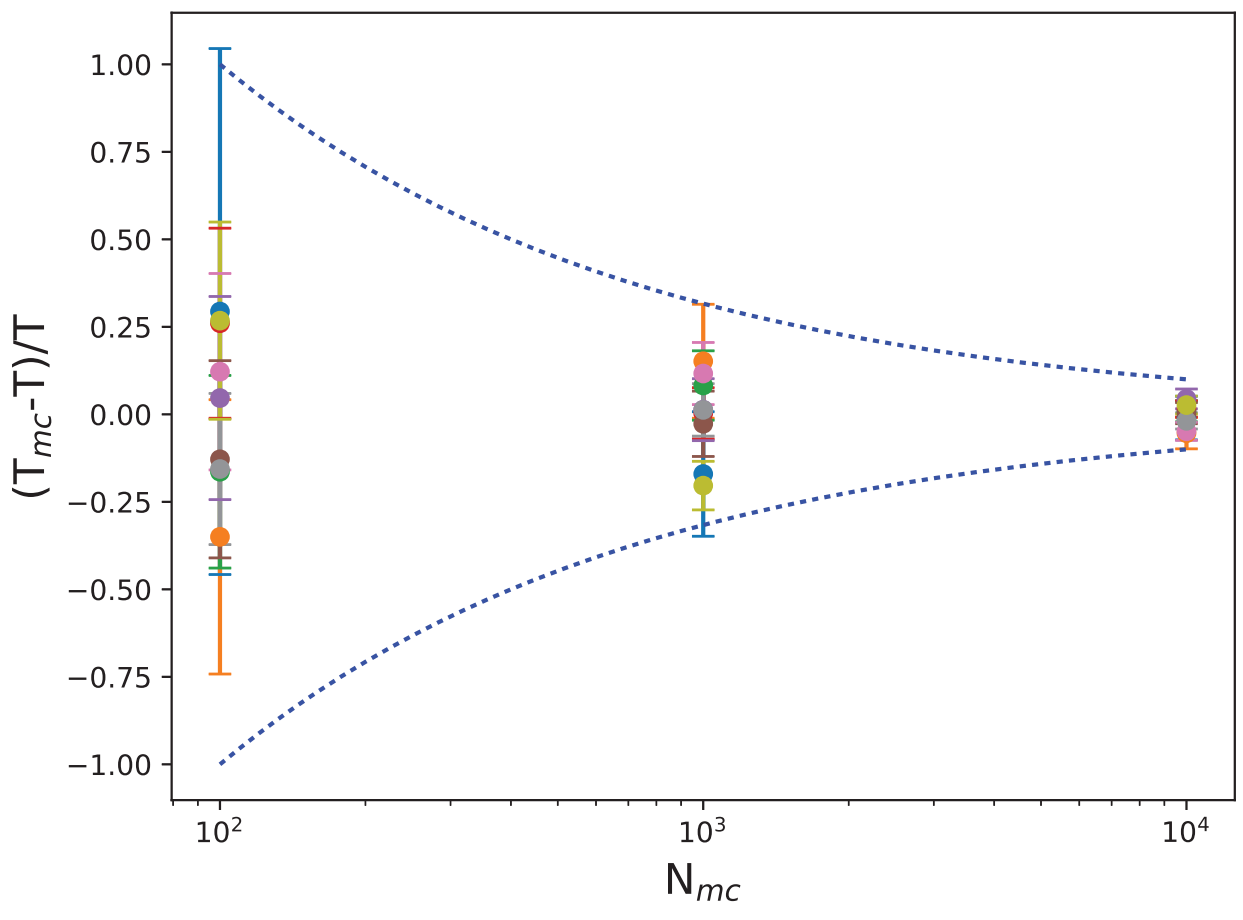

(b) Convergence of Monte Carlo estimates

Figure 8.19: (a) Transmission function $T(E)$. The solid blue line shows the transmission function of the quantum dot, calculated by the direct inversion approach. The dashed orange line shows the transmission function of the same quantum dot without scattering. The green red and purple error bars respectively show Monte Carlo estimates obtained with 100, 1000 and 10000 samples. (b) Relative error $(T_{mc} - T)/T$ as function of the number $n$ of Monte Carlo samples. The nine different colors represent the nine energies in (a). The two dashed blue lines show the functions $\pm 10/\sqrt{n}$.

Figure 8.20 shows the conductance of the quantum dot. The Monte Carlo simulation was set to accumulate samples for 1000 s (17 min) for each value of $\mu$, which resulted in approximately 20000 samples per value. As can be seen from Figure 8.20b, this results in a relative standard deviation $\Delta G/G$ in the range of 1-4 percent. The Monte Carlo results are compared to the direct inversion method, which is integrated over a regular energy grid of 13 points, in the range $\mu \pm 0.25$ eV. The interpolated method referenced in this figure and the succeeding ones will be explained below.

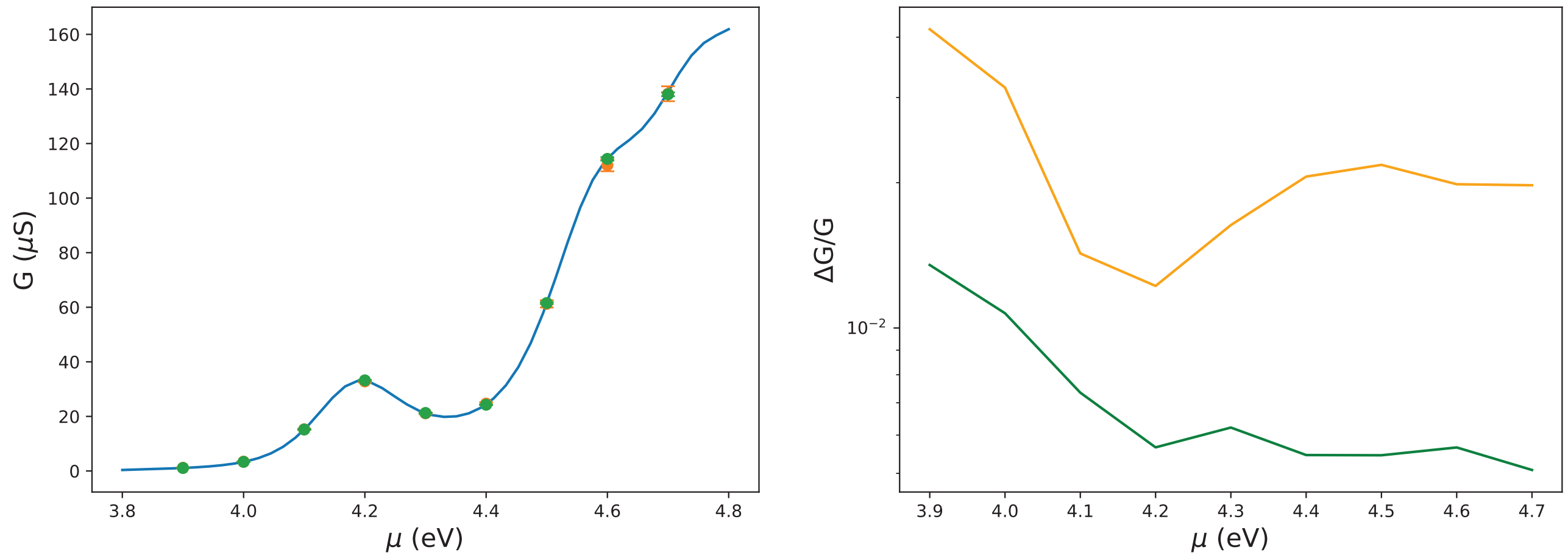

(a) Conductance (b) Relative error estimate

Figure 8.20: (a) Conductance in Siemens, as function of chemical potential $\mu$ in eV. The solid blue line shows the result of the direct inversion approach. The orange and green error bars show results of the Monte Carlo simulations, respectively employing the straight forward and the interpolated method. All of the Monte Carlo results were obtained with an accumulation time of 1000 s. (b) Error estimate of the Monte Carlo calculations in terms of the relative standard deviation $\Delta G/G$. The orange line represents the straight forward method, while the green line represents the interpolated method.

To obtain the other thermoelectric transport coefficients we make use of (2.24)-(2.26), and define

$$\mathcal{S}_n = \int (E-\mu)^n T(E)\mathrm{Th}(E)\mathrm{d}E. \tag{8.45}$$

The thermoelectric coefficients $A$ and $B$ are obtained from $\mathcal{S}_1$, while the coefficient $C$ is obtained from $\mathcal{S}_2$. In the Monte Carlo simulation these coefficients are obtained simultaneously with the conductance calculation, by simply weighting the result respectively with $(E-\mu)$ and $(E-\mu)^2$, $E$ being the energy of the transmitted electron. The thermal conductance $k$ of the quantum dot, and its Seebeck coefficient $\alpha$ can be calculated from $A$ and $C$ using (2.27)-(2.29). In terms of $\mathcal{S}_1$ and $\mathcal{S}_2$, the resulting expressions are

$$\alpha = -\frac{2e\mathcal{S}_1}{hTG}, \quad \text{and} \tag{8.46}$$

$$k_e = \frac{2\mathcal{S}_2}{hT} - \frac{4e^2\mathcal{S}_1^2}{h^2TG}. \tag{8.47}$$

By making use of the general formula

$$\Delta F^2 \approx \sum_i \left(\frac{\partial F}{\partial x_i}\right)^2 \Delta x_i^2, \tag{8.48}$$

we can estimate the standard deviation of the Monte Carlo calculations of $\alpha$ and $k$ as

$$\Delta\alpha = \frac{2e}{hT}\sqrt{\frac{\Delta\mathcal{S}_1^2}{G^2} + \frac{\mathcal{S}_1^2\Delta G^2}{G^4}}, \quad \text{and} \tag{8.49}$$

$$\Delta k = \frac{2}{hT}\sqrt{\Delta\mathcal{S}_2^2 + \left(\frac{2e^2}{h}\right)^2\left(\frac{4\mathcal{S}_1^2\Delta\mathcal{S}_1^2}{G^2} + \frac{\mathcal{S}_1^4\Delta G^2}{G^4}\right)}. \tag{8.50}$$

Monte Carlo calculations of $\alpha$ and $k$ are compared to the direct inversion approach respectively in figures 8.21 and 8.22.

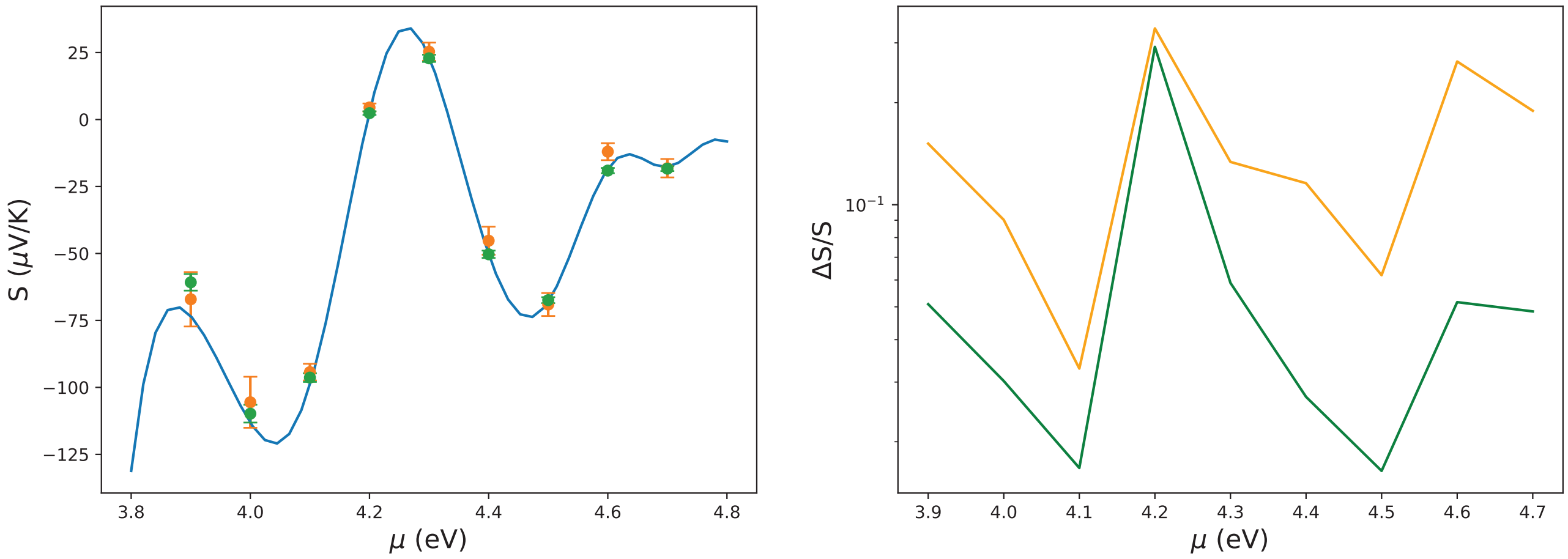

(a) Seebeck coefficient

(b) Relative error estimate

Figure 8.21: (a) Seebeck coefficient $S = V/\Delta T$ in $\mu$V/K. The solid blue line shows the result of the direct inversion approach. The orange and green error bars show results of the Monte Carlo simulations, respectively employing the straight forward and the interpolated method. (b) Error estimate of the Monte Carlo calculations in terms of the relative standard deviation $\Delta S/S$. The orange line represents the straight forward method, while the green line represents the interpolated method.

#### 8.3.1.1 Interpolated method

While the transmission function shown in Figure 8.19 is quite fast to calculate using the Monte Carlo technique, calculation of transport coefficients is much slower. This is because in these calculations the electron energy is different in every Monte Carlo sample, which means $G^r$ will have to be recalculated for every such sample. Since calculation of $G^r$ is the most demanding part of the procedure, one could potentially gain a substantial speed improvement if the amount of such recalculations were reduced. One way of obtaining this would be interpolation. That is, instead of always recalculating the scattering probabilities $p_{ij}$ from the Green's function, we sometimes interpolate between previously calculated values of these functions at different energies. One such interpolation scheme is tested here.

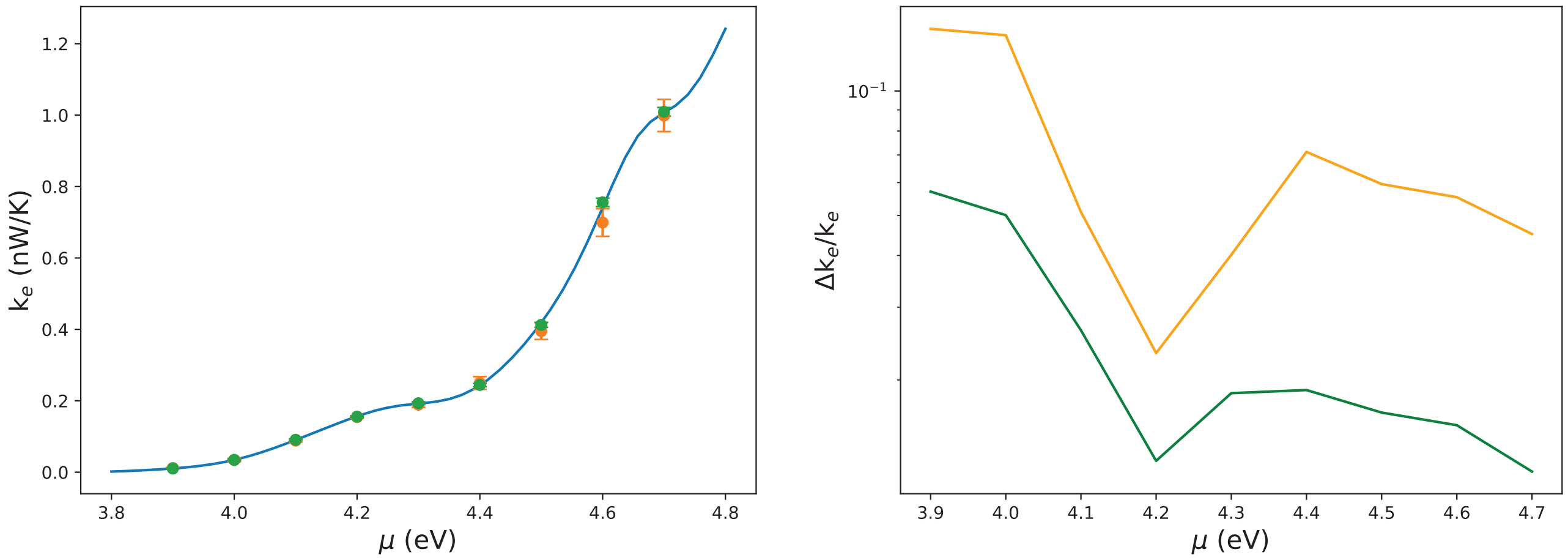

(a) Thermal conductance

(b) Relative error estimate

Figure 8.22: (a) Thermal conductance in W/K. The solid blue line shows the result of the direct inversion approach. The orange and green error bars show results of the Monte Carlo simulations, respectively employing the straight forward and the interpolated method. (b) Error estimate of the Monte Carlo calculations in terms of the relative standard deviation $\Delta k/k$. The orange line represents the straight forward method, while the green line represents the interpolated method.

This interpolation scheme is implemented as follows: Whenever some column of the Green's function is calculated, all corresponding scattering probabilities $p_{ij}(E) = \gamma_i\gamma_j|G_{ij}(E)|^2$ are stored in a table together with their respective energy. Then, whenever transmission functions are required at an energy $E$ between two energies $E_1$ and $E_2$ where these probabilities have previously been calculated, we test whether the relevant probabilities differ by more than 5 percent on average between $E_1$ and $E_2$. If this is not the case, $p_{ij}(E)$ is interpolated linearly between $p_{ij}(E_1)$ and $p_{ij}(E_2)$. Otherwise, the relevant sections of the Green's function is recalculated, and $p_{ij}(E)$ is obtained from this.

The figures 8.20-8.22 also contain results from the interpolated method. These calculations were executed for 1000 s (17 min), and resulted on average in approximately 200 000 Monte Carlo samples. Relative to the uninterpolated method this is a tenfold improvement. As can be seen from the figures, this has resulted in the relative standard deviation being reduced by approximately a factor of $\sqrt{10} = 3.16$. Figure 8.23 compares the sampling rates of the two methods in a single simulation. The sampling rate of the uninterpolated method stays approximately fixed at 60 samples per second, while the interpolated method starts out at this frequency, but then becomes progressively faster as the interpolation table is filled in. This process saturates after approximately 10000 samples, after which the interpolated method stays fixed at 600-700 samples per second.

Thus, the sampling rate of the interpolated method is seen to saturate at approximately ten times that of the uninterpolated one. However, since the Monte

Carlo method is implemented in Python, it seems likely that this ratio is limited primarily by the Python overhead, and that for instance a C-implementation of the interpolation scheme would be significantly faster still.

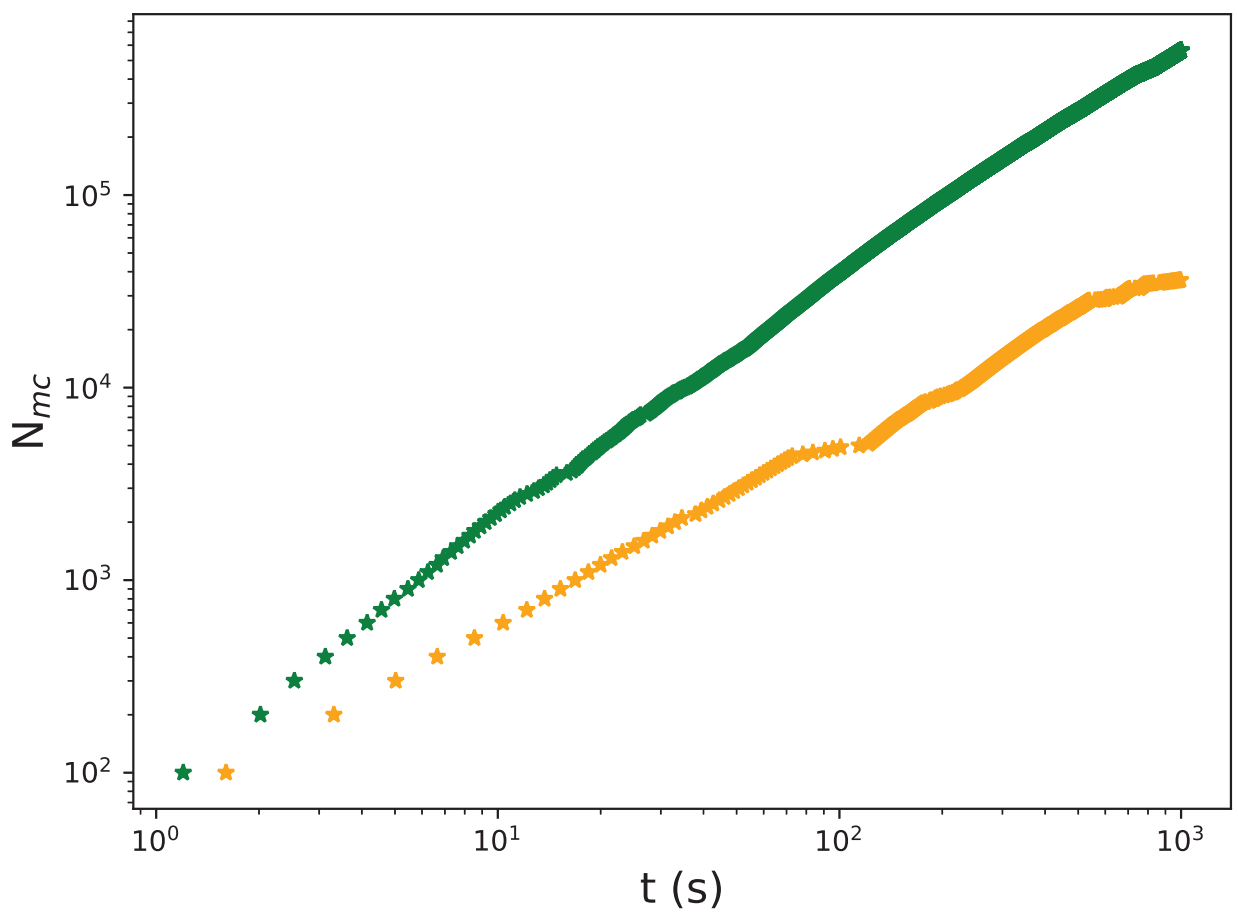

(a) Number of acquired Monte Carlo samples

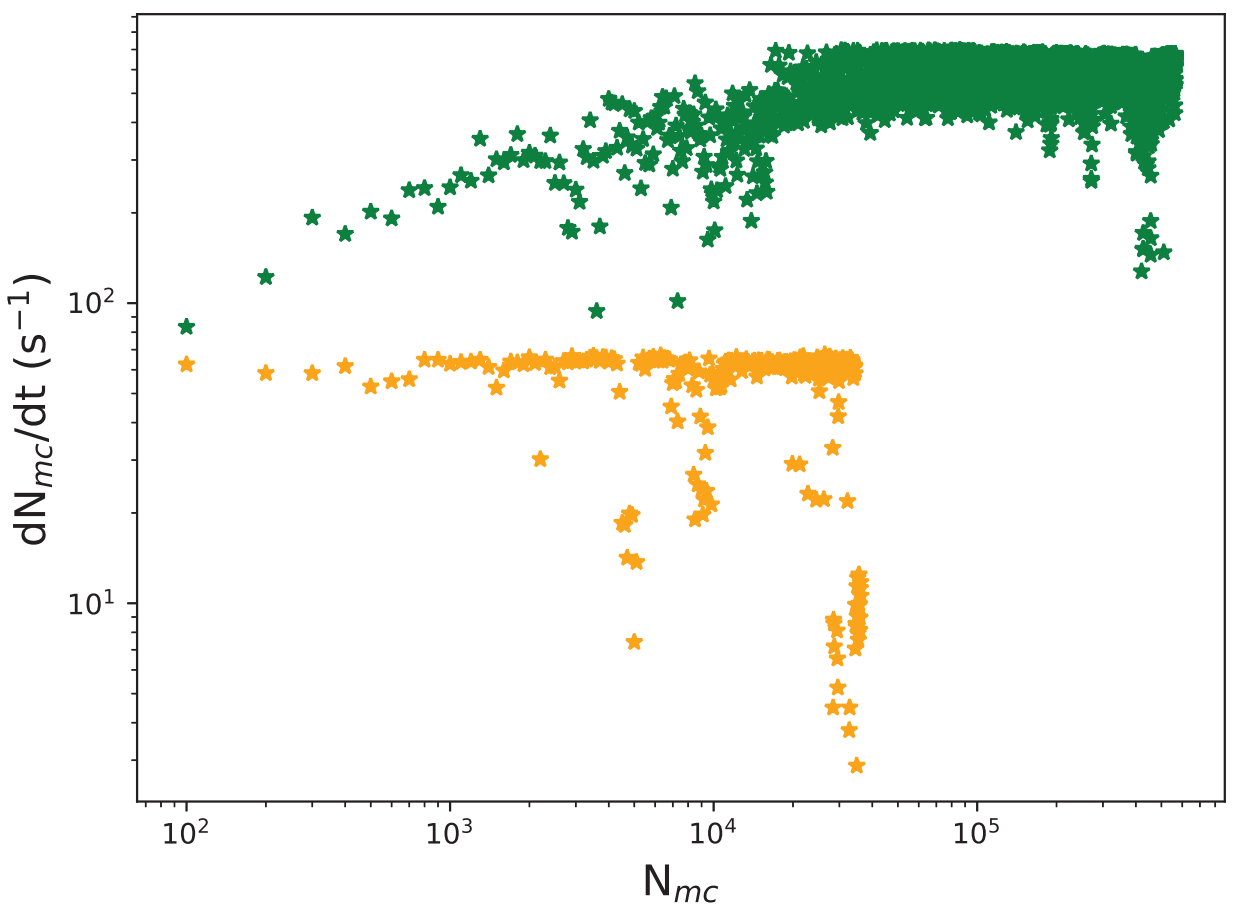

(b) Sampling rate as a function of the number of samples already acquired

Figure 8.23: Comparison of sampling rate. The uninterpolated method is shown in orange, while the interpolated method is shown in green.

### 8.3.2 Thin Film

This section includes results of transport calculations in a CdTe thin film with a thickness of two unit cells. Both the CdTe Hamiltonian, the scattering model, and the coupling to the leads make use of the same models as in the previous section. However, this time we set $\Gamma = \Gamma_l = 0.02$ eV.

A major focus of this section is to compare the computational efficiency of the Monte Carlo method to that of alternative approaches, when these have to be integrated over a k-grid. Thus, some effort is put into a study of how the computation time and accuracy of these other methods scale with the grid resolution. We study both the direct solution approach mentioned in the previous section, and in addition an iterative approach. These methods solve the same set of equations, and their only difference is that the first method solves the equations by direct linear inversion, while the second method solves them iteratively. The particular iterative method employed is the Scipy implementation of the gmres method. In both methods, integration over energy is performed over a grid with a quite high resolution of 50 points in the range $\mu \pm 0.25$ eV. This is to assure minimal contribution to the error from energy integration.

#### 8.3.2.1 Results

In Figure 8.24 the direct and iterative methods, employing various k-grid resolutions, are compared to Monte Carlo results. The calculations are performed both at a chemical potential $\mu = 4.3$ eV, which is quite low in the conduction band, and at $\mu = 5.1$ eV which lies deeply into the conduction band, where the entire Brillouin zone contributes to conduction. In addition, the calculations are performed with two different scattering models. In one of these, scattering does not couple the k-points, i.e. the k-value of an electron is not changed after scattering. In the other model, the electrons are scattered homogeneously between all k-points, i.e. when an electron is scattered, it may end up with any value of $\boldsymbol{k}$ with equal probability.

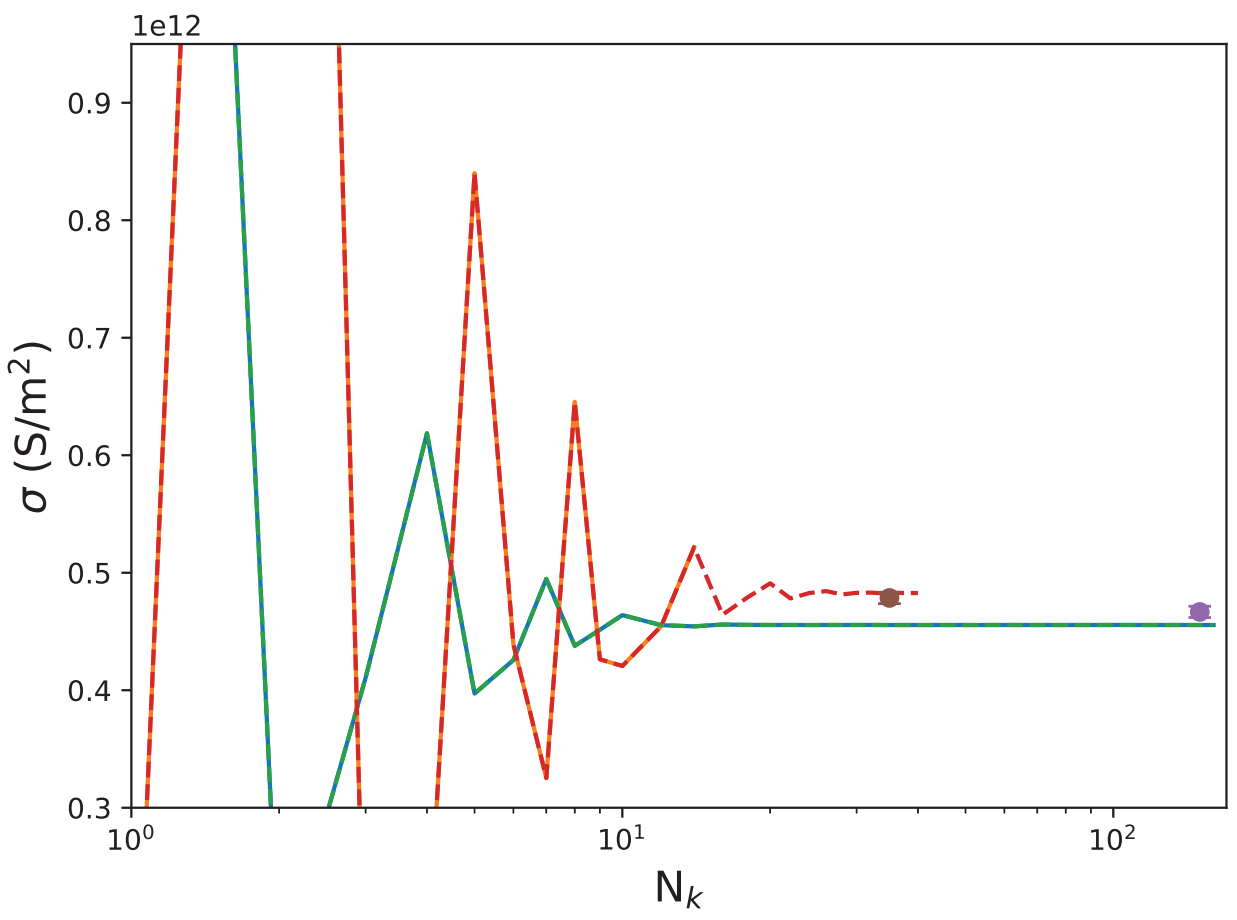

(a) $\mu = 4.3$ eV

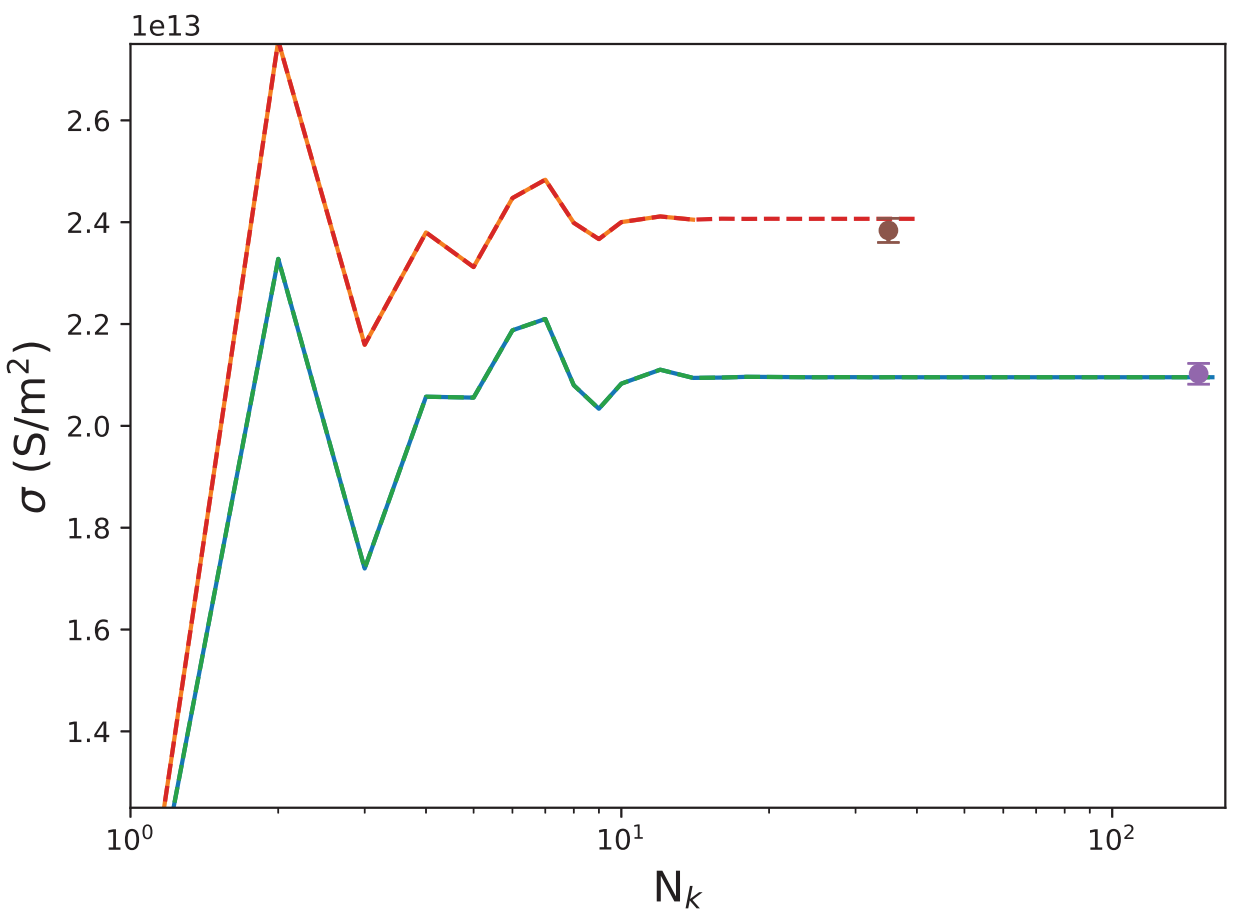

(b) $\mu = 5.1$ eV

Figure 8.24: Conductance of thin film in Siemens/m$^2$, as a function of k-grid resolution. Solid lines employ direct inversion, dashed lines employ the iterative approach, while the error bars represent the Monte Carlo results. Among these, the results shown in green blue and purple employ the k-conserving scattering model, while the results shown in red, orange and brown employ the k-coupling scattering model. The Monte Carlo results do not employ a k-grid, and their position along the horizontal axis is determined simply so as to be convenient in comparing them to the other results.

At the high value om $\mu$, integration is performed over the entire Brillouin zone $k_x, k_y \in [-\frac{1}{2}, \frac{1}{2}]$, wile at the low value of $\mu$ it is only performed over the range $k_x, k_y \in [-\frac{1}{4}, \frac{1}{4}]$. This is because the outer extents of the Brillouin zone have no significant contribution to transport this low in the conduction band. In addition, because of the symmetry of the Brillouin zone, values need only be calculated in one quarter of the integration region, for instance that where $k_x, k_y > 0$. The grid resolution $N_k$ referenced in Figure 8.24 and in the following figures, refers to the number of points along one dimension, in the region where values are calculated. That is, in the region $k_x, k_y \in [0, \frac{1}{2}]$ when $\mu = 4.3$ eV, and $k_x, k_y \in [0, \frac{1}{4}]$ when $\mu = 5.1$ eV.

When the k-conserving scattering model is employed, the transport problem can be separated into isolated problems for each k-point. When the k-coupling model is employed however, this is not possible, and as a result the computational demands of the alternative approaches become drastically larger. When employing the k-conserving model, it was possible to take the grid resolution all the way up to $N_k = 160$ without trouble, but with the k-coupling model, we ran into problems of too little memory at $N_k > 14$ with the direct inversion approach. With the iterative approach such problems did not arise, but after $N_k > 40$ the calculations were just too time consuming.

The Monte Carlo calculations were allowed to accumulate samples until the result was determined with a relative standard deviance of approximately one percent. Relevant information about the results is summarized in table 8.2. The required computation time is determined by the sampling rate, and by the variance $\Delta X^2$ of the estimator. We observe that the sampling rate is roughly 3 times higher when the k-conserving model is employed, compared to the k-coupling model. This is because when the k-coupling model is employed, the diagonal of the Green's function must be recalculated every time an electron is moved, whereas when the k-conserving model is used, this needs only be calculated once per sample.

| $\mu$ (eV)<br>scattering model | 4.3<br>k-conserving | 4.3<br>k-coupling | 5.1<br>k-conserving | 5.1<br>k-coupling |
|---|---|---|---|---|
| $G$ (S/m$^2$) | $4.5552{\cdot}10^{11}$ | $4.8270{\cdot}10^{11}$ | $2.0955{\cdot}10^{13}$ | $2.4066{\cdot}10^{13}$ |
| $G_{mc}$ (S/m$^2$) | $4.6667{\cdot}10^{11}$ | $4.7856{\cdot}10^{11}$ | $2.1022{\cdot}10^{13}$ | $2.3839{\cdot}10^{13}$ |
| $\Delta G_{mc}$ (S/m$^2$) | $4.7897{\cdot}10^{9}$ | $4.8657{\cdot}10^{9}$ | $2.0424{\cdot}10^{11}$ | $2.3668{\cdot}10^{11}$ |
| $\Delta X$ (S/m$^2$) | $2.9387{\cdot}10^{12}$ | $6.6054{\cdot}10^{12}$ | $8.9526{\cdot}10^{13}$ | $9.3588{\cdot}10^{13}$ |
| num. samples | 376451 | 1842945 | 192140 | 156351 |
| comp. time (s) | 6411.1 | 102406. | 2565.9 | 6405.8 |
| $t_s$ (s) | 0.01703 | 0.05557 | 0.01335 | 0.04097 |

Table 8.2: Monte Carlo results. $G$ is the result of the standard methods, $G_{mc}$ is the Monte Carlo result, $\Delta G_{mc}$ is the standard deviance of the Monte Carlo result, $\Delta X$ is the standard deviance of the estimator $X$ which is sampled to calculate $G_{mc} = \bar{X}$, and $t_s$ is the computation time per sample.

In addition, we observe that the ratio $\Delta X/G$ is higher at the low value of $\mu$. This is because a large fraction of the samples are then drawn from k-points which do not contribute significantly to conduction. In the k-conserving case, this problem is mitigated by drawing the k-points from a probability distribution proportional to the k-projected density of states of bulk CdTe, so that most of them are drawn from the region contributing to conduction. In the k-coupled case however, this technique is not effective, and in fact leads to a significant increase in the variance of $X$. Accordingly, we have not made use of this technique in the k-coupled case.

#### 8.3.2.2 Analysis of computational scaling

In Figure 8.25a we study how the relative error of the direct solution and iterative method depends on the k-grid resolution. The relative error is estimated as $\Delta G/G$, where $\Delta G$ is calculated as the difference between each result and the same result with maximal grid size. The first thing to note is that for the k-conserving $\mu = 4.3$ eV case, the direct solution and iterative results are coinciding until an error of less than about $10^{-6}$ is reached. We thus conclude that the separation point $10^{-6}$ represents the accuracy of the iterative method. Similar conclusions hold for the other three cases, so in order to simplify the figure we include only the direct inversion result in the other k-conserving case, and only the iterative results for the two k-coupling cases.

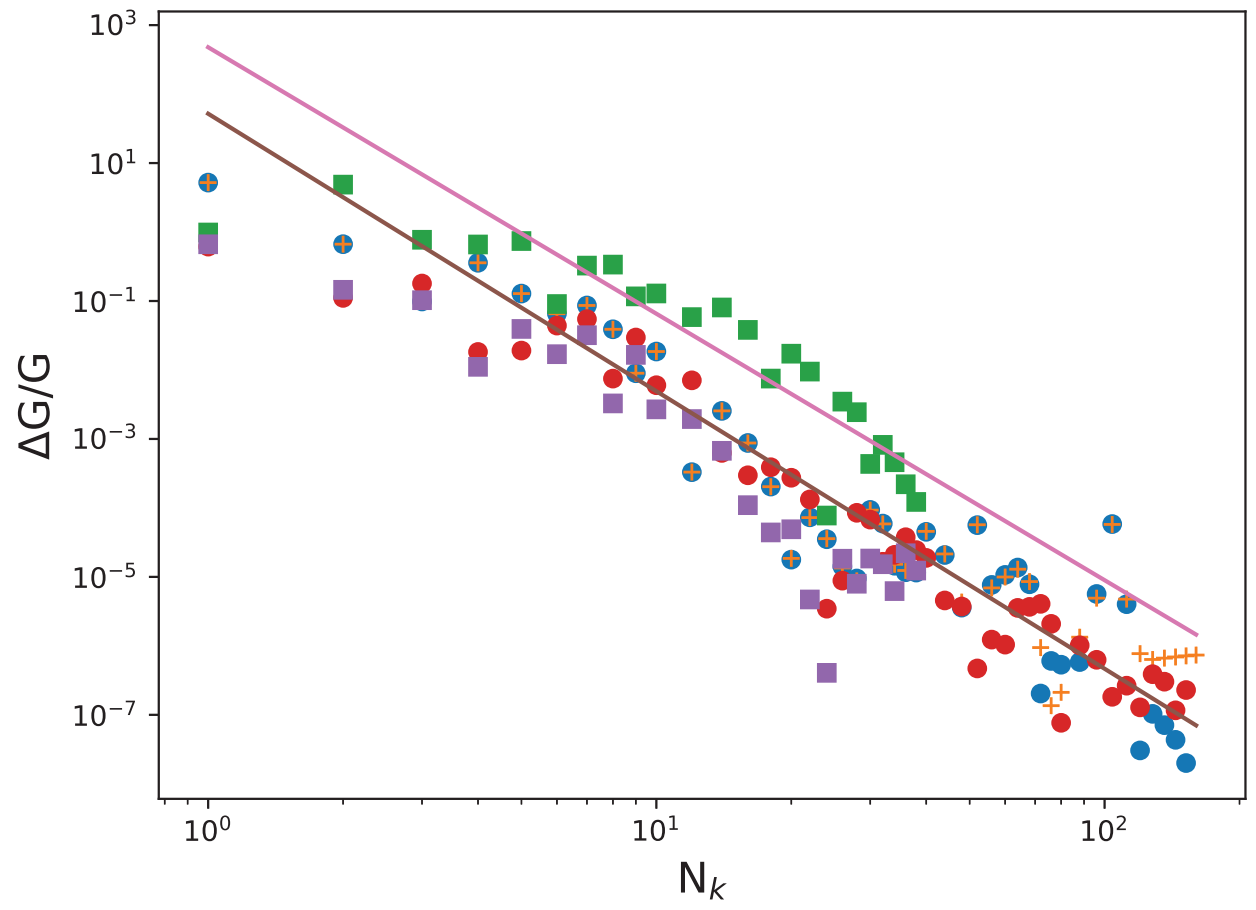

(a) Accuracy

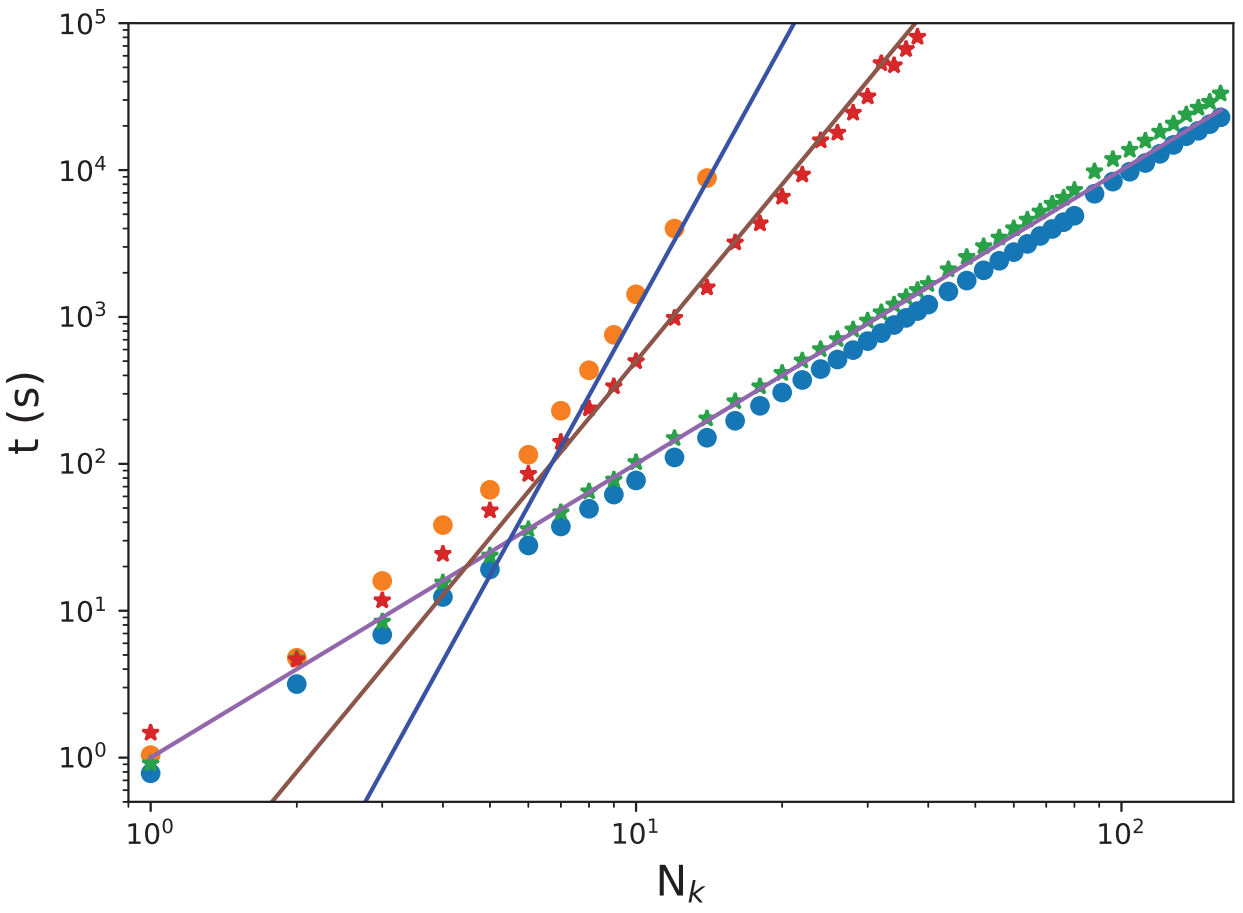

(b) Computation time

Figure 8.25: Accuracy and computation time of direct and iterative solution. (a) Error estimate as function of grid size. Circles represent results with k-conserving scattering, employing the direct solution approach. Squares represent results with k-coupling scattering, employing the iterative approach. The blue circles and green squares are obtained with $\mu = 4.3$ eV, while the red circles and purple squares are obtained with $\mu = 5.1$. The orange crosses are obtained with the same parameters as the blue circles, but using the iterative approach. The brown and pink line are power functions fitted respectively to the k-conserving and k-coupling results at $\mu = 4.3$ eV. (b) Computation time as function of grid size. Direct inversion results are shown as circles, and iterative results as stars. The blue circles and green stars were obtained using the k-conserving scattering model, while the orange circles and red stars were obtained using the k-coupling model. The results are from the calculation at $\mu = 4.3$ eV, but the results at $\mu = 5.1$ eV are almost identical. The solid lines in purple, brown and blue show respectively the power functions $N_k^2$, $0.05 \cdot N_k^4$ and $0.00111 \cdot N_k^6$.

The solid line shown in brown in Figure 8.25a represents a power function fitted to the k-conserving $\mu = 4.3$ eV case. The first six points are ignored in this fit, and the result is $\epsilon = 51.931 \cdot N_k^{-4.0235}$. We observe that the asymptotic behavior of the two cases with $\mu = 5.1$ eV also agree fairly well with this function. The k-coupling $\mu = 4.3$ eV case however, lies a little higher than the other results, and we thus make a separate fit of that case. The result is $\epsilon = 477.44 \cdot N_k^{-3.8633}$, and is shown as a solid line in pink in Figure 8.25a. Both of the fitted functions have an exponent fairly close to 4, and we thus conclude that the integration accuracy is of fourth order in these cases.

In Figure 8.25b we study how the computation time of the direct solution and iterative methods scale with k-grid resolution. The results are compared to appropriate power functions. We see that in the k-conserving case, both the direct inversion and iterative approach lies close to the function $t = N_k^2$. This is reasonable, since in that case the problem can be divided into isolated problems for each k-point, so that the total computation time should be proportional to the total number of k-points $N_k^2$.

In the k-coupled case, we see that the iterative computation time asymptotically approaches the function $t = 0.05 \cdot N_k^4$, while the direct inversion time approaches $0.00111 \cdot N_k^6$. This is also reasonable, since we are then solving a linear equation set with a number of equations proportional to the number of k-points. It is well known that direct inversion scales as $N^3$ and an iterative method as $N_{itt} \cdot N^2$, where $N$ is the number of equations and $N_{itt}$ the required number of iterations.

At small grid sizes all of the methods lie close to the function $N_k^2$. We conclude that this function describes the time required to calculate $G^r$, which will always be proportional to the number of k-points.

Summarizing, the error $\epsilon$ of both the direct solution and the iterative method scales with $N_k^{-4}$, the computation time $t$ of the k-conserving cases with $N_k^2$, while in the k-coupled cases the computation time of the iterative method scales with $N_k^4$, and the computation time of direct inversion method with $N_k^6$. Combining all this, we find that the error scales with $t^{-2}$ in the k-conserving cases, while in the k-coupled cases the iterative and direct inversion errors scale respectively as $t^{-1}$ and $t^{-\frac{2}{3}}$.

For the case of the Monte Carlo calculations, corresponding expressions for the relative deviation $\epsilon = \Delta G/G$, with $\Delta G$ now being the standard deviation, can be found from the general relation

$$\epsilon = \frac{\Delta X}{G\sqrt{N_s - 1}} \approx \frac{\sqrt{t_s}\Delta X}{G} t^{-\frac{1}{2}}, \tag{8.51}$$

where $t_s$ is the average computation time per Monte Carlo sample. Thus, the error of the Monte Carlo method always scales as $t^{-\frac{1}{2}}$. Concrete expressions for the relation between $\epsilon$ and $t$ in the four cases we have tested can be found by inserting the numbers from table 8.2 into (8.51). In the k-conserving cases, we obtain respectively $\epsilon \approx 0.84 \cdot t^{-\frac{1}{2}}$ and $\epsilon \approx 0.94 \cdot t^{-\frac{1}{2}}$ in the $\mu = 4.3$ eV and $\mu = 5.1$ eV cases, while in the k-coupling cases, we obtain respectively $\epsilon \approx 3.2 \cdot t^{-\frac{1}{2}}$ and $\epsilon \approx 0.79 \cdot t^{-\frac{1}{2}}$ in the same two cases. All of these scaling relations are compared in Figure 8.26, where the error and computation time of the actual results is also included. We observe that even the worst scaling of the alternative approaches, the k-coupled direct solution, has an error which scales better with computation time than the Monte Carlo method.

However, it remains a question how general the fourth order scaling of the error seen in Figure 8.25a actually is. Further, the Monte Carlo method can be parallelized much more efficiently than the other methods, and has smaller memory requirements. It should also be mentioned that the k-coupled Monte

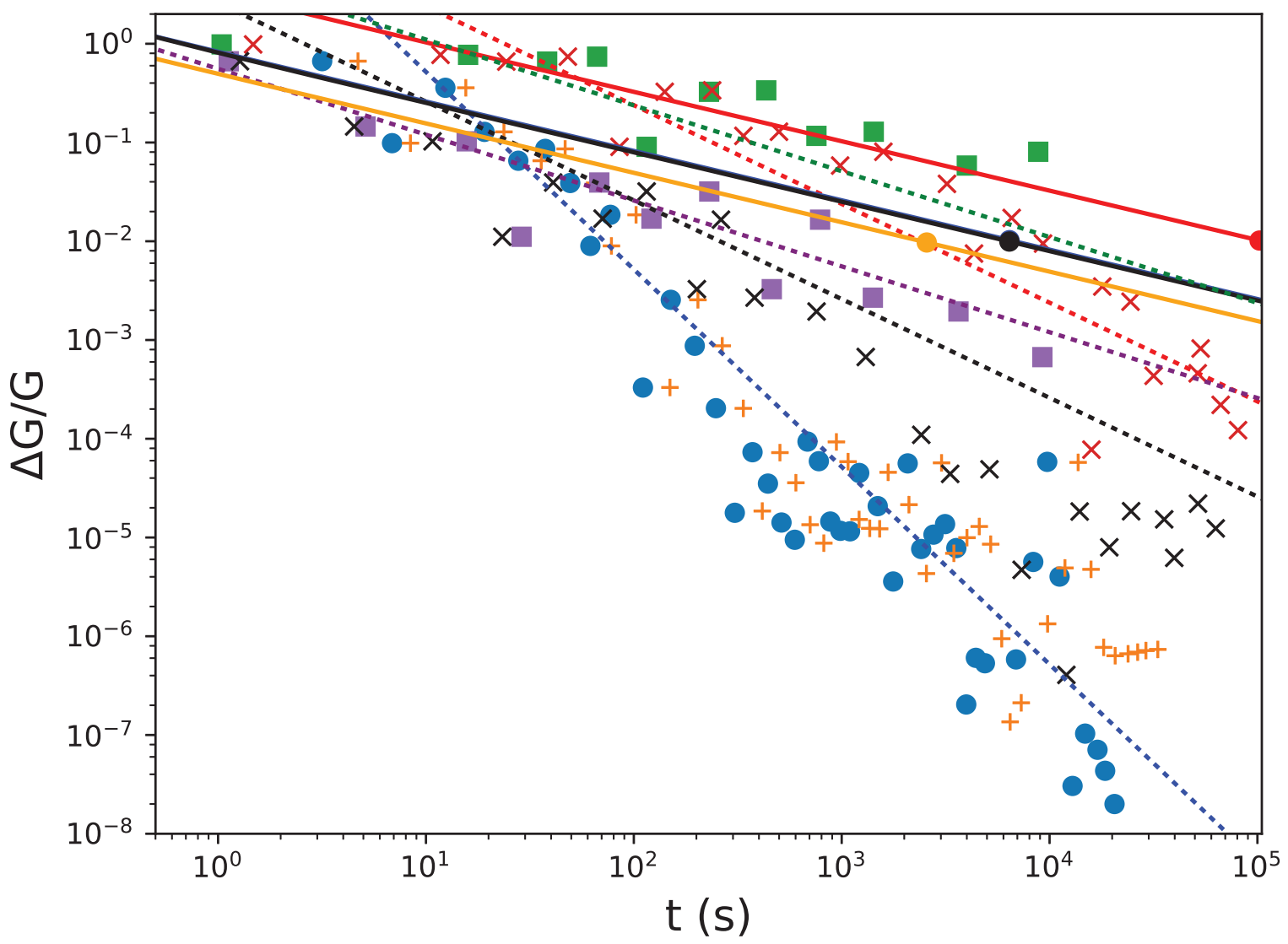

Figure 8.26: Relative error as a function of computation time in seconds. The blue circles and orange crosses show respectively the direct and iterative solutions in the k-conserving $\mu = 4.3$ eV case. The other results assume k-coupled scattering. Among these, the green and purple squares show respectively the direct solution in the $\mu = 4.3$ eV and $\mu = 5.1$ eV cases, while the red and black crosses show respectively the iterative solutions for the same values of $\mu$. The dotted lines show scaling estimates obtained in the text. These are in the same colors as the results they are obtained from. The solid lines show the scaling of the relative deviation in the Monte Carlo calculations. The blue and red lines represent respectively the k-conserving and k-coupled $\mu = 4.3$ eV cases, while the orange and black lines represent respectively the k-conserving and k-coupled $\mu = 5.1$ eV cases. The Monte Carlo results from our simulations obtained a relative deviation of $10^{-2}$. These points are marked on the scaling relation as overlying circles.

Carlo calculations are slowed down quite significantly by the unnatural scattering mechanism, where all k-points are drawn with equal probability. This causes the electrons to spend most of the simulation in regions of the Brillouin zone where they are almost immobile, and accordingly causes the calculation to be very inefficient. In more realistic scattering models, the probability of scattering to a particular k-point is proportional to the k-projected density of states, meaning that the electrons would spend most of the simulation time in regions contributing more to the conductivity. This could significantly improve the efficiency of the simulation.

Chapter 9

# Conclusions

During the course of this thesis, I have pursued a series of different methods for performing transport calculations. All of these have been lacking important aspects, by not employing sufficiently accurate models of band structure and scattering, and by ignoring important physical effects. Because of this, none of my results are useful as predictions concerning values of transport coefficients in real materials. Nevertheless, my experimentation with these methods has allowed me to make a few conclusions concerning the appropriate way forward for implementing a general transport framework. In particular, I would like to share some conclusions concerning how the appropriate method varies between different regimes of heterostructure transport.

The first parameter I will use to distinguish different regimes, is the typical size scale of structures in the heterostructure. In a superlattice, this size scale would typically be the superlattice period, while in a disordered three dimensional structure it could be the average grain size, or some similar parameter. Whenever this size scale is considerably larger than the coherence length of the transported particles, which is usually similar to their mean free path, the transport problem is well described by the Boltzmann equation, and the Boltzmann Monte Carlo method seems best suited to solve this. In high field applications, one must use the method described in Section 8.1.1, while for the calculation of linear transport coefficients, the method of Section 8.1.2 seems much better suited, assuming it is appropriately generalized.

A second important parameter which distinguishes different regimes, is the effective dimensionality of the problem. In particular, I will distinguish between quasi-one-dimensional systems, and systems that have fully three-dimensional structures. All of my own calculations have been limited to the quasi-one-dimensional case, so my discussion of the fully three-dimensional case will be limited to a discussion merely of what seems intuitive. In fully three-dimensional systems, there seems to have been some success in modeling heterostructure transport using the Boltzmann equation, even when the heterostructure size scale is small. In a sense, it also seems intuitive that this approach should work better in such systems than in quasi-one-dimensional systems, since failure of the Boltzmann equation is due to interference effects, which will be reduced by the inverse square law in fully three-dimensional systems.

On the other hand, peculiar transport phenomena like Anderson localization have been predicted to occur also in three-dimensionally disordered systems. Since such effects are not predicted by the Boltzmann equation, it is clear that Boltzmann based approaches are not always applicable to three-dimensional heterostructures. All in all this indicates that in the fully three-dimensional case, more study is required in order to understand precisely which cases are treatable by the Boltzmann equation, and which ones are not. In the mean time, Boltzmann Monte Carlo methods seems like a good starting point for doing practical calculations. Again, the choice between the methods described in 8.1.1 and 8.1.2 will depend on whether the problem is in the linear or high field regimes.

In the quasi-one-dimensional case, we must distinguish between the periodic case of superlattices, and the non-periodic case, which is sometimes referred to as disordered superlattices. The regimes of high field transport in superlattices is well described by Wacker[28], so the discussion here will be limited to the regimes of linear transport, which was investigated in our second publication (Chapter II). Assuming a simple superlattice, with a single well and barrier layer, our results show that semiclassical approaches are applicable when the superlattice period is ether considerably larger, or considerably smaller than the coherence length. However, when the period is large, one must use the band structure of the composing materials, while in the regime of small superlattice period, the minibands of the superlattice should be used. In the first of these cases, the Monte Carlo method of Section 8.1.2 seems like the best suited method. In the second case, the Boltzmann equation can be solved directly as long as the scattering model is not too complicated. Given a more complicated scattering model, the method of Section 8.1.2 again seems like the appropriate choice.

The results in our second publication also show that one can expect the existence of an intermediate regime, where neither of the two semiclassical approaches work particularly well. Based upon the discussion of Section 6.4, the failure of the miniband approach in this regime is due to the step from (6.82) to (6.88) not being justified. Thus, the fairly specialized regime of superlattices with a period comparable to the coherence length, can probably be handled by applying (6.82) directly.

The final regime to be discussed, is that of non-periodic quasi-one-dimensional heterostructures. It is probably in this regime that the Monte Carlo method described in our third paper (Chapter III) is most useful. This is both because there are few other methods capable of handling this regime, at least when accurate scattering models are required, and also because it is precisely in quasi-one-dimensional systems where one can expect Monte Carlo methods to have a major advantage, since they do not require explicit integration over the crystal momentum.

Thus, the NEGF Monte Carlo method could have potential applications within several different areas, among which our original application of thermoelectric effect in quasi-one-dimensional heterostructures is only one. However, before this can be realized, considerable amounts of additional testing is required to determine the actual utility of the method. In particular, we must perform more appropriate tests of how competitive the method is performance wise, and perform tests with more realistic scattering mechanisms. In addition, additional investigations are needed into the linear limit, and the range of applicability of the positivity assumption.

Finally, in retrospect it seems that the choice to base our calculations on the NEGF formalism rather than the Kubo relations was a mistake. The main reason for this has to do with the numerical derivative of the NEGF currents. As it happens, all methods for solving the NEGF equations we have encountered, have some problems with numerical noise. Often, this noise will be insignificant in the currents themselves, but the numerical derivative can severely magnify the effect. Thus, to obtain stable results, it is best to calculate these derivatives analytically, in which case we will in any case obtain expressions similar to the Kubo relations. In addition to this, there is no reason to assume four point functions to be considerably much harder to handle in a perturbative framework, where everything is in any case expressed in terms of unperturbed Green's functions. Thus, the perceived advantage of the NEGF formalism, in that one does not need to handle four point functions, seems not to be that relevant.